\documentclass[offcenter]{UIUC07aastex_ARtitlepagefix}
\pdfoutput=1
\usepackage{natbib}
\usepackage{longtable,lscape}
\usepackage{chngcntr}

\def\geqsim{\lower.73ex\hbox{$\sim$}\llap{\raise.4ex\hbox{$>$}}$\,$}
\def\leqsim{\lower.73ex\hbox{$\sim$}\llap{\raise.4ex\hbox{$<$}}$\,$}

\def\Mpch {~h^{-1}~{\rm Mpc}}
\def\Mpchseventy {~h_{70}^{-1}~{\rm Mpc}}
\def\kpch {~h^{-1}~{\rm kpc}}
\def\kpchseventy {~h_{70}^{-1}~{\rm kpc}}
\def\kms{~{\rm km~s^{-1}}}
\def\LOIII{L_{\rm [OIII]}}
\def\MBH{M_{\rm BH}}
\def\Hbeta{\rm H_{\beta}}

\def\Lstar{L*}

\counterwithin{figure}{chapter}
\counterwithin{table}{chapter}

\begin{document} 
\title{The Relationship of Active Galactic Nuclei \& Quasars \\with Their Local Galaxy Environment} 
\author{Natalie Erin Strand} 
\department{Physics} 
\schools{B.S., The Pennsylvania State University, 2003\\ 
M.S., University of Illinois at Urbana-Champaign, 2005} 
\phdthesis 
\advisor{Robert J. Brunner} 
\committee{Professor John Stack, Chair\\ Associate Professor Robert Brunner, Director of Research\\ Associate Professor Brian Fields \\ Assistant Professor Leslie Looney}
\degreeyear{2009} 

\maketitle 

\frontmatter 
\begin{abstract} 
We explore how the local environment is related to properties of active galactic nuclei (AGNs) of various luminosities.  Recent simulations and observations are converging on the view that the extreme luminosity of quasars, the brightest of AGNs, is fueled in major mergers of gas-rich galaxies.  In such a picture, quasars, the highest luminosity AGNs, are expected to be located in regions with a higher density of galaxies on small scales where mergers are more likely to take place.  However, in this picture, the activity observed in low-luminosity AGNs is due to secular processes that are less dependent on the local galaxy density.  To test this hypothesis, we compare the local photometric galaxy density on kiloparsec scales around spectroscopic type I and type II quasars to the local density around lower-luminosity spectroscopic type I and type II AGNs.  To minimize projection effects and evolution in the photometric galaxy sample we use to characterize AGN environments, we place our random control sample at the same redshift as our AGNs and impose a narrow redshift window around both the AGNs and control targets.  Our results support these merger models for bright AGN origins.  We find that the brightest sources have overdensities that increase on the smallest scales compared to dimmer sources.  In addition, we investigate the nature of the quasar and AGN environments themselves and find that the increased overdensity of early-type galaxies in the environments of bright type I sources suggests that they are located in richer cluster environments than dim sources.  We measure increased environment overdensity with increased quasar black hole mass, consistent with the well-known $M_{\rm DMH}-\MBH$ relationship, and find evidence for quenching in the environments of high accretion efficiency type I quasars.  
\end{abstract} 
\begin{dedication} 
to my husband Joel\\
and\\
to my dad
\end{dedication} 

\chapter*{Acknowledgments} 

Many thanks are due to the people who have been a part of this project:

My advisor, Robert Brunner, who supported and challenged me throughout the process.  

My group mates, especially Ashley Ross, Adam Myers, and Britt Lundgren, with whom I had conversations that provoked insight and understanding.  

Reinabelle Reyes and Lei Hao, who generously provided data and willingly answered questions about it.  

My husband Joel, my family, and my friends, who encouraged me, supported me, and prayed for me continually.  

Most importantly, I praise my Rock and my Redeemer, who set me on this path and sustained me along the way.\\

Additionally, I acknowledge support from Microsoft Research, the University of Illinois, and NASA through grants NNG06GH156 and NB 2006-02049. I made extensive use of the storage and computing facilities at the National Center for Supercomputing Applications and thank the technical staff for their assistance in enabling this work. 

Funding for the creation and distribution of the SDSS Archive has been provided by the Alfred P. Sloan Foundation, the Participating Institutions, the National Aeronautics and Space Administration, the National Science Foundation, the U.S. Department of Energy, the Japanese Monbukagakusho, and the Max Planck Society. The SDSS Web site is http://www.sdss.org/.

The SDSS is managed by the Astrophysical Research Consortium (ARC) for the Participating Institutions. The Participating Institutions are The University of Chicago, Fermilab, the Institute for Advanced Study, the Japan Participation Group, The Johns Hopkins University, the Korean Scientist Group, Los Alamos National Laboratory, the Max-Planck-Institute for Astronomy (MPIA), the Max-Planck-Institute for Astrophysics (MPA), New Mexico State University, University of Pittsburgh, University of Portsmouth, Princeton University, the United States Naval Observatory, and the University of Washington.

%
\tableofcontents 

\mainmatter 
\chapter{Introduction}

\section[A BRIEF HISTORY OF STRUCTURE FORMATION]{A BRIEF HISTORY OF STRUCTURE FORMATION  or A Bit of Cosmic Background}

In the early universe, tiny density fluctuations appeared and continued to increase when massive particles began to govern the expansion of the universe (``matter era") instead of the background blackbody radiation (``radiation era").  As the matter era progressed and the universe continued to expand and cool, massive particles began to be attracted to these fluctuations, increasing the density.  When the variations in density were measured in the cosmic microwave background by such surveys as WMAP and its predecessors, it became clear that visible baryonic matter alone is not enough to explain the sizes of the density fluctuations.  The presence of a significant fraction of nonbaryonic cold dark matter in the universe is the currently accepted solution to this problem \citep[e.g.,][]{Spergel}.  The interaction of this dark matter with radiation is thought to be negligible, therefore, the dark matter could have begun clumping during the radiation era.  Thus, by the time the matter era began, the clumps of dark matter could have reached larger densities than were possible for baryons alone.  Baryonic matter fell into the potential wells caused by these enhanced density regions of dark matter, known as halos, eventually collapsing into the first globular clusters, galaxies, and galaxy clusters \citep[e.g.,][]{WhiteRees}.  The relative clustering strength of visible baryonic matter (e.g., in the form of galaxies) to the underlying dark matter distribution is known as bias.  

In the cold dark matter paradigm, which accurately describes the observed universe, the physical structures that we see were formed hierarchically:  the smallest structures combining to form larger structures.  Gas clouds condense to form stars, and collections of stars form galaxies.  Galaxies tend to clump together to form groups (less than about 50 galaxies with scales of $\lesssim1$~Mpc) and clusters (larger numbers of galaxies in regions with scales $\gtrsim1$~Mpc); the abundance of galaxies in a cluster determines its richness \citep{Abell}.  

The first stars (known as Population III stars) are proposed to have been zero-metallicity and very massive, on the order of hundreds of solar masses, forming in regions of enhanced density.  The reactions that fueled these stars enriched the interstellar medium with heavier metals, which became the building blocks for later (Population II and I) stars.  When Population III stars collapsed, they retained a large fraction of their mass to form massive black holes, which are thought to be the progenitors of supermassive black holes at the centers of galaxies.  Mergers of the massive halos containing the black holes and their surrounding gas resulted in the building up of larger and larger halos.  The massive black holes sank to the center, combining to form supermassive black holes at the centers of the newly-forming galaxies \citep[see, e.g.,][and references therein]{Ferrara, MadauRees, Volonteri}.  

\section{ACTIVE GALACTIC NUCLEI}
It is now generally accepted that every massive galaxy harbors a supermassive black hole \citep[e.g.][]{KormendyRichstone, Richstone}.  Some of these supermassive black holes are in the process of actively accreting material \citep{LyndenBell}, thereby causing the central regions of the galaxy to emit large amounts of radiation across the electromagnetic spectrum.  These active galactic nuclei (AGNs) can easily be classified by luminosity; Seyfert galaxies are among the dimmest variety of AGNs, while the most luminous AGNs are known as quasars.  The Unified Model of an AGN \citep[see, e.g.,][for a detailed review]{Antonucci, UrryPadovani} consists of a central supermassive black hole surrounded by an accretion disk which is itself surrounded by dusty clouds or a dusty torus.  The object also may have jets visible in X-ray, optical, and/or radio wavelengths that emerge perpendicular to the plane of the torus.  \citet{Antonucci} claims that properties of observed AGNs of all luminosities differ primarily due to the orientation angle of the quasar to the viewer.  In this popular model, broad emission lines are seen in the spectrum of an object viewed at an angle that allows a more direct line-of-sight to the central engine.  Objects viewed at a more oblique angle exhibit narrow emission lines, since radiation from the central source is being re-emitted by the surrounding dust along the direction of observation.  

However, this observationally-driven model does not explicitly address the origin or evolution of AGNs.  Currently proposed theories \citep[see, e.g.,][]{Hopkins2005b} suggest that the brightest AGNs are fueled by dust and gas driven toward the galactic nucleus by mergers or interactions of galaxies, while secular mechanisms cause the activity in lower luminosity AGNs.  Therefore, In this model, the amount of fuel available to the AGN plays a major role in the properties that we observe.  The aim of this dissertation is to use studies of AGN environments to constrain the validity of theories for the activation and fueling of AGNs.  

\section{AGN ENVIRONMENTS}
     The local environments of AGNs and quasars provide valuable insights into the formation history and evolution of those sources \citep[e.g.,][]{Ellingson}.  Quasar environments were first studied by \citet{Bahcall}, who used a sample of five quasars and concluded that quasars are associated with galaxy clusters.  \citet{YeeGreen} found that quasars reside in regions with higher galaxy density, and more recent work has confirmed that quasars are found in regions with densities consistent with galaxy groups or clusters of poor to moderate richness \citep{BahcallChokshi, Fisher, McLureDunlop, Wold, Coldwell2002, Barr}.  Although studies of several X-ray- and radio-selected samples have found evidence for a relationship between environment and AGN activity \citep[e.g.,][]{Wurtz, Best, Sochting}, the Sloan Digital Sky Survey (SDSS) is the first survey to allow meaningful studies of quasar environments, because it samples large numbers of both quasars and galaxies for redshifts $z\lesssim0.5$.  Using SDSS data, \citet{Serber} concluded that the density of photometric galaxies around quasars increases with decreasing angular scale, but is independent of redshift for $z\leq0.4$.  They also provided evidence for a higher density of galaxies around more luminous quasars at scales less than $100\kpch$, while at larger angular scales, the density appears to be largely independent of luminosity \citep[c.f.,][]{PorcianiNorberg, daAngela, Myers07a}.  

     The results of \citet{Serber} agree with other studies showing enhanced clustering of quasars on small scales.  \citet{Djorgovski} first linked the excess of quasar clustering on small scales to galaxy interactions.  Studies of the small-scale clustering of quasars (e.g., binary and triplet quasars) also support the hypothesis that there is excess quasar clustering on scales of $\lesssim100\kpch$ \citep{Kochanek, Mortlock, Hennawi, Djorgovski2007, Myers07a}.
    
    An excess of quasar pairs on small scales naturally follows from a merger origin for quasar activity, whether these pairs simply trace biased groups where mergers are likely to occur \citep{Hopkins2008} or are being excited in merging galaxies \citep{Djorgovski, Myers07b}.  \citet{Hopkins2006} have developed a unified, merger-driven framework that naturally predicts that quasar environments should be highly biased \citep{Hopkins2008}.  These simulations show that major mergers between gas-rich galaxies are the likely mechanisms to trigger bright quasar activity, and that this activity is a phase in the evolution of massive spheroidal galaxies \citep{Hopkins2005b, Hopkins2008}.  In contrast, secular mechanisms (e.g. disk instabilities that feed cold gas onto the central black hole)                                                                                                                                                      may fuel the activity in most low-luminosity AGN, implying that the small-scale environments of these objects should have a smaller bias \citep{HopkinsHernquist, Hopkins2008}.  Therefore, we would expect that objects driven by major mergers will have biased environments on small scales, whereas objects fueled by secular means will reside in less rich environments.  Such a simplification hides many subtleties, however, as secular mechanisms such as harassment can probably only occur in slightly overdense environments.  Further, for objects whose observed characteristics differ purely because of viewing angle or internal structure \citep{Antonucci, Elvis}, there should be no particular difference in local environment.  This, of course, would only be the case if that structure is not correlated with fueling, as could occur, for instance, if more luminous quasars had strong winds.  Therefore, it is important to understand the relationship between the physical properties of AGN and their local environment, which will in turn provide insight into what aspects of AGN properties are explained by formation history, fueling, or simply by structure and orientation.  
    
    In this dissertation, we address this merger hypothesis by studying the relationships between different types of AGNs and their environments.  Our work improves upon previous studies of AGN environments in several ways.  First, we use larger samples of background photometric galaxies, as well as more cleanly defined samples of spectroscopic AGNs of various luminosities.  Additionally, we include cuts in photometric redshift space around spectroscopic targets and the random positions to which they are compared to minimize interloping foreground or background objects, as well as to marginalize any redshift evolution of the photometric galaxy sample.  By using photometric redshift cuts, we obtain more realistic overdensity estimates and errors, and we are able to extend the study of quasar environments in the SDSS to $z\approx0.5$.  With our large samples, we are able to isolate specific characteristics of the AGNs to quantify physical relationships between the AGN and its local environment.  

We assume a concordance cosmology $\Omega_{M} = 0.3$, $\Omega_{\Lambda} = 0.7$, $H_{0} = 70$ km s$^{-1}$ Mpc$^{-1}$ (with $h = 0.7$) in order to compare to results from previous studies.  Our spectroscopic data is divided into four target samples:  Type I and Type II quasars (e.g., AGNs with the highest intrinsic luminosity) and lower-luminosity Type I and Type II AGNs.  The spectra of Type I AGNs and quasars are characterized by broad emission lines \citep[FWHM $> 1000~\rm{km~s^{-1}}$; e.g.,][]{Haoa, Schneider}, while the spectra of Type II AGN and quasars exhibit narrow emission lines \citep[e.g.,][]{Haoa, Zakamska}.  From this point forward, for simplicity, we will generally use ``AGN" to describe lower-luminosity objects and ``quasar" to describe higher-luminosity objects.

\chapter{The Data}

\section{OVERVIEW}
We study the environments of spectroscopic AGN targets by counting photometric galaxies within a 2.0$\Mpchseventy$ projected comoving distance of the target center (i.e., we consider a conical slice around the target rather than a spherical volume).  The samples of Type I quasars (QIs), Type II quasars (QIIs), lower luminosity AGNs (AIs and AIIs), spectroscopic and photometric galaxies selected from the SDSS are described below.  We mask the spectroscopic samples to eliminate objects within 2.0$\Mpchseventy$ of any of the following:  the survey edge, an area masked out by SDSS\footnote{See http://www.sdss.org/dr5/products/images/use\_masks.html; negative random positions (see Section~\ref{sort_negran}) provided by A. D. Myers, private communication}, an area with seeing $>1.5\arcsec$, or an area with $r$-band reddening $A_{r}\geq0.2$ \citep{Scranton2002, Ross2006}. 

\begin{figure}
\plotone{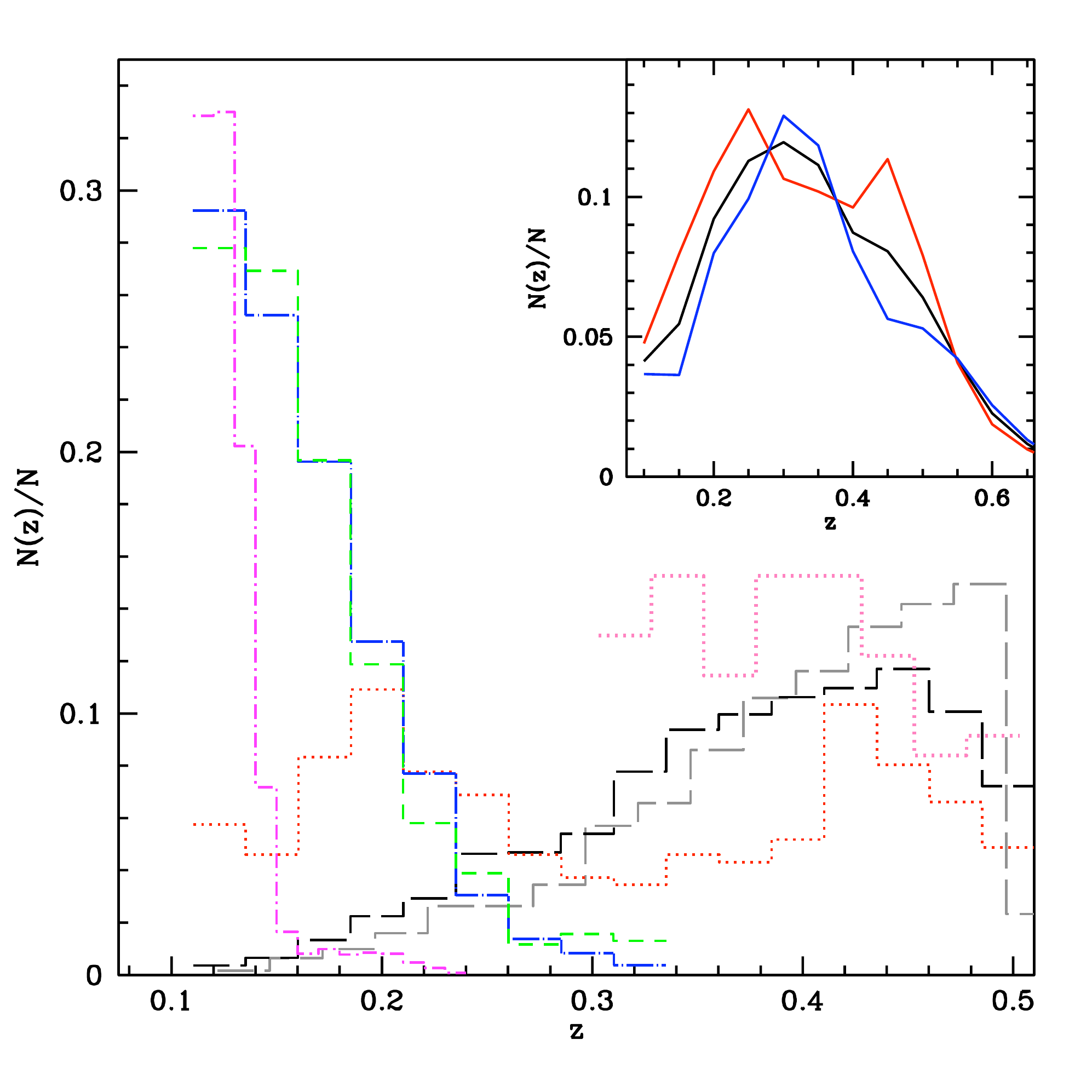}
\caption{Normalized redshift distributions of each of the spectroscopic target samples.  QI\_pt:  long gray dashed (2,314).  QI:  long black dashed (3,793).  QII\_Z:  pink dotted (131).  QII\_R:  red dotted (348).  AI:  short green dashed (1,464).  AII:  long blue dash dotted (3,329). $\Lstar$ galaxies: short magenta dash dotted (8,618).  \emph{Inset:} Normalized redshift distribution of photometric galaxies.  The solid black curve shows the distribution of all 28,851,353 galaxies, the solid red curve shows the distribution of the 12,150,909 early-type galaxies, and the solid blue curve shows the normalized of the 16,700,444 late-type galaxies. 
\label{data_Nofz}}
\end{figure}

Since our spectroscopic data are compiled from several sources, as detailed below, we have eliminated duplicate objects between the samples.  If an object appears in both a lower-luminosity AGN sample and a quasar sample, the object is classified as a quasar (and thus removed from the AGN sample), since the quasar samples have lower luminosity limits imposed.  If an object appears in both the QI and QII sample, we classify it as a QII, since \citet{Reyes} have imposed strict line width cuts on objects that enter their sample.   Figure~\ref{data_Nofz} shows the redshift distributions for the spectroscopic target samples, which are also summarized in Table~\ref{table_datasummary}.    

\section{SPECTROSCOPIC TARGETS}\label{targetsection}
\subsection{Active Targets}
\subsubsection{Quasars}
Our samples of spectroscopic Type I quasar targets are drawn from the SDSS Fifth Data Release \citep[DR5; ][]{DR5paper} Quasar Catalog \citep{Schneider}, which includes $K$-corrected absolute $i$-band magnitudes for each object.  The $0.11\leq z\leq0.5$ QI samples have $-26.4\leq M_{i}\leq-22.0$.  
In addition to absolute magnitude information, the catalog provides a flag to distinguish between resolved and point source objects.  The measured absolute broad-band luminosity of resolved sources will likely be contaminated by starlight from the host galaxy, so initially, when we divide the sample by absolute luminosity, we exclude those quasars with extended morphology for a sample (QI\_pt) that contains $2,314$ objects.  In our analyses which use the [OIII] emission line luminosity and black hole mass measurements (Chapters~\ref{OIIIlumchapter}ff), we re-introduce the extended sources, because $\LOIII$ is due to AGN activity.  The combined point + extended sample of QIs with good [OIII] and $\MBH$ measurements contains $3,793$ objects.  

For our initial analysis, we draw Type II quasar targets from the sample presented by \citet{Zakamska}.  After cutting the sample to match the high redshift limit of the main quasar sample and masking this sample as described above, we have $131$ QII\_Z targets with $0.3\leq z\leq 0.5$.  In the subsequent stages of our analysis, we extend our sample by using QIIs from \citet{Reyes} (QII), which includes over 90\% of the sources in the QII\_Z sample.  After masking there are $348$ QII targets with $0.11\leq z\leq0.5$.  

\subsubsection{Lower-Luminosity AGNs}
We compare the quasar targets to lower-luminosity Type I and Type II AGN (AI and AII, respectively) from Hao (private communication) selected from SDSS DR5 spectroscopic galaxies according to the criteria laid out in \citet{Haoa}.  The classification of these galaxies as AGN depends on the strengths of the [O III] and H$\beta$ lines \citep{Haoa, Kauffmann}; therefore our low-redshift limit is set to $z = 0.11$, as this is where the [O III] ($\lambda$$\lambda$4959, 5007) lines enter the $r$-band \citep{Kauffmann}, resulting in a more uniform classification.  After masking as above, there are $1,464$ Type I AGN and $3,329$ Type II AGN following the criteria of \citet{Kewley}.  While we could have adopted the less stringent criteria of \citet{Kauffmann}, we wish to be conservative in our sample selection in this analysis and minimize the contribution from non-accretion luminosity sources \citep{Haoa}.  The lower-luminosity AGN samples have a redshift range of $0.11\leq z\leq0.33$, with the majority of sources at $z < 0.15$.  

\subsubsection{Combined Samples}
We have matched all of the targets with Galactic extinction-corrected [OIII] line fluxes (in addition, QI targets set been matched with $\Hbeta$ line and 5100\AA~continuum fluxes from which black hole masses and Eddington accretion ratios have been calculated; see Chapter~\ref{MBHchapter}).  Because we have [OIII] flux measurements for all of our active spectroscopic targets, we can investigate the relationship between $\LOIII$ and environment overdensity for Type I AGN samples and Type II AGN samples.  We convert the observed [OIII] flux to luminosity by first correcting for extinction using Galactic dust maps \citep{Schlegel} with an $R_V=3.1$ extinction law \citep{CCM} to find the extinction correction for the object's coordinates.  
The corrected flux is then converted to luminosity using the target's spectroscopic redshift and our assumed cosmology.  

\begin{figure}
\plotone{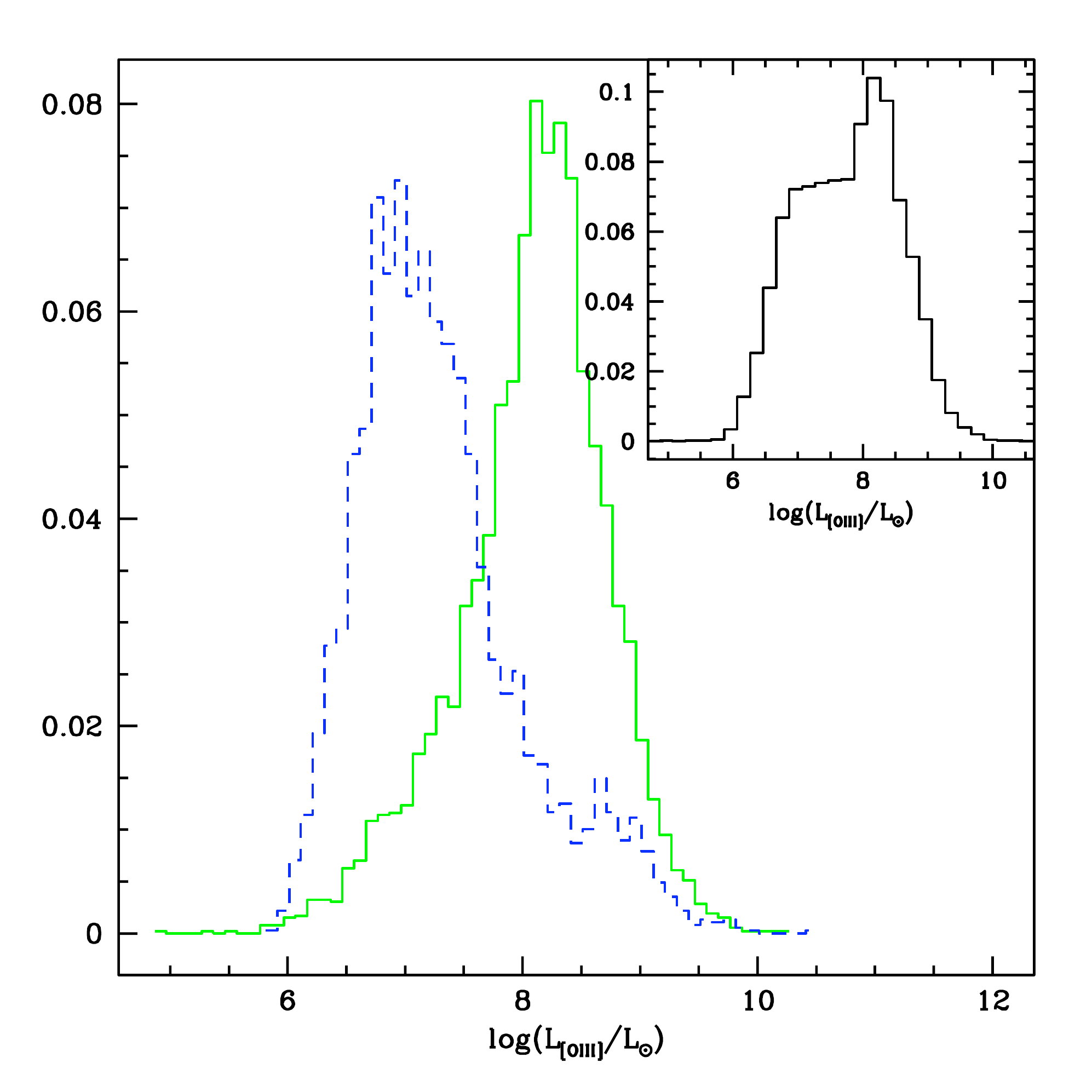}
\caption{Normalized $\LOIII$ distribution for 5,257 Type I (green solid line) and 3,677 Type II (blue dashed line) targets.   $\LOIII$ values have been corrected for Galactic extinction.  
\emph{Inset:} Normalized $\LOIII$ distribution for all types of spectroscopic targets combined (8,934 objects).  
\label{OIIIhistogram_types_extcorr_smbins_remdup_inset}}
\end{figure}

The Type I sample has $5,257$ objects (QIs + AIs), and the Type II sample has $3,677$ objects (QIIs + AIIs; the low redshift, lower luminosity AGNs dominate this sample) for a total of $8,934$ targets when the samples are combined.  We plot the $\LOIII$ distribution of the combined Type I and Type II samples in the inset of Figure~\ref{OIIIhistogram_types_extcorr_smbins_remdup_inset}.  Upon comparing the $\LOIII$ distributions of the two samples in Figure~\ref{OIIIhistogram_types_extcorr_smbins_remdup_inset}, we see that the whole distribution composed of distinctly separate peaks for the two types.  We plot the redshift distributions and show $\LOIII$ vs. redshift in Figure~\ref{zhistogramOIIIvsz_types_extcorr_remdup}.  

\subsubsection{Note:  Slight Discrepancy in AI and QI Flux Measurements}
When we compare the [OIII] measurements for AI and QI objects that are present in both samples (from Hao and Reyes, respectively), we see in the upper panels of Figure~\ref{compareOIII_AIvsQI_duplicates_shift} that there is a systematic offset where the QI measurement is brighter than the AI measurement.  The offset is not due to differences in redshift.  According to Reyes (private communication, 2009), the difference is likely due to an improvement in the spectrophotometric flux scale calibration between the SDSS data analyzed by \citet{Haoa} and the later SDSS data analyzed by \citet{Reyes}.  In this later data release (i.e., DR6), spectra are calibrated relative to the PSF magnitudes of reference stars on each plate rather than their fiber magnitudes.  \citet{Reyes} comment on this change, noting that the measured fluxes using the new calibration are on average 38\% higher.  

Rather than attempting to recalibrate the \citet{Haoa} data, we instead make a ``correction" based on the offset between AI and QI measured fluxes in the lower left panel of Figure~\ref{compareOIII_AIvsQI_duplicates_shift}.  We estimate the correction by calculating a linear fit to the objects not on the QI flux = AI flux black dashed line, which is shown as a magenta dashed line.  The correction is $\rm{AI_{flux, corrected}}=\frac{11}{7}(\rm{AI_{flux}}-45.45)$.  Using this correction, we recalculate the extinction-corrected [OIII] luminosity values for the AIs and plot them against the QI luminosity values in the lower right panel of Figure~\ref{compareOIII_AIvsQI_duplicates_shift}.  

While there is definite improvement in the agreement of the luminosities, it is not a very accurate correction.  However, it is sufficient to estimate the extent to which this difference might affect our environment analysis (specifically in the circumstances where we combine AI and QI data to make the TI sample or all four samples to make the TI+TII sample).  When the correction is applied to the AI dataset used for our analysis (duplicates removed), there are now 288 objects with log($\LOIII$/$L_{\odot}$)$\geq8.0$, whereas before the correction, there were 201 objects above that volume limit value.  
Assuming that the same correction can be made to the AIIs, we similarly recalculate the luminosity values for the AII sample and find 303 AIIs with log($\LOIII$/$L_{\odot}$)$\geq8.0$, whereas before the correction, there were 208 objects above that volume limit value.  

For both samples, the changes in the measured overdensities (using the technique described in Chapter~\ref{techniquechapter}) with this very rough correction are well within the error bars (which only include errors due to variation in the environment overdensity, not error in the measurement of the [OIII] line).  We estimate that there may be a few percent change in our measurements due to these differences.  For the purposes of this work, we will combine the AI and QI (AII and QII) samples without applying a flux correction, but when we tabulate the results, we will additionally report only QI values where appropriate.  

\subsection{Normal (Quiescent) Galaxies}\label{specgals}
In order to compare a representative non-active galaxy population to our AGN samples, we have constructed a spectroscopic galaxy target sample.  We select primary SDSS DR5 objects that are spectroscopically classified as galaxies with extinction-corrected $i$-band magnitudes $<18.5$ using the following query:  

\begin{verbatim}    
  SELECT p.objID, s.ra, s.dec, s.z, 
     (p.modelMag_u-p.extinction_u) AS uMag, 
     p.modelMagErr_u AS uMagErr, 
     (p.modelMag_g-p.extinction_g) AS gMag, 
     p.modelMagErr_g AS gMagErr, 
     (p.modelMag_r-p.extinction_r) AS rMag, 
     p.modelMagErr_r AS rMagErr, 
     (p.modelMag_i-p.extinction_i) AS iMag, 
     p.modelMagErr_i AS iMagErr, 
     (p.modelMag_z-p.extinction_z) AS zMag, 
     p.modelMagErr_z AS zMagErr, p.flags, p.insideMask
  FROM PhotoTag as p, specObj as s
  WHERE s.SpecClass = 2 AND p.Objid = s.bestObjID AND 
      (p.modelMag_i - p.extinction_i)  < 18.5 AND
      s.z BETWEEN 0.077 and 0.41
\end{verbatim}
that returns $380,678$ rows.  

We calculate absolute magnitudes for the spectroscopic galaxy sample, where the absolute magnitude in that band is given by
\begin{equation}\label{eqn_Absmag}
M_{band} = m_{band} - DM - K_{band}
\end{equation}
where $m_{band}$ is the extinction-corrected apparent magnitude and DM is the distance modulus, which is found from the luminosity distance $D_{L}$ in units of Mpc:
\begin{equation}\label{eqn_DM}
DM = 5\log_{10}(D_{L})+25
\end{equation}

The quantity $K_{band}$ is the K-correction in the particular band for which the magnitude is calculated.  It corrects for the fact that when sources are observed at different redshifts, the broad-band magnitude is sampling flux from different features in the galaxies' rest frame spectra \citep{Hogg2002}.  Thus, the K-correction depends on both redshift and the spectral type for each galaxy \citep[e.g.,][]{Ellis}.  We use K-correction software ``Kcorrect" created by \citet{Blanton}.  This software fits galaxy spectral energy distribution (SED) templates to the photometric information for an individual galaxy and uses the best fit SED to calculate the K-correction.  

We used the standalone C-code provided in the Kcorrect package to calculate the template fitting coefficients and reconstruct magnitudes from these coefficients from the observed bandpasses to rest-frame bandpasses shifted by 0.1, since the SDSS main galaxy sample is primarily at $z=0.1$ \citep[see][for more details]{Blanton}.  
The K-correction in a particular band is given by\footnote{see http://cosmo.nyu.edu/blanton/kcorrect/}
\begin{equation}\label{eqn_Kcorr}
K_{band} = -2.5\log_{10}(\frac{maggies_{band}}{maggies.z0_{band}})
\end{equation} 
where ``maggie" is referring to a linear measure of flux, related to an AB magnitude by $maggie = 10^{-0.4*\rm{m}}$ \citep[note that SDSS magnitudes are asinh magnitudes, or ``luptitudes," and their relationship to more traditional astronomical magnitudes is described in detail in][]{LuptonGunnSzalay}.  

The galaxy targets are then masked in the same manner as the AGN targets.  Additionally, in order to ensure that we have only non-active galaxies in this sample, we remove the galaxies that are classified as AGNs in the DR5 samples described previously as well as in the \citet{Kauffmann} DR4 AGN catalog\footnote{The AGNs in the \citet{Kauffmann} DR4 AGN catalog are selected on slightly less stringent criteria than the \citet{Kewley} criteria used by \citet{Haoa}.  We use the most conservative AGN classification to define our low-luminosity active galaxy samples, but here we eliminate a broader set of galaxies which could be considered active for our non-active galaxy sample.}.  In order to define a sample of $\Lstar$ galaxies, we select only those galaxies with $i$-band magnitudes within $0.25$ magnitudes of $M_{i}*= -21.59$ \citep{Blanton2003}, resulting in a final sample of $8,618$ spectroscopic galaxies with $0.11\leq z\leq0.24$.   

\section{PHOTOMETRIC GALAXIES}\label{photogalsection}
We use the local number of photometric galaxies to characterize the environments of the spectroscopic targets described previously.  The photometric galaxies are drawn from the SDSS DR5 database by selecting all primary objects photometrically classified as galaxies with $r$-band extinction corrected magnitude in the range $14.0\leq r\leq21.0$\footnote{Note that \citet{Serber} used an $r$-band limit on their photometric galaxies rather than $i$-band as stated in their paper (W. Serber and R. Scranton, private communication).}.  All of these objects have been assigned photometric redshifts via a template-fitting technique \citep{Csabai}.  
    The photometric galaxy sample including photometric redshifts and galaxy type information is acquired from the DR5 database using the following query:    
\begin{verbatim}
  SELECT p.objID, p.ra, p.dec, 
     p.u, p.Err_u, p.g, p.Err_g, p.r, p.Err_r, p.i, p.Err_i, 
     p.z, p.Err_z, p.extinction_u, p.extinction_g, 
     p.extinction_r, p.extinction_i, p.extinction_z,
     p.type, p.type_u, p.type_g, p.type_r, p.type_i, p.type_z,
     p.flags, p.flags_u, p.flags_g, p.flags_r, 
     p.flags_i, p.flags_z, p.psfMagErr_g, p.psfMagErr_r, 
     p.psfMagErr_i, p.insideMask,
     z.chiSq, z.z, z.zErr, z.t, z.terr, z.kcorr_r,
     z.rest_ug, z.rest_gr, z.absMag_r
  FROM PhotoPrimary AS p LEFT OUTER JOIN 
     photoz as z ON p.objID = z.objID
  WHERE ((p.dered_g < 23.0) OR (p.dered_r < 23.0) 
     OR (p.dered_i < 23.0))
     AND ((p.type_g = 3) OR (p.type_r = 3) OR (p.type_i = 3))
\end{verbatim}
which returned 163,507,385 objects.  

We accept galaxies in the clean \emph{r}-band galaxy sample if they pass the following requirements:
\begin{list}{$\bullet$}{}
\item \emph{type}: type\_\emph{r} = 3 (classified as a galaxy by the DR5 photometric pipeline)
\item \emph{extinction corrected magnitude}: $14.0\leq r - \rm{extinction}\_r\leq21.0$
\item \emph{flags}$\footnote{as defined by http://cas.sdss.org/astro/en/help/docs/realquery.asp\#flags}$: 
\begin{verbatim}
  ((flag_r & BINNED1)!=0) and
  ((flag_r & NOPROFILE)==0) and
  ((flag_r & PEAKCENTER)==0) and
  ((flag_r & NOTCHECKED)==0) and
  ((flag_r & PSF_FLUX_INTERP)==0) and
  ((flag_r & SATURATED)==0) and
  ((flag_r & BAD_COUNTS_ERROR)==0) and
  ((flag_r & BRIGHT)==0) and 
  (((flag_r & DEBLEND_NOPEAK)==0))or
      (float(psfErr_r)<=0.2)) and
  (((flag_r & INTERP_CENTER)==0)or
      ((flag_r & COSMIC_RAY)==0)) and
  ((flag_r & DEBLENDED_AS_MOVING)==0) 
\end{verbatim}
\end{list}
The full $r$-band galaxy sample contains $28,856,324$ objects.  While we do not make explicit redshift cuts on the photometric galaxy sample, our technique effectively limits the sample to $0.06\leq z\leq0.55$.  


\begin{table}\centering  \begin{minipage}{140mm}
\caption[Summary of spectroscopic target data]{Summary of spectroscopic target data sets used for AGN environment analysis \label{table_datasummary}}
\begin{tabular}{c c c c c}\hline\hline
Name&Type&\# Objects&Range&Reference\\ \hline
QI\_pt\footnotemark[1]&Type I quasars&2314&$0.11\leq z\leq0.5$&\citet{Schneider}\\
QI\footnotemark[2]&Type I quasars&3793&$0.11\leq z\leq0.5$&\citet{Schneider}\\ 
QII\_Z\footnotemark[4]&Type II quasars&131&$0.3\leq z\leq0.5$&\citet{Zakamska}\\
QII\footnotemark[3]&Type II quasars&348&$0.11\leq z\leq0.5$&\citet{Reyes}\\
AI\footnotemark[5]&Type I AGNs&1464&$0.11\leq z\leq0.33$&\citet{Haoa}\\
AII\footnotemark[6]&Type II AGNs&3329&$0.11\leq z\leq0.33$&\citet{Haoa}\\
$\Lstar$\footnotemark[7]&$\Lstar$ galaxies&8618&$0.11\leq z\leq0.24$&SDSS CAS\\
\footnotetext{$^{1}$ point sources only; used in Paper I}
\footnotetext{$^{2}$ point + extended sources; [OIII], $\Hbeta$, continuum luminosity measurements from Reyes, private communication (2008); includes only objects with both good [OIII] and $\MBH$ measurements; duplicates removed}
\footnotetext{$^{4}$ used in Paper I}
\footnotetext{$^{5}$ selected from SDSS DR5 galaxies; [OIII] measurements from Hao, private communication (2008); duplicates removed}
\footnotetext{$^{6}$ selected from SDSS DR5 galaxies; [OIII] measurements from Hao, private communication (2008); \citet{Kewley} AGN definition; duplicates removed}
\footnotetext{$^{7}$ AGN matches removed}
    \end{tabular} \end{minipage}
    \end{table}


\clearpage

\clearpage

\clearpage
\begin{figure}
\plotone{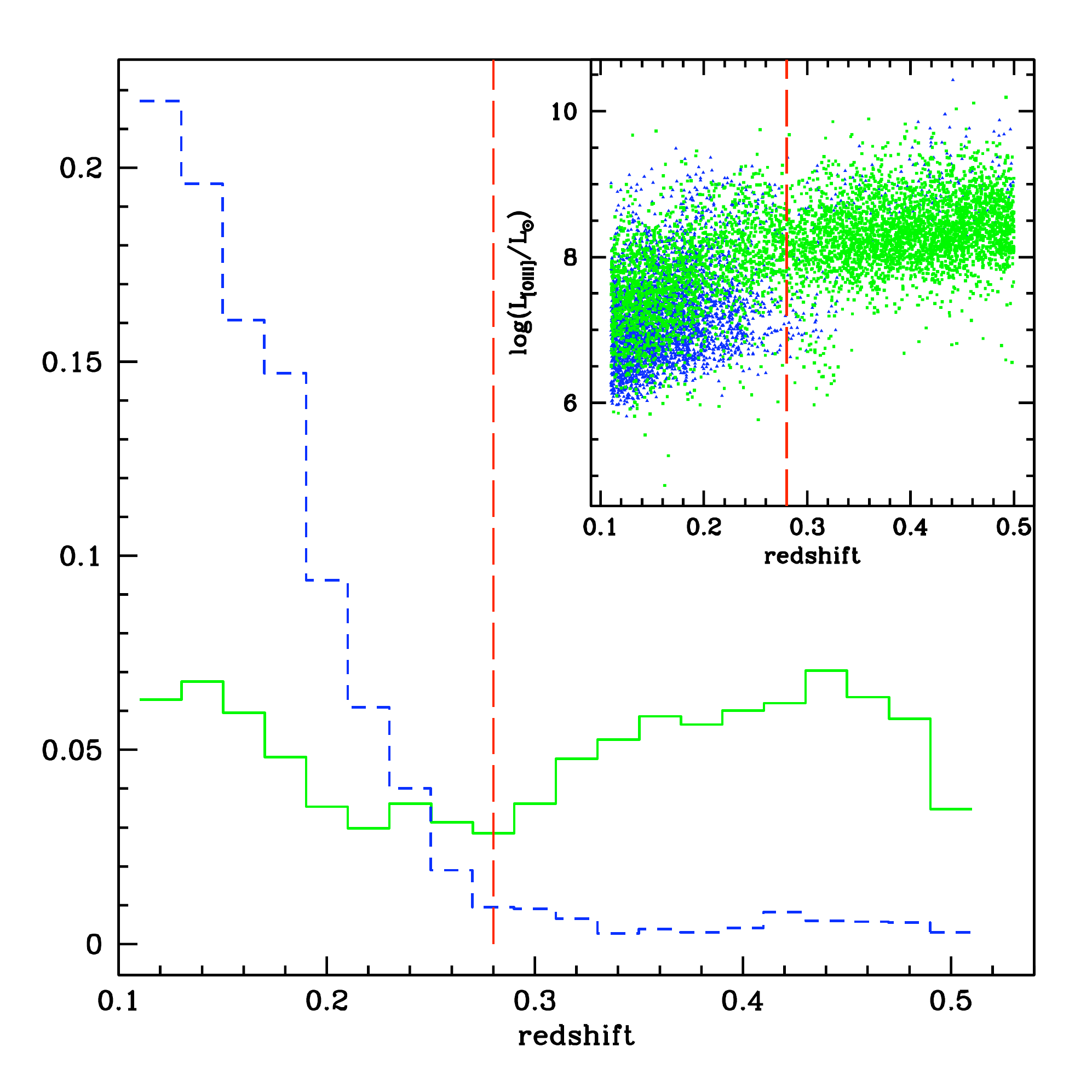}
\caption{Normalized redshift distribution of the 5,257 Type I (green solid line) and 3,677 Type II (blue dashed line) targets.   \emph{Inset:} $\LOIII$ vs. $z$ for Type I (green squares) and Type II (blue triangles) target samples.  The dashed red vertical line in both plots shows $z=0.28$, which divides the two main clusters of points.  
\label{zhistogramOIIIvsz_types_extcorr_remdup}}
\end{figure}
\clearpage
\begin{figure}
\plotone{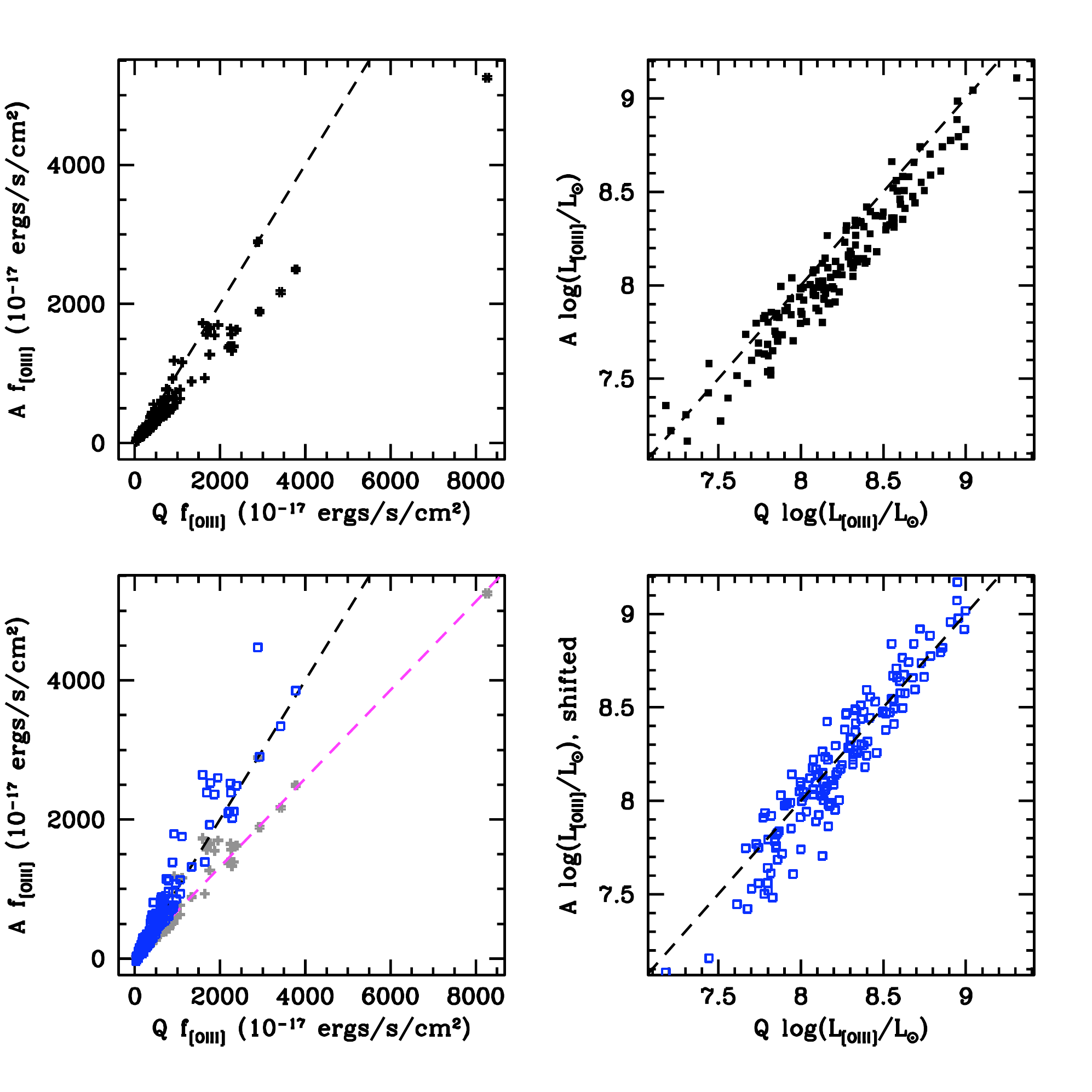}
\caption{\emph{Upper panels:}  Comparison of $f_{[OIII]}$ (left) and $\LOIII$ (right) for QI objects that had duplicates in the AI sample (note: these objects are removed from the AI sample in all other figures).  The measured fluxes are systematically lower in the AI catalog \citep[measured by][]{Haoa} compared to the QI catalog \citep[measured by][]{Reyes}, and thus the luminosities calculated for the AIs are lower. \emph{Lower left:} A correction of the form $\rm{AI_{flux, corrected}}=\frac{11}{7}(\rm{AI_{flux}}-45.45)$ (magenta dashed line) is applied to original AI fluxes.  Blue points show the shifted AI fluxes based on this correction; gray points are the original flux values.  \emph{Lower right:} Blue points show the extinction-corrected luminosities calculated from the ``corrected" fluxes in the lower left panel.  
\label{compareOIII_AIvsQI_duplicates_shift}}
\end{figure}



\chapter{The Technique}\label{techniquechapter}

\section{COUNTING GALAXIES AROUND TARGETS}
We count the number of photometric galaxies within a comoving radius of $2.0 \Mpchseventy$ of each spectroscopic target (e.g., spectroscopic quasar, AGN, or spectroscopic galaxy), excluding any galaxies that are within $25 \kpchseventy$ of the target.  At $z < 0.4$, $25 \kpchseventy$ corresponds to an angular size of $> 3.3\arcsec$, which is approximately twice the average seeing in DR5 \citep{DR5paper}.  At angular scales smaller than this, deblending begins to complicate the reliable detection and measurement of faint galaxies.  

We generate a large number of random positions in the DR5 footprint area for each redshift increment of $0.001$ in our redshift range.  We mask these random positions in the same manner as we masked the spectroscopic targets, requiring at least $1,000$ random positions that are more than $2.0 \Mpchseventy$ away from the survey edge or a masked area for each redshift value.  
We count the number of photometric galaxies within a designated comoving distance around random positions and calculate the mean cumulative number of counts for that redshift increment as

\begin{equation}\label{randomeqn}
R_{i} = \frac{\sum_{z}R_{z}}{N_{z}}
\end{equation} 
Here, $N_{z}$ is the number of random positions $R_{z}$ at a given redshift increment $z$.  We calculate the error corresponding to the mean random counts as
\begin{equation}\label{randomerreqn}
e_{R_{i}}^{2} = \frac{N_{z}}{N_{z}-1}(\overline{R_{i}^{2}}-\overline{R_{i}}^{2}) = \sigma_{R_i}^2
\end{equation}
which is the variance on the mean random counts at a given redshift increment $z$.  

The cumulative bincounts \emph{C$_{i}$} around spectroscopic targets are matched with the mean cumulative random bincounts $R_{i}$ (and error $e_{R_{i}}$) at the redshift increment closest to the target's redshift.  We calculate a mean overdensity $\delta_{bin}$ in a particular scale, redshift, or absolute magnitude bin as
\begin{equation}\label{avgoverdensityeqn}
\delta_{bin} = \frac{\frac{1}{N}\sum_{i}^{N}C_i}{\frac{1}{N}\sum_{i}^{N}R_i} - 1 = \frac{\overline{C_{bin}}}{\overline{R_{bin}}} - 1 = \frac{C_{bin}}{R_{bin}} - 1 
\end{equation}
where \emph{C$_{i}$} is the counts around each target in the bin, \emph{R$_{i}$} is the mean counts around random positions at the corresponding redshift, and there are \emph{N} total targets in the bin.  The error on the overdensity is determined via error propagation:
\begin{equation}\label{overdensityerreqn}
e_{\delta_{bin}}^2 = \frac{e_{C_{bin}}^2}{R_{bin}^2} + \frac{C_{bin}^2}{R_{bin}^4}e_{R_{bin}}^2
\end{equation}
where $e_{C_{bin}}$ = $\sqrt{C_{bin}}$ and $e_{R_{bin}}^2$ = $\sum_{i}^{N}e_{R_{i}}^2$.  

We will refer to the quantity of $\frac{C_{bin}}{R_{bin}}$ as the \emph{mean density}; this quantity is used to compare our results to those of \citet{Serber}.  In order to compare the environments of our various spectroscopic populations, however, we use the \emph{mean overdensity}, $\frac{C_{bin}}{R_{bin}}-1$, which is related to the underlying dark matter distribution and can be more directly related to correlation analyses \citep[e.g.,][]{Padmanabhan}.  

Appendix~\ref{bincountDoc} summarizes the C code written to perform the bincounting and masking.  

\section{APPLYING A PHOTOMETRIC REDSHIFT CUT}\label{deltazcutsection}
One of the difficulties in using photometric galaxy samples for overdensity measurements is the issue of projection effects, where foreground or background objects contaminate a measurement.  We use photometric redshifts that are assigned to the photometric galaxies to minimize this complication.  We apply a photometric redshift cut on the galaxies so that only those galaxies which satisfy $|z_{target} - z_{photogal}|\leq\delta z$ are counted in each bin.  
Crucially, the same $\delta z$ cut is applied to both the spectroscopic targets and the random positions to which they are compared, which, as noted above, are also placed at the same redshift as the spectroscopic targets.  We therefore minimize contamination by most galaxies outside of the $\delta z$ interval.  Additionally, by calculating our spectroscopic-photometric and random-photometric counts in the same $z\pm\delta z$ bin, we marginalize redshift evolution in the photometric galaxy sample outside of that $z\pm\delta z$ bin.  We make the reasonable assumption that there is no redshift evolution in the photometric galaxy sample over this small $\delta z$ interval.  We have not accounted for the changes in photometric redshift accuracy as a function of magnitude and redshift, which we also assume are negligible over these redshift ranges.  

We verify that the projection effect issue is mitigated by the $\delta z$ cut without introducing systematics by calculating overdensities for random positions with the same redshift distribution as the QI\_pt sample.  We find that the overdensities of photometric galaxies around random positions is consistent with zero on all scales with and without the photometric redshift cut.  We use the value $\delta z=0.05$, which is large enough to encompass the effective rms error of the photometric redshifts \citep[$\Delta z_{rms} = 0.04$ for $r < 18$;][]{Budavari}, for all further analysis.  We have tested other values of $\delta z$ and find that they give consistent results, albeit with larger uncertainties for narrower cuts, which is consistent with the expectations of Poissonian sampling.  

To compare directly to the results of \citet{Serber}, who did not apply any such redshift cut, we calculate the mean density of photometric galaxies around QI\_pt targets with $-24.2\leq M_{i}\leq-22.0$ and $0.08\leq z\leq0.4$.  At a scale of $250\kpchseventy$, the density of photometric galaxies around quasars is $1.41\pm0.033$ and around $L*$ galaxies is $1.15\pm0.005$ without the $\delta z$ cut.  However, applying the $\delta z$ cut decreases the random background noise, and with this cut we measure an environment density of $2.11\pm0.096$ around quasars and $1.74\pm0.020$ around $\Lstar$ galaxies at the same scale.  In order to confirm that we have not added any systematics by using the $\delta z$ cut, we compare the relative densities of QI\_pt environments to $L*$ galaxy environments.  The relative density of photometric galaxies around quasars to that around $\Lstar$ galaxies is $1.22\pm0.029$ without the $\delta z$ cut.  The relative density does not appreciably change when the $\delta z$ cut is used, and we find the relative density to be $1.22\pm0.057$.  

The true physical effect of the $\delta z$ cut is shown in the comparison of mean $over$densities.  At the same scale of $250 \kpchseventy$, the relative overdensity around quasars compared to around $\Lstar$ galaxies is $2.67\pm0.236$ without the $\delta z$ cut, but is $1.51\pm0.137$ with the $\delta z$ cut.  Because we have removed projection effects, the relative overdensities are lower when the $\delta z$ cut is used; however, the errors on the mean densities with the $\delta z$ cut have increased.  We believe these larger errors are more physically relevant: with no $\delta z$ cut, objects not actually correlated with the target will reduce Poissonian error estimates.  Therefore, all subsequent analysis and figures include the $\delta z=0.05$ cut.  

\section{QUANTIFYING REDSHIFT UPPER LIMITS}\label{techniquechapter_redshiftlimits}
\begin{figure}
\plotone{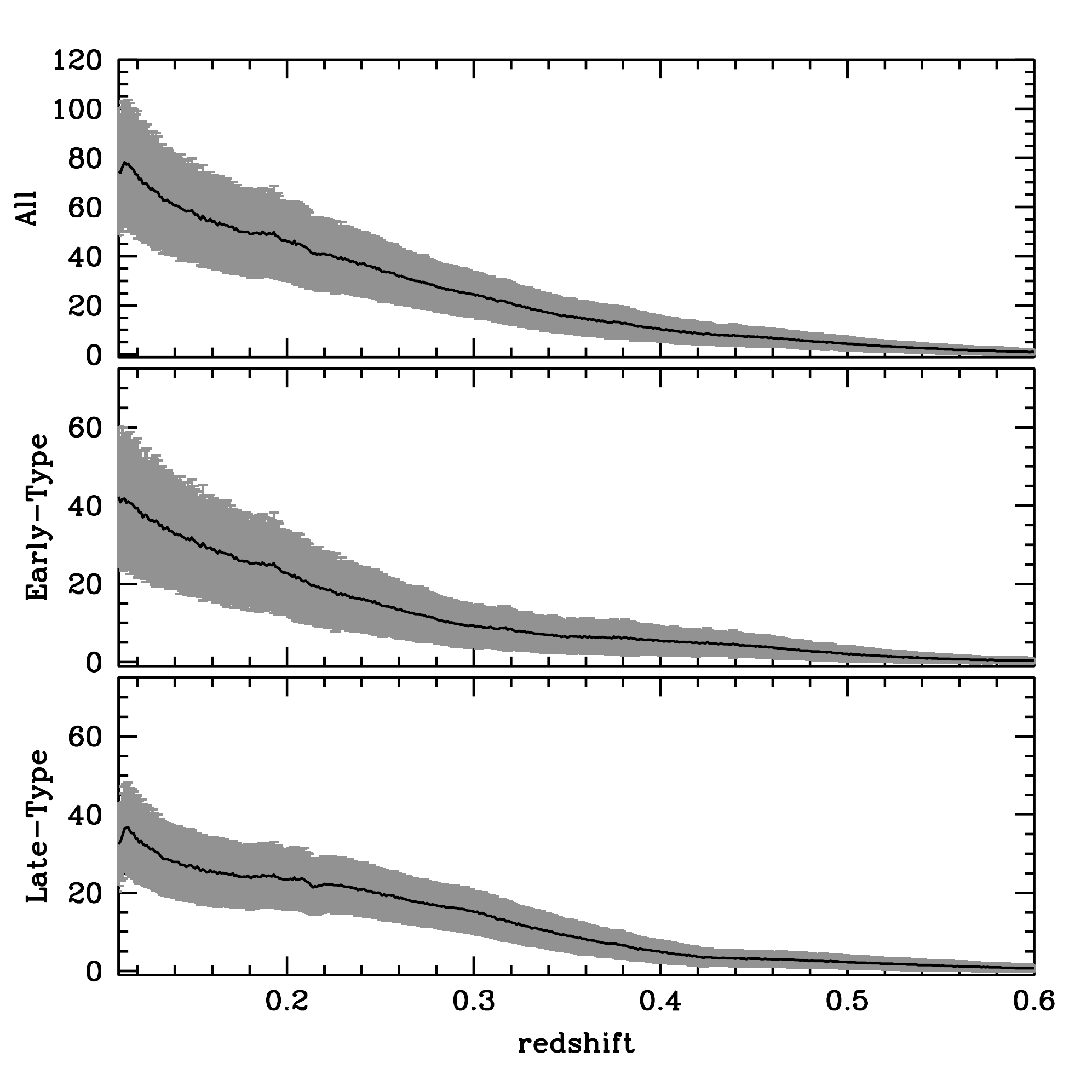}
\caption{Expected galaxy counts in $2.0\Mpchseventy$ radius vs. redshift. The black line corresponds to the mean number of counts around random targets, the grey area corresponds to the standard deviation from this mean ($Note-$ at least 1000 random targets contribute to the mean at each 0.001 in redshift).  
\label{expectedNum}}
\end{figure}

\begin{figure}
\plotone{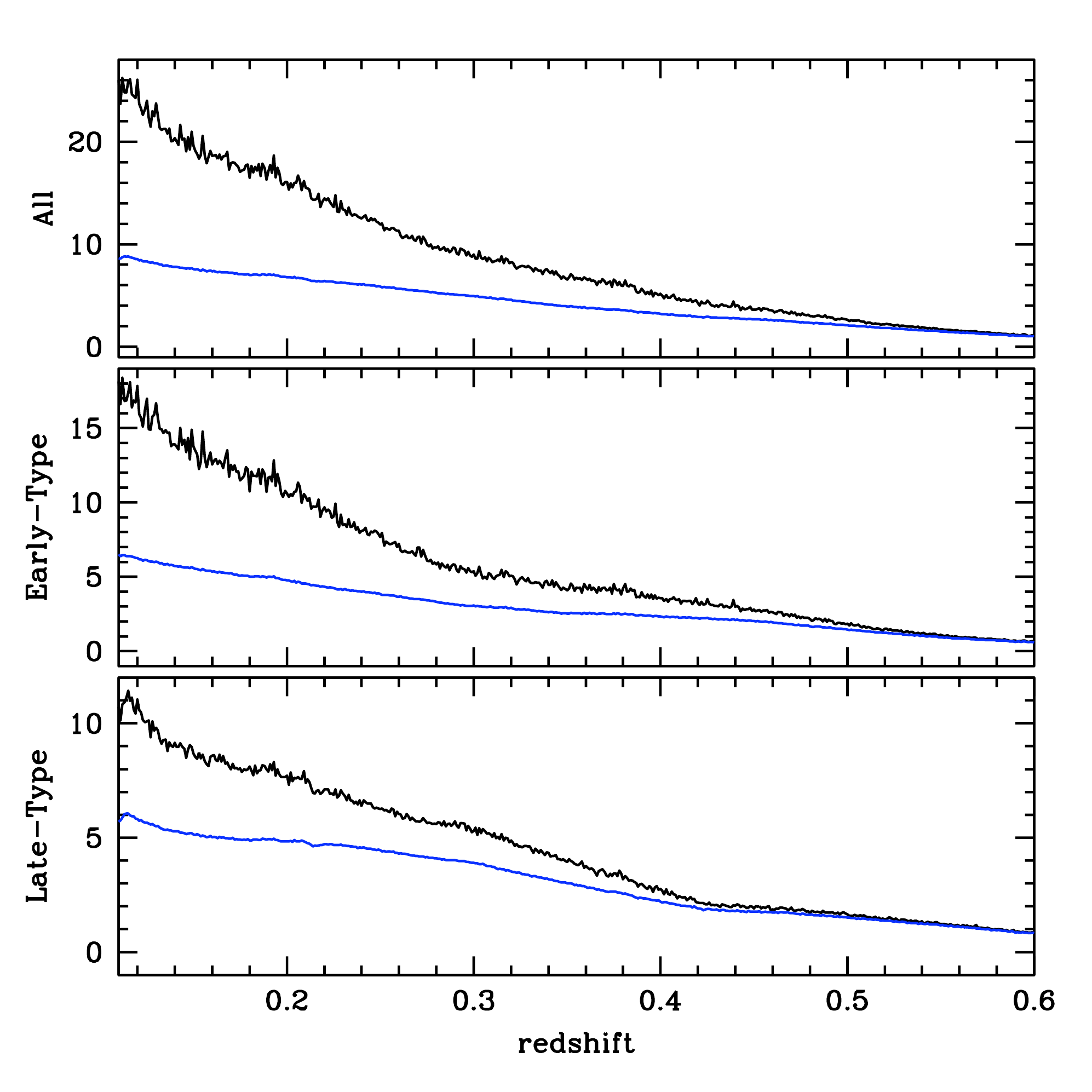}
\caption{Comparison of the standard deviaion on the mean number of counts around random targets (black) to the Poisson error ($=\sqrt{\rm{mean~counts}}$; blue).  
\label{expectedNumErr}}
\end{figure}

We use the $\delta z$ cut to extend our redshift range to include spectroscopic targets with redshifts $z\geq0.4$ without concern that foreground objects will contaminate the overdensity measurements; however caution must be exercised at these higher redshifts because the number of photometric galaxies at these higher redshifts is falling off quickly.  In Figure~\ref{expectedNum}, we present the mean counts around random positions at each redshift increment of 0.001 in $0.11\leq z\leq0.6$ including the $\delta z$ cut.  The number of galaxies expected to be counted within $2.0\Mpchseventy$ and the $\delta z$ cut decreases quickly as the angular size of this radius shrinks with redshift and the number of photometric galaxies in the entire sample decreases.  If we assume that the photometric galaxies follow a Poisson distribution, the expected signal-to-noise ratio within the $2.0\Mpchseventy$ area is $\sqrt{R_{i}}$, where $R_{i}$ is the mean number of galaxies counted within that area at a particular redshift.  We compare the expected signal-to-noise ratio to the standard deviation on the mean counts with redshift in Figure~\ref{expectedNumErr}.  At low redshifts it is clear that the expected error (due to true physical variations in the galaxy distribution) overwhelms the Poisson error, but at higher redshifts, the Poisson error has a larger contribution.  We require a signal-to-noise of at least $2$, which means that it is necessary to set an upper limit on redshift of $z=0.5$ when the photometric galaxy sample is used as a whole.  However, we also subdivide the photometric galaxy sample by type, and in order to keep our signal-to-noise above $2$, we must decrease the upper redshift limit when smaller background samples are used.  Table~\ref{PoissonSNtable} lists the redshift limits and expected signal-to-noise values for each of the background galaxy samples that we use in our analysis.  

\begin{figure}
\plotone{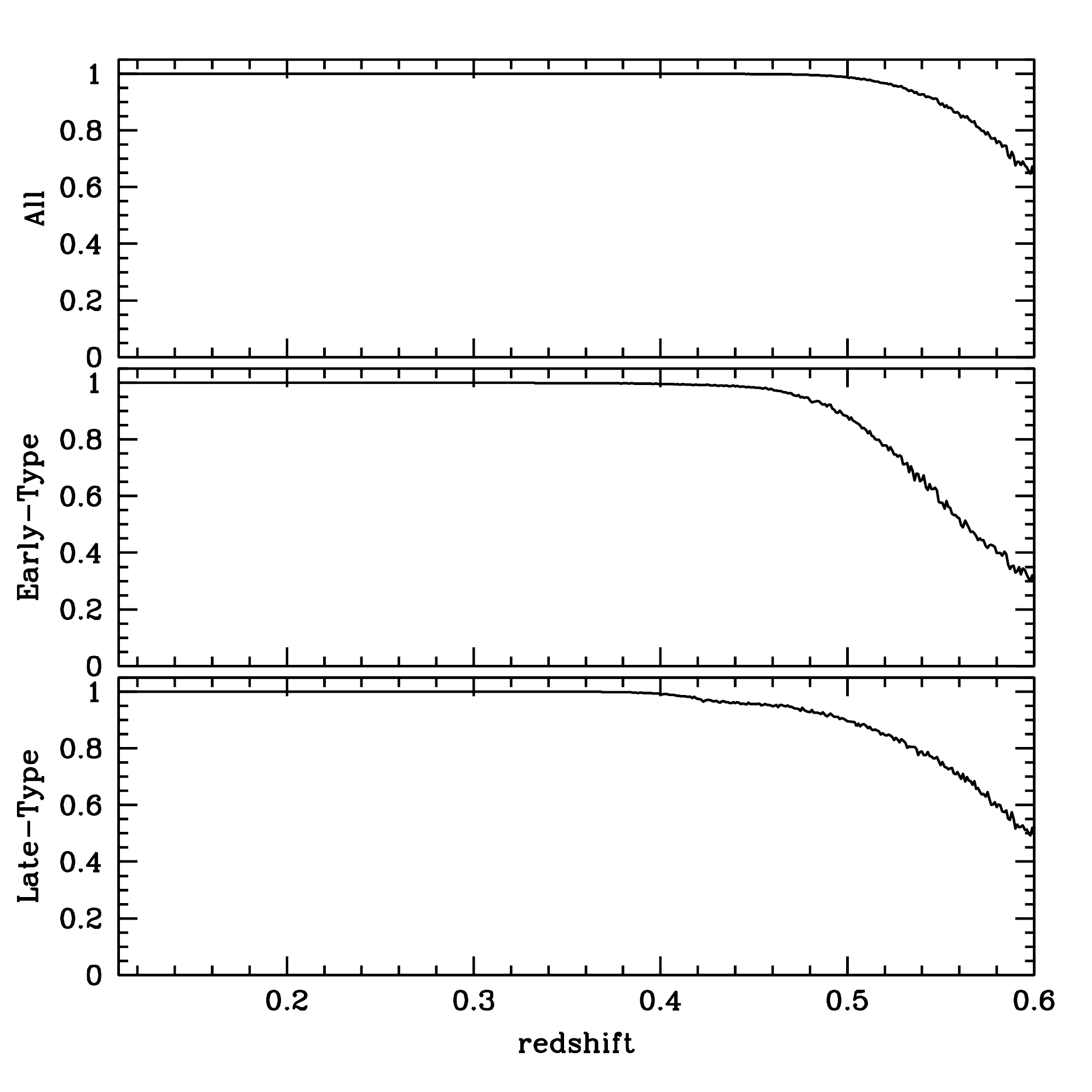}
\caption{Probability of finding more than zero counts if the environment galaxies can be approximated as having a Poissonian distribution.  
\label{probIFpoisson}}
\end{figure}

In Figure~\ref{probIFpoisson}, we confirm the necessity of an upper limit by determining the probability of counting at least one galaxy in $2\Mpchseventy$ if the environment galaxies are assumed to follow a Poisson distribution.  At $z=0.4$, it is certain that at least one environment galaxy of any type will be detected, and probability of finding an early-type (late-type) galaxy is 0.996 (0.992).  At $z=0.5$, the probability of finding at least one environment galaxy is 0.988; probability of finding an early-type (late-type) galaxy is 0.881 (0.896).  At $z=0.6$, the probability of finding at least one environment galaxy is much lower at 0.649; the probabilities of finding an early-type or late-type galaxy are further reduced to 0.300 and 0.498, respectively.  

\section{VOLUME-LIMITING}
The spectroscopic and photometric data samples used in this analysis have been selected from magnitude-limited surveys and parent samples.  Therefore, because the apparent magnitude limit is constant for all redshifts, there is a systematic increase in minimum (intrinsic) luminosity of objects detected with redshift due to the inverse-square law for electromagnetic radiation.  To remove this evolution in minimum luminosity, one selects a volume-limited sample by determining the intrinsic luminosity corresponding to the (apparent) magnitude limit at the maximum redshift and allowing only those sources that are brighter than this luminosity into the sample.  Such a sample minimizes systematics due to redshift-dependent properties such as galaxy type \citep[e.g.,][]{Budavari}

\subsection{Spectroscopic Target Samples}
Primarily, we will use the extinction-corrected $\LOIII$ measurement to create volume-limited samples of spectroscopic targets.  Based on Figure~\ref{zhistogramOIIIvsz_types_extcorr_remdup}, we select targets with log($\LOIII$/$L_{\odot}$)$\geq8.0$ for a volume-limited sample to $z=0.5$.  When we restrict the redshift to $z\leq0.28$ ($z\leq0.3$), we lower the luminosity limit to log($\LOIII$/$L_{\odot}$)$\geq6.75$.  Table~\ref{vollimSpectargtable} summarizes the various volume-limits that will be implemented throughout the analysis.  

\subsection{Photometric Environment Galaxy Sample}
We test the usefulness of a volume-limited photometric galaxy sample by creating a sample with $z\leq0.45$ and $M_r>-20.55$ (hereafter referred to as the V4 volume limit).  

\begin{figure}
\plotone{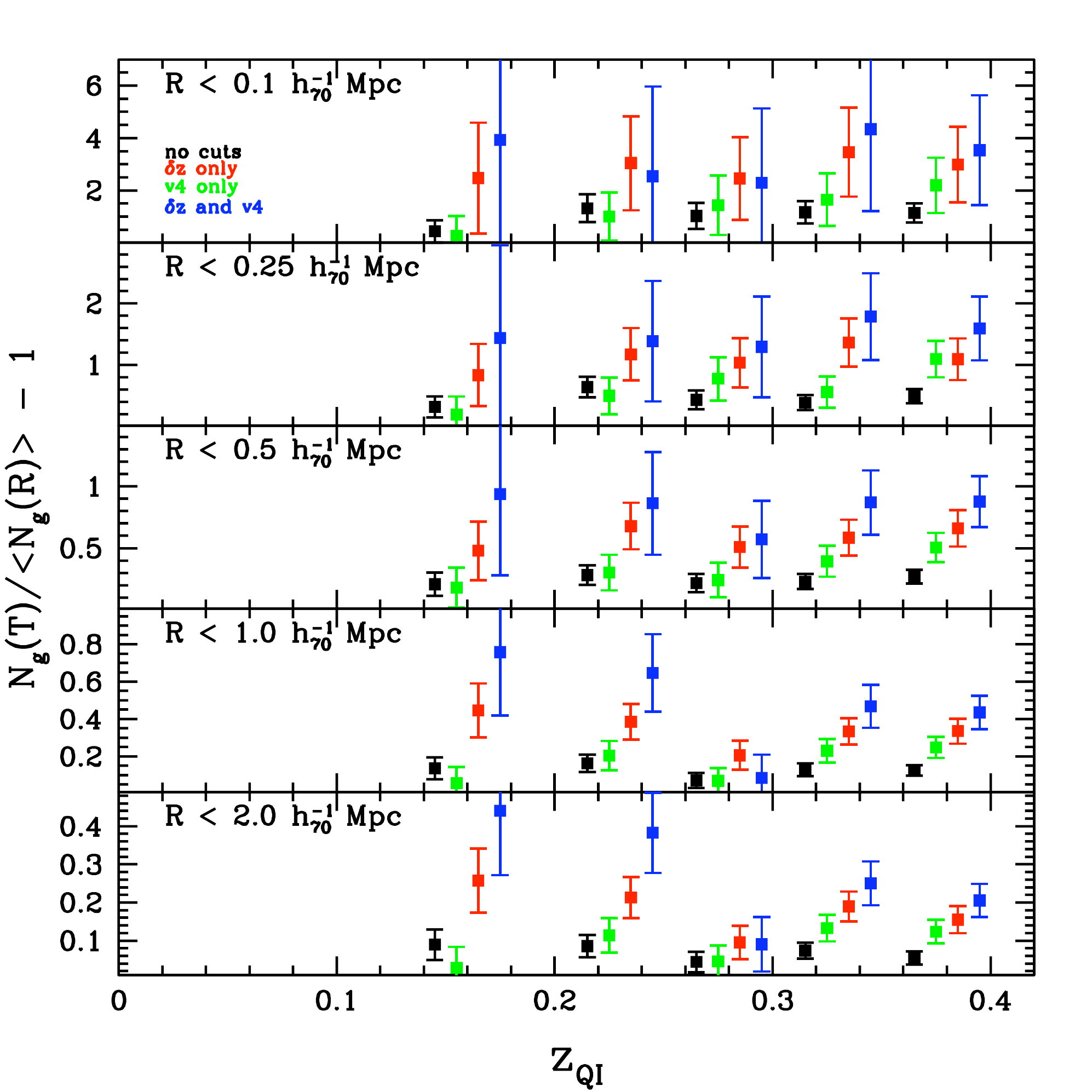}
\caption{Cumulative overdensity vs. redshift for $0.11\leq z\leq0.4$ QI\_pt sources with $-24.2 \leq M_{i} \leq -22.0$ using no cuts (black), only $\delta z = 0.05$ cut (red), only V4 sample photometric galaxy sample (green), and both V4 photometric galaxy sample and $\delta z = 0.05$ (blue).  
\label{redshift_comparecutsQI}}
\end{figure}

We compare the cumulative overdensity of targets as a function of redshift using no photometric galaxy cuts, the $\delta z = 0.05$ cut, the V4 limit, and both the V4 and $\delta z$ cuts (see Figure~\ref{redshift_comparecutsQI}, which uses QI\_pt sources with $-24.2 \leq M_{i} \leq -22.0$).
We find that applying just the V4 volume-limit has a negligible effect on the measured overdensity of the target environments.  Applying the $\delta z = 0.05$ cut has a strong effect as it eliminates most of the projection effects of background and foreground galaxies.  Combining the $\delta z = 0.05$ cut with the V4 background sample increases the overdensities measured but also increases the measurement noise.  The increase of overdensity when using the V4 background sample is due to the fact that only brighter galaxies are being counted, and it is well known that brighter galaxies cluster more strongly \citep[e.g.][]{Norberg, Hogg2003, Blanton2005}.  

Our goal is to marginalize over the photometric redshift distribution of the background galaxy sample, which is accomplished with the $\delta z$ cut; adding the volume limit increases the noise without improving the overdensity measurement.  Therefore, in general, we will not use volume-limited background galaxy samples.  

    \begin{table} \centering \begin{minipage}{140mm}
    \caption{Redshift limits based on Poisson signal-to-noise.  \label{PoissonSNtable}}
    \begin{tabular}{c c c c}\hline\hline
    Galaxy Sample&Redshift Limit&Expected S/N&$P>0$ counts\\ \hline
    all&0.5&2.09&0.988\\ \hline
    early&0.4&2.33&0.996\\
    late&0.4&2.21&0.992\\ \hline
    Ell&0.28&2.48&0.998\\
    Sbc&0.28&2.20&0.992\\
    Scd&0.28&2.14&0.990\\
    Irr&0.28&3.50&1.0
    \end{tabular} \end{minipage}
    \end{table}

    \begin{table} \centering \begin{minipage}{140mm}
    \caption{Volume-limited spectroscopic samples.  \label{vollimSpectargtable}}
    \begin{tabular}{c c c c}\hline\hline
    Sample&Redshift&Luminosity&\# Objects\\ \hline
    TI+TII&$0.11\leq z\leq0.5$&log($\LOIII$/$L_{\odot}$)$\geq8.0$&3781\\
    TI&$0.11\leq z\leq0.5$&log($\LOIII$/$L_{\odot}$)$\geq8.0$&3229\\
    TII&$0.11\leq z\leq0.5$&log($\LOIII$/$L_{\odot}$)$\geq8.0$&552\\ \hline
    TI+TII&$0.11\leq z\leq0.28$&$6.75\leq$log($\LOIII$/$L_{\odot}$)$\geq8.0$&2944\\
    TI&$0.11\leq z\leq0.28$&$6.75\leq$log($\LOIII$/$L_{\odot}$)$\geq8.0$&894\\
    TII&$0.11\leq z\leq0.28$&$6.75\leq$log($\LOIII$/$L_{\odot}$)$\geq8.0$&2050\\ \hline 
    TI+TII&$0.11\leq z\leq0.4$&log($\LOIII$/$L_{\odot}$)$\geq8.0$&2229\\
    TI&$0.11\leq z\leq0.4$&log($\LOIII$/$L_{\odot}$)$\geq8.0$&1791\\
    TII&$0.11\leq z\leq0.4$&log($\LOIII$/$L_{\odot}$)$\geq8.0$&438\\ \hline
    QI&$0.11\leq z\leq0.5$&log($\LOIII$/$L_{\odot}$)$\geq8.0$&3030\\
    QI&$0.11\leq z\leq0.4$&log($\LOIII$/$L_{\odot}$)$\geq8.0$&1592\\ 
    AI&$0.11\leq z\leq0.15$&log($\LOIII$/$L_{\odot}$)$\geq6.5$&621\\
    AII&$0.11\leq z\leq0.15$&log($\LOIII$/$L_{\odot}$)$\geq6.5$&1222
    \end{tabular} \end{minipage}
    \end{table}

\clearpage

\clearpage

\chapter{The Dependence on Redshift, Type, \& Broad-Band Luminosity}\label{paper1chapter}

\section{OVERVIEW}
We now begin our analysis of the environments of AGNs and quasars using the galaxy counting technique established in the previous chapter.  Here we explore how the local environment of the targets is related to their redshift, type, and broad-band luminosity.  This work was originally published as \citet{Paper1}; since its publication, however, we have modified our upper limits in redshift (see Section~\ref{techniquechapter_redshiftlimits}).    The updated results, which are qualitatively unchanged, are presented in this chapter.  

In Figure~\ref{scale_spectargs}, we present the mean cumulative overdensity of photometric galaxies as a function of scale for the QI\_pt, QII\_Z, AI, and AII spectroscopic samples.  There are clear differences: 
Type II objects are in more overdense environments than Type I objects for both higher luminosity AGN (i.e. quasars) and lower luminosity AGN, and quasars are in the most overdense environments at all scales.  Within a scale of $\approx150\kpchseventy$, QII\_Zs have an environment 1.3 times more overdense than that of QI\_pts, albeit with large errors, while the AIIs have an environment 1.4 times more overdense than AIs.  At the same scale, the QI\_pts have environments more overdense than AIs by a factor of 1.8, and QII\_Zs have environments 1.6 times more overdense than AIIs.  Moving out to the scale of $\approx 1\Mpchseventy$, the QII\_Zs again have an environment 1.3 more overdense than the environment of QI\_pts, and the QI\_pts are in environments 1.4 times more overdense than AIs.  

The differences between the target samples' environment overdensities could be an effect of AGN type, however, and the intertwined effects of AGN luminosity and redshift will certainly play into these differences.  In this chapter, we explore how type, redshift and broad-band luminosity influence the measured differences in AGN environments. 

\section{TARGET REDSHIFT}\label{redshiftsubsec}
We first isolate the effects of redshift and investigate the redshift dependence of AGN environments.  Figure~\ref{redshift_spectargs} shows the mean cumulative overdensity as a function of redshift for different spectroscopic targets, which provides marginal evidence for redshift evolution in the environment overdensity of  QI\_pts.  The magenta dashed line shows the linear weighted least-squares fit to the QI\_pt environment overdensity data with redshift at different maximum radii;  the fitting parameters for these lines is given in Table~\ref{table_compareFit}.  The rightmost column of the table gives the $\chi^2$ probability for each fit using the relevant degrees of freedom.  While these fits indicate a slight redshift dependence, we also try a zero-slope linear fit and find that the zero-slope fit, i.e., no redshift dependence, is also a good fit to the data and in the case of the $1\Mpchseventy$ scale, slightly more likely.  In contrast, weighted least-squares fits for AIs and AIIs (which are not shown in Figure~\ref{redshift_spectargs}, but the parameters are listed in Table~\ref{table_compareFit}) show that there is evolution with redshift (especially for AIIs).  This conclusion is strengthened by the fact that zero-slope fits are increasingly poor characterizations of the overdensity with redshift relationship as the scale decreases.  The average QII\_Z environment overdensity, which is placed at the average redshift of the sample, is consistent with the QI\_pt overdensity values at all scales.  
We note that both larger samples and higher redshift measurements will be necessary to place strong constraints on the functional form of QI environment overdensity evolution with redshift.  
However, if the QI environment overdensity is indeed independent of redshift, this implies that the significant differences in environment seen in Figure~\ref{scale_spectargs} are caused primarily by luminosity and type effects, rather than the influence of redshift evolution.

In the top panel of Figure~\ref{scale_spectargs_z}, we show the evolution of the mean cumulative overdensity of photometric galaxies in the environments of QI\_pt, AI and AII samples.  It is important to recall that we have placed the random points at the same redshift as the spectroscopic targets, and that we have imposed $\delta z$ cuts on the photometric galaxies (as described in Section~\ref{deltazcutsection}) in order to minimize the effect of redshift evolution in the photometric galaxy sample.  Therefore we can compare objects in different redshift bins.  Figure~\ref{scale_spectargs_z} demonstrates that higher redshift QI\_pts are in environments 1.29 times more overdense than the lower redshift quasars on scales $\lesssim500\kpchseventy$, while at larger scales, there appears to be little-to-no redshift evolution.  However, there is scale-dependent redshift evolution evident on scales $\lesssim1.0\Mpchseventy$ for the AIIs, shown in the lowest panel.  The AIs begin to exhibit more noticeable redshift evolution at scales $\lesssim300\kpchseventy$, where the environments of lower redshift AIs are 1.16 times less dense than those of the higher redshift AIs. 

We see therefore that there is some evidence for a change in local environment as a function of redshift, all else being held constant.  However, we have not yet taken AGN luminosity into account.  Even in the same redshift range, selection effects due to the magnitude-limited samples may come into play, which we investigate in Sections~\ref{absmagsubsec} and \ref{allsubsec}.  

\section{TARGET TYPE}\label{typesubsec}
In Figure~\ref{scale_spectargs_type}, we identify three redshift ranges where there is overlap between our AGN samples and explore whether differences in type are reflected in the relative overdensity.  The top panel shows the overdensity as a function of scale for both types of higher-luminosity AGNs (i.e. quasars) in the range $0.3\leq z\leq0.5$, and for both types of lower-luminosity AGNs in two redshift ranges, $0.11\leq z\leq0.15$ and $0.15<z\leq0.33$.  The dividing redshift value of $z=0.15$ is chosen to roughly equalize the number of lower-luminosity AGNs in each redshift range.  The lower three panels show the ratio of Type II environment overdensity to Type I environment overdensity in the three redshift ranges.  Again, we are able to compare objects in different redshift ranges because we have imposed $\delta z$ cuts on the photometric galaxies around both the spectroscopic targets and the random positions to which they are compared in order to account for any redshift evolution in the photometric galaxy sample and to minimize projection effects.  

QII\_Zs have higher overdensity environments than QI\_pts with little scale dependence: at $R\approx1.0\Mpchseventy$, the overdensity of QII\_Z environments is a factor of 1.4 greater than the overdensity of QI\_pt environments, and at the smaller scale of $R\approx250\kpchseventy$, the QII\_Z environments have 1.2 times the overdensity of QI\_pt environments.   However, the large errors due to the small number of QII\_Zs prevent us from drawing strong conclusions.  

Type II environment overdensity is again consistently about a factor of 1.3 higher on all scales than the Type I environment overdensity for AIs and AIIs in the redshift range $0.15<z\leq0.33$.  In the lower redshift range of $0.11\leq z\leq0.15$, however, the AII and AI environment overdensities have a ratio consistent with unity until scales $R<200\kpchseventy$, where the ratio increases to 1.3.  This result agrees with previous work that concluded that Type II Seyfert galaxies are more likely to have close neighbors than Type I Seyferts at very low redshifts \citep{Koulouridis}.  

\begin{figure}
\plotone{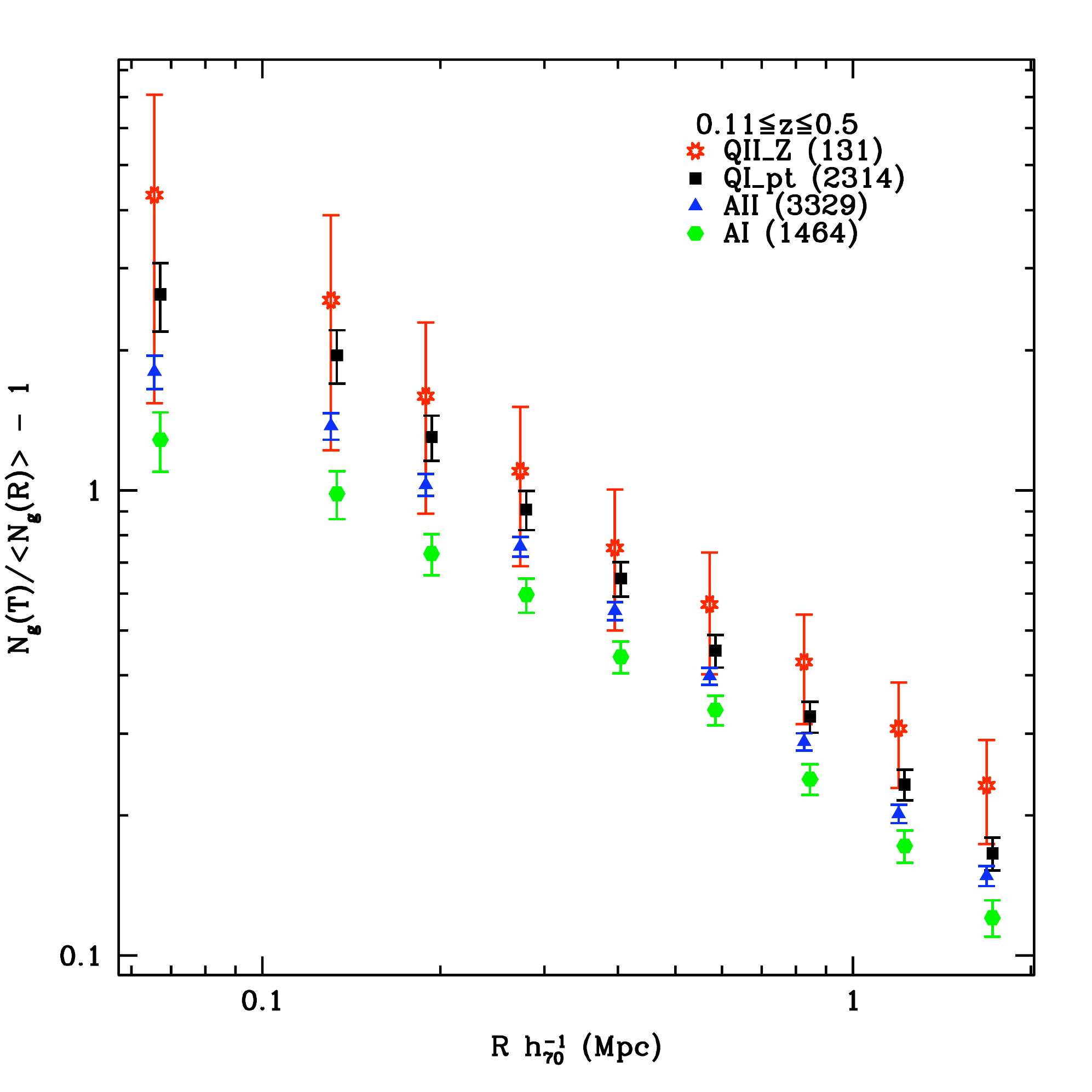}
\caption{Mean cumulative overdensity of photometric galaxies as a function of comoving scale around spectroscopic targets.  Solid black squares represent QI\_pts, open red starred points represent QII\_Zs, solid blue triangles represent AIIs, and solid green hexagons represent AIIs.   Points have been slightly offset horizontally for clarity.  
\label{scale_spectargs}}
\end{figure}

\begin{figure}
\plotone{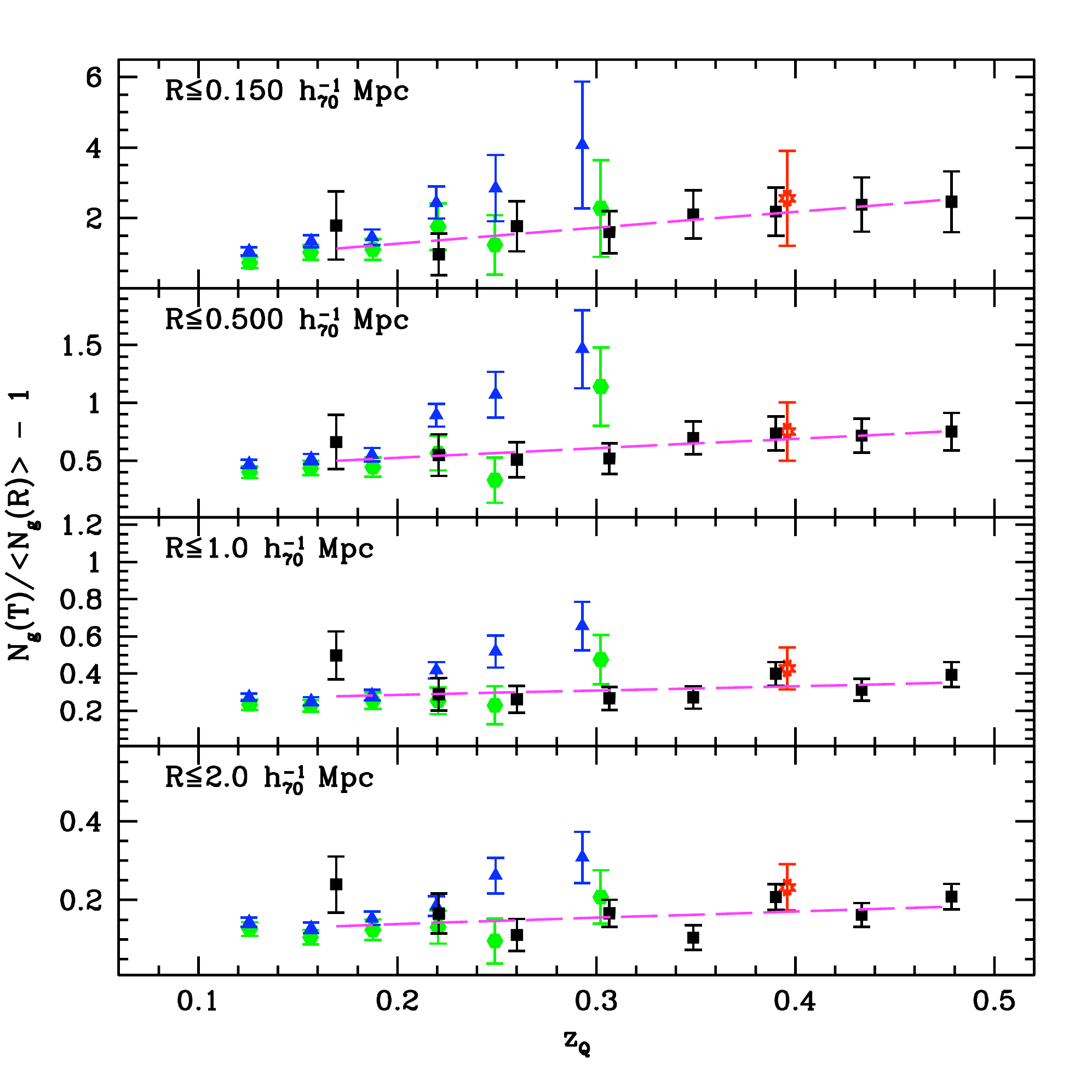}
\caption{Mean cumulative overdensity vs. redshift for spectroscopic targets.  Symbols correspond to those used in Figure~\ref{scale_spectargs}.  The magenta dashed lines are linear weighted least-squares fits for the QI\_pt sample; the parameters for these lines are given in Table~\ref{table_compareFit}.  
\label{redshift_spectargs}}
\end{figure}

\begin{figure}
\plotone{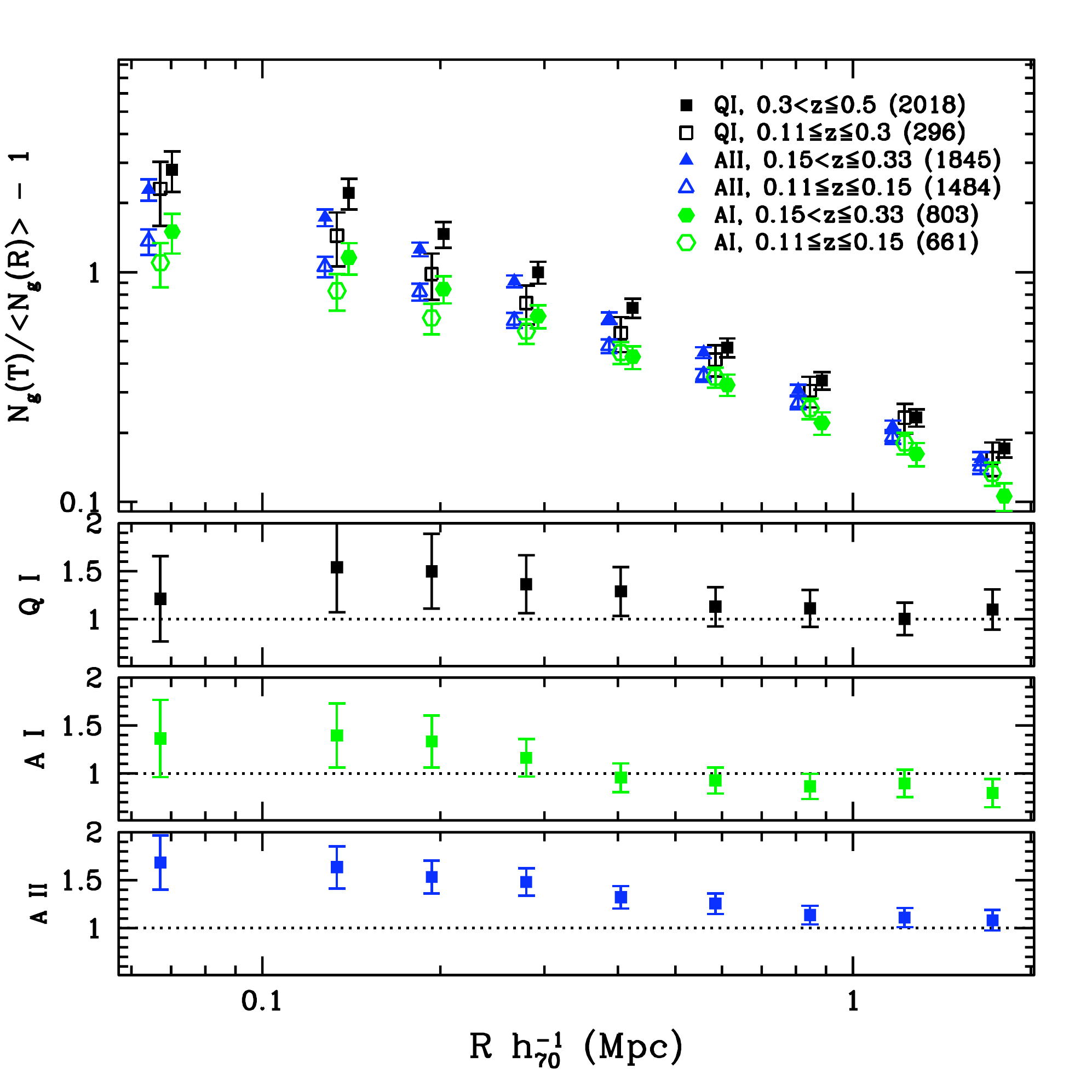}
\caption{Upper panel:  Mean cumulative overdensity of photometric galaxies around quasars and lower-luminosity AGN as a function of scale and redshift.  Points have been slightly offset horizontally for clarity.  Top lower panel:  Ratio of environment overdensity of higher-redshift QI\_pts to that of lower-redshift QI\_pts.  Middle lower panel:   Ratio of environment overdensity of higher-redshift AIs to that of lower-redshift AIs.  Bottom lower panel:  Ratio of environment overdensity of higher-redshift AIIs to that of lower-redshift AIIs.    
\label{scale_spectargs_z}}
\end{figure}

Because we see increased overdensity for AIIs compared to AIs on small scales in overlapping redshift ranges, we can conclude that in Figure~\ref{scale_spectargs}, the differences seen in environment overdensity between the AGN types are not primarily due to redshift evolution.  However, we have not ruled out the effects of AGN luminosity.  The AI and AII samples are selected from magnitude-limited spectroscopic galaxy samples, which will be dominated by intrinsically more luminous sources at higher redshift.  The AI sample could be more affected by this magnitude limit, as the broad emission lines contribute more significantly to the overall flux in a given band and therefore the two AGN populations could have different average intrinsic luminosities.  

\section{TARGET BROAD-BAND LUMINOSITY}\label{absmagsubsec}

Unlike Type I quasars, which are targeted largely based on their strong nuclear luminosity \citep{Schneider}, the lower luminosity AGN we use were selected from objects classified as galaxies by the SDSS selection algorithms \citep{Haoa}.  The broad-band flux of these sources will be dominated by host galaxy starlight and/or flux from star formation, etc., which has little or no association with the nuclear luminosity.  Therefore, we first focus on the point-source QIs only (QI\_pt) for our analysis of the relationship between absolute magnitude and environment, as the QI sample spans the entire redshift range we study, and with this long redshift baseline we are best able to disentangle redshift and luminosity effects on environment overdensity.  

In order to verify that the observed evidence for evolution of environment overdensity is not due to the $i\leq19.1$ ($z\lesssim3.0$) limit imposed on QI selection in the SDSS \citep{Schneider}, we performed several tests in which we vary the apparent magnitude limit of the data.  We considered two quasar samples limited to $i\leq18.9$ and to the $i\leq19.1$ SDSS limit.  
The two magnitude-limited samples were each subdivided into two luminosity bins.  We first compared environment overdensity measurements of bright or dim quasars in each of the magnitude-limited samples and found no appreciable difference.  Additionally, no difference was observed when different absolute magnitude values were used to define the bright and dim samples.  In order to ensure that there is no difference between environments of quasars with $i > 19.1$, which were selected by the high-redshift targeting algorithm, and the rest of the apparent magnitude-selected sample, we performed similar tests comparing the environment overdensity of the entire quasar sample to that of the subset of quasars with $i\leq18.9$ or $i>19.1$.  In all cases, there was no appreciable change in the observed overdensity. 

We compare the environment overdensities of QI\_pt in two luminosity bins to the other target samples without redshift cuts in Figure~\ref{scale_spectargs_M}.  The threshold value $M_{i}=-23.0$ is chosen to give roughly equal numbers of QI\_pt sources in each luminosity bin: there are 1,136 (1,178) quasars with $-26.4\leq M_{i} \leq-23.0$ ($-23.0<M_{i}\leq-22.0$).  The average magnitude of the brighter (fainter) bin is $\overline{M_{i}}=-23.60$ ($\overline{M_{i}}=-22.60$).  

QII\_Zs and the brighter QI\_pt sources are located in similarly overdense environments consistently at all scales, while the dimmer QI\_pts are located in environments slightly less overdense than the QII\_Zs.  At a scale $R\approx500\kpchseventy$, the cumulative overdensity of QII\_Z environment is 1.14 times that of the brighter QI\_pts, but 1.18 times as the dimmer QI\_pts.  At the scale of $R\approx1.0\Mpchseventy$, QII\_Zs have environment overdensities 1.3 times the environment overdensity of brighter QI\_pts but 1.3 times that of dimmer QI\_pts.  Again we note that the large error bars nearly overlap with unity and prevent strong conclusions.  

The more luminous QI\_pts are located in environments more overdense than AIs, while there is less difference in the overdensities of dimmer QI\_pts and AIs.  The environment overdensity ratio increases with decreasing scale for both brighter and dimmer QI\_pts.  At a scale $R\approx500\kpchseventy$, brighter QI\_pt environments have an overdensity 1.5 times the overdensity of AI environments with significance $2,2\sigma$, and dimmer QI\_pt environments have an overdensity 1.4 times the overdensity of AI environments with significance $2.2\sigma$.  At $R\approx150\kpchseventy$, the environments of brighter QI\_pts are 2.2 times as overdense ($2.4\sigma$), and the environments of dimmer QI\_pts are 1.9 times as overdense as the environments of AIs ($2.2\sigma$).  

The ratio of QI\_pts to AIIs increases for both bright and dim quasars with decreasing scale, but less dramatically as the ratio to AIs.  The ratio between dimmer QI\_pts and AIIs is approximately consistent with unity for scales $150\kpchseventy<R\leq2.0\Mpchseventy$; the ratio between brighter QI\_pts and AIIs is 1.2 ($1.2\sigma$) for scales $R\approx500\kpchseventy$.  On smaller scales, both ratios increase.  At scales $R\approx150\kpchseventy$, the ratio of brighter QI\_pts to AIIs is 1.6 ($1.7\sigma$), and the ratio of dimmer QI\_pts to AIIs is 1.3 ($1.3\sigma$).  This scale dependency could be evidence for the merger origin of quasars, since one would expect to see a higher density of environment galaxies at small scales where merger events are likely to take place \citep{Hopkins2008}. 

\section{BROAD-BAND LUMINOSITY, REDSHIFT, AND TYPE}\label{allsubsec}
We combine our analysis of type, redshift and broad-band luminosity effects on environment overdensity in Figures~\ref{scale_spectargs_Mandz_higherz}, \ref{scale_spectargs_Mandz_lowerz}, and \ref{scale_qso_newMzcompare}.  Our $\delta z$ cuts on the photometric galaxies around the spectroscopic targets as well as around the random positions to which they are compared (as described in Section~\ref{deltazcutsection}) allow us to make meaningful comparisons of objects in different redshift ranges.  In Figure~\ref{scale_spectargs_Mandz_higherz}, QII\_Zs are compared to QI\_pts in the redshift range $0.3 \leq z \leq 0.5$.  
We divide the QI\_pts into bright ($1,039$; $\overline{M_{i}}=-23.60$) and dim ($980$; $\overline{M_{i}}=-22.60$, about 2.5 times fainter) samples of roughly equal numbers at $M_{i}=-23.0$.  

\begin{figure}
\plotone{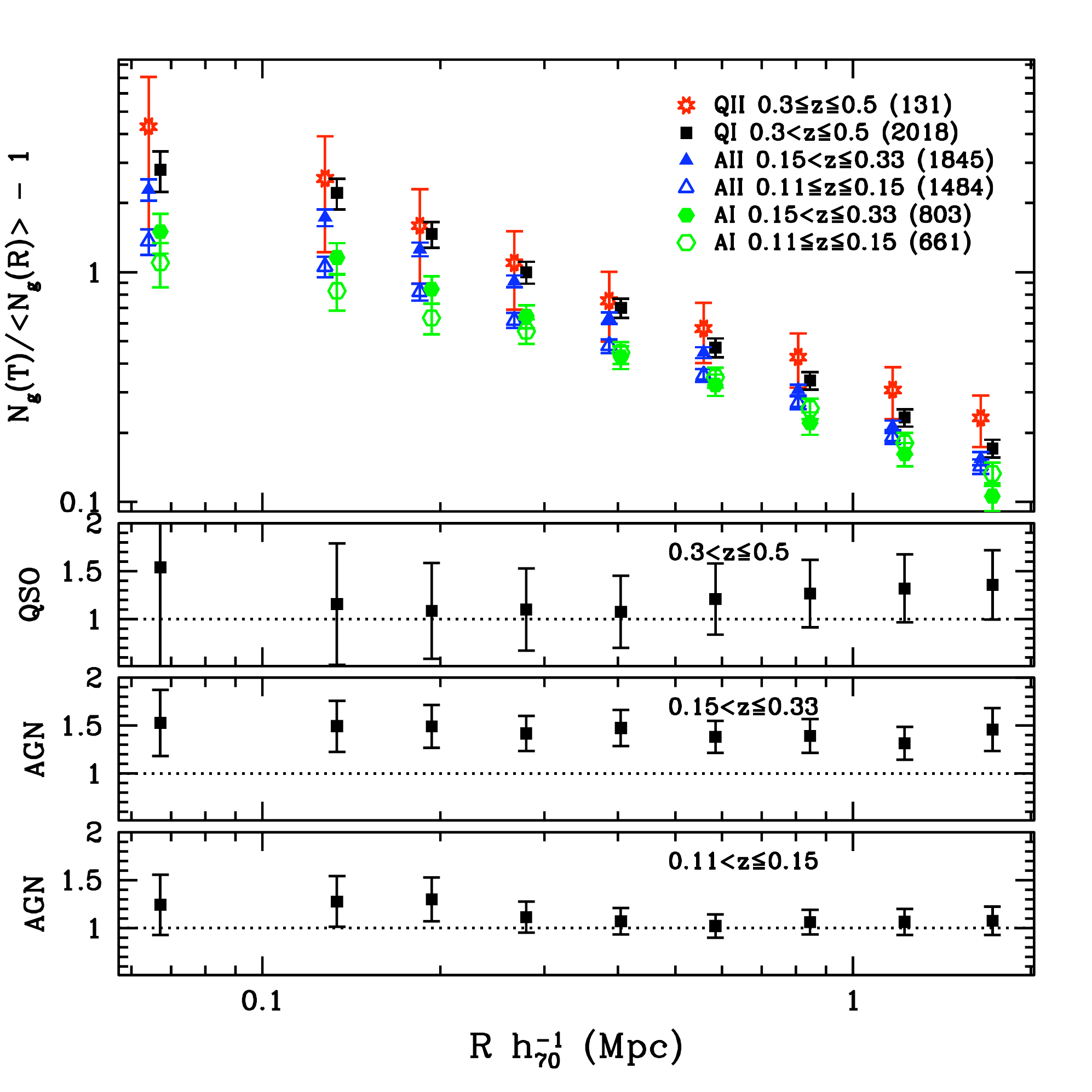}
\caption{Upper panel:  Mean cumulative overdensity of photometric galaxies around quasars and lower-luminosity AGN as a function of scale and redshift.  Points have been slightly offset horizontally for clarity.   Top lower panel:  Ratio of environment overdensity of QII\_Zs to that of QI\_pts in the redshift range $0.3 \leqslant z \leqslant 0.5$.  Middle lower panel:   Ratio of environment overdensity of AIIs to that of AIs in the redshift range $0.15 < z \leqslant 0.33$.  Bottom lower panel:  Ratio of environment overdensity of AIIs to that of AIs in the redshift range $0.11 < z \leqslant 0.15$.   
\label{scale_spectargs_type}}
\end{figure}

\begin{figure}
\plotone{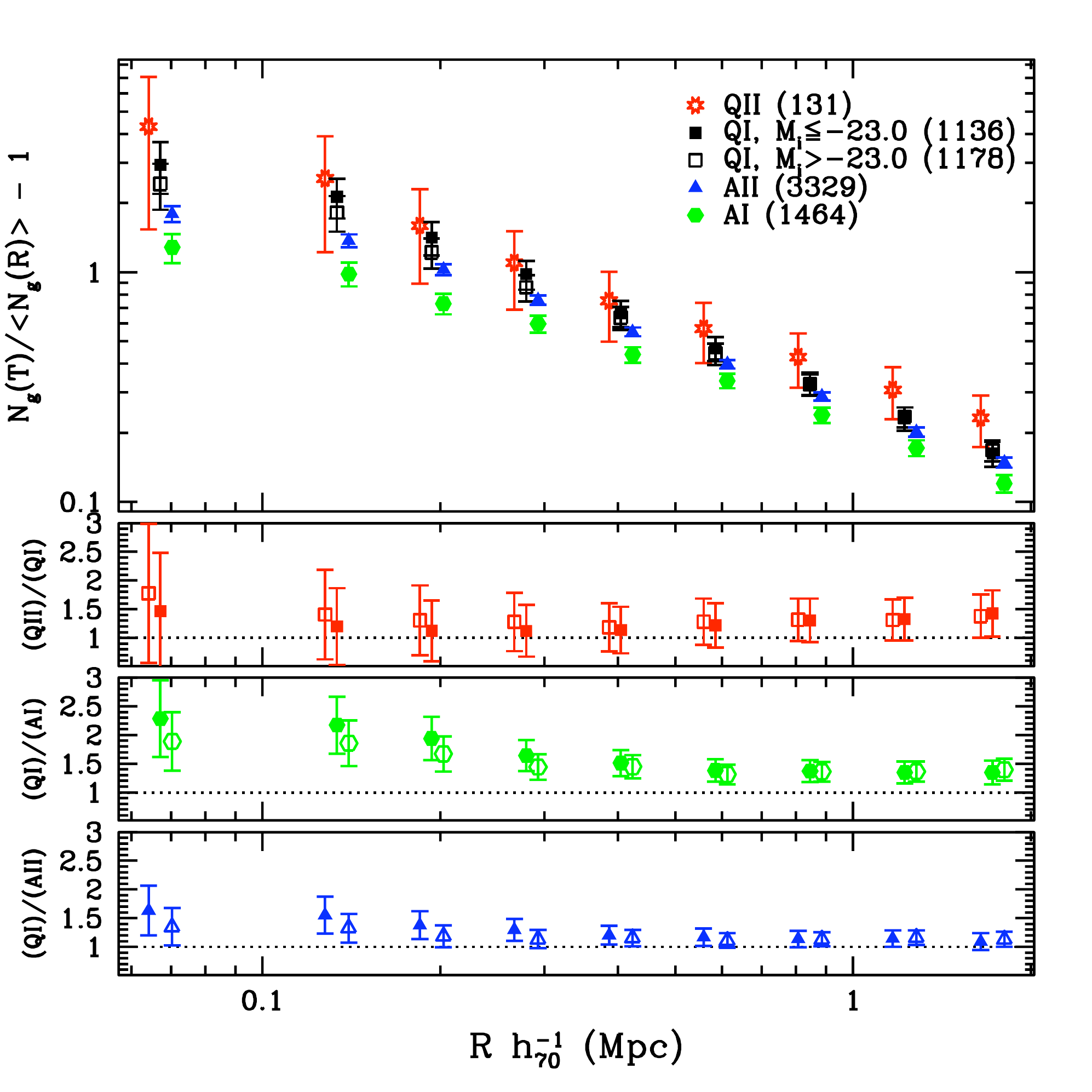}
\caption{Top panel:  Mean cumulative overdensity of photometric galaxies around QI\_pts split by luminosity; low luminosity AGN; and QII\_Zs.  Bright quasars have absolute magnitude $-26.4 \leq M_{i} \leq -23.0$ and dim quasars have absolute magnitude $-23.0 < M_{i} \leq -22.0$.  Top lower panel:  ratio of QII\_Zs environment overdensity to bright (solid points) and dim (open points) QI\_pt environment overdensities.  Middle lower panel:  ratio of bright (solid points) and dim (open points) QI\_pt environment overdensities to AI environment overdensity.  Bottom lower panel: ratio of bright (solid points) and dim (open points) QI\_pt environment overdensities to AII environment overdensity.  No redshift limits have been imposed.  
\label{scale_spectargs_M}}
\end{figure}

\begin{figure}
\plotone{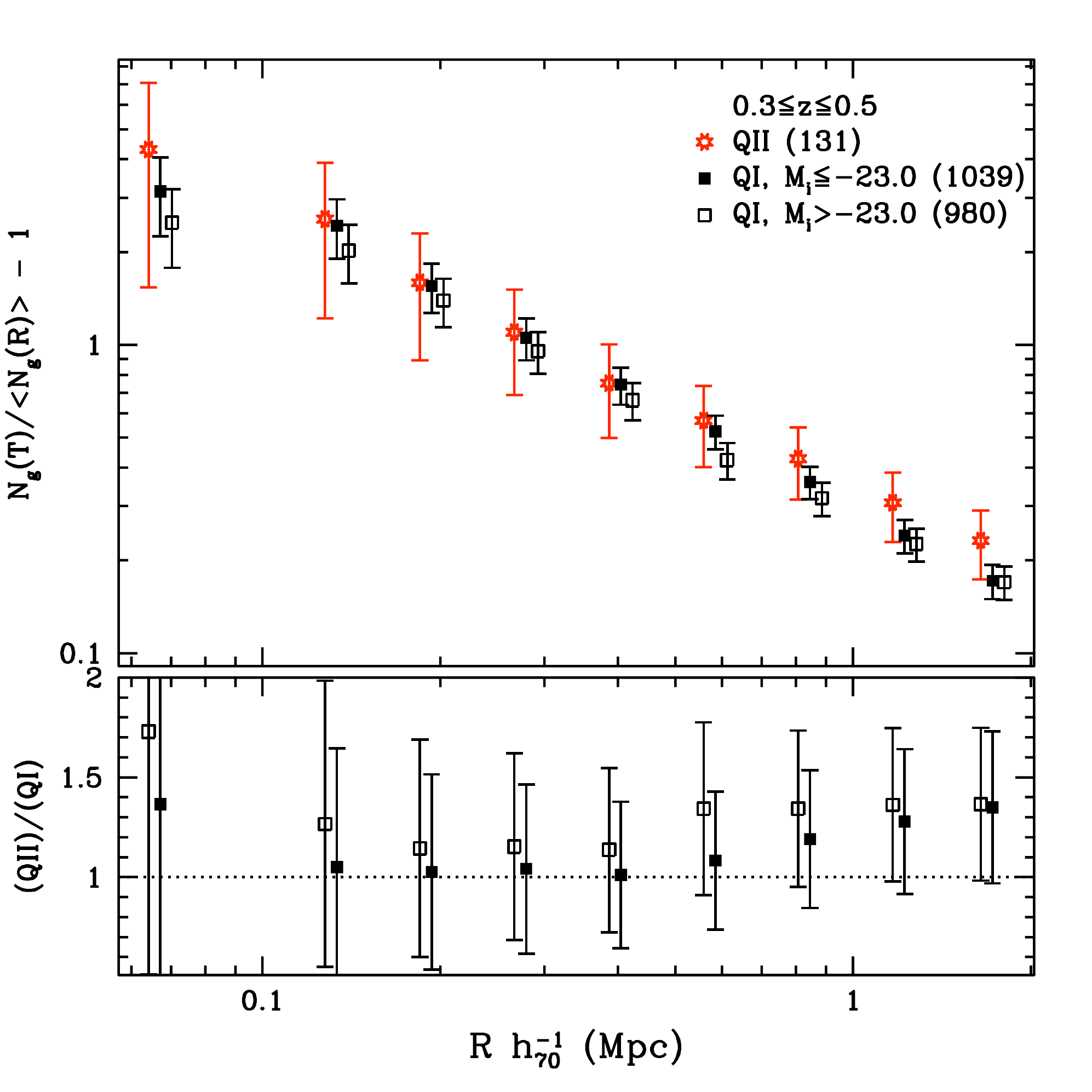}
\caption{Upper panel:  Mean cumulative overdensity of photometric galaxies around QII\_Zs and QI\_pts in the redshift range $0.3\leq z\leq0.5$.  The QI\_pts in this redshift range have been divided at $M_{i}=-23.0$ so that the luminosity bins contain approximately equal numbers of QI\_pts.  Lower panel:  Ratio of environment overdensities of QII\_Zs to brighter and dimmer QI\_pts in this redshift range.  Points in both panels have been slightly offset horizontally for clarity.  
\label{scale_spectargs_Mandz_higherz}}
\end{figure}

\begin{figure}
\plotone{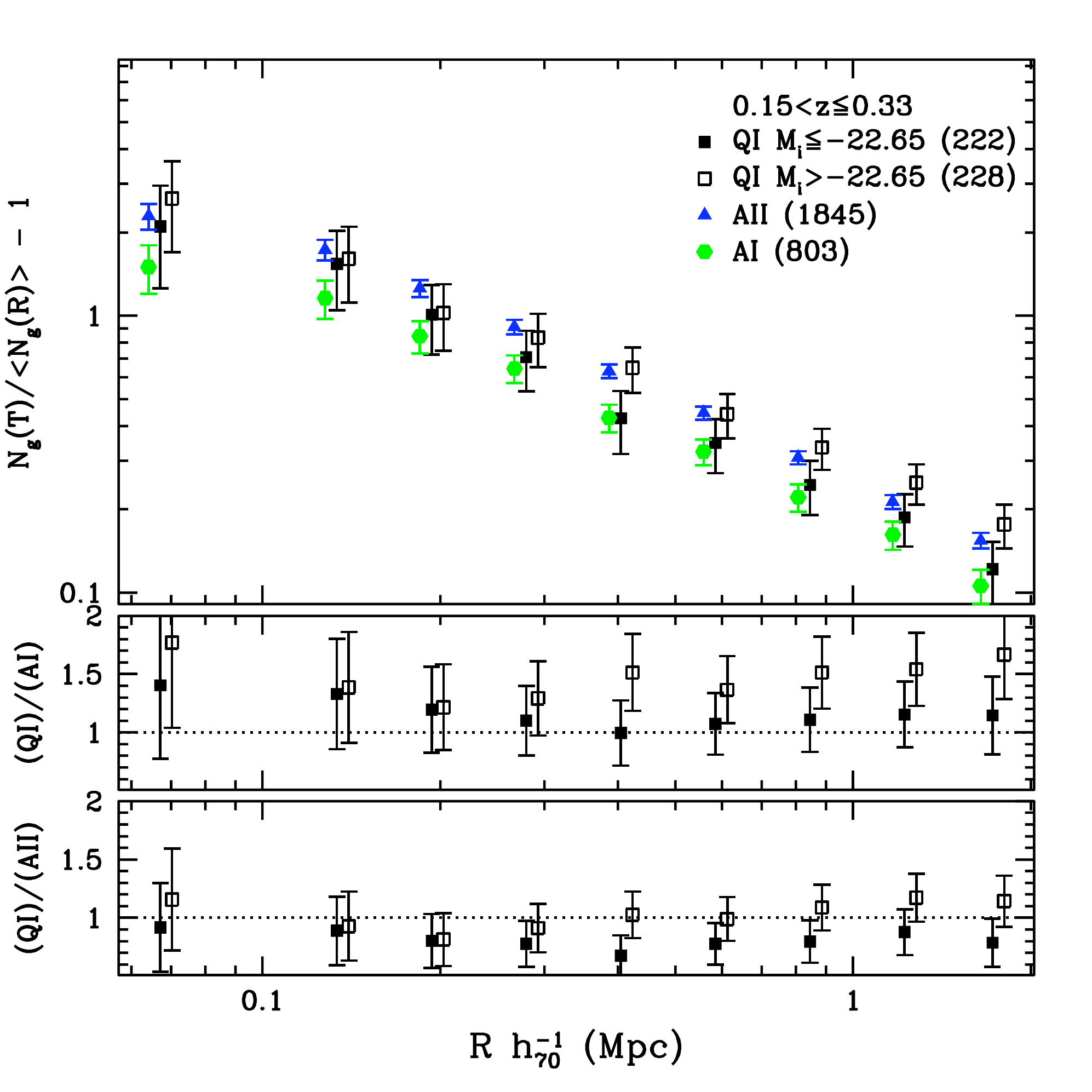}
\caption{Upper panel:  Mean cumulative overdensity of photometric galaxies around QI\_pts and AIs and AIIs in the redshift range $0.15 < z \leq 0.33$.  The QI\_pts in this redshift range have been divided at $M_{i}=-22.65$ so that the luminosity bins contain approximately equal numbers of QI\_pts.  Lower panels:  Ratio of environment overdensities of brighter and dimmer QI\_pts to AIs and AIIs in this redshift range.  Points in both panels have been slightly offset horizontally for clarity.  
\label{scale_spectargs_Mandz_lowerz}}
\end{figure}

\begin{figure}
\plotone{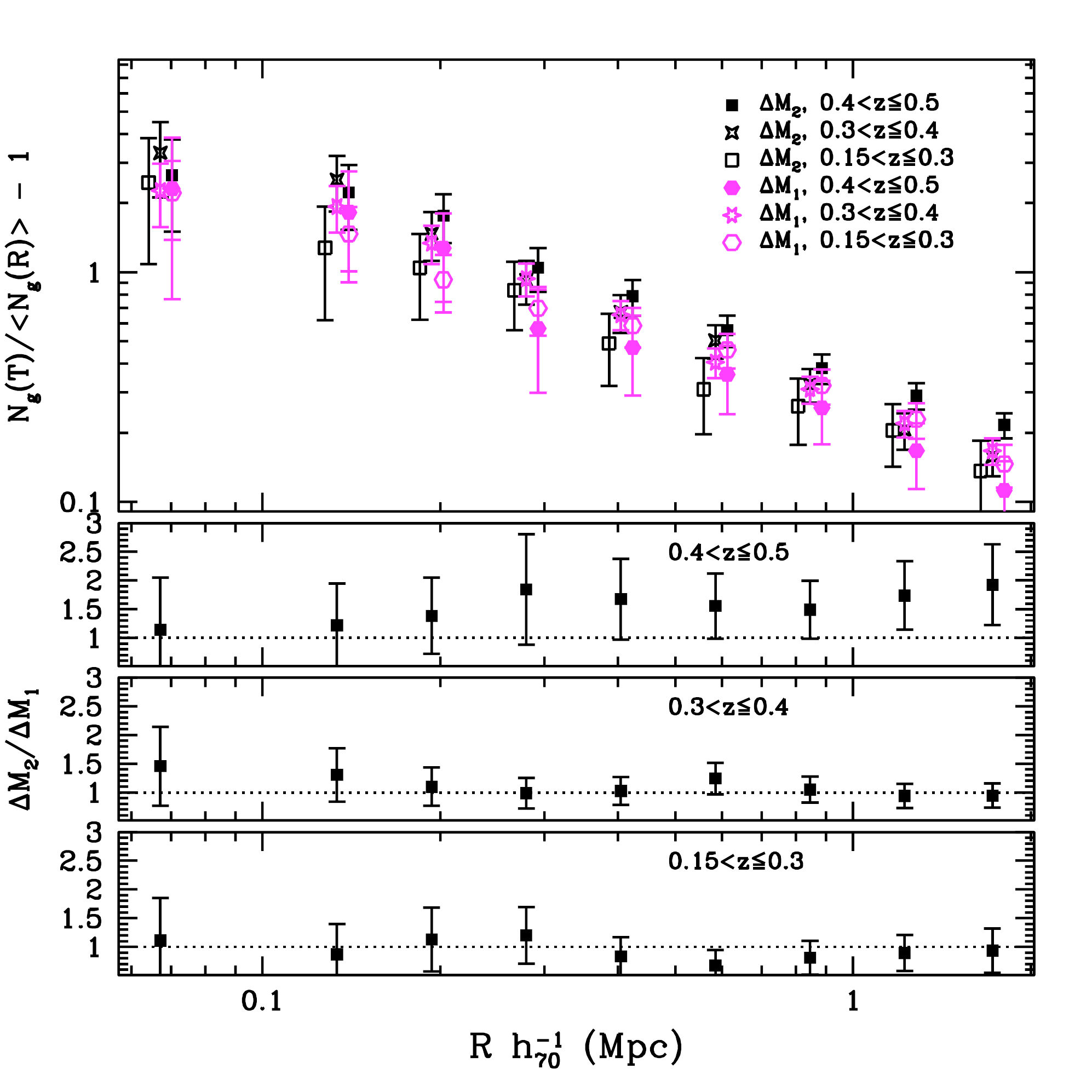}
\caption{Mean cumulative overdensity of photometric galaxies around QI\_pts with redshift for two luminosity bins, where $\Delta M_{1} : -23.0 < M_{i} \leq -22.0$ and $\Delta M_{2} : -24.0 < M_{i} \leq -23.0$.  Lower panels:  Ratio of $\Delta M_{2}$ to $\Delta M_{1}$ quasar environment overdensities in the three redshift ranges.  
\label{scale_qso_newMzcompare}}
\end{figure}

Comparing the lower panel of Figure~\ref{scale_spectargs_Mandz_higherz} to the top ratio panel of Figure~\ref{scale_spectargs_type} shows the dramatic part luminosity plays compared to evolution alone.  The environment of QII\_Zs is similar to the signature of brighter QI\_pts for the smallest scales.  The similarity of environments at small scales suggests that the differences observed between brighter Type I quasars and Type II quasars are due to a non-environmentally driven mechanism such as orientation or internal structure effects.  This in turn implies that Type II quasars are not a different cosmological population from these brighter Type I quasars.  

However, the QII\_Z environments are slightly more overdense, albeit with large errors, than those of dimmer QI\_pts on scales $R\leq2.0\Mpchseventy$ that we measure.  The different characteristics of the environments of the dimmer QI\_pt population from the QII\_Z population are most likely due to intrinsic luminosity differences rather than redshift differences.  
We see consistent overdensity ratios on all scales and do not see small scale effects, therefore we conclude that the difference in environment overdensity between the brighter and dimmer quasars is primarily due to mass effects.  More luminous AGN are expected to have higher mass black holes \citep[e.g.,][]{Magorrian, MarconiHunt}, which are in turn correlated with more massive dark matter halos \citep[e.g.,][]{FerrMerr, Gebhardt, Tremaine}.  
Selection effects in the magnitude-limited photometric galaxy sample could also play into the difference in overdensity between the brighter and dimmer QI\_pts.  The redshift distribution of QI\_pts in the brighter ($M_{i}\leq-23.25$) bin is slightly different from that of the dimmer ($M_{i}>-23.25$) bin even over the redshift range of $0.3 \leq z \leq 0.5$ (the brighter quasars have a mean redshift of 0.43, and the dimmer quasars have a mean redshift of 0.40).  The galaxies seen in the environments of brighter (higher redshift) quasars will tend themselves to be brighter, and consequently more massive, and therefore cluster more strongly than the dimmer environment galaxies \citep{Maddox, Zehavi}.  This, however, should not be a major effect.  

Figure~\ref{scale_spectargs_Mandz_lowerz} compares the environments of QI\_pts in two luminosity bins to the environments of AIs and AIIs in the redshift range $0.15<z\leq 0.33$.  We use $M_{i}=-22.65$ as the threshold value for brighter and dimmer quasars in this lower redshift range to equalize the number in each luminosity bin.  The $222$ brighter ($228$ dimmer) QI\_pts have a mean magnitude of  $\overline{M_{i}}=-23.39$ ($\overline{M_{i}}=-22.32$, about 2.4 times fainter than the brighter sample).  We note that these QI\_pt samples are more than four times smaller than the QI\_pt samples in the higher redshift range, thus the measurements (and resulting interpretation) will be less precise.  

The top panel of Figure~\ref{scale_spectargs_Mandz_lowerz} shows that at all scales, the environments of dimmer QI\_pts are more overdense than those of brighter QI\_pts.  It appears that the situation has been reversed from Figure~\ref{scale_spectargs_Mandz_higherz}, where the environments of dimmer quasars were less overdense than the environments of brighter quasars.  However, the range of luminosity at this lower redshift range of $0.15<z\leq0.33$ is much smaller than for the higher range of $0.3\leq z\leq 0.5$.  The overall absolute magnitude distribution of these QI\_pts is skewed toward the faint end, thus the dividing value $M_{i}=-22.65$ is very close to the quasar-Seyfert divide of $M_{i} \approx-22.5$ as defined by \citet{Haob}.   Significant variation in overdensities was seen when different magnitude cuts were imposed, with the dimmer quasars consistently having higher overdensities by varying margins.  The dramatic sensitivity of results on the bright/dim dividing value emphasize that for low luminosites and redshifts, broad-band absolute magnitudes are a poor proxy for AGN luminosity.  The measured flux is more likely to be affected by galaxy starlight, star formation, etc. at this faint end.  Therefore, any attempt to use broadband magnitudes to correlate nuclear luminosity with environment will be skewed.  

With these caveats in mind, we compare the QI\_pts to lower-luminosity AIs and AIIs in the lower two panels of the figure.  Dimmer QI\_pt environments have overdensities greater than the AIs, but the environments of brighter QI\_pts and the AIs have about the same amplitude on all scales.  The lower ratio panel shows the ratio of bright and dim QI\_pts to AIIs.  The brighter quasars have environments with slightly lower overdensity than AIIs;  the environments of dimmer QI\_pts are only slightly more overdense than the environments of AIIs, but are consistent within the error bars.  

In Figure~\ref{scale_qso_newMzcompare} we focus on QI\_pts alone to investigate the evolution of the environment overdensity of brighter and dimmer objects.   We have chosen two luminosity intervals of one magnitude in width and compare the environment overdensities of brighter to dimmer objects in three redshift intervals. $\Delta M_{1}$ corresponds to the dimmer luminosity interval of $-23.0<M_i\leq-22.0$ and contains 1,174 QI\_pts, and the brighter luminosity interval $\Delta M_2$ is $-24.0<M_{i}\leq-23.0$, containing 924 QI\_pts.  Table~\ref{table_Fig9Details} gives the number of quasars as well as the mean magnitude in each redshift and magnitude bin.  

In the two lower redshift bins $0.15<z\leq0.3$ and $0.3<z\leq0.4$, there is little difference in the environment overdensity of brighter and dimmer quasars with little-to-no scale dependence.  
However, in the highest redshift interval of $0.4 < z \leq 0.5$, brighter quasars are shown to be  located in slightly more overdense environments than the dimmer quasars.  At scale of $R\approx1.0\Mpchseventy$, the brighter quasars are located in environments with overdensity 1.5 times that of the dimmer quasars.  The brighter quasars have environments with overdensity 1.4 times the overdensity of dimmer quasar environments at a scale of $R\approx250\kpchseventy$, and then the ratio begins to drop toward unity at the innermost scales.  However, the large errors are nearly consistent with unity on all scales we measure.  

It appears, therefore, that there is again slight evidence for some redshift evolution of QI\_pt environments, but it is mainly manifested at the highest redshift range, which also has the largest error bars.  This emphasizes the need for additional studies of the environments around higher redshift Type I quasars.  We caution that the increased overdensity at higher redshift be affected by the change in mean luminosity of the dimmer quasar sample with increasing redshift (see Table~\ref{table_Fig9Details}).  While the bright quasar luminosity hardly changes between the three redshift intervals, the mean dim quasar luminosity changes by 0.23 magnitudes.  Therefore we cannot draw strong conclusions, but reiterate the need for higher precision and higher redshift measurements of quasar environments.   

\section{CONCLUSIONS}
We have shown in this chapter that 
QII\_Zs are shown to have similar environments as brighter Type I quasars in the same redshift range on all scales that we study, which suggests the observational differences in Type I and Type II quasars are driven by orientation and/or structure and not by cosmological evolution.  
Evidence that dimmer quasars and lower-luminosity AGN are located in environments with similar overdensity might suggest that dimmer quasars could be a transition population between low-luminosity AGN (likely fueled in dry mergers, close encounters, or secular processes) and high-luminosity AGN (likely fueled in major mergers).  Rather than disparate populations of merger-fueled and secularly fueled AGN, there may be a continuum of galaxy interactions from major mergers to close encounters or harassment that cause AGN luminosity differences.  Alternatively, a mix of mergers and secular processes could drive the AGN population near the quasar-Seyfert divide \citep[$M_{i}\approx-22.5$;][]{Haob}.  We have compared the AGN samples without redshift cuts, but we note that in Section~\ref{redshiftsubsec} we demonstrated that evolution of quasar environments with redshift is negligible.  

The significant difference in the environments of bright QI\_pts and the environments of both AIs and AIIs could imply that these populations have different fueling mechanisms.  A weak link between nearby neighbors of narrow-line AGN and their nuclear activity \citep{Li2006, Li2008} implies that it is likely internal mechanisms rather than merger activity that gives rise to the AGN activity in the low-luminosity sources.  The scale dependency in the relative environment overdensities of bright QI\_pt overdensities and lower-luminosity AGN could be evidence for the merger origin of bright QIs, since mergers are expected to be more likely in regions with a higher local density of galaxies \citep{Hopkins2008}.  

Finally, there is marginal evidence for redshift evolution of Type I quasar environments on all scales, especially for $0.4 \leqslant z \leqslant 0.5$, not noted in previous studies.   However, this evolution is not the primary explanation for the environment overdensity differences seen between Type I quasars and Type II quasars, and between Type I AGN and Type II AGN.  In order to place strong constraints on the functional form of this redshift evolution,  it is necessary to acquire higher precision measurements and higher redshift measurements.

\begin{table}\centering \begin{minipage}{140mm}
\caption{Linear weighted least-squares fit parameters for QI\_pt, AI, and AII environment data in Figure~\ref{redshift_spectargs}.  \label{table_compareFit}}
\begin{tabular}{c c c c c}\hline\hline
$R_{max}$&slope&intercept&$\chi^{2}$&P($\chi^{2}$,$\nu$)\\ \hline
\multicolumn{5}{c}{QI\_pt}\\ \hline
$2.0\Mpchseventy$&$0.161\pm0.152$&$0.107\pm0.057$&$8.91$&0.1787\\
&0.0&$0.165\pm0.013$&$10.02$&0.1874\\
$1.0\Mpchseventy$&$0.237\pm0.286$&$0.237\pm0.104$&$6.14$&0.4077\\
&0.0&$0.321\pm0.024$&$6.83$&0.4468\\
$0.5\Mpchseventy$&$0.830\pm0.622$&$0.358\pm0.219$&$1.43$&0.9640\\
&0.0&$0.641\pm0.055$&$3.21$&0.8649\\
$0.15\Mpchseventy$&$4.49\pm2.79$&$0.382\pm0.927$&$1.15$&0.9793\\
&0.0&$1.82\pm0.249$&$3.75$&0.8081\\
\hline
\multicolumn{5}{c}{AI}\\ \hline
$2.0\Mpchseventy$&$0.152\pm0.268$&$0.096\pm0.045$&$2.32$&0.6771\\
&0.0&$0.121\pm0.011$&$2.64$&0.7553\\
$1.0\Mpchseventy$&$0.555\pm0.472$&$1.52\pm0.077$&$2.13$&0.7119\\
&0.0&$0.240\pm0.018$&$3.51$&0.6219\\
$0.5\Mpchseventy$&$1.35\pm0.964$&$0.222\pm0.155$&$3.93$&0.4156\\
&0.0&$0.434\pm0.034$&$5.90$&0.3161\\
$0.15\Mpchseventy$&$7.28\pm3.60$&$-0.164\pm0.551$&$0.661$&0.9560\\
&0.0&$0.929\pm0.114$&$4.77$&0.4446\\
\hline
\multicolumn{5}{c}{AII}\\ \hline
$2.0\Mpchseventy$&$0.610\pm0.198$&$0.053\pm0.032$&$7.20$&0.1257\\
&0.0&$0.150\pm0.007$&$16.7$&0.0051\\
$1.0\Mpchseventy$&$1.37\pm0.353$&$0.072\pm0.057$&$11.7$&0.0197\\
&0.0&$0.288\pm0.012$&$26.9$&$<0.0001$\\
$0.5\Mpchseventy$&$3.50\pm0.743$&$-3.32\pm0.117$&$8.90$&0.0636\\
&0.0&$0.537\pm0.024$&$31.1$&$<0.0001$\\
$0.15\Mpchseventy$&$10.8\pm2.98$&$-0.332\pm0.451$&$2.71$&0.6075\\
&0.0&$1.27\pm0.087$&$15.9$&0.0071\\
\end{tabular}
\end{minipage}
\end{table}

\begin{table}\centering \begin{minipage}{140mm}
\caption{Details for data used in Figure~\ref{scale_qso_newMzcompare}. \label{table_Fig9Details}}
\begin{tabular}{c | c c c | c c c}\hline\hline
\multicolumn{1}{c}{Redshift}&\multicolumn{3}{c}{$-24.0<M_{i}\leq-23.0$}&\multicolumn{3}{c}{$-23.0<M_{i}\leq-22.0$}\\ 
Range&\#&$\overline{M_{i}}$&$\overline{z}$&\#&$\overline{M_{i}}$&$\overline{z}$\\ \hline
$0.15<z\leq0.3$&82&$-23.38$&0.241&195&$-22.45$&0.248\\
$0.3<z\leq0.4$&233&$-23.40$&0.361&472&$-22.51$&0.355\\
$0.4<z\leq0.5$&609&$-23.38$&0.457&507&$-22.68$&0.447
\end{tabular} \end{minipage}
\end{table}

\clearpage

\chapter{The Dependence on [OIII] Luminosity}\label{OIIIlumchapter}

\section{OVERVIEW}
In the previous chapter, we discussed that broad-band magnitudes were not a good proxy for the nuclear luminosity of AGNs and quasars, especially for lower-luminosity AGNs, due to possible contamination by stellar light from the host galaxy.  In this section, we use the observed luminosity of the [OIII]$\lambda$5007 emission line ($\LOIII$) to better quantify the AGN power \citep{Kauffmann, Heckman2004, Haoa}.  The [OIII]$\lambda$5007 line arises from the narrow-line region of the AGN \citep{Reyes, Li2008} and is negligibly effected by emission due to star-formation \citep{Kauffmann}.  Because the line will be present in both Type I and Type II AGNs and quasars, we can more effectively compare the nuclear luminosity of the different types of quasars and AGNs.  

\begin{figure}
\plotone{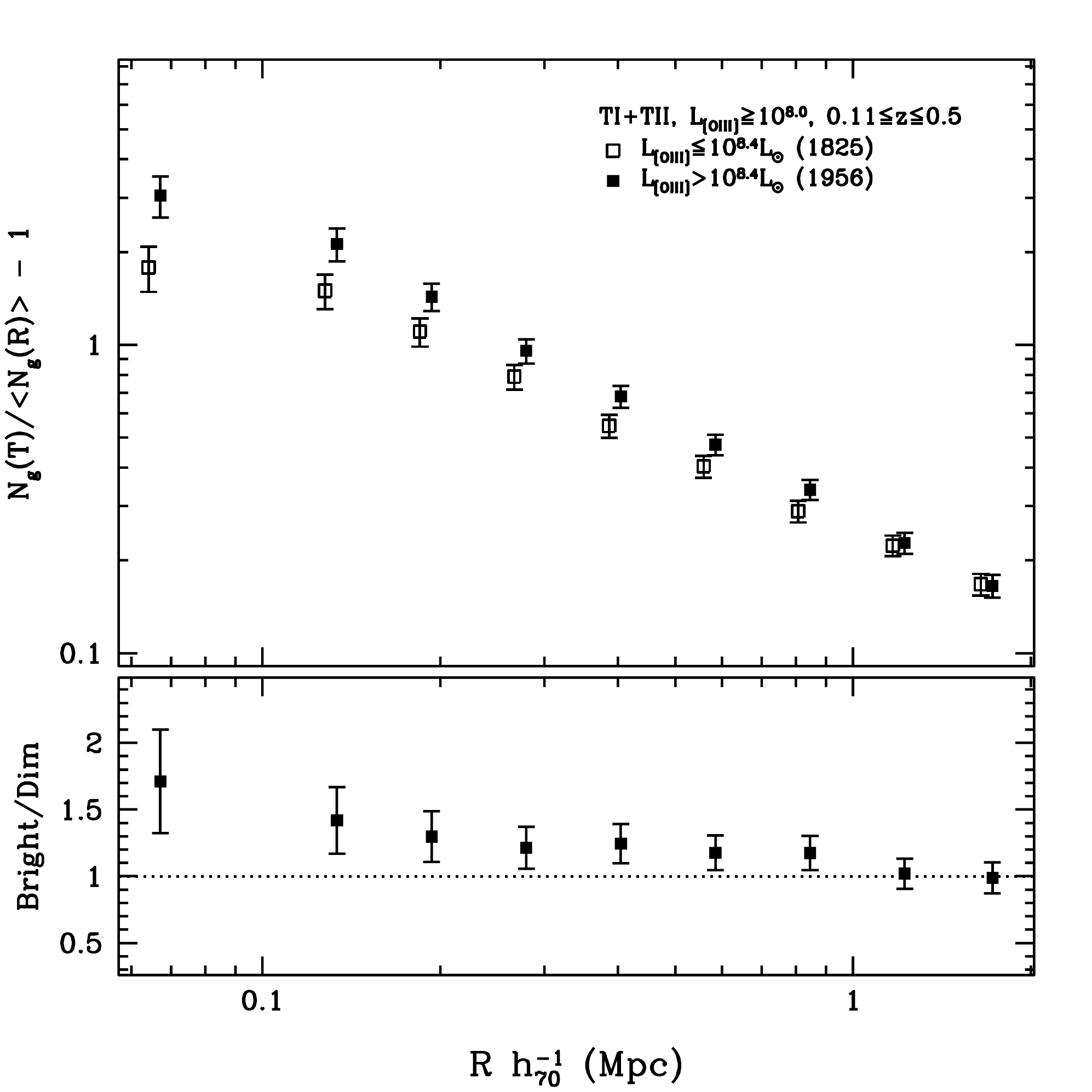}
\caption{\emph{Upper panel:} Mean cumulative environment overdensity vs. scale and $\LOIII$ for TI+TII targets with log($\LOIII$/$L_{\odot}$)$\geq8.0$.  \emph{Lower panel:} Ratio of high luminosity AGN environment overdensity to low luminosity AGN environment overdensity.
\label{scale_OIII_2bins_OIIIvollim8_extcorr_remdup}}
\end{figure}

In Figure~\ref{scale_OIII_2bins_OIIIvollim8_extcorr_remdup}, we limit the combined $0.11\leq z\leq0.5$ AGN sample to log($\LOIII$/$L_\odot$)$\geq8.0$ to make one large, volume-limited sample of $3,781$ sources.  We divide the sample into two luminosity bins at log($\LOIII$/$L_\odot$)$\geq8.4$, which gives an approximately equal number of objects in each luminosity bin.  It is clear that at scales $R\leq1.0\Mpchseventy$, targets with higher [OIII] luminosity are located in environments more overdense than targets with lower [OIII] luminosity (upper panel).  Table~\ref{table_overdensityratios_LOIII} summarizes the relative environment overdensity for bright and dim targets at $R\approx500\kpchseventy$.  At a scale of $R\approx500\kpchseventy$, the brighter AGNs reside in environments that have overdensity 1.24 times the overdensity of the dimmer AGNs, with a significance of $1.7\sigma$; at $R\approx2\Mpchseventy$, the ratio is $0.99$ with significance $<1\sigma$, and at $R\approx150\kpchseventy$, the ratio is 1.4 with significance $1.7\sigma$.  From this we see evidence that the ratio of bright AGN environment overdensity to dim AGN environment overdensity is dependent on scale.  
Throughout this chapter, we will explore in greater detail the relationship between environment overdensity and [OIII] luminosity.  

\section{TARGET TYPE}\label{OIIIlumchapter_typesection}
We divide the $\LOIII$ volume-limited sample into Type I and Type II sources (recall that an object is characterized as Type I or Type II based on the width of its emission lines).  There are 3,229 Type I (TI) sources and 552 Type II (TII) sources with $0.11\leq z\leq0.5$ and log($\LOIII$/$L_{\odot}$)$\geq8.0$.  We show overdensity vs. scale and $\LOIII$ for TIs and TIIs in Figure~\ref{scale_OIIITypeI_2bins_OIIIvollim8_extcorr_remdup} and~\ref{scale_OIIITypeII_2bins_OIIIvollim8_extcorr_remdup}, respectively.    
It is noteworthy that the scale dependency in the ratio of overdensities (which can also be interpreted as a measurement of the relative bias) remains evident for the TIs,  but from Figure~\ref{scale_OIIITypeII_2bins_OIIIvollim8_extcorr_remdup}, it appears that the scale dependency is not present for the TIIs. 

We quantify the scale dependency of the overdensity-$\LOIII$ relationship by calculating a linear least-squares fit to cumulative overdensity vs. $\LOIII$ at several fixed scales (see Figure~\ref{vsOIII_combined_4bins_OIIIvollim8_extcorr_remdup}).  The fit parameters for overdensity vs. $\LOIII$ at $150\kpchseventy$, $500\kpchseventy$, $1\Mpchseventy$, and $2\Mpchseventy$ are summarized in Table~\ref{table_overdensityVSLOIII_4scales}.   As expected, there is an increase in slope at smaller scales for the TI target sample.  Though less dramatic than for TIs, we still see evidence for a slight scale dependency of TII environment overdensity.   


The log($\LOIII$/$L_{\odot}$)$\geq8.0$ volume limit is a fairly bright one and removes the majority of our lower-luminosity AGNs (AI and AIIs) from consideration.  If we instead select targets with $z\leq0.28$ (see dividing line in Figure~\ref{zhistogramOIIIvsz_types_extcorr_remdup}), we can use a lower limit of log($\LOIII$/$L_{\odot}$)$>6.75$ to create the volume-limited samples; we additionally restrict the objects in the sample to have log($\LOIII$/$L_{\odot}$)$\leq8.0$, resulting in 894 TIs and 2,050 TIIs.  At these low redshifts and luminosities, the AI and AIIs dominate the samples.  We again calculate least-squares fits to the overdensity vs. $\LOIII$ relationship at four scales (Table~\ref{table_overdensityVSLOIII_4scales}) and find that there is very little change in slope with scale for both samples.  In fact, with the exception of the $R=150\kpchseventy$ measurement for TIs (where the error bars are very large and a zero-slope line would be an acceptable fit), the slope at any given scale is close to zero. 

In contrast, if we instead look at the high-luminosity (log($\LOIII$/$L_{\odot}$)$>8.0$) targets at these low redshifts, we see that the scale dependence appears again especially for the TI sources.  When we select TI and TII targets with log($\LOIII$/$L_{\odot}$)$\geq8.0$ and $z>0.28$ (4317 and 271 objects, respectively), the QI and QII objects dominate the samples.  Due to the very small number of TII objects in this sample, it is difficult to draw strong conclusions, but any scale dependency is small in the relationship between overdensity and luminosity for TIIs.  Overdensity vs. $\LOIII$ fits for this sample are presented in Table~\ref{table_overdensityVSLOIII_4scales}, and the increase in slope with decreasing scale is again seen for TIs.  Thus we can conclude that there is a scale-dependent relationship between environment overdensity and luminosity for the brightest sources, especially those classified as Type I.  

\section{TARGET REDSHIFT}
Using the $\LOIII$ volume-limited samples and without making any additional luminosity cuts, we investigate the effect of redshift on environment overdensity by plotting cumulative overdensity vs. redshift and calculating fits.  Table~\ref{table_overdensityVSredshift_4scales} gives the least-squares fitting parameters corresponding to the fits in Figure~\ref{vsredshift_combined_4bins_OIIIvollim8_extcorr_remdup} (only the fit for the combined TI and TII sample is shown in the figure).  It appears that there is some redshift evolution of overdensity, and that this evolution depends on scale, especially for TIs.  The small sample size of the TIIs gives rise to very large error bars, so we cannot draw conclusions for this sample.  

\begin{figure}
\plotone{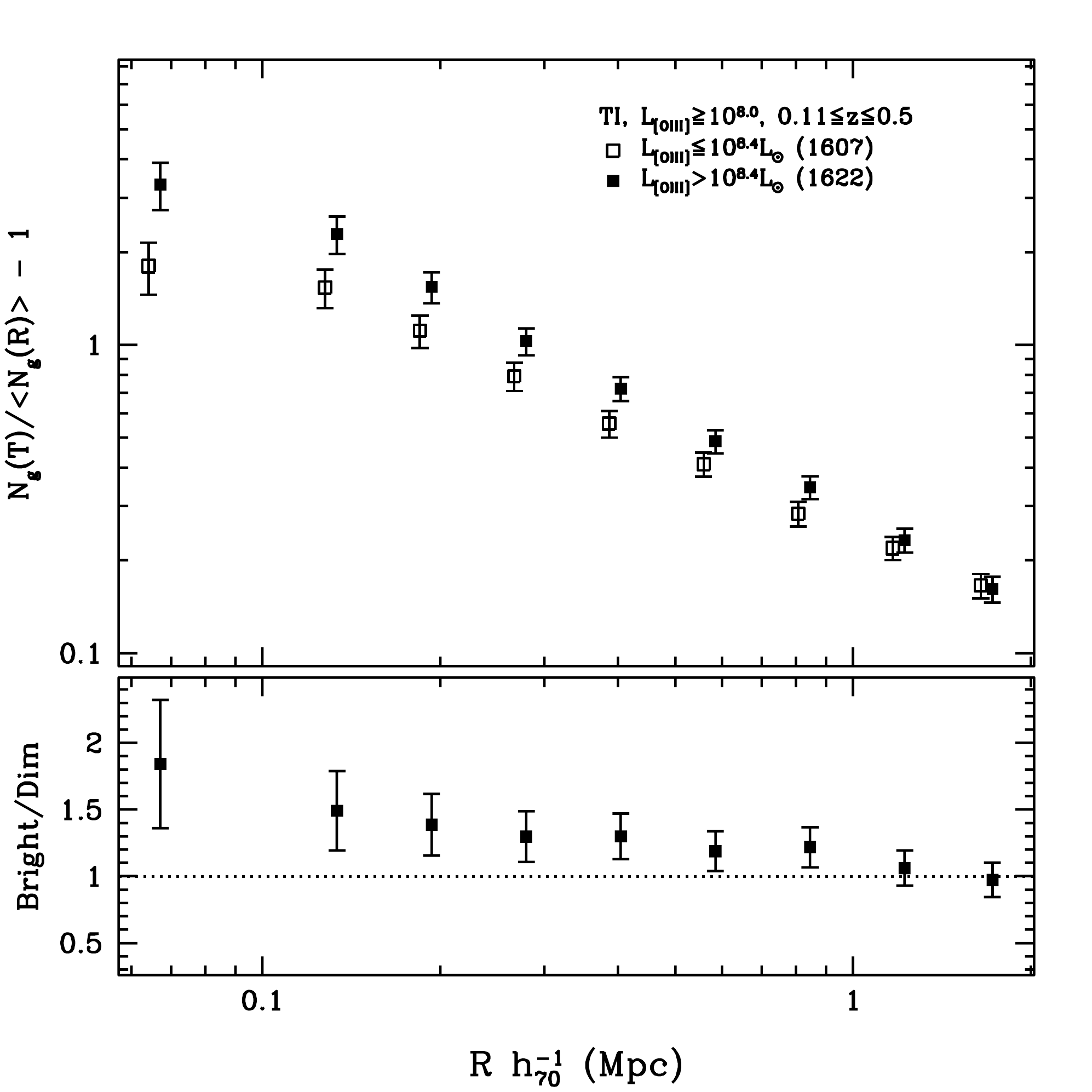}
\caption{\emph{Upper panel:} Mean cumulative environment overdensity vs. scale and $\LOIII$ for TI targets with log($\LOIII$/$L_{\odot}$)$\geq8.0$.  \emph{Lower panel:} Ratio of high luminosity AGN environment overdensity to low luminosity AGN environment overdensity.
\label{scale_OIIITypeI_2bins_OIIIvollim8_extcorr_remdup}}
\end{figure}

\begin{figure}
\plotone{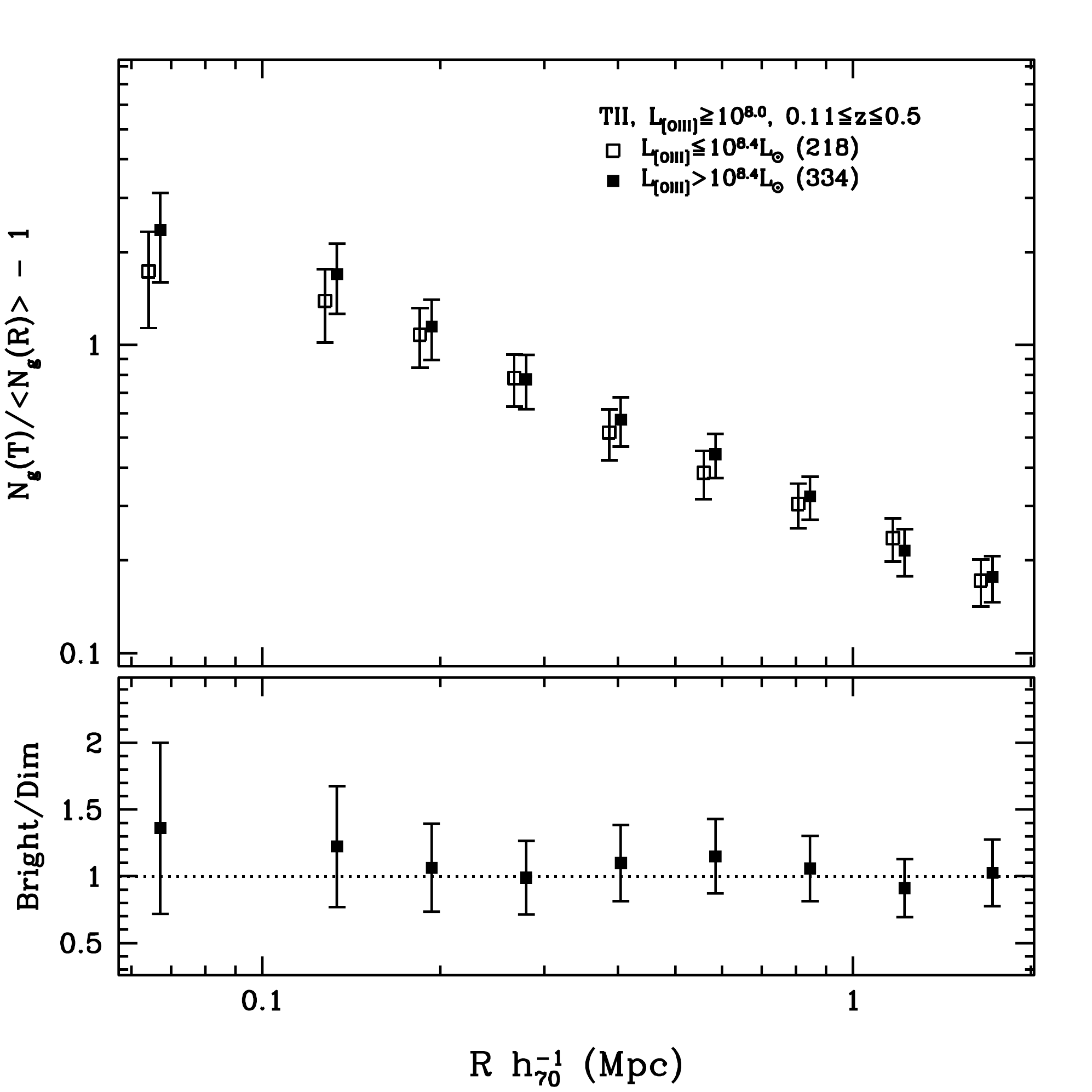}
\caption{\emph{Upper panel:} Mean cumulative environment overdensity vs. scale and $\LOIII$ for TII targets with log($\LOIII$/$L_{\odot}$)$\geq8.0$.  \emph{Lower panel:} Ratio of high luminosity AGN environment overdensity to low luminosity AGN environment overdensity.
\label{scale_OIIITypeII_2bins_OIIIvollim8_extcorr_remdup}}
\end{figure}

\begin{figure}
\plotone{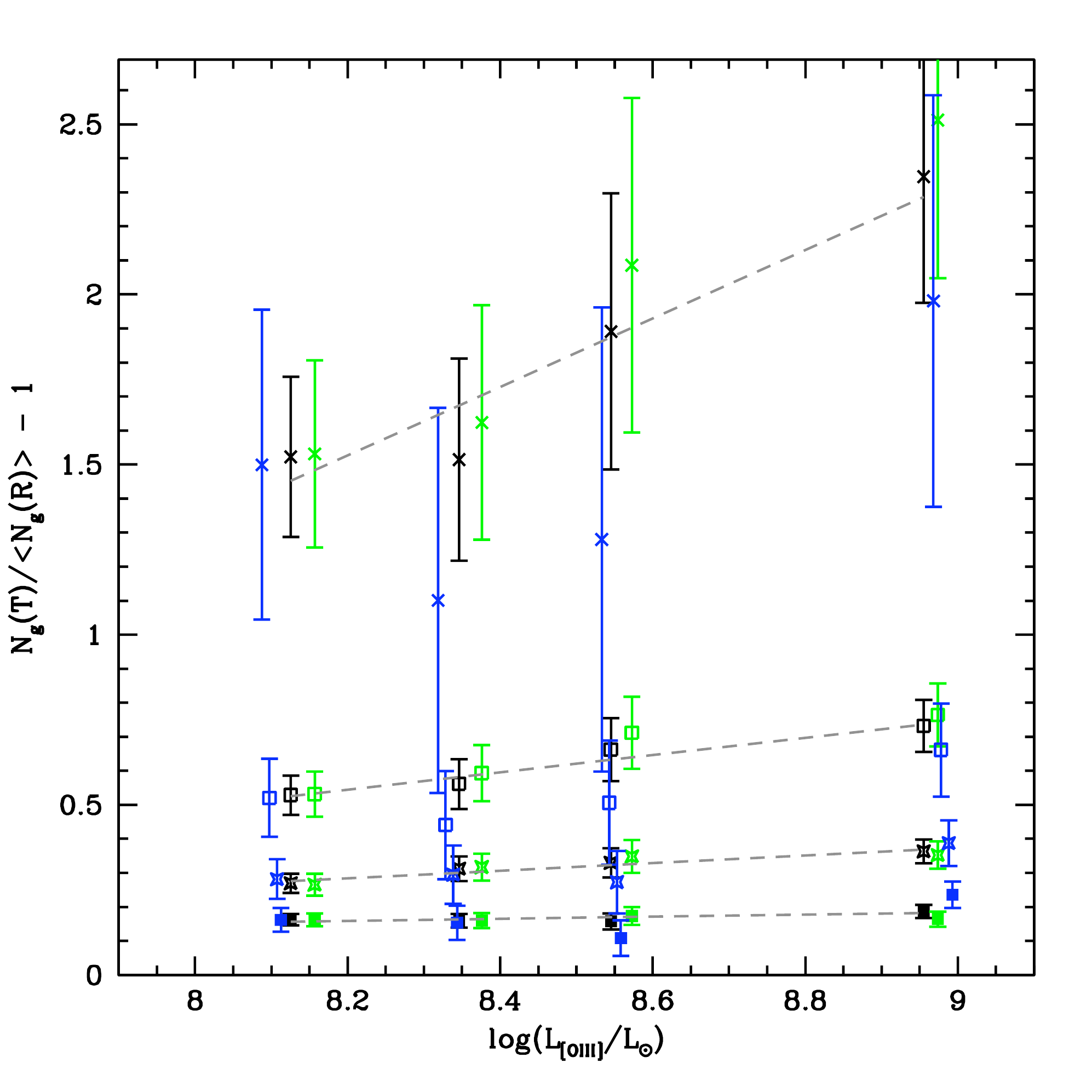}
\caption{Mean cumulative environment overdensity vs. redshift and scale for all targets (black points), Type I targets (green points), and Type II targets (blue points) with $0.11\leq z\leq0.5$ and log($\LOIII$/$L_{\odot}$)$\geq8.0$.  The solid square points show overdensity vs. redshift for $R\approx2\Mpchseventy$, starred points for $R\approx1\Mpchseventy$, open squares for $R\approx500\kpchseventy$, and crosses for $R\approx150\kpchseventy$.  The linear least-squares fits to the combined target sample data for each of these scales are given by the dashed lines; The fit parameters for each of the samples are summarized in Table~\ref{table_overdensityVSLOIII_4scales}.  Points are shifted from their average $\LOIII$ bin value (horizontal axis) by a small amount for clarity.  
\label{vsOIII_combined_4bins_OIIIvollim8_extcorr_remdup}}
\end{figure}

\begin{figure}
\plotone{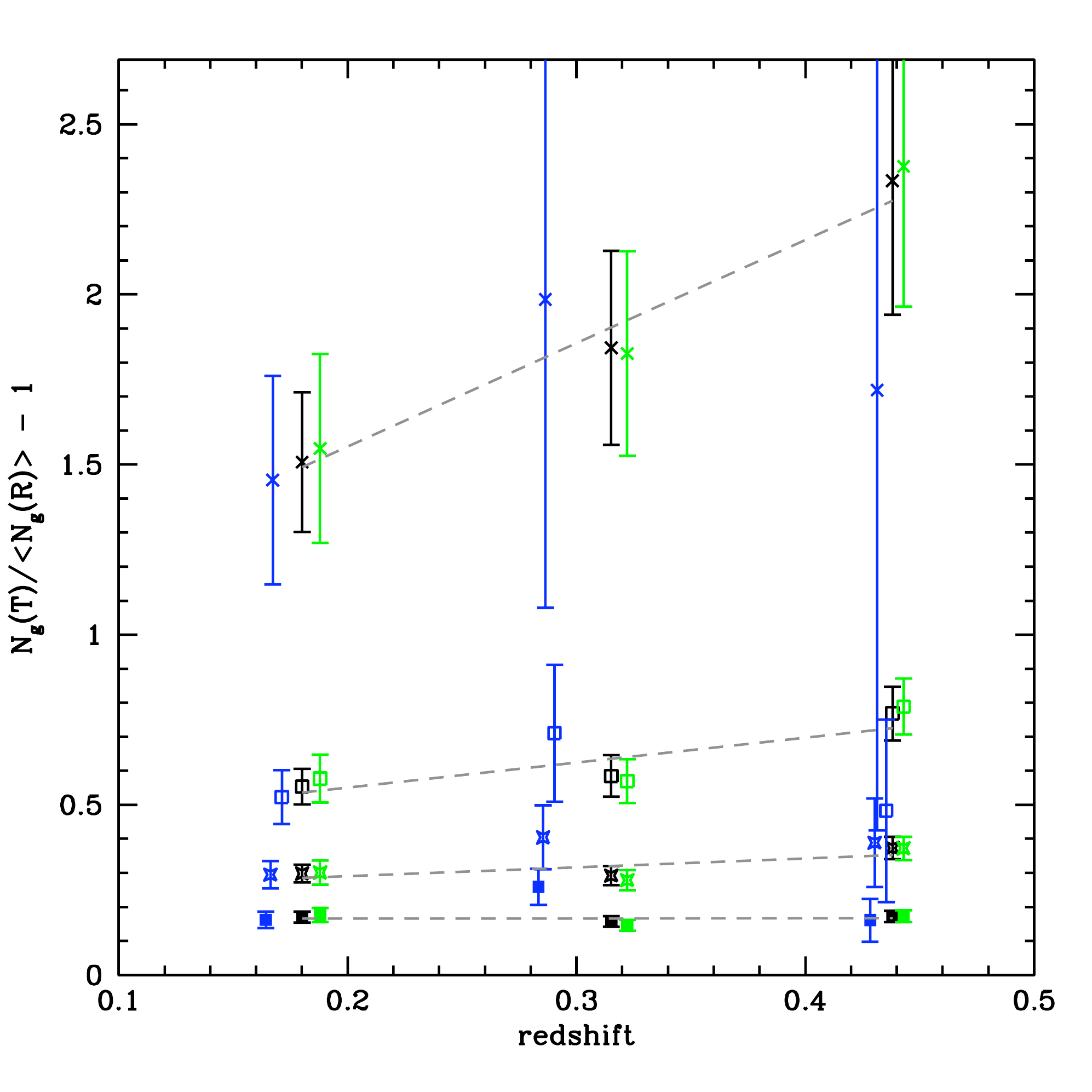}
\caption{Mean cumulative environment overdensity vs. redshift and scale for all targets (black points), Type I targets (green points), and Type II targets (blue points) with log($\LOIII$/$L_{\odot}$)$\geq8.0$.  The solid square points show overdensity vs. redshift for $R\approx2\Mpchseventy$, starred points for $R\approx1\Mpchseventy$, open squares for $R\approx500\kpchseventy$, and Xes for $R\approx150\kpchseventy$.  The linear least-squares fits to the combined target sample data for each of these scales are given by the dashed lines; The fit parameters for each of the samples are summarized in Table~\ref{table_overdensityVSredshift_4scales}.  Points are shifted from their average $z$ bin value (x-axis) by a small amount for clarity.  
\label{vsredshift_combined_4bins_OIIIvollim8_extcorr_remdup}}
\end{figure}

\begin{figure}
\plotone{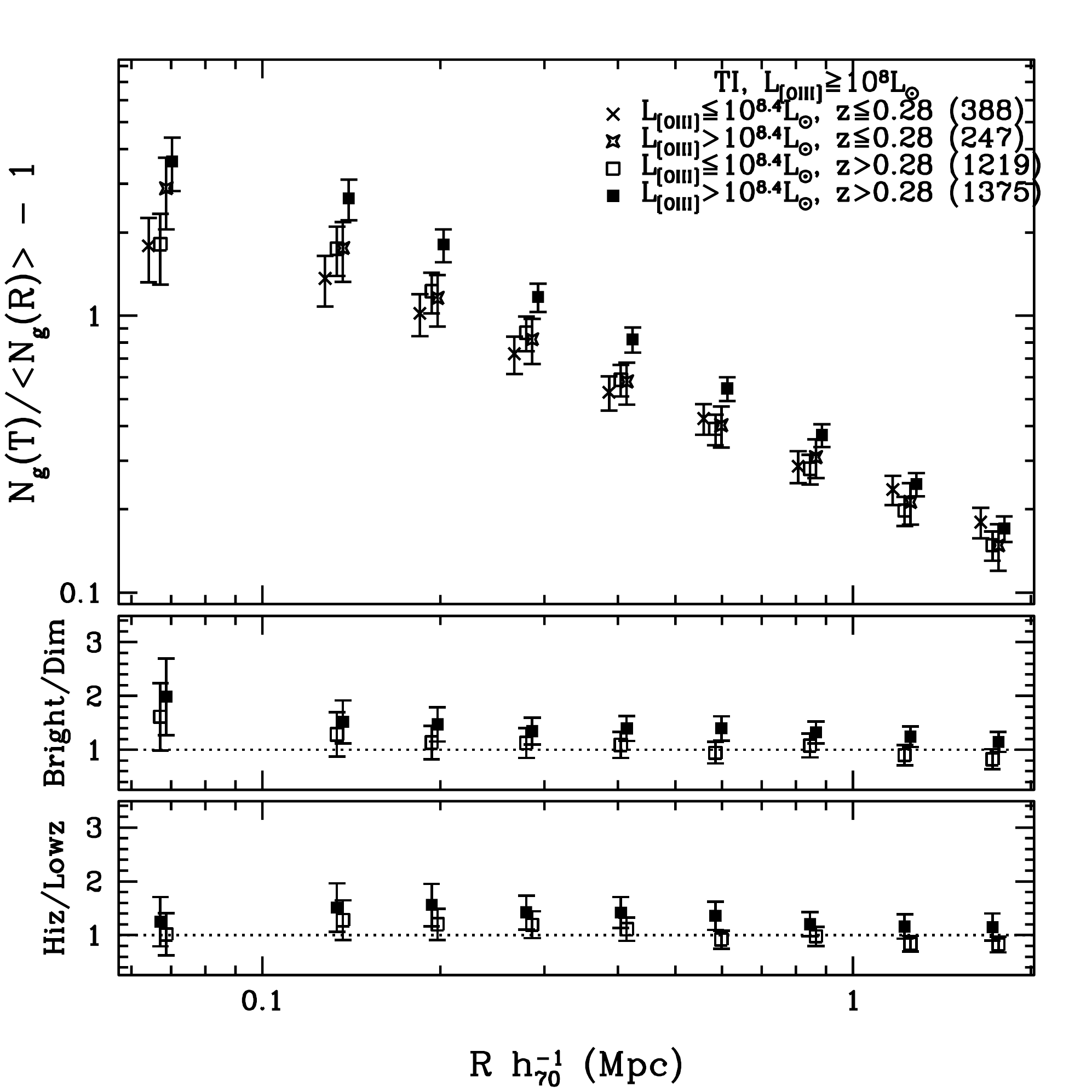}
\caption{Mean cumulative environment overdensity vs. scale, $\LOIII$, and redshift for TI targets with log($\LOIII$/$L_{\odot}$)$\geq8.0$.  \emph{Middle panel:}  ratio of environment overdensities of bright TIs to dim TIs for $z\leq0.28$ (open points), and $z>0.28$ (filled points) targets.  \emph{Lower panel:}  ratio of overdensities for high-redshift TIs to low-redshift TIs for bright (filled points) and dim (open points) targets.  
\label{scale_OIIITypeI_2bins_zevol_OIIIvollim8_extcorr_remdup}}
\end{figure}

Because we have a larger sample of TIs with the log($\LOIII$/$L_\odot$)$\geq8.0$ volume-limit, we can subdivide each luminosity bin from Figure~\ref{scale_OIIITypeI_2bins_OIIIvollim8_extcorr_remdup} into two redshift bins at $z=0.28$.  Figure~\ref{scale_OIIITypeI_2bins_zevol_OIIIvollim8_extcorr_remdup} shows the overdensity vs. scale, $\LOIII$, and redshift for TIs.  It appears that there is slight evidence for redshift evolution in the brighter objects.  The scale dependency seen in the ratio of environment overdensity of bright to dim TIs is seen in the $z>0.28$ redshift bin 
but is harder to discern in the low redshift bin, where the ratio is consistent with unity on all scales.  However, the subset of TIs with log($\LOIII$/$L_\odot$)$\geq8.0$ and $0.11\leq z\leq0.28$ is small, so we must be cautious in our conclusions.  The scale dependency at high redshifts and not low redshifts could signal mass evolution with redshift in our sample which we have not yet accounted for.  Chapter~\ref{MBHchapter} will discuss the relationship between black hole mass and QI environment overdensity.  

\section{CONCLUSIONS}
To summarize, we see that the environments of the brightest sources (i.e., those with at least log($\LOIII$/$L_{\odot}$)$\geq8.4$) have greater overdensity with greater luminosity compared to dimmer sources (with at least log($\LOIII$/$L_{\odot}$)$\geq8.0$), and that there is a slight scale dependency observed, especially for bright TIs at scales $R\leq150\kpchseventy$.  Since at $\sim$Mpc scales we are seeing approximately equal environment overdensity for bright and dim TIs, we can conclude that the halo mass is not dependent on luminosity.  However, on smaller scales, higher luminosity sources have higher overdensity, implying that the halos in which they reside are occupied differently than the lower-luminosity (but still log($\LOIII$/$L_{\odot}$)$\geq8.0$)) sources.  

The weak or absent luminosity dependence of environment overdensity on $\gtrsim$Mpc scales provides support for the results of simulations by \citet{Lidz}, which showed that galaxy merger models predict that there is a weak dependency of clustering of quasars on luminosity at larger scales.  The luminosity independence of quasar clustering has implications for understanding the lifetime and luminosity ``cycle" of quasars \citep[e.g.,][]{Hopkins2005a, AdelbergerSteidel, Lidz, Myers2006}. While the brightest observed quasars are likely radiating at their peak, the dimmer end of the quasar luminosity function is a combination of quasars that are building up to their peak luminosity (and possibly obscured by the material that powers them) or fading away as the material available for accretion is ejected by outflows from the central engine.  Thus, bright and dim sources can have similar clustering at larger scales because halo mass is not necessarily dependent the luminosity, which changes over the lifetime of the quasar with accretion efficiency.  
However, at smaller scales, we measure a higher galaxy overdensity in the environments of bright sources compared to dimmer sources which increases with decreasing scale.  The scale-dependency of the environments on these small scales is tell-tale evidence for the merger origin of these bright AGNs \citep[e.g.,][]{Djorgovski, Hennawi, Hopkins2008, Myers07b}, as mergers are expected to occur with greater frequency in regions with higher galaxy density \citep[e.g.,][]{LaceyCole}.  

For AGNs with lower $\LOIII$ (i.e., $6.75\leq$log($\LOIII$/$L_{\odot}$)$\leq8.0$), we do not see scale dependency or even much increase in overdensity with luminosity even at the smallest scales, which supports the predictions for less biased environments of low-luminosity AGNs in a secular-fueling model \citep[e.g.,][]{HopkinsHernquist, Hopkins2008}.

\begin{table} \begin{minipage}{140mm}
\caption[Overdensity ratios: $\LOIII$]{Overdensity ratios comparing environments of bright (log($\LOIII$/$L_\odot$)$>8.4$) and dim ($8.0\leq$log($\LOIII$/$L_\odot$)$\leq8.4$) targets at scale $R\approx500\kpchseventy$.  \label{table_overdensityratios_LOIII}}
\begin{tabular}{c c c}\hline\hline
Sample&bright/dim&significance\\ \hline
TI+TII&$1.24\pm0.147$&$1.66\sigma$\\
TI&$1.30\pm0.172$&$1.74\sigma$\\
TI, $0.11\leq z\leq0.28$&$1.09\pm0.244$&$<1\sigma$\\
TI, $0.28<z\leq0.5$&$1.40\pm0.232$&$1.71\sigma$\\
TII&$1.10\pm0.287$&$<1\sigma$\\
QI&$1.25\pm0.179$&$1.41\sigma$\\
QI, $0.11\leq z\leq0.28$&$0.971\pm0.269$&$<1\sigma$\\
QI, $0.28<z\leq0.5$&$1.40\pm0.232$&$1.71\sigma$\\
\end{tabular} \end{minipage}
\end{table}

\begin{landscape}
\begin{table} 
\begin{minipage}{8in}
\caption[Fit parameters for overdensity vs. $\LOIII$ at four different scales]{Linear least-squares fit parameters for overdensity vs. $\LOIII$ at four different scales using all AGNs combined, TIs, and TIIs.  (See also Figure~\ref{vsOIII_combined_4bins_OIIIvollim8_extcorr_remdup}).  \label{table_overdensityVSLOIII_4scales}}
\begin{tabular}{c | c c c | c c c | c c c}\hline\hline
\multicolumn{1}{c}{scale}&\multicolumn{3}{c}{All AGNs}&\multicolumn{3}{c}{Type I AGNs}&\multicolumn{3}{c}{Type II AGNs} \\
($\Mpchseventy$)&slope&intercept&$\chi^2$&slope&intercept&$\chi^2$&slope&intercept&$\chi^2$\\ \hline
\multicolumn{10}{c}{log($\LOIII$/$L_{\odot}$)$\geq8.0$, $0.11\leq z\leq0.28$}\\ \hline
$2.0$&$0.027\pm0.012$&$-0.063\pm0.102$&$0.404$&$0.004\pm0.009$&$0.133\pm0.072$&$0.152$&$0.074\pm0.053$&$-0.462\pm0.449$&$2.18$\\
$1.0$&$0.098\pm0.019$&$-0.518\pm0.159$&$0.317$&$0.094\pm0.037$&$-0.487\pm0.311$&$0.869$&$0.104\pm0.033$&$-0.572\pm0.278$&$0.286$\\
$0.5$&$0.226\pm0.035$&$-1.31\pm0.296$&$0.241$&$0.266\pm0.054$&$-1.62\pm0.458$&$0.389$&$0.155\pm0.080$&$-0.782\pm0.679$&$0.427$\\
$0.15$&$0.911\pm0.189$&$-5.95\pm0.159$&$0.332$&$1.12\pm0.182$&$-7.61\pm1.53$&$0.195$&$0.526\pm0.433$&$-2.99\pm3.67$&$0.703$\\ \hline
\multicolumn{10}{c}{$6.75<$log($\LOIII$/$L_{\odot}$)$\leq8.0$, $0.11\leq z\leq0.28$}\\ \hline
$2.0$&$0.011\pm0.012$&$0.063\pm0.091$&$10.1$&$0.042\pm0.011$&$-0.195\pm0.088$&$3.16$&$0.001\pm0.014$&$0.140\pm0.100$&$6.51$\\
$1.0$&$0.002\pm0.016$&$0.256\pm0.115$&$5.67$&$0.052\pm0.023$&$-0.155\pm0.079$&$4.17$&$-0.011\pm0.021$&$0.368\pm0.155$&$5.64$\\
$0.5$&$-0.010\pm0.031$&$0.584\pm0.230$&$5.82$&$0.082\pm0.035$&$-0.176\pm0.268$&$2.76$&$-0.030\pm0.031$&$0.758\pm0.222$&$2.99$\\
$0.15$&$0.050\pm0.108$&$0.873\pm0.798$&$5.12$&$0.292\pm0.141$&$-1.21\pm1.07$&$3.70$&$0.049\pm0.077$&$1.01\pm0.553$&$1.30$\\ \hline
\multicolumn{10}{c}{log($\LOIII$/$L_{\odot}$)$>8.0$, $0.11\leq z\leq0.28$}\\ \hline
$2.0$&$0.033\pm0.012$&$-0.107\pm0.088$&$2.04$&$0.052\pm0.017$&$-0.281\pm0.133$&$1.81$&$0.027\pm0.025$&$-0.050\pm0.183$&$4.41$\\
$1.0$&$0.037\pm0.018$&$-0.003\pm0.140$&$1.77$&$0.074\pm0.025$&$-0.332\pm0.195$&$1.33$&$0.026\pm0.025$&$0.093\pm0.185$&$1.58$\\
$0.5$&$0.045\pm0.027$&$0.174\pm0.203$&$0.955$&$0.139\pm0.054$&$-0.627\pm0.418$&$1.57$&$0.006\pm0.038$&$0.500\pm0.279$&$0.929$\\
$0.15$&$0.318\pm0.082$&$-1.14\pm0.614$&$0.545$&$0.681\pm0.116$&$-4.24\pm0.895$&$0.468$&$0.142\pm0.179$&$0.313\pm1.32$&$1.30$\\ \hline
\multicolumn{10}{c}{log($\LOIII$/$L_{\odot}$)$>8.0$, $0.28<z\leq0.5$}\\ \hline 
$2.0$&$-0.003\pm0.020$&$1.95\pm0.169$&$2.25$&$0.007\pm0.016$&$0.107\pm0.134$&$1.17$&$-0.082\pm0.027$&$0.918\pm0.238$&$0.523$\\
$1.0$&$0.040\pm0.047$&$-0.003\pm0.394$&$3.31$&$0.066\pm0.038$&$-0.232\pm0.314$&$1.73$&$-0.146\pm0.062$&$1.68\pm0.533$&$0.696$\\
$0.5$&$0.126\pm0.139$&$-0.366\pm1.16$&$5.07$&$0.221\pm0.129$&$-1.17\pm1.07$&$3.54$&$-0.400\pm0.309$&$4.17\pm2.70$&$2.82$\\
$0.15$&$0.731\pm0.426$&$-4.11\pm3.35$&$2.03$&$1.18\pm0.417$&$-7.86\pm3.43$&$1.62$&$-1.07\pm0.994$&$11.01\pm8.80$&$1.21$
\end{tabular} \end{minipage}
\end{table}
\end{landscape}

\begin{table} \begin{minipage}{140mm}
\caption[Fit parameters for overdensity vs. redshift at four different scales]{Linear least-squares fit parameters for overdensity vs. redshift at four different scales using all AGNs combined, TIs, and TIIs with log($\LOIII$/$L_{\odot}$)$\geq8.0$.  (See also Figure~\ref{vsredshift_combined_4bins_OIIIvollim8_extcorr_remdup}).  \label{table_overdensityVSredshift_4scales}}
\begin{tabular}{c | c c c | c c c | c c c}\hline\hline
\multicolumn{1}{c}{scale}&\multicolumn{3}{c}{All AGNs}&\multicolumn{3}{c}{Type I AGNs}&\multicolumn{3}{c}{Type II AGNs} \\
($\Mpchseventy$)&slope&intercept&$\chi^2$&slope&intercept&$\chi^2$&slope&intercept&$\chi^2$\\ \hline
$2.0$&$0.005$&$0.165$&$0.4714$&$0.002$&$0.162$&$1.860$&$0.143$&$0.145$&$2.551$\\
$1.0$&$0.259$&$0.239$&$1.656$&$0.278$&$0.226$&$2.431$&$0.490$&$0.213$&$0.3509$\\
$0.5$&$0.734$&$0.404$&$1.087$&$0.751$&$0.400$&$1.974$&$0.243$&$0.492$&$0.7431$\\
$0.15$&$3.03$&$0.946$&$0.0703$&$3.04$&$0.954$&$0.1613$&$1.92$&$1.13$&$0.1421$\\
\end{tabular} \end{minipage}
\end{table}

\clearpage

\clearpage

\clearpage

\clearpage

\clearpage

\chapter{The Dependence on Environment Galaxy Type}\label{envtgalchapter}
\section{OVERVIEW}
We now focus on the characteristics of the galaxies in the environment of AGNs to provide additional insight into the relationship between the AGN and its environment.  The composition of AGN environments can be related to galaxy formation models for clues about their location in or proximity to groups or clusters.  The photometric galaxy catalog from which our environment galaxy sample is derived contains a parameter $t$ that provides an estimate of the galaxy spectral type.  The type classification is based on the spectral energy distribution of the template galaxy used to estimate the photometric redshift.  Following \citet{Budavari}, we first use the separation that $t<0.3$ is an early-type (red) galaxy, and late-type (blue) galaxies have $t\geq0.3$ \citep[in Section~\ref{envtgalchapter_4envtgaltypes}, we will further subdivide the environment galaxies into][hereafter CWW, spectral types]{CWW}.   As discussed in Section~\ref{techniquechapter_redshiftlimits}, subdividing our environment galaxy sample limits the maximum redshift at which we have a reasonable probability of counting galaxies.  Thus, when we divide the environment sample into two environment galaxy types, we use a redshift range of $0.11\leq z\leq0.4$.  

\begin{figure}
\plotone{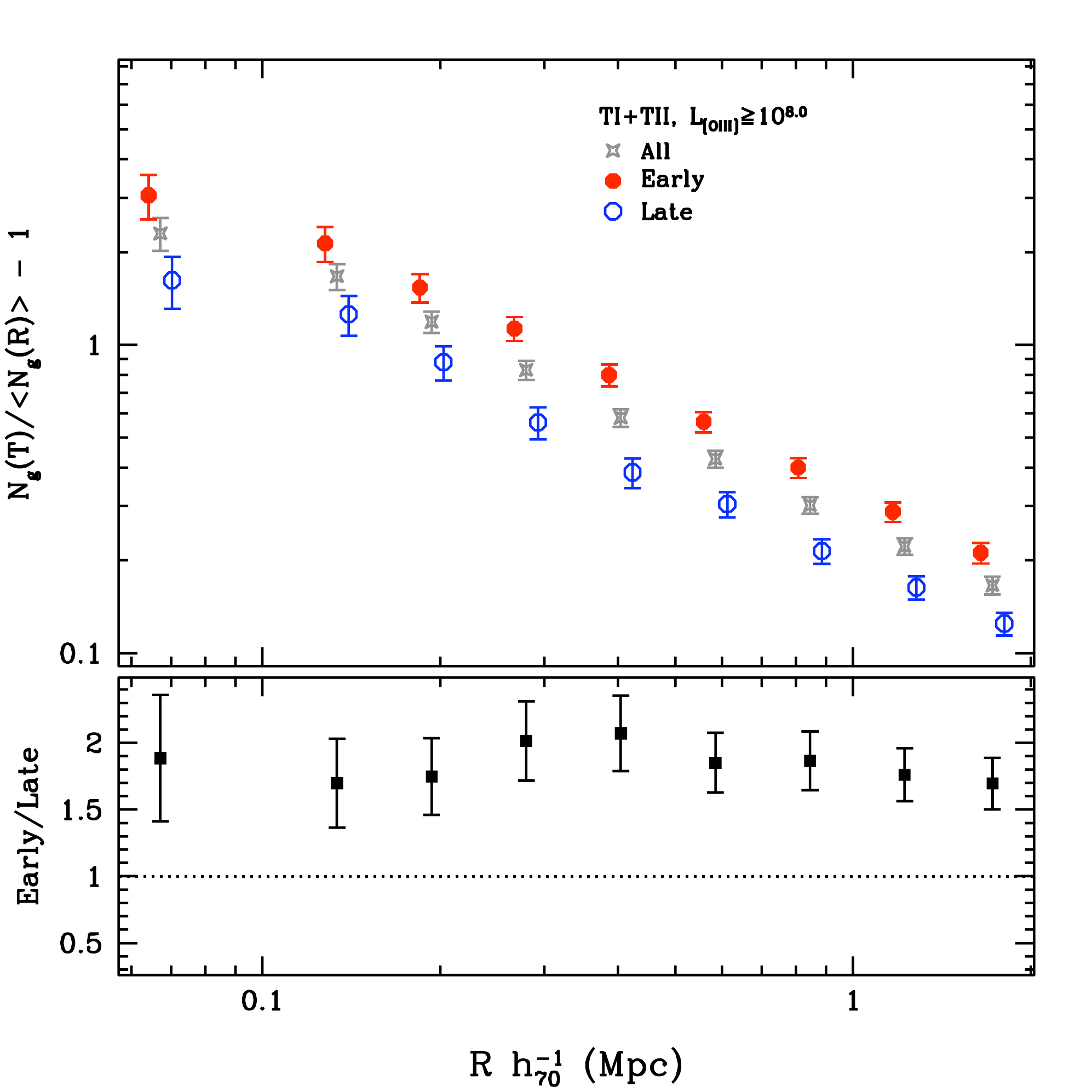}
\caption{Mean cumulative environment overdensity vs. scale and environment galaxy type for all targets with $0.11\leq z\leq0.4$ and log($\LOIII/L_{\odot}$)$\geq8.0$.  \emph{Lower panel:} Ratio of early-type environment overdensity to late-type environment overdensity. 
\label{scale_OIII_1bin_earlylate_OIIIvollim8_extcorr_remdup}}
\end{figure}

We use the same log($\LOIII$/$L_\odot$)$\geq8.0$ combined sample of AGNs in Figure~\ref{scale_OIII_1bin_earlylate_OIIIvollim8_extcorr_remdup}, where we compare the overdensity of early-type environments to the overdensity of late-type environments.  At all scales, we see that the AGNs reside in environments more densely populated with early-type environment galaxies.  In Table~\ref{table_overdensityratios_envtgaltype}, we give the ratio of early-type environment galaxy overdensity to late-type environment galaxy overdensity at a scale $R\approx500\kpchseventy$.  Throughout this chapter, we will expand on the details of the relationship between overdensity of different types of galaxies and AGN type, luminosity, and redshift.  

\section{TARGET TYPE}\label{envtgalchapter_targettype}
\begin{figure}
\plotone{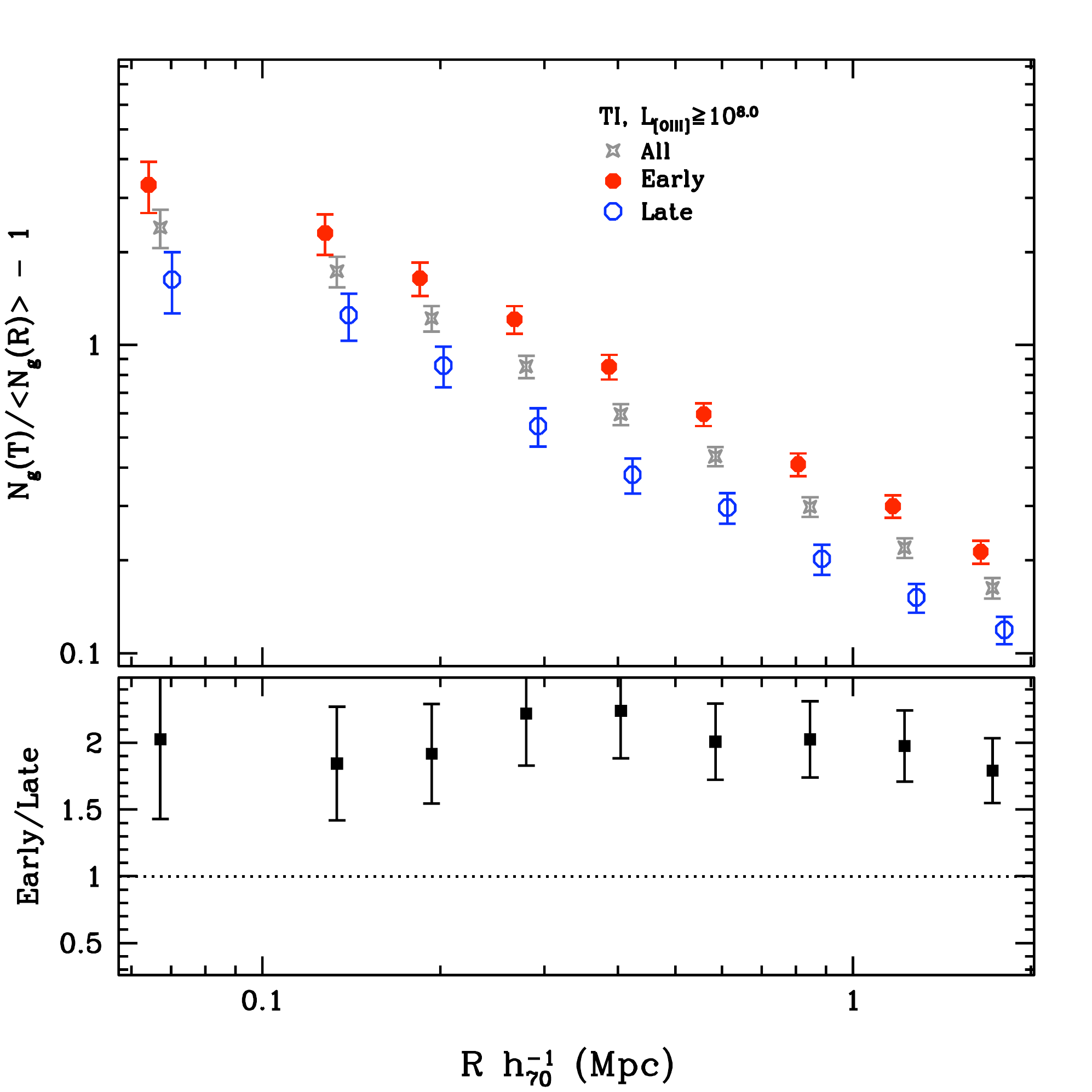}
\caption{Mean cumulative environment overdensity vs. scale and environment galaxy type for TIs with $0.11\leq z\leq0.4$ and log($\LOIII/L_{\odot}$)$\geq8.0$.  \emph{Lower panel:} Ratio of early-type environment overdensity to late-type environment overdensity. 
\label{scale_OIIITypeI_1bin_earlylate_OIIIvollim8_extcorr_remdup}}
\end{figure}

\begin{figure}
\plotone{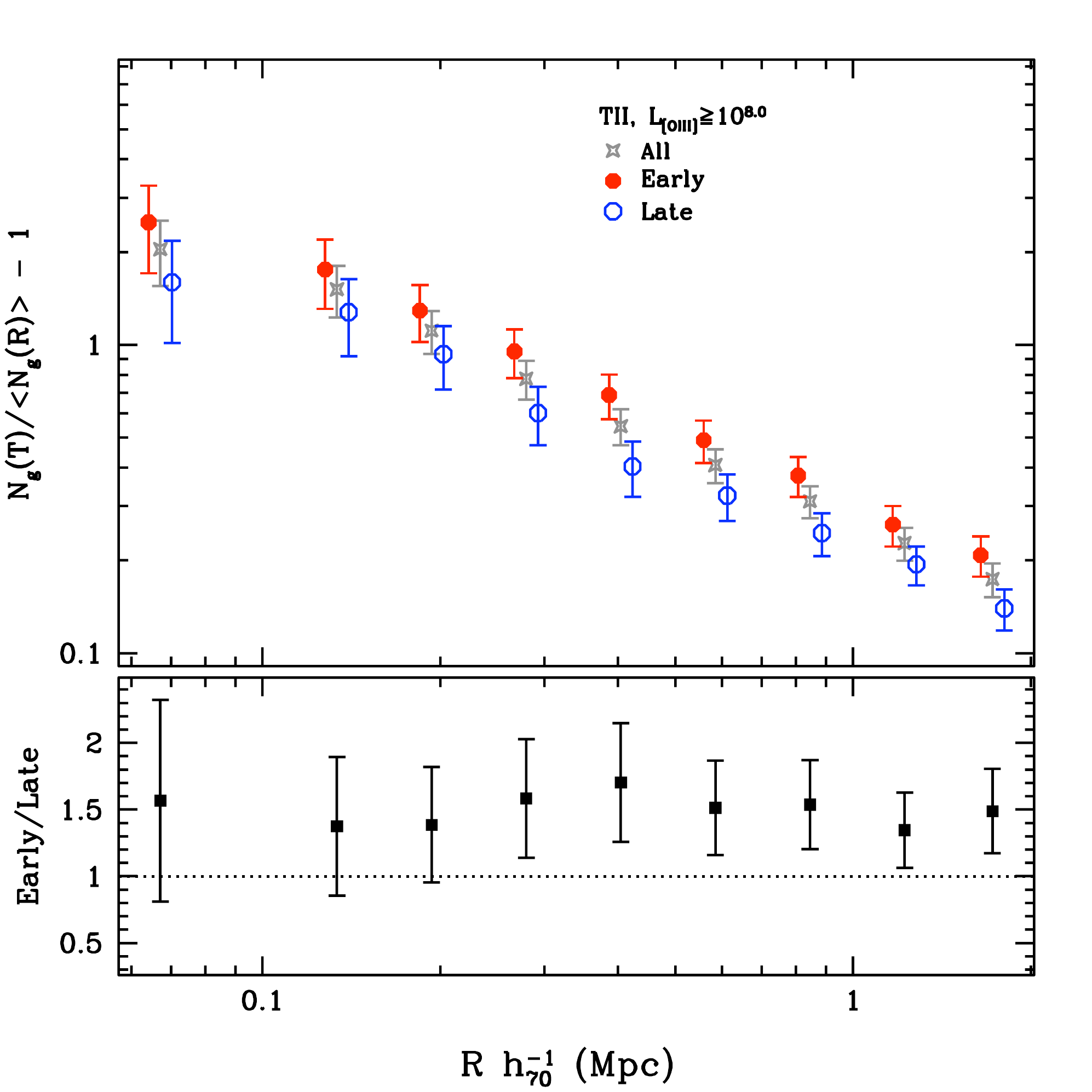}
\caption{Mean cumulative environment overdensity vs. scale and environment galaxy type for TIIs with $0.11\leq z\leq0.4$ and log($\LOIII/L_{\odot}$)$\geq8.0$.  \emph{Lower panel:} Ratio of early-type environment overdensity to late-type environment overdensity. 
\label{scale_OIIITypeII_1bin_earlylate_OIIIvollim8_extcorr_remdup}}
\end{figure}

First, we explore whether the type of the target is related to the composition of its environment.  Figures~\ref{scale_OIIITypeI_1bin_earlylate_OIIIvollim8_extcorr_remdup} and~\ref{scale_OIIITypeII_1bin_earlylate_OIIIvollim8_extcorr_remdup} show the overdensity vs. scale and environment galaxy type for the TI and TII samples with log($\LOIII/L_{\odot}$)$\geq8.0$, respectively.  
By comparing the overdensity values for the two types (see Table~\ref{table_overdensitycompare_envtgaltype}), we see that the lower early/late ratio for TIIs is due to the fact that the difference between early-type overdensity between TI and TII AGNs increases with decreasing scale:  at $1.0\Mpchseventy$, the early-type overdensity around TI (TII) targets is $0.41\pm0.04$ ($0.38\pm0.06$), and at $500\kpchseventy$, the overdensity has more than doubled for TIs ($0.85\pm0.08$), but has not quite doubled for TIIs ($0.69\pm0.11$).  The late-type overdensities show only slight differences between TI and TIIs; thus the difference in overdensity ratios is due to lower overdensities of early-type galaxies in the environments of TIIs.  

\subsection{Comparison to $\Lstar$ Galaxies}
\begin{figure}
\plotone{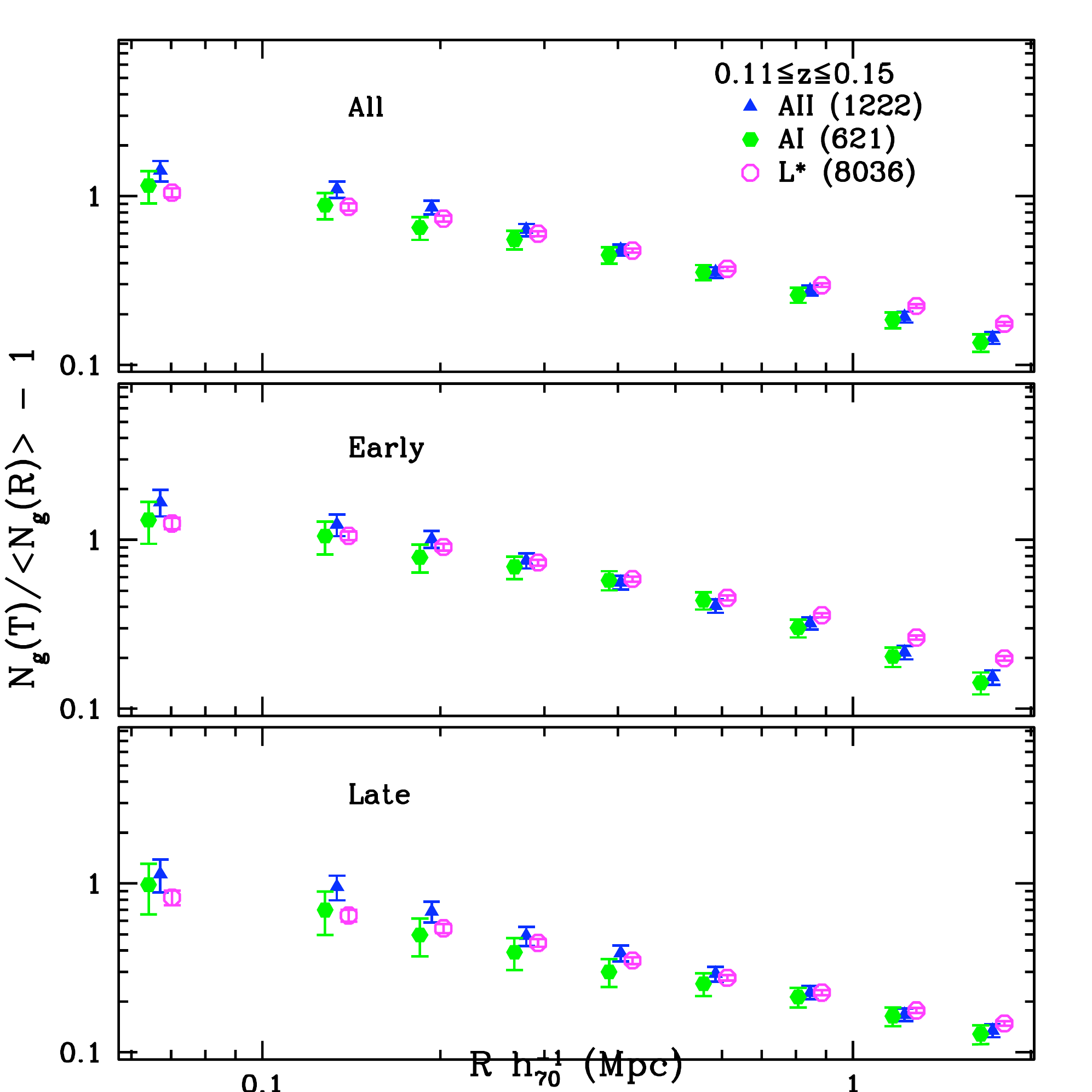}
\caption{Mean cumulative environment overdensity vs. scale for AIs, AIIs, and $\Lstar$ galaxies with all environment galaxies; and early- and late-type environment galaxies alone.  All targets have $0.11\leq z\leq0.15$, and the AGN targets have log($\LOIII/L_{\odot}$)$\geq6.5$.  See Table~\ref{table_overdensitycompare_AIAIILstar} for numerical overdensity values and comparisons.  
\label{scale_spectargs_compareLstar_11z15_remdup}}
\end{figure}

At low redshifts,  we can compare the environments of log($\LOIII/L_{\odot}$)$\geq6.5$ AIs and AIIs to the environments of $\Lstar$ galaxies:  it is in the range $0.11\leq z\leq0.15$ that the vast majority of our $\Lstar$ galaxies are located.  Figure~\ref{scale_spectargs_compareLstar_11z15_remdup} and Table~\ref{table_overdensitycompare_AIAIILstar} shows the comparison of environments using all environment galaxies, and early- and late-type environment galaxies individually.  We find that these lower-luminosity AGN environments are approximately consistent with the $\Lstar$ galaxy environments, but there is a slight excess at the smallest scales, which is seen especially the comparison of late-type environment galaxy overdensities.  The late-type excess is especially noticeable in the smallest scales of AII environments where at $150\kpchseventy$, the ratio of late-type overdensity around AIIs to $\Lstar$ galaxies is $1.45\pm0.28$ with a significance of $\sim2\sigma$.  The slight excess of gas-rich late-type galaxies in the nearby environment of AGNs compared to the $\Lstar$ galaxies could be evidence that the AGNs are affecting star formation in the nearby galaxies, or that they may accrete gas from outside their host galaxies \citep{Coldwell2003, Coldwell2006, Koulouridis}.  

\section{TARGET LUMINOSITY}
\begin{figure}
\plotone{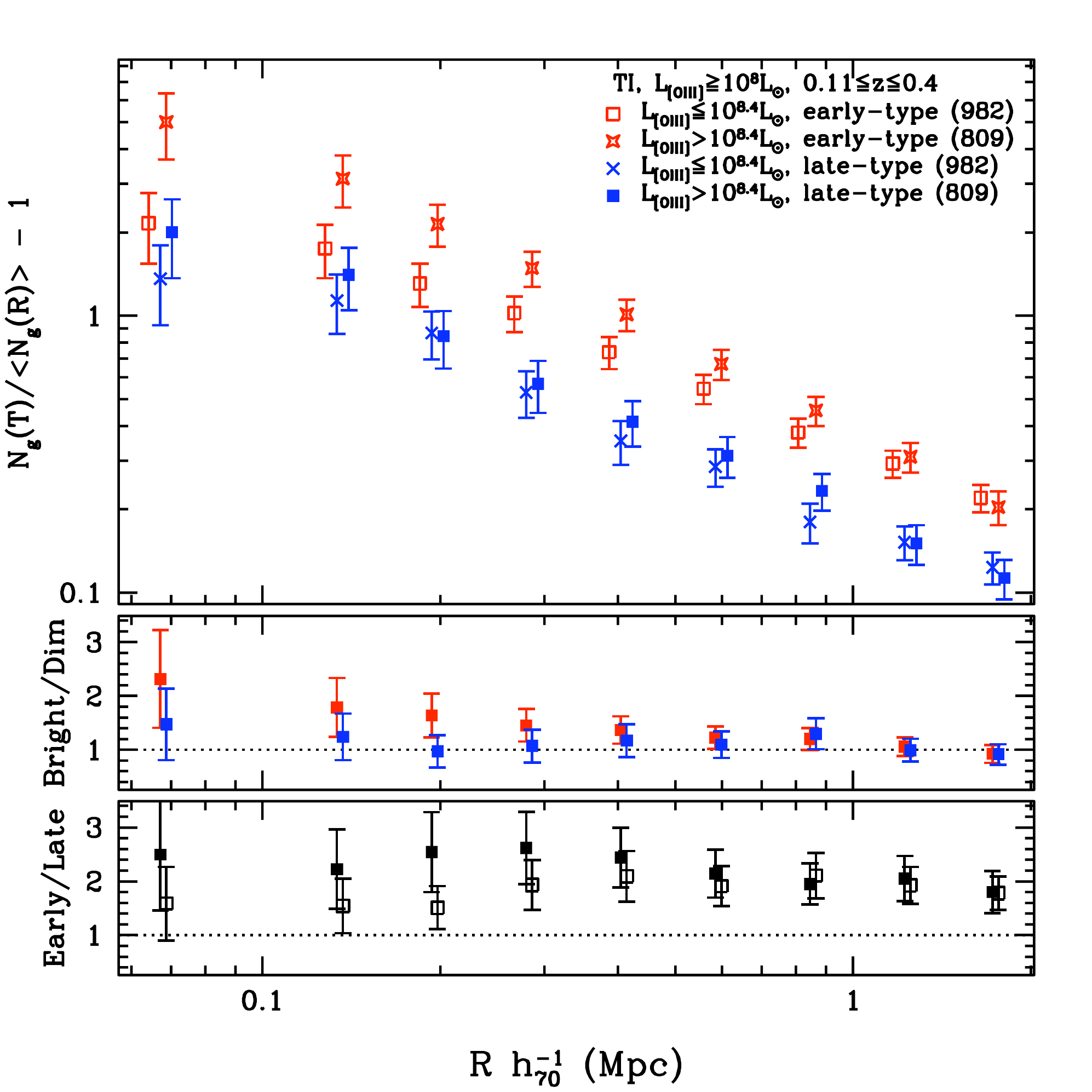}
\caption{\emph{Upper panel:} Mean cumulative environment overdensity vs. scale, $\LOIII$ and environment galaxy type for $0.11\leq z\leq0.4$ TI targets with log($\LOIII/L_{\odot}$)$\geq8.0$.  
\emph{Middle panel:} Ratio of early- or late-type environment overdensity around bright targets to that around dim targets (red, blue points respectively).  
\emph{Lower panel:} Ratio of early-type overdensity to late-type overdensity around bright (solid points) and dim (open points) targets.  
\label{scale_TI_earlylate_LOIII2bins_OIIIvollim8_extcorr_remdup}}
\end{figure}

Now that we have established that there is a difference in environment galaxy composition due to target type, we investigate the relationship between target luminosity and the types of galaxies in the environment in order to further understand the luminosity dependency we saw in Chapter~\ref{OIIIlumchapter}.  In Figure~\ref{scale_TI_earlylate_LOIII2bins_OIIIvollim8_extcorr_remdup}, we divide the ($\LOIII$ volume-limited) TI sample by luminosity at log($\LOIII/L_{\odot}$)$=8.4$ to quantify the relationship of the early- or late-type environment galaxy overdensity to the luminosity of the targets.  We observe a scale dependency in the early-type environment galaxy overdensity for bright TIs to dim TIs (red points in the middle panel of Figure~\ref{scale_TI_earlylate_LOIII2bins_OIIIvollim8_extcorr_remdup}).  However, the bright/dim overdensity ratio for the late-type environment galaxies has virtually no scale dependency.  Additionally, the ratios of early-type to late-type overdensity for bright or dim TIs (lower panel of Figure~\ref{scale_TI_earlylate_LOIII2bins_OIIIvollim8_extcorr_remdup}) have no notable change with scale, and the ratios at scale $R\approx500\kpchseventy$ are shown in Table~\ref{table_overdensityratios_envtgaltype}.  

\begin{figure}
\plotone{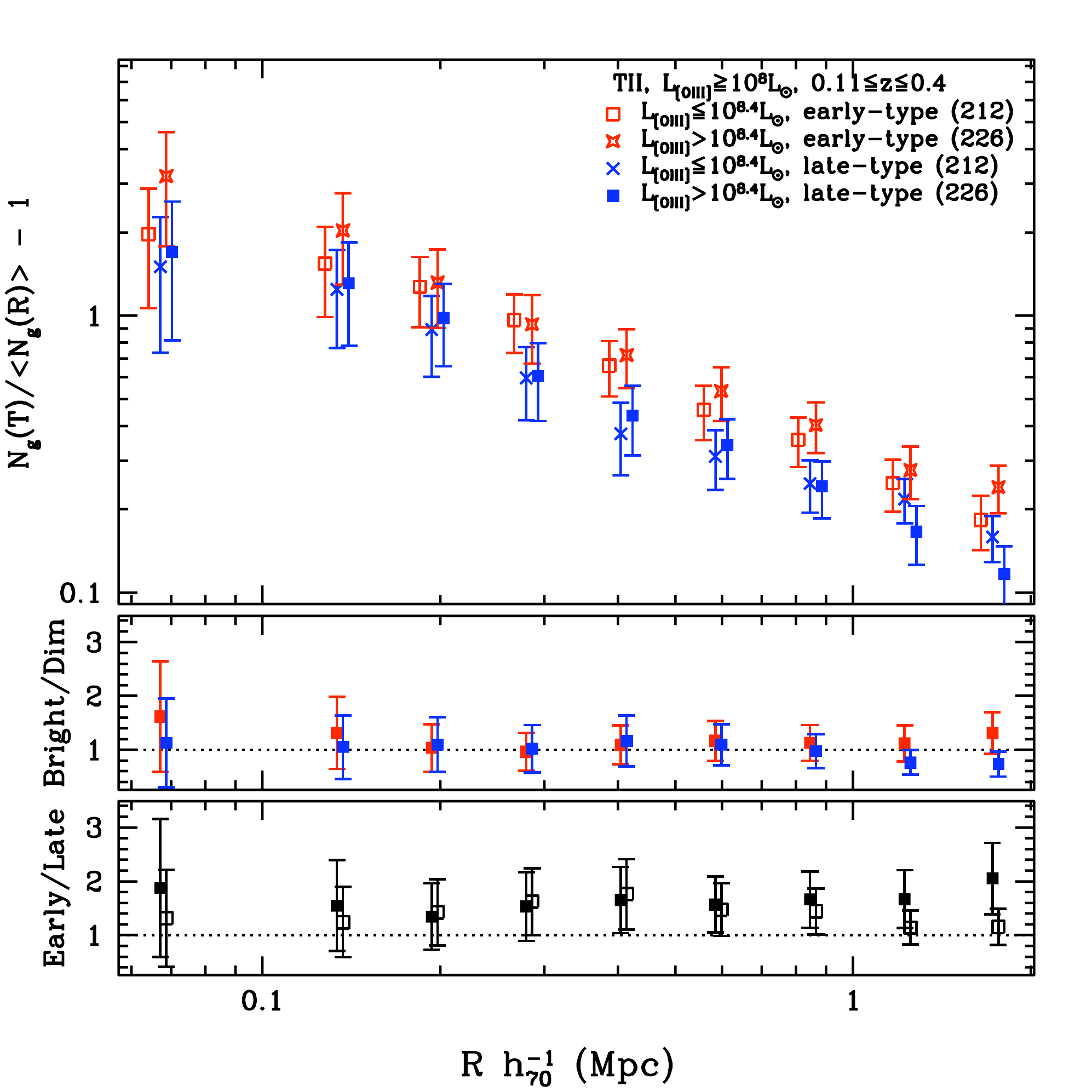}
\caption{\emph{Upper panel:} Mean cumulative environment overdensity vs. scale, $\LOIII$ and environment galaxy type for $0.11\leq z\leq0.4$ TII targets with log($\LOIII/L_{\odot}$)$\geq8.0$.  
\emph{Middle panel:} Ratio of early- or late-type environment overdensity around bright targets to that around dim targets (red, blue points respectively).  
\emph{Lower panel:} Ratio of early-type overdensity to late-type overdensity around bright (solid points) and dim (open points) targets.  
\label{scale_TII_earlylate_LOIII2bins_OIIIvollim8_extcorr_remdup}}
\end{figure}

Though the overdensity errors are larger due to the smaller sample size, we make the same luminosity division for TIIs in Figure~\ref{scale_TII_earlylate_LOIII2bins_OIIIvollim8_extcorr_remdup}.  For these targets, there are no obvious scale dependencies in any of the ratios, and in fact, the early- to late-type overdensity ratios are virtually the same for bright and dim TII targets.  

We compare the early- and late-type overdensity values for bright and dim TIs and TIIs in Table~\ref{table_overdensitycompare_brightdim_envtgaltype}.  For both the bright and dim samples, TIs and TIIs have very similar late-type galaxy overdensities in their environments.  However, the bright TIs have increasing early-type galaxy overdensities with decreasing scale.  Thus the difference that we have observed (e.g., in Chapter~\ref{OIIIlumchapter}) in the TI and TII environments is most likely caused by this increase in early-type overdensity around bright targets at smaller scales.  

\section{TARGET REDSHIFT}
\begin{figure}
\plotone{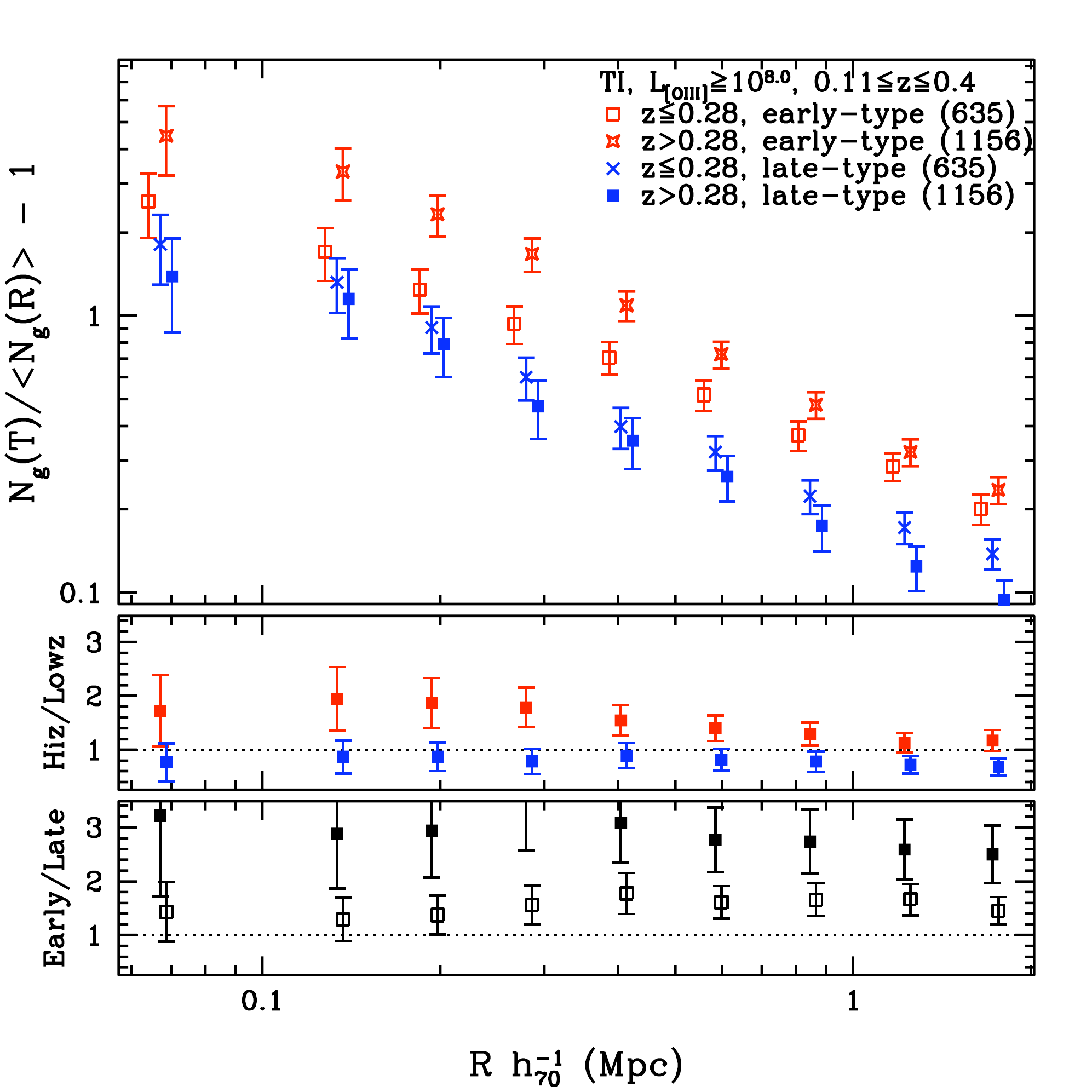}
\caption{Mean cumulative environment overdensity vs. scale, redshift and environment galaxy type for TI targets with log($\LOIII/L_{\odot}$)$\geq8.0$.  \emph{Middle panel:}  ratio of high redshift environment overdensity to low-redshift environment overdensity of early-type galaxies (red points) and late-type galaxies (blue points).  \emph{Lower panel:}  ratio of early-type galaxy overdensity to late-type galaxy overdensity for high redshift (solid points) and low redshift (open points) TIs.  
\label{scale_TI_earlylate_zevol_OIIIvollim8_extcorr_remdup}}
\end{figure}

Next, we examine whether there are redshift effects influencing the composition of TI source environments, recalling that in Chapter~\ref{paper1chapter} we saw some evidence for redshift evolution in QI environments, but that the redshift effects were minor compared to those of luminosity.  In Figure~\ref{scale_TI_earlylate_zevol_OIIIvollim8_extcorr_remdup}, we divide the TI sample at $z=0.28$ (see Figure~\ref{zhistogramOIIIvsz_types_extcorr_remdup}) and find that there is evidence for redshift evolution in the early-type overdensity in the environment of TIs, especially on the smallest scales (red points in the middle panel).  In contrast, there is virtually no redshift evolution in the late-type overdensity.  Redshift evolution in the background galaxy sample is marginalized because we are comparing counts around targets to counts around random at the same redshift with the $\delta z$ cut.  Therefore, the variation is in the composition of the small-scale environment of AGNs with redshift, specifically that there is a larger percent of early-type galaxies in higher redshift environments.

\begin{figure}
\plotone{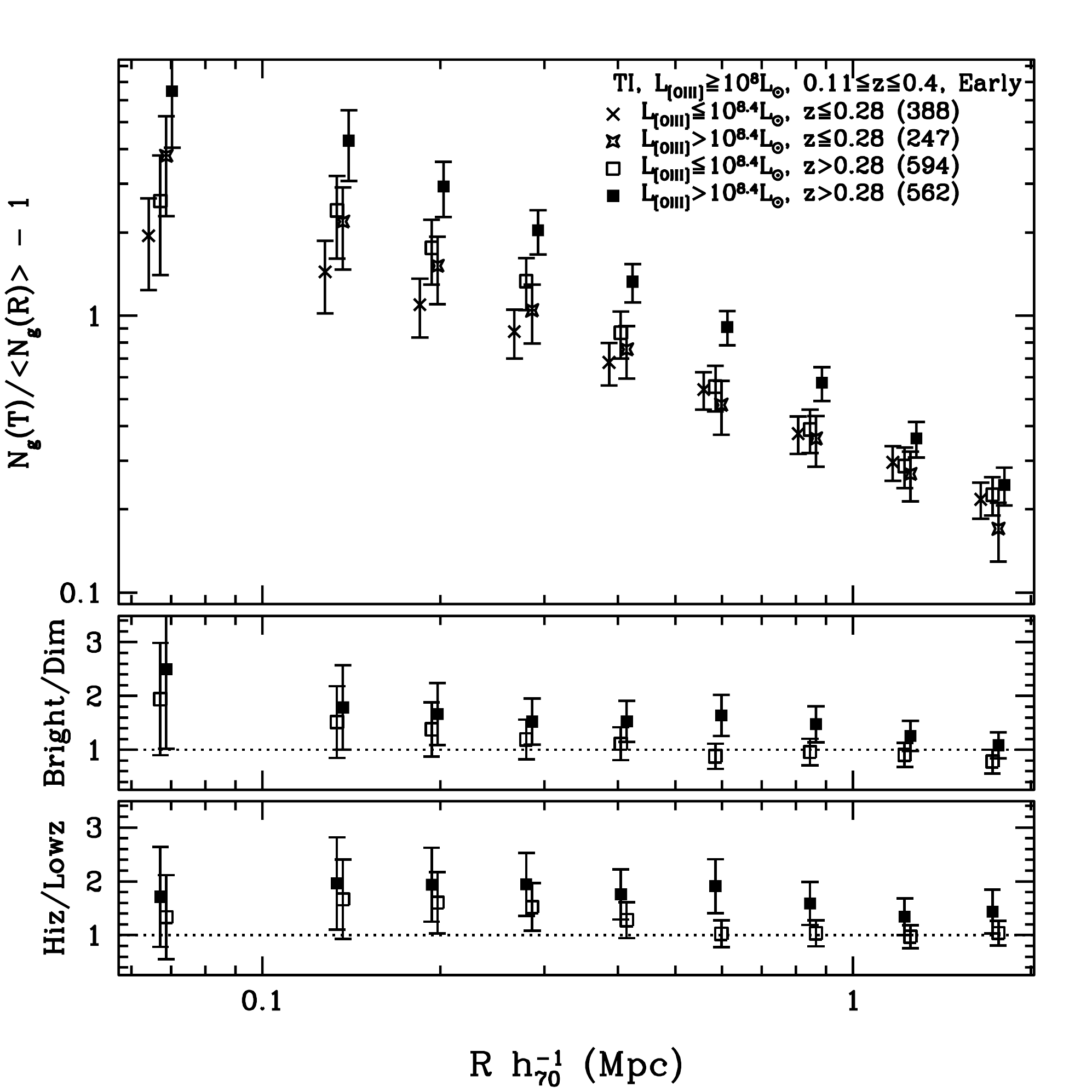}
\caption{Mean cumulative environment overdensity of early-type galaxies vs. scale, $\LOIII$, and redshift for TI targets with log($\LOIII$/$L_{\odot}$)$\geq8.0$.  \emph{Middle panel:}  ratio of environment overdensities of bright TIs to dim TIs for $z\leq0.28$ (open points), and $z>0.28$ (filled points) targets.  \emph{Lower panel:}  ratio of overdensities for high-redshift TIs to low-redshift TIs for bright (filled points) and dim (open points) targets.  
\label{scale_OIIITypeI_2bins_zevol_early_OIIIvollim8_extcorr_remdup}}
\end{figure}

\begin{figure}
\plotone{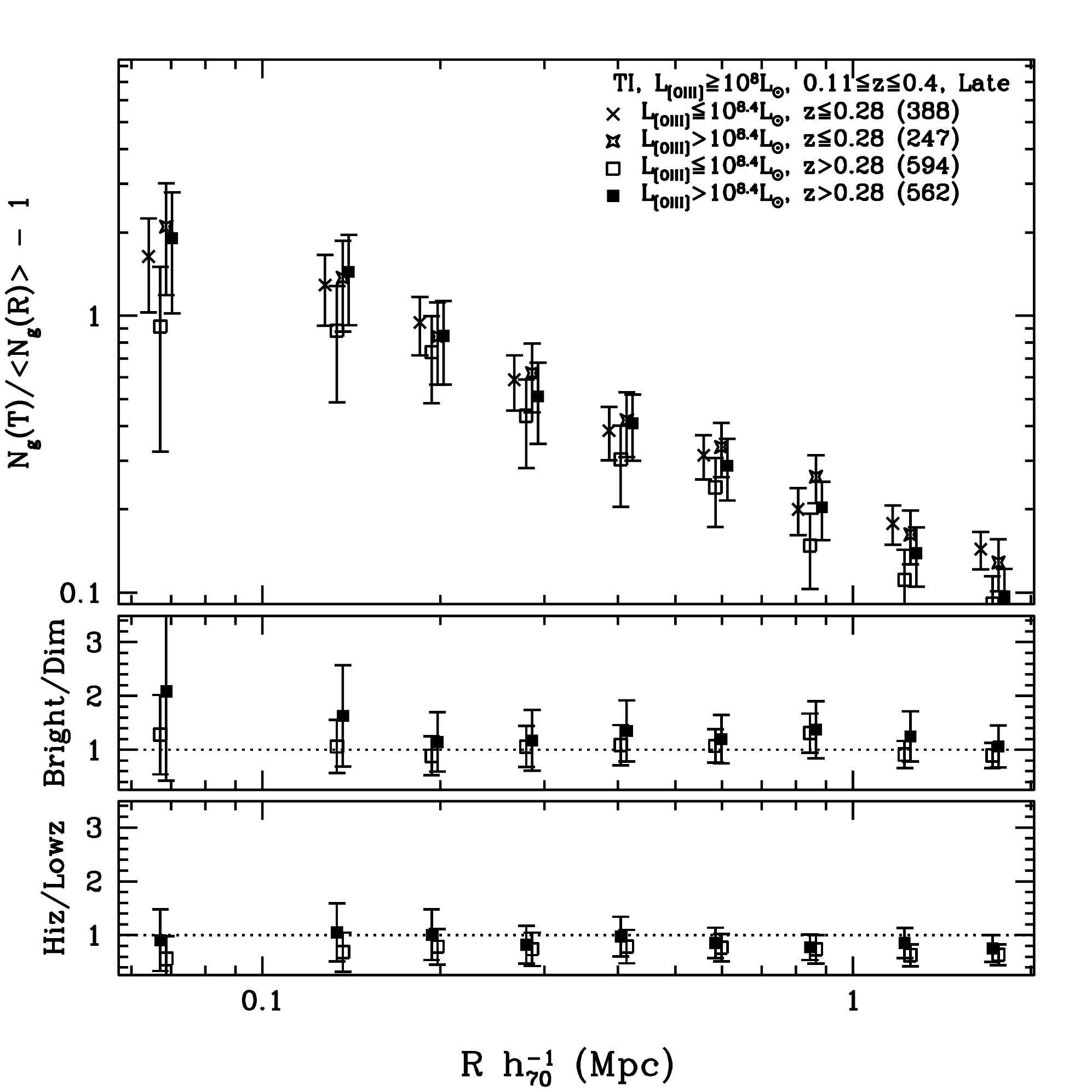}
\caption{Mean cumulative environment overdensity of late-type galaxies vs. scale, $\LOIII$, and redshift for TI targets with log($\LOIII$/$L_{\odot}$)$\geq8.0$.  \emph{Middle panel:}  ratio of environment overdensities of bright TIs to dim TIs for $z\leq0.28$ (open points), and $z>0.28$ (filled points) targets.  \emph{Lower panel:}  ratio of overdensities for high-redshift TIs to low-redshift TIs for bright (filled points) and dim (open points) targets.  
\label{scale_OIIITypeI_2bins_zevol_late_OIIIvollim8_extcorr_remdup}}
\end{figure}

There is sufficient data in the volume-limited TI sample for us to make cuts on both redshift and luminosity and measure the early- and late-type overdensities in the subsamples.  We plot this information in two figures for readability:  Figure~\ref{scale_OIIITypeI_2bins_zevol_early_OIIIvollim8_extcorr_remdup} shows the effects on early-type overdensity when the TI sample is divided on luminosity and redshift, and Figure~\ref{scale_OIIITypeI_2bins_zevol_late_OIIIvollim8_extcorr_remdup} shows the effects on late-type overdensity using the same redshift and luminosity divisions.  Additionally, the overdensities are compared in Table~\ref{table_overdensitycompare_brightdimzevol_envtgaltype}.  In the lower panel of Figure~\ref{scale_OIIITypeI_2bins_zevol_early_OIIIvollim8_extcorr_remdup}, we see that there is redshift evolution in the early-type overdensity for both bright and dim TIs, but the redshift effects are slightly more pronounced for brighter TIs.  At $R\approx500\kpchseventy$, the environment overdensity ratio of bright high redshift to bright low redshift TIs is $1.76\pm0.47$ (significance $1.6\sigma$), and for dim TIs, the ratio is $1.28\pm0.33$ (significance $\sim1\sigma$).  In contrast, we see that there is essentially no redshift evolution in the late-type overdensity around both bright ($0.98\pm0.37$, significance $>1\sigma$) and dim ($0.79\pm0.31$, significance $>1\sigma$) TIs in the lower panel of Figure~\ref{scale_OIIITypeI_2bins_zevol_late_OIIIvollim8_extcorr_remdup}.  However, in the middle panels of both figures, we find that the bright overdensity to dim overdensity ratio has similar behavior for both high- and low-redshift TIs (including some evidence for scale-dependency for the early-type environment galaxies similar to what is seen in Figure~\ref{scale_TI_earlylate_LOIII2bins_OIIIvollim8_extcorr_remdup}.  Therefore, we again conclude that although there is redshift evolution present in the early-type environment overdensities, the effect of target luminosity is more important.  

\section{FOUR ENVIRONMENT GALAXY TYPES}\label{envtgalchapter_4envtgaltypes}
\begin{figure}
\plotone{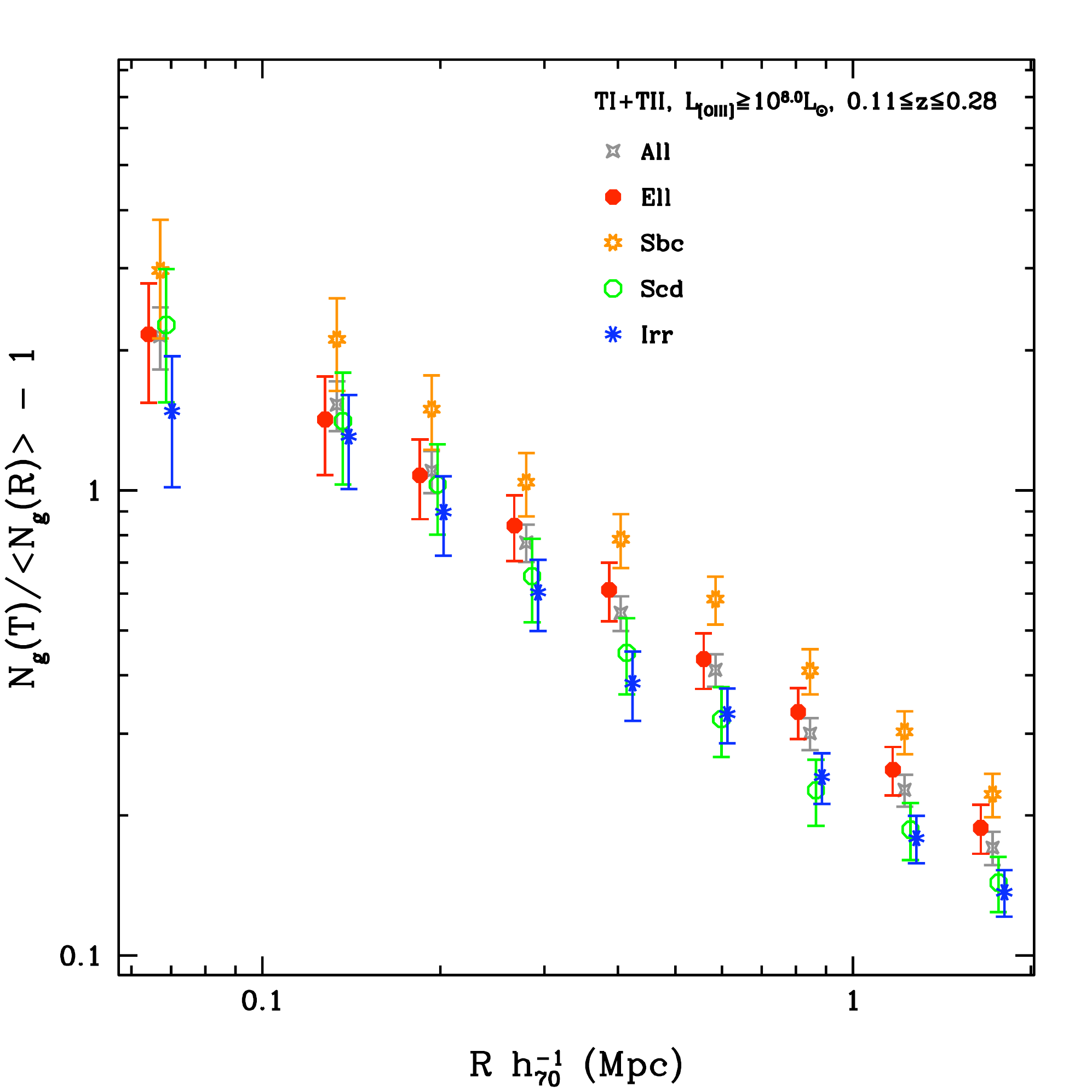}
\caption{Mean cumulative environment overdensity vs. scale and environment galaxy type for all targets with $0.11\leq z\leq0.28$ and log($\LOIII/L_{\odot}$)$\geq8.0$.  
\label{scale_typeIandII_OIII_1bin_4types_OIIIvollim8_extcorr_remdup}}
\end{figure}

\begin{figure}
\plotone{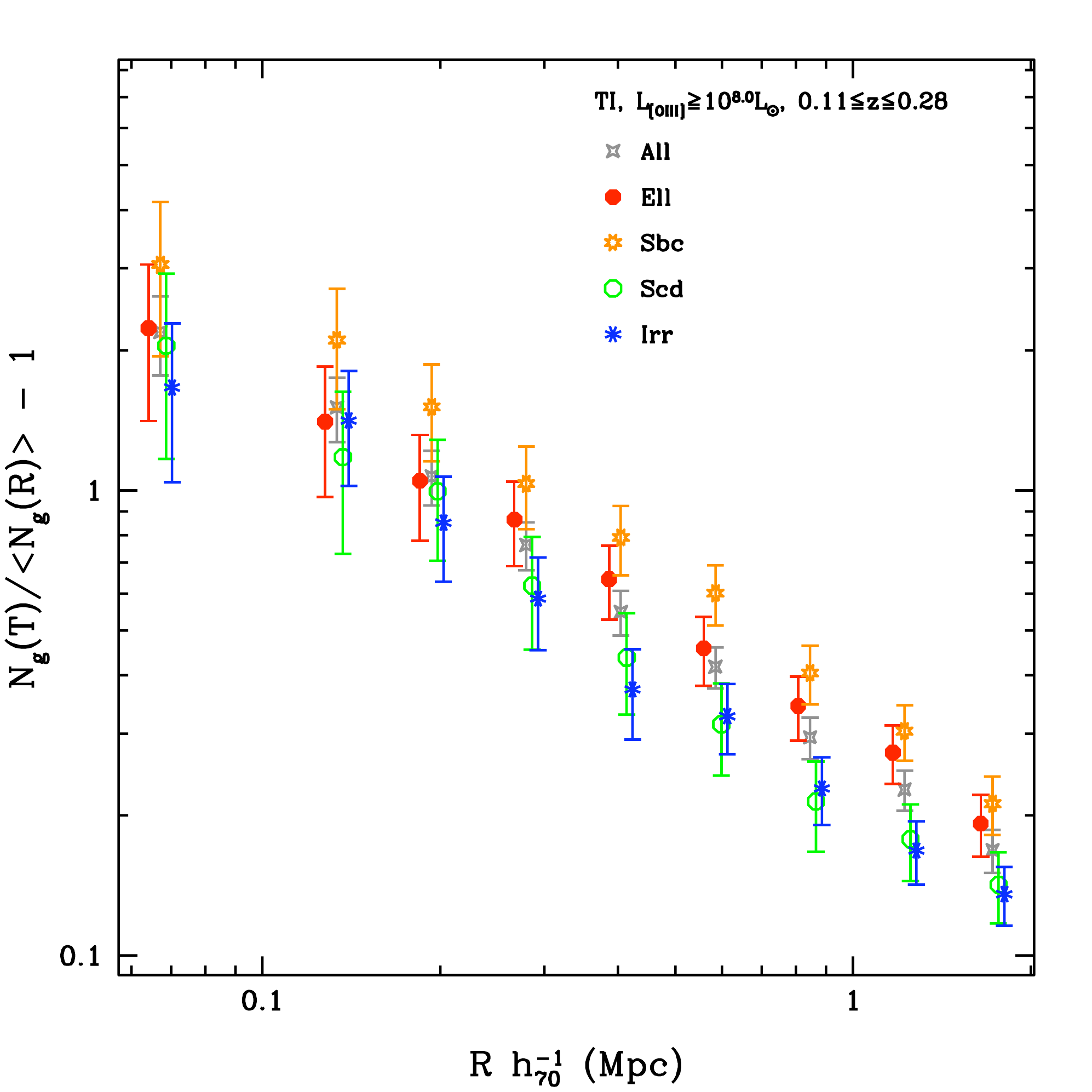}
\caption{Mean cumulative environment overdensity vs. scale and environment galaxy type for TIs with $0.11\leq z\leq0.28$ and log($\LOIII/L_{\odot}$)$\geq8.0$.  
\label{scale_typeI_OIII_1bin_4types_OIIIvollim8_extcorr_remdup}}
\end{figure}

\begin{figure}
\plotone{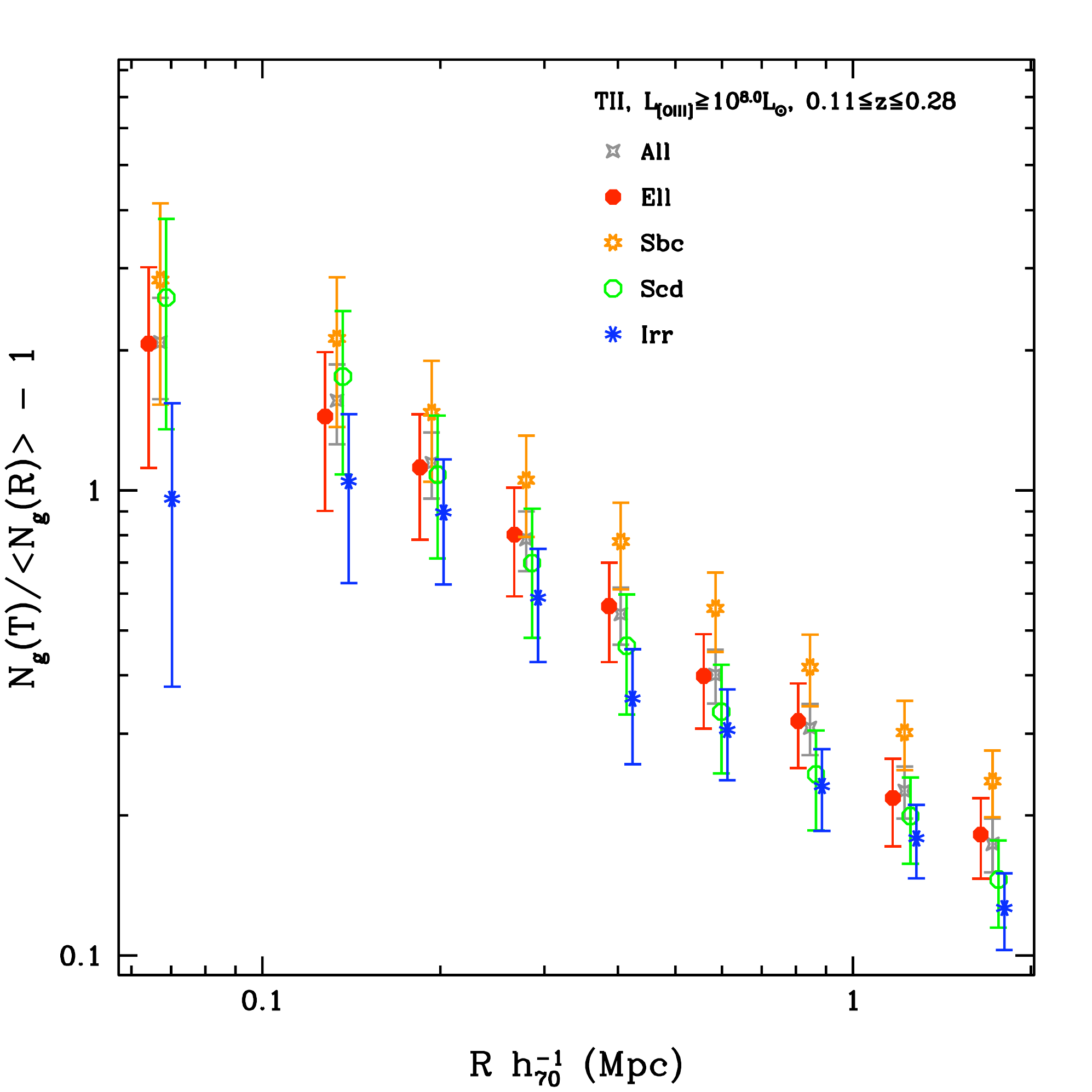}
\caption{Mean cumulative environment overdensity vs. scale and environment galaxy type for Type II AGNs with $0.11\leq z\leq0.28$ and log($\LOIII/L_{\odot}$)$\geq8.0$.  
\label{scale_typeII_OIII_1bin_4types_OIIIvollim8_extcorr_remdup}}
\end{figure}

Finally, we take advantage of additional subdivisions using the photometric galaxy type parameter $t$ that correspond to four galaxy types based on the CWW spectral templates.  Following the divisions chosen by \citet{Budavari}, we define $t<0.02$ to be Ell type, $0.02\leq t<0.3$ is approximately CWW Sbc type, $0.3\leq t<0.65$ is approximately CWW Scd type, and $t\geq0.65$ is approximately CWW Irregular type.  Because we are further subdividing the environment galaxies, the average expected counts of any given type will be lower within $2\Mpchseventy$, so it is best to use a maximum redshift of $z=0.28$ (see Section~\ref{techniquechapter_redshiftlimits}).  Figures~\ref{scale_typeIandII_OIII_1bin_4types_OIIIvollim8_extcorr_remdup},~\ref{scale_typeI_OIII_1bin_4types_OIIIvollim8_extcorr_remdup},~and~\ref{scale_typeII_OIII_1bin_4types_OIIIvollim8_extcorr_remdup} show the overdensity with scale and environment galaxy types for TI+TIIs, TIs, and TIs, respectively.  The overdensities at $R\approx500\kpchseventy$ are compared in Table~\ref{table_overdensitycompare_4envtgaltype}.  We especially note that at these low redshifts, the Sbc type galaxies dominate the environments.  Although the CWW templates are based on observations of nearby galaxies classified by morphology \citep[e.g., for the Sbc type: M51, NGC 470, NGC 1659, and NGC 2903;][]{CWW} we interpret the calculated spectral energy distributions as containing information about the star formation history of galaxies in a particular class.  Galaxies with more young stars are bluer, while galaxies with older stars appear redder.  The Sbc type galaxy is representative of a galaxy with less recent star formation than Scd or Irr type galaxies and is therefore considered ``early-type" in the broader definition we use in our work.  


\section{CONCLUSIONS}
Given our results, we speculate that bright TIs are located nearby to or in galaxy clusters \citep[see also, e.g.,][]{Barr, Sochting} because of the high overdensity of early-type galaxies in their environments.  Early-type galaxies are known to cluster more strongly than late-type galaxies \citep[e.g.,][]{Willmer, Zehavi}, to be located closer to the centers of galaxy clusters \citep[e.g.,][]{Dressler, Aguerri}, and compose a predominant fraction of the galaxies in more massive halos \citep[e.g.][]{Zehavi2005}.   

Bright and dim TIIs are found to be in environments with fewer early-type galaxies.  If the difference between TIs and TIIs is merely observation angle, the difference in their environments could be evidence that we are observing these narrow-line targets at a different stage of evolution, closer to the interaction of gas-rich (late-type) galaxies that triggered the activity in the nucleus, perhaps in the outskirts of interacting galaxy clusters \citep{Sochting}.  The lower early-type overdensity could also be evidence that the clusters in which the TIIs are located are poorer.  

According to the merger model, the active galactic nucleus is shrouded in gas and dust (only narrow-line emission will be observed) immediately following the interaction of massive gas rich galaxies, which also may trigger local star formation activity \citet{Li2008}.  In time, however, the feedback from the active nucleus blows out the gas and dust, revealing its bright center (now seen as a Type I source) and eventually quenching the nuclear activity and local star formation \citep[e.g.,][]{DiMatteo, Hopkins2008}.  Our observation of a higher overdensity of Sbc type galaxies around our sources, especially TIs, could be evidence that this feedback also affects nearby galaxies outside of the host galaxy, quenching or suppressing star formation in nearby galaxies \citep[e.g.,][]{Croton}.   


\begin{table} \centering \begin{minipage}{5.5in}
\caption[Overdensity ratios: Early/Late]{Ratio of early-type environment galaxy overdensity to late-type environment galaxy overdensity at scale $R\approx500\kpchseventy$.  Targets have $0.11\leq z\leq0.4$ and log($\LOIII/L_{\odot}$)$\geq8.0$.  \label{table_overdensityratios_envtgaltype}}
\begin{tabular}{c c c}\hline\hline
Sample&early/late&significance\\ \hline
TI+TII&$2.07\pm0.282$&$3.80\sigma$\\
TI&$2.24\pm0.358$&$3.47\sigma$\\
TII&$1.70\pm0.446$&$1.58\sigma$\\
QI&$2.24\pm0.393$&$3.16\sigma$\\
TI, log($\LOIII/L_{\odot}$)$\leq8.4$&$2.09\pm0.468$&$2.33\sigma$\\
TI, log($\LOIII/L_{\odot}$)$>8.4$&$2.44\pm0.556$&$2.59\sigma$\\
TII, log($\LOIII/L_{\odot}$)$\leq8.4$&$1.76\pm0.653$&$1.17\sigma$\\
TII, log($\LOIII/L_{\odot}$)$>8.4$&$1.65\pm0.612$&$1.07\sigma$\\
QI, log($\LOIII/L_{\odot}$)$\leq8.4$&$2.06\pm0.529$&$2.00\sigma$\\
QI, log($\LOIII/L_{\odot}$)$>8.4$&$2.42\pm0.580$&$2.45\sigma$\\
\end{tabular} \end{minipage}
\end{table}

\begin{landscape}
\begin{table} \centering \begin{minipage}{7.5in}
\caption[Overdensity comparisons: Early-, Late-type environment galaxies]{Comparison of early- and late-type environment galaxy overdensities at several scales (see also Figures~\ref{scale_OIIITypeI_1bin_earlylate_OIIIvollim8_extcorr_remdup}~and~\ref{scale_OIIITypeII_1bin_earlylate_OIIIvollim8_extcorr_remdup}).  \label{table_overdensitycompare_envtgaltype}}
\begin{tabular}{c | c c c | c c c}\hline\hline
\multicolumn{1}{c}{scale}&\multicolumn{3}{c}{Early-Type}&\multicolumn{3}{c}{Late-Type}\\
($\Mpchseventy$)&TI&TII&TI/TII&TI&TII&TI/TII\\ \hline
$2.0$&$0.213\pm0.019$&$0.208\pm0.031$&$1.027\pm0.176$&$0.119\pm0.012$&$0.140\pm0.021$&$0.853\pm0.157$\\
$1.0$&$0.410\pm0.035$&$0.377\pm0.055$&$1.088\pm0.184$&$0.202\pm0.023$&$0.245\pm0.039$&$0.825\pm0.160$\\
$0.5$&$0.850\pm0.078$&$0.687\pm0.113$&$1.237\pm0.234$&$0.379\pm0.049$&$0.403\pm0.082$&$0.940\pm0.227$\\
$0.15$&$2.304\pm0.346$&$1.755\pm0.447$&$1.313\pm0.388$&$1.249\pm0.219$&$1.277\pm0.357$&$0.978\pm0.323$\\
\end{tabular} \end{minipage}
\end{table}
\end{landscape}

\begin{table} \centering \begin{minipage}{140mm}
\caption[Overdensity comparisons: AGNs and $\Lstar$ galaxies]{Comparison environment overdensities of $0.11\leq z\leq0.15$ AI, AII, and $\Lstar$ galaxies at several scales (see also Figure~\ref{scale_spectargs_compareLstar_11z15_remdup}).  \label{table_overdensitycompare_AIAIILstar}}
\begin{tabular}{c | c c c | c c }\hline\hline
\multicolumn{1}{c}{Scale}&\multicolumn{5}{c}{All Environment Galaxies}\\
($\Mpchseventy$)&AI &AII&Lstar&AI/Lstar&AII/Lstar\\ \hline
$2.0$&$0.136\pm0.016$&$0.145\pm0.012$&$0.175\pm0.005$&$0.775\pm0.095$&$0.828\pm0.070$\\
$1.0$&$0.260\pm0.027$&$0.277\pm0.019$&$0.297\pm0.008$&$0.876\pm0.093$&$0.933\pm0.069$\\
$0.50$&$0.448\pm0.050$&$0.481\pm0.036$&$0.477\pm0.014$&$0.941\pm0.108 $&$1.009\pm0.081$\\
$0.15$&$0.885\pm0.157$&$1.101\pm0.123$&$0.865\pm0.043$&$1.023\pm0.189$&$1.273\pm0.155$\\
\hline
\multicolumn{1}{c}{}&\multicolumn{5}{c}{Early-Type Environment Galaxies}\\ \hline
$2.0$&$0.142\pm0.021$&$0.154\pm0.015$&$0.198\pm0.006$&$0.717\pm0.108$&$0.775\pm0.080$\\
$1.0$&$0.301\pm0.037$&$0.321\pm0.027$&$ 0.358\pm0.011$&$0.841\pm0.107$&$0.897\pm0.079$\\
$0.5$&$0.576\pm0.074$&$0.560\pm0.052$&$0.586\pm0.020$&$0.984\pm0.130$&$0.956\pm0.094$\\
$0.15$&$1.051\pm0.233$&$1.230\pm0.178$&$1.057\pm0.064$&$0.994\pm0.228$&$1.164\pm0.183$\\
\hline
\multicolumn{1}{c}{}&\multicolumn{5}{c}{Late-Type Environment Galaxies}\\ \hline
$2.0$&$0.128\pm0.016$&$0.135\pm0.012$&$0.148\pm0.005$&$0.864\pm0.113$&$0.910\pm0.084$\\
$1.0$&$0.213\pm0.028$&$0.226\pm0.020$&$0.226\pm0.008$&$0.941\pm0.130$&$1.000\pm0.097$\\
$0.5$&$0.300\pm0.056$&$0.388\pm0.042$&$0.349\pm0.016$&$0.858\pm0.166$&$1.111\pm0.131$\\
$0.15$&$0.696\pm0.200$&$0.954\pm0.160$&$0.647\pm0.053$&$1.077\pm0.322$&$1.475\pm0.275$\\
\hline
number&621&1222&8036&&
\end{tabular} \end{minipage}
\end{table}

\begin{landscape}
\begin{table} \centering \begin{minipage}{7.5in}
\caption[Overdensity comparisons: Early-, Late-type environment galaxies, bright and dim TI, TII targets]{Comparison of early- and late-type environment galaxy overdensities at several scales for bright and dim TI and TII targets (see also Figures~\ref{scale_TI_earlylate_LOIII2bins_OIIIvollim8_extcorr_remdup}~and~\ref{scale_TII_earlylate_LOIII2bins_OIIIvollim8_extcorr_remdup}).  \label{table_overdensitycompare_brightdim_envtgaltype}}
\begin{tabular}{c | c c c | c c c}\hline\hline
\multicolumn{1}{c}{scale}&\multicolumn{3}{c}{Early-Type}&\multicolumn{3}{c}{Late-Type}\\
($\Mpchseventy$)&TI&TII&TI/TII&TI&TII&TI/TII\\ \hline
\multicolumn{7}{c}{$8.0\leq$log($\LOIII/L_{\odot}$)$\leq8.4$}\\ \hline
$2.0$&$0.220\pm0.025$&$0.183\pm0.041$&$1.201\pm0.299$&$0.123\pm0.016$&$0.159\pm0.030$&$0.777\pm 0.179$\\
$1.0$&$0.379\pm0.045$&$0.357\pm0.073$&$1.062\pm0.252$&$0.180\pm0.029$&$0.248\pm0.054$&$0.726\pm0.197$\\
$0.5$&$0.740\pm0.097$&$0.661\pm0.150$&$1.120\pm0.293$&$0.354\pm0.064$&$0.375\pm0.110$&$0.942\pm0.325$\\
$0.15$&$1.751\pm0.381$&$1.544\pm0.555$&$1.134\pm0.477$&$1.135\pm0.274$&$1.246\pm0.481$&$0.911\pm0.415$\\
\hline
\multicolumn{7}{c}{log($\LOIII/L_{\odot}$)$>8.4$}\\ \hline
$2.0$&$0.204\pm0.028$&$0.240\pm0.047$&$0.847\pm0.203$&$0.113\pm0.019$&$0.117\pm0.030$&$0.966\pm0.297$\\
$1.0$&$0.455\pm0.055$&$0.403\pm0.084$&$1.130\pm0.272$&$0.233\pm0.036$&$0.242\pm0.057$&$0.963\pm0.269$\\
$0.5$&$1.012\pm0.131$&$ 0.721\pm0.174$&$1.403\pm0.383$&$0.415\pm0.078$&$0.436\pm0.123$&$0.950\pm0.321$\\
$0.15$&$3.130\pm0.666$&$2.033\pm0.738$&$1.540\pm0.648$&$1.407\pm0.359$&$1.312\pm 0.533$&$1.072\pm0.514$
\end{tabular} \end{minipage}
\end{table}
\end{landscape}

\begin{landscape}
\begin{table} \centering \begin{minipage}{7.5in}
\caption[Overdensity comparisons: Early-, Late-type environment galaxies, bright and dim TI targets, $z$-evolution]{Comparison of early- and late-type environment galaxy overdensities at several scales for bright and dim TI targets in two redshift ranges (see also Figures~\ref{scale_OIIITypeI_2bins_zevol_early_OIIIvollim8_extcorr_remdup}~and~\ref{scale_OIIITypeI_2bins_zevol_late_OIIIvollim8_extcorr_remdup}).  \label{table_overdensitycompare_brightdimzevol_envtgaltype}}
\begin{tabular}{c | c c c | c c c}\hline\hline
\multicolumn{1}{c}{scale}&\multicolumn{3}{c}{Early-Type}&\multicolumn{3}{c}{Late-Type}\\
($\Mpchseventy$)&$0.11\leq z\leq0.28$&$0.28<z\leq0.4$&highz/lowz&$0.11\leq z\leq0.28$&$0.28<z\leq0.4$&highz/lowz\\ \hline
\multicolumn{7}{c}{$8.0\leq$log($\LOIII/L_{\odot}$)$\leq8.4$}\\ \hline
$2.0$&$0.217\pm0.032$&$0.226\pm0.036$&$1.039\pm0.227$&$0.144\pm0.022$&$0.091\pm0.024$&$0.635\pm0.192$\\
$1.0$&$0.375\pm0.058$&$0.388\pm0.069$&$1.035\pm0.244$&$0.200\pm0.039$&$0.148\pm0.045$&$0.739\pm0.266$\\
$0.5$&$0.679\pm0.118$&$0.869\pm0.168$&$1.280\pm0.333$&$0.385\pm0.084$&$0.303\pm0.099$&$0.787\pm0.308$\\
$0.15$&$1.442\pm0.421$&$2.404\pm0.795$&$1.667\pm0.735$&$1.291\pm0.371$&$0.885\pm0.398$&$0.685\pm0.366$\\
number&388&594&&388&594&\\ \hline
\multicolumn{7}{c}{log($\LOIII/L_{\odot}$)$>8.4$}\\ \hline
$2.0$&$0.170\pm0.041$&$0.245\pm0.038$&$1.437\pm0.409$&$0.129\pm0.028$&$0.097\pm0.025$&$0.753\pm0.252$\\
$1.0$&$0.361\pm0.075$&$0.573\pm0.080$&$1.590\pm0.398$&$0.262\pm0.052$&$0.203\pm0.049$&$0.776\pm0.241$\\
$0.5$&$0.756\pm0.162$&$1.330\pm0.210$&$1.758\pm0.468$&$0.420\pm0.110$&$0.409\pm0.110$&$0.975\pm0.366$\\
$0.15$&$2.189\pm0.719$&$4.296\pm1.231$&$1.963\pm0.855$&$1.372\pm0.496$&$1.443\pm0.519$&$1.052\pm0.536$\\
number&247&562&&247&562&
\end{tabular} \end{minipage}
\end{table}
\end{landscape}


\begin{table} \centering \begin{minipage}{5.5in}
\caption[Overdensity comparison: Four Environment Galaxy Types]{Comparison of four types (CWW) of environment galaxy overdensities at scale $R\approx500\kpchseventy$.  Targets have log($\LOIII/L_{\odot}$)$\geq8.0$ and $0.11\leq z\leq0.28$ (see also Figures~\ref{scale_typeIandII_OIII_1bin_4types_OIIIvollim8_extcorr_remdup} through~\ref{scale_typeII_OIII_1bin_4types_OIIIvollim8_extcorr_remdup}). 
\label{table_overdensitycompare_4envtgaltype}}
\begin{tabular}{c c c c}\hline\hline
Galaxy Type&TI+TII&TI&TII\\ \hline
Ell&$0.611\pm0.088$&$0.644\pm0.116$&$0.563\pm0.136$\\
Sbc&$0.785\pm0.104$&$0.791\pm0.134$&$0.776\pm0.165$\\
Scd&$0.447\pm0.084$&$0.437\pm0.107$&$0.463\pm0.134$\\
Irr&$0.385\pm0.065$&$0.373\pm0.082$&$0.357\pm0.099$
\end{tabular} \end{minipage}
\end{table}

\clearpage

\clearpage

\clearpage

\clearpage

\clearpage

\clearpage

\clearpage

\clearpage

\clearpage


\clearpage

\clearpage

\clearpage


\chapter{The Dependence on Black Hole Mass}\label{MBHchapter}

\section{OVERVIEW}
We calculate black hole masses $\MBH$ for QIs using Equation 1 from \citet{Shen}:
\begin{equation}
{\rm log}\left(\frac{M_{\rm BH, vir}}{M_{\odot}}\right) = a + b~{\rm log}\left(\frac{\lambda L_{\lambda}}{10^{44}~{\rm ergs~s^{-1}}}\right)+2~{\rm log}\left(\frac{\rm FWHM}{\rm km~s^{-1}}\right)
\end{equation}
where FWHM is the full-width at half-maximum measurement of the $\Hbeta$ emission line and $L_{\lambda}$ is the luminosity of the quasar continuum spectrum at $\lambda=5100\rm\AA$ \citep[see][for details on line measurement procedures; line measurements provided by R. Reyes, private communication, 2008]{Shen}.   
We see no evolution of $\Hbeta$ FWHM and continuum $5100\rm\AA$~luminosity measurements in our redshift range of $0.11\leq z\leq0.5$, therefore the black hole mass distribution for the sample is also independent of redshift.  As in previous chapters, we will apply a volume-limit to ensure that the objects we consider have log($\LOIII$/$L_{\odot}$)$\geq8.0$.  For simplicity, we will use ``mass" interchangeably with $\MBH$ throughout the chapter, unless explicitly noted otherwise.  

Cumulative overdensity vs. scale and $\MBH$ is plotted for this $\LOIII$ volume-limited QI sample in Figure~\ref{scale_MBH_OIIIgte8_QIptext_extcorr} (see also Table~\ref{table_overdensityratios_MBH} for the ratio of high-mass black hole environment overdensity to low-mass black hole environment overdensity at $R\approx500\kpchseventy$).  The dividing value of log($\MBH/M_{\odot})=8.25$ is chosen to give roughly equal subsamples of quasars in each mass bin.  It is quite clear that at scales $R\gtrsim100\kpchseventy$, the environments of quasars with more massive black holes are more overdense than the less massive ones, which reflects the well-known correlation between black hole mass and dark matter halo mass, the so-called $M-\sigma$ or $M_{\rm DMH}-M_{\rm BH}$ relation \citep[e.g.,][]{FerrMerr, Gebhardt, Tremaine}.  At scales $\lesssim100\kpchseventy$, small numbers combined with an decreasing angular radius with redshift make strong conclusions difficult.    
The ratio of environment overdensity of high mass QIs to low mass QIs increases with increasing radius:  at $R\approx150\kpchseventy$, the ratio of environment overdensity for higher mass QIs to lower mass QIs is $1.24\pm0.27$, at $R\approx500\kpchseventy$ the ratio is $1.30\pm0.19$, and at $R\approx1.0\Mpchseventy$ it is $1.61\pm0.22$.  Thus, higher mass QIs are in environments that are more overdense than those of their lower mass counterparts to larger radii.  

We quantify the relationship between overdensity and $\MBH$ in Figure~\ref{vsMBH_4bins_OIIIgte8_extcorr_remdup}, where the linear least-squares fits for each of the scales are given in Table~\ref{table_overdensityVSMBH_4scales}.  Here we see that at small scales, QI environment overdensity increases with increasing $\MBH$ as shown by the increasingly nonzero slope in the relationship of overdensity with $\MBH$.  At the scale $\approx150\kpchseventy$, the relationship has the form $\rm{overdensity}=0.59$log$(\MBH/M_{\odot})-2.9$,~while at the largest scale of $2\Mpchseventy$, the slope is close to zero, meaning that there is very little variation in overdensity with $\MBH$.  In the following sections, we will isolate specific physical attributes of the QIs to identify their influence on the relationship between QI $\MBH$ and environment overdensity.  

\section{QI LUMINOSITY}\label{QIlumsection}
\begin{figure}
\plotone{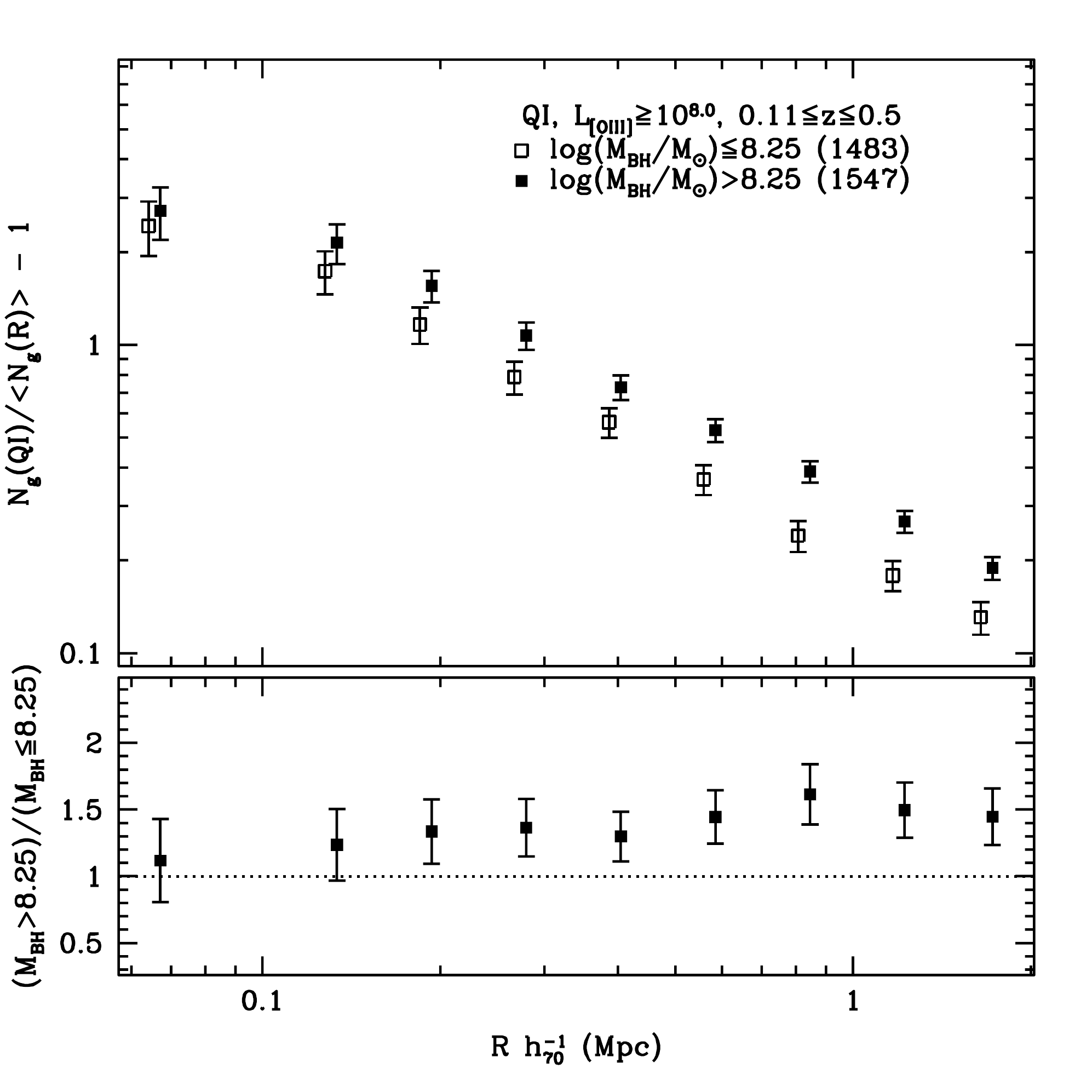}
\caption{\emph{Upper panel:}  Mean cumulative overdensity vs. scale, $\MBH$ for QIs with log($\LOIII$/$L_{\odot}$)$\geq8.0$.  \emph{Lower panel:}  Environment overdensity ratio for high mass QIs to low mass QIs.  
\label{scale_MBH_OIIIgte8_QIptext_extcorr}}
\end{figure}

We first subdivide the mass bins by $\LOIII$ in Figure~\ref{scale_MBH_OIIIsplit_OIIIgte8_QIptext_extcorr}.  We see that the ratio of overdensity around high mass to low mass QIs increases with increasing scale in the lower panel similar to what is seen in the lower panel of Figure~\ref{scale_MBH_OIIIgte8_QIptext_extcorr}, though the effect is not quite as strong for the dim quasars.  The middle panel shows the ratio of bright QI environment overdensities to dim QI environment overdensities for high mass and low mass targets.  The scale-dependency seen here is consistent with our results for TIs in Chapter~\ref{OIIIlumchapter}.  That the ratios for the two mass bins have similar values at all scales is evidence that the effect of $\LOIII$ dominates $\MBH$ in the relationship between the QI and its environment.  

\begin{figure}
\plotone{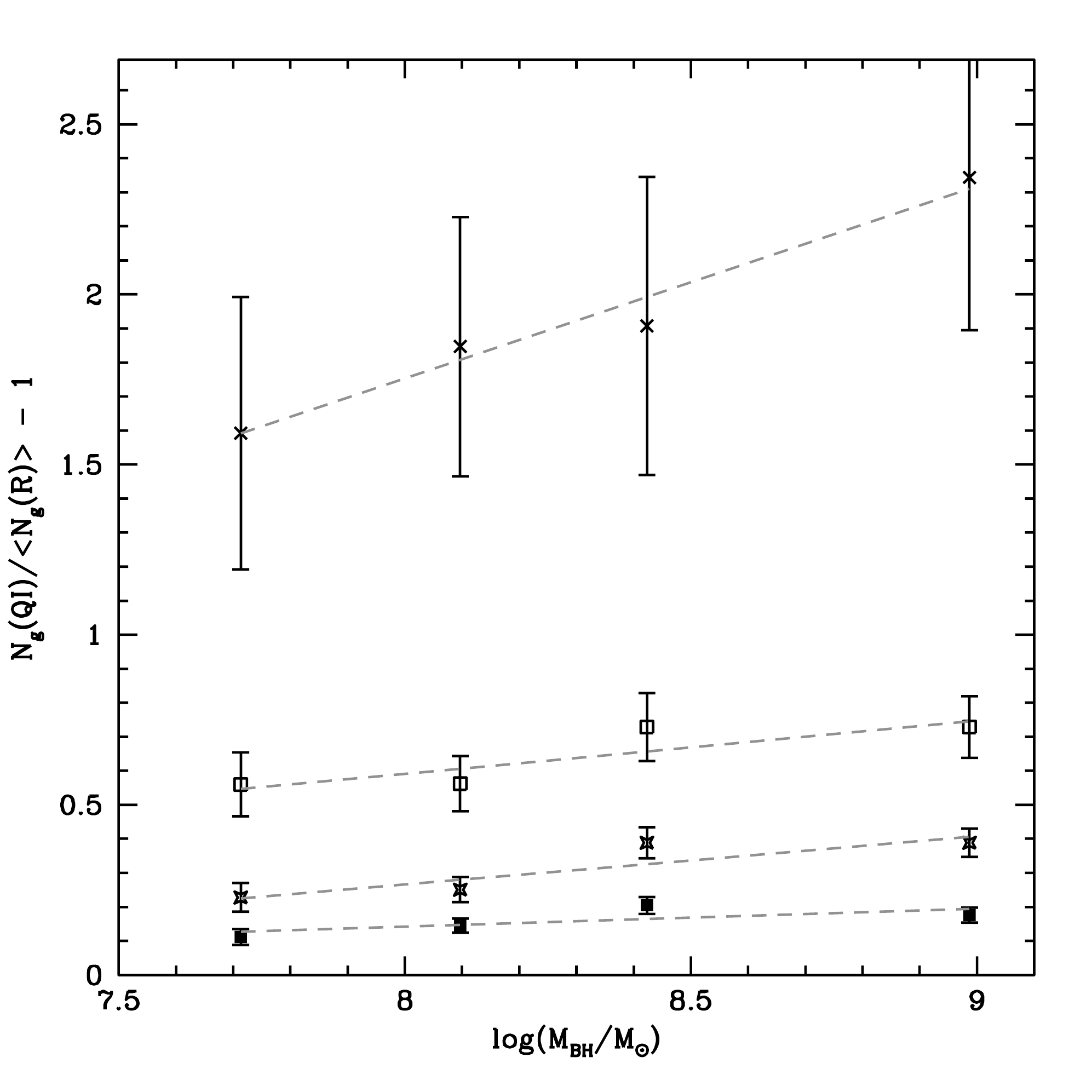}
\caption{Mean cumulative environment overdensity vs. $\MBH$ for QIs with $0.11\leq z\leq0.5$ and log($\LOIII$/$L_{\odot}$)$\geq8.0$ at four scales:  $R\approx2\Mpchseventy$ (solid squares), $R\approx1\Mpchseventy$ (starred squares), $R\approx500\kpchseventy$ (open squares), $R\approx150\kpchseventy$ (crosses).  The linear least-squares fits to the combined target sample data for each of these scales are given by the dashed lines. The fit parameters are summarized in Table~\ref{table_overdensityVSMBH_4scales}.  
\label{vsMBH_4bins_OIIIgte8_extcorr_remdup}}
\end{figure}

\begin{figure}
\plotone{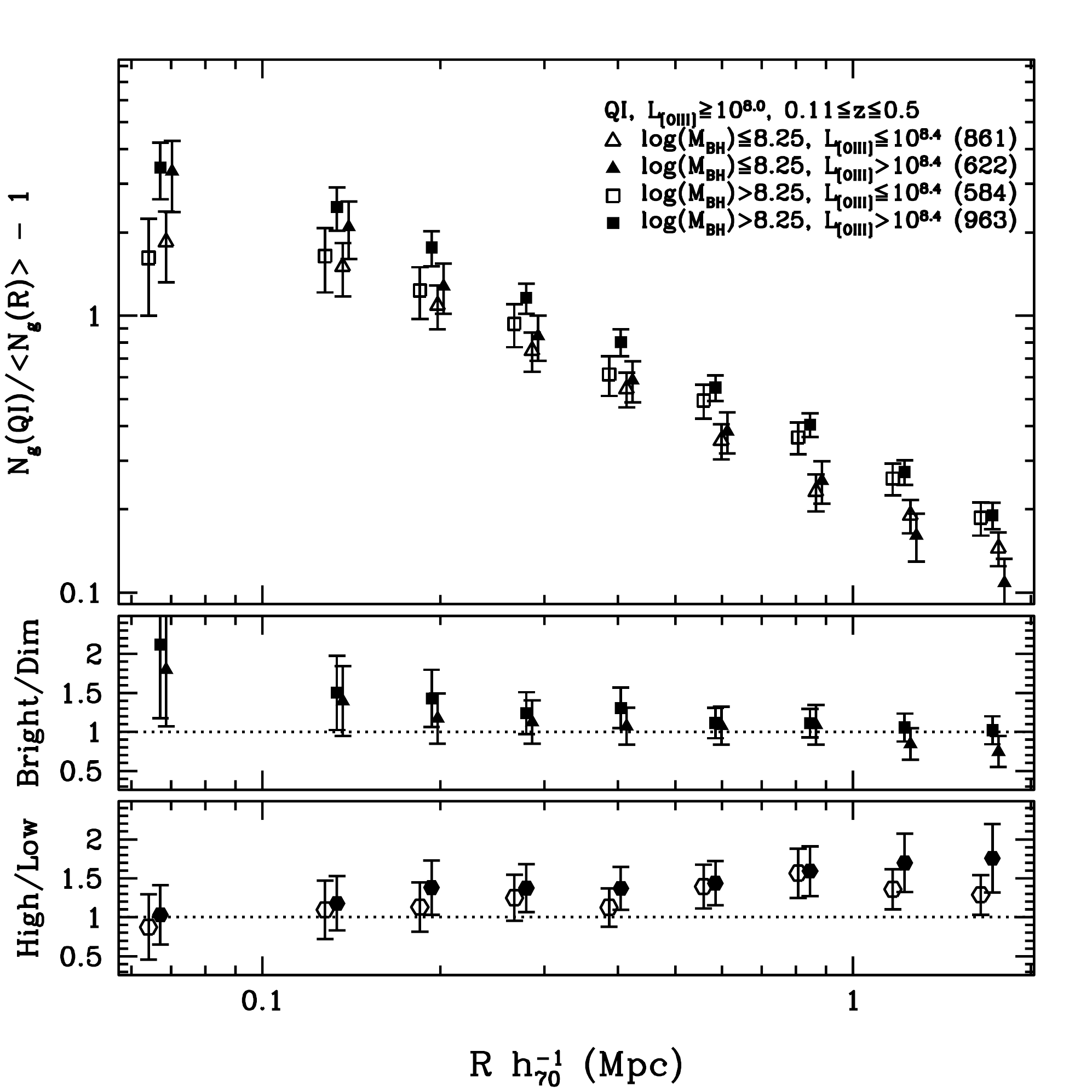}
\caption{\emph{Upper panel:}  Mean cumulative overdensity vs. scale, $\MBH$ and $\LOIII$ for QIs with $0.11\leq z\leq0.5$ and log($\LOIII$/$L_{\odot}$)$\geq8.0$.  \emph{Middle panel:} Ratio of bright QI environment to dim QI environment for high mass (squares) and low mass (triangles) QIs.  \emph{Lower panel:} Ratio of  high mass QI environment to low mass QI environment for bright (solid points) and dim (open points) QIs.  
\label{scale_MBH_OIIIsplit_OIIIgte8_QIptext_extcorr}}
\end{figure}

\begin{figure}
\plotone{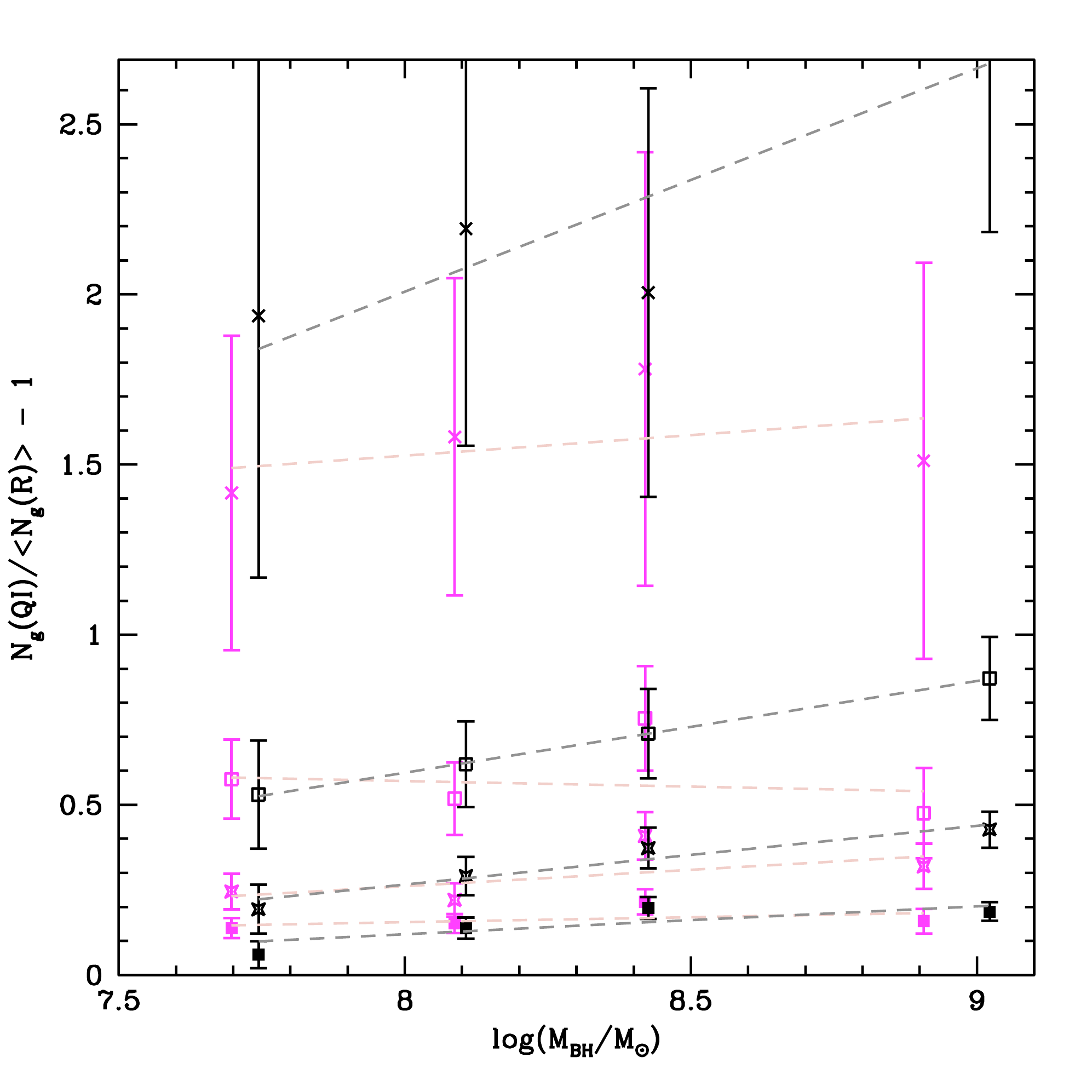}
\caption{Mean cumulative environment overdensity vs. $\MBH$ for bright (black points) and dim (magenta points) QIs with $0.11\leq z\leq0.5$ and log($\LOIII$/$L_{\odot}$)$\geq8.0$ at four scales:  $R\approx2\Mpchseventy$ (solid squares), $R\approx1\Mpchseventy$ (starred squares), $R\approx500\kpchseventy$ (open squares), $R\approx150\kpchseventy$ (crosses).  The linear least-squares fits to the combined target sample data for each of these scales are given by the dashed lines.  The fit parameters are summarized in Table~\ref{table_overdensityVSMBH_4scales}.  
\label{vsMBH_4bins_OIIIsplit_OIIIgte8_extcorr_remdup}}
\end{figure}

Again, we quantify the relationship of overdensity with $\MBH$ and $\LOIII$ in Figure~\ref{vsMBH_4bins_OIIIsplit_OIIIgte8_extcorr_remdup}.  Table~\ref{table_overdensityVSMBH_4scales} lists the linear least-squares fits for the two samples at four different scales.  The brighter QIs show the dependency of increasing environment overdensity for increasing scale that was evident in Figure~\ref{vsMBH_4bins_OIIIgte8_extcorr_remdup}, but the dimmer QIs show a much less dramatic scale dependency, where the change in slope between scales of $2.0\Mpchseventy$ to $150\kpchseventy$ is only about $0.1$ compared to $0.6$ for the brighter QIs.  

\section{QI REDSHIFT}
\begin{figure}
\plotone{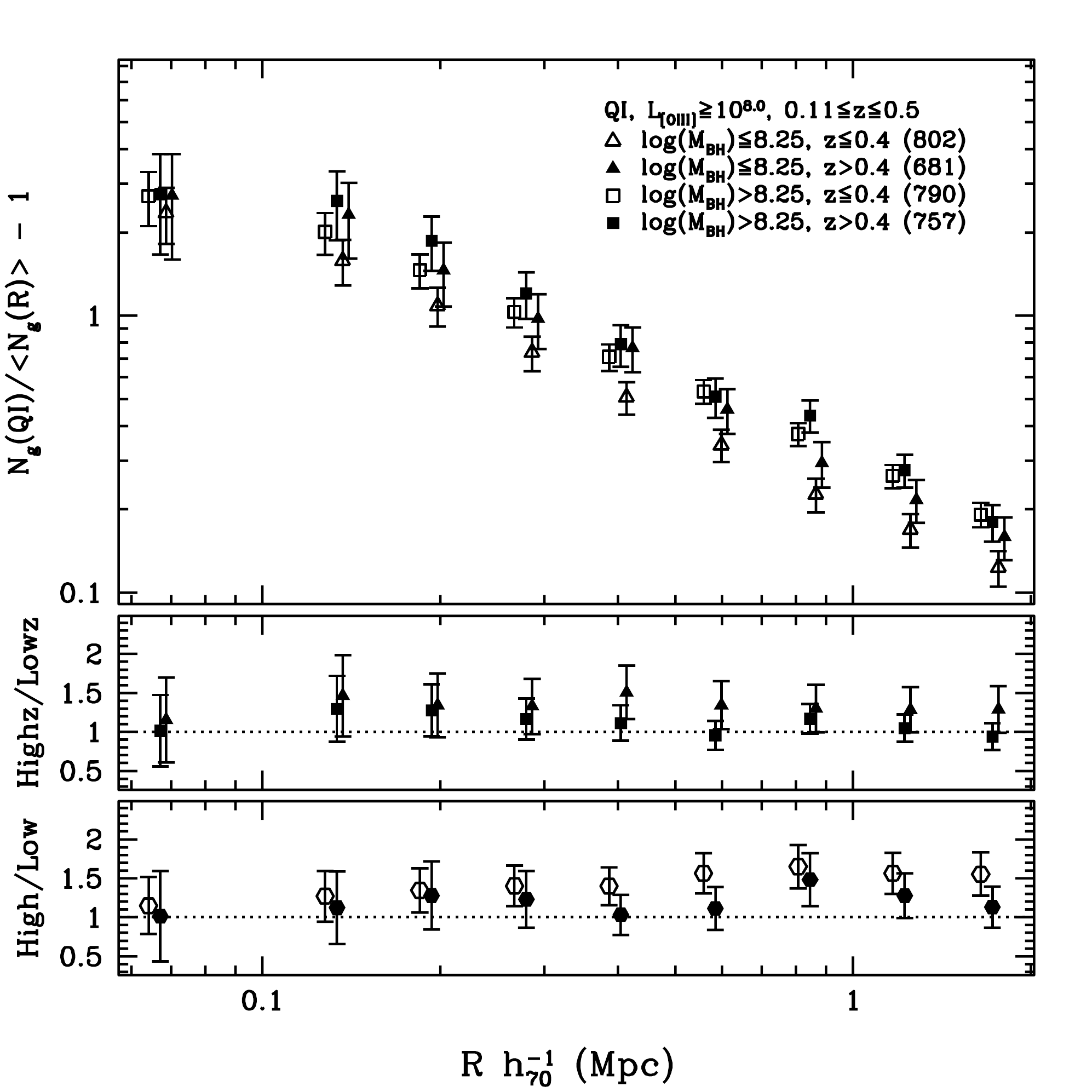}
\caption{\emph{Upper panel:}  Mean cumulative overdensity vs. scale, $\MBH$ and redshift for QIs with $0.11\leq z\leq0.5$ and log($\LOIII$/$L_{\odot}$)$\geq8.0$.  \emph{Middle panel:} Ratio of high-redshift QI environment to low-redshift QI environment for high mass (squares) and low mass (triangles) QIs.  \emph{Lower panel:} Ratio of high mass QI environment to low mass QI environment for high-redshift (solid points) and low-redshift (open points) QIs.  
\label{scale_MBH_zsplit_OIIIgte8_QIptext_extcorr}}
\end{figure}

The redshift distributions of the high and low mass QI samples are approximately the same, thus we can conclude that the overdensity differences we have observed are not an effect of a systematic difference in the two samples.  However, we now investigate whether those environment differences are an effect of redshift evolution.  We divide each of the two mass bins into two redshift bins; the dividing redshift value of $z=0.4$ gives roughly equal numbers of QIs in each redshift bin before dividing by mass.  As shown in Figure~\ref{scale_MBH_zsplit_OIIIgte8_QIptext_extcorr}, the redshift evolution is not strong for quasars with high- or low-mass black holes: on all scales the ratio of environment overdensity for higher redshift QIs to lower redshift QIs is consistent with unity (middle panel), which supports the redshift independence of the $M-\sigma$ relationship \citep[e.g.,][]{Shields}.  However, in the lower panel, it is interesting to note that the overdensity ratio for high-mass QIs to low-mass QIs at low-redshifts increases slightly with increasing scale, while the high-redshift quasars have a high-mass to low-mass overdensity ratio that is consistent with unity for nearly all scales (see also Table~\ref{table_overdensityratios_MBH}).  

\section{ENVIRONMENT GALAXY TYPE}
\begin{figure}
\plotone{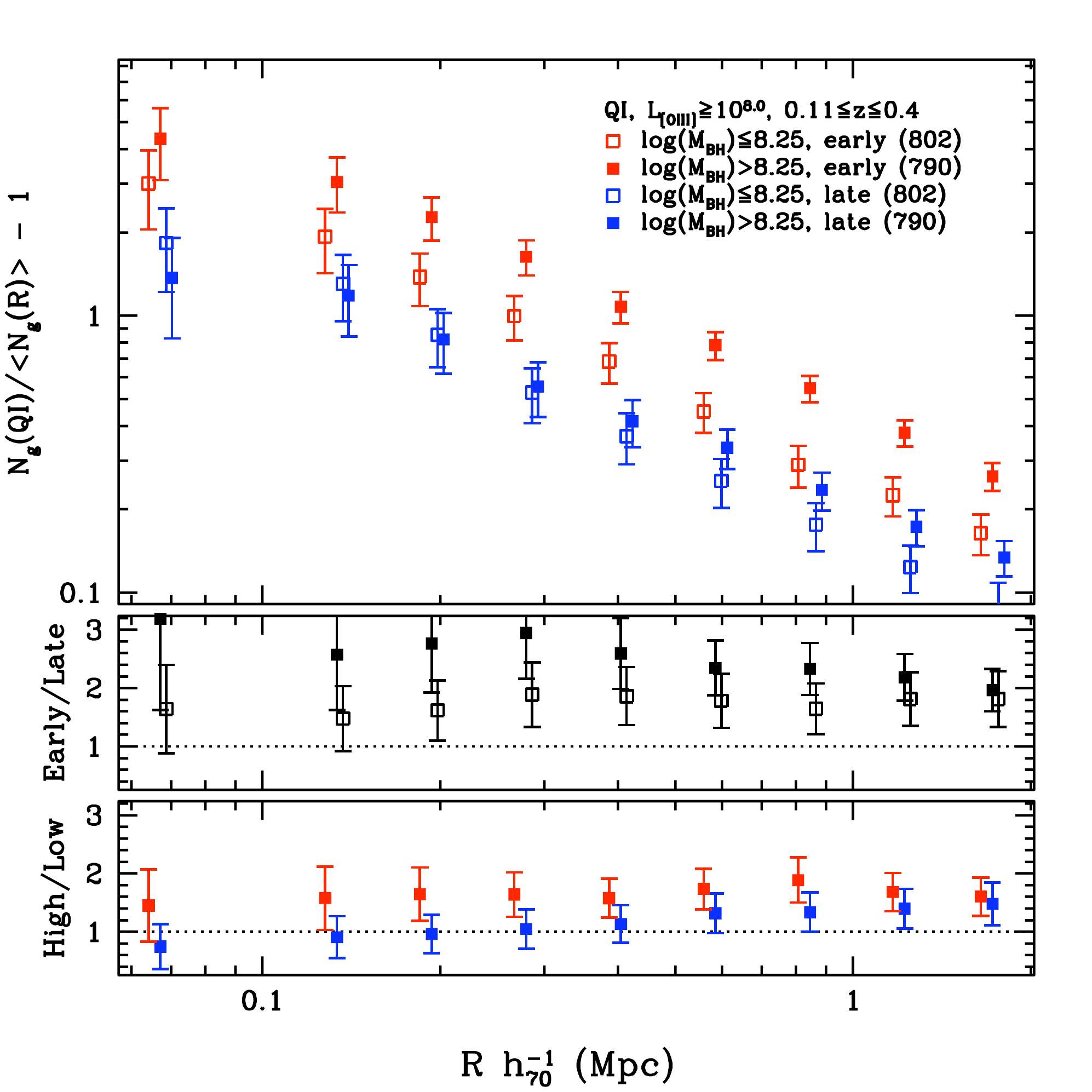}
\caption{\emph{Upper panel:}  Mean cumulative overdensity vs. scale, $\MBH$ and environment galaxy type for QIs with $0.11\leq z\leq0.4$ and log($\LOIII$/$L_{\odot}$)$\geq8.0$.  \emph{Middle panel:} Ratio of early-type environment overdensity to late-type environment overdensity for high mass (solid) and low mass (open) QIs.  \emph{Lower panel:} Ratio of  high mass QI environment to low mass QI environment for early-type (red) and late-type (blue) environments
\label{scale_MBH_earlylate_OIIIgte8_QIptext_extcorr}}
\end{figure}

Finally, we explore the nature of the environment galaxies themselves.  Figure~\ref{scale_MBH_earlylate_OIIIgte8_QIptext_extcorr} shows the environment overdensity of early- and late-type galaxies around QIs in the redshift range $0.11\leq z\leq0.4$ (note that the redshift range has been restricted as discussed in Section~\ref{techniquechapter_redshiftlimits}) divided into two mass bins.  High mass QIs have a noticeably higher overdensity of early-type galaxies in their environments compared to low mass QIs.  The early/late overdensity ratio increases with decreasing scale for high-mass quasars from a ratio value of $2.0\pm0.4$ ($2.7\sigma$) at $R\approx2.0\Mpchseventy$ to $2.6\pm0.9$ ($1.6\sigma$) at $R\approx150\kpchseventy$, but for low-mass quasars, this ratio is scale-independent (middle panel).  In the lower panel, the high-mass to low-mass overdensity ratio for early-type galaxies remains scale independent with higher-mass QIs in an environment at $R\approx500\kpchseventy$ with $1.58\pm0.33$ times more early-type galaxies than the environment of lower-mass QIs (with significance $1.7\sigma$). However, the late-type overdensity for high-mass QIs seems to increase compared to the late-type overdensity for low-mass QIs with increasing scale, though not with large statistical significance.  Thus the increase in high mass/low mass ratio with increasing scale may be due to a slight change in late-type overdensity in the environments of QIs.  

\section{ACCRETION EFFICIENCY}
Following \citet{Li2008}, we use the ratio $\LOIII$/$\MBH$ as an indicator of the AGN accretion efficiency of the supermassive black hole at the center of the QI.  We see no evolution of the distribution of $\LOIII$/$\MBH$ over our redshift range, and as before we apply a volume limit and use only QIs with log($L_{[OIII]}$/$L_{\odot}$)$\geq8.0$ 

\begin{figure}
\plotone{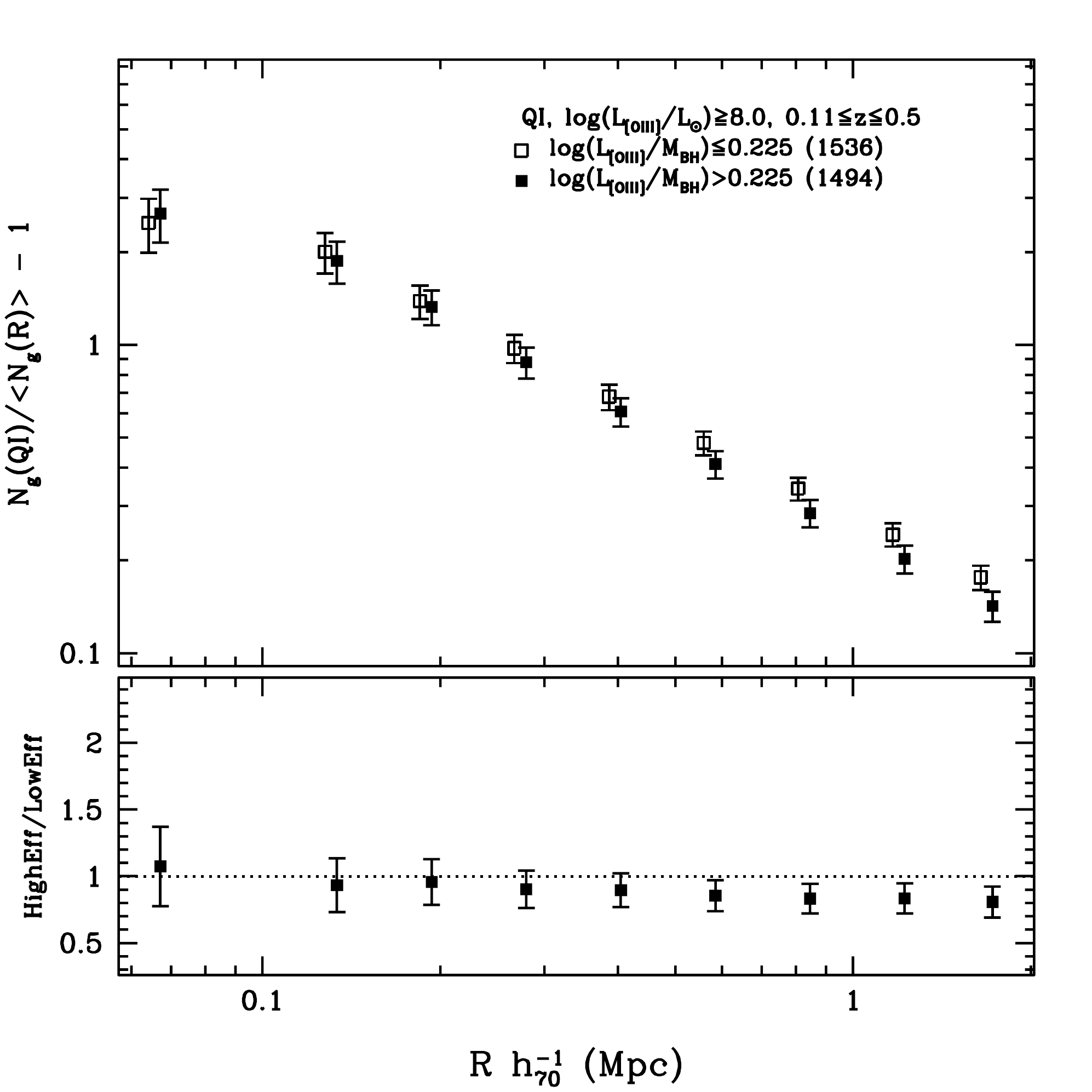}
\caption{\emph{Upper panel:} Mean cumulative overdensity vs. scale and $\LOIII$/$\MBH$ ratio for QIs with log($L_{[OIII]}$/$L_{\odot}$)$\geq8.0$.  \emph{Lower panel:} Ratio of environment overdensity for higher efficiency QIs to the environment overdensity for lower efficiency QIs.  
\label{scale_OIIIMBHratio_QIptext_extcorr}}
\end{figure}

\begin{figure}
\plotone{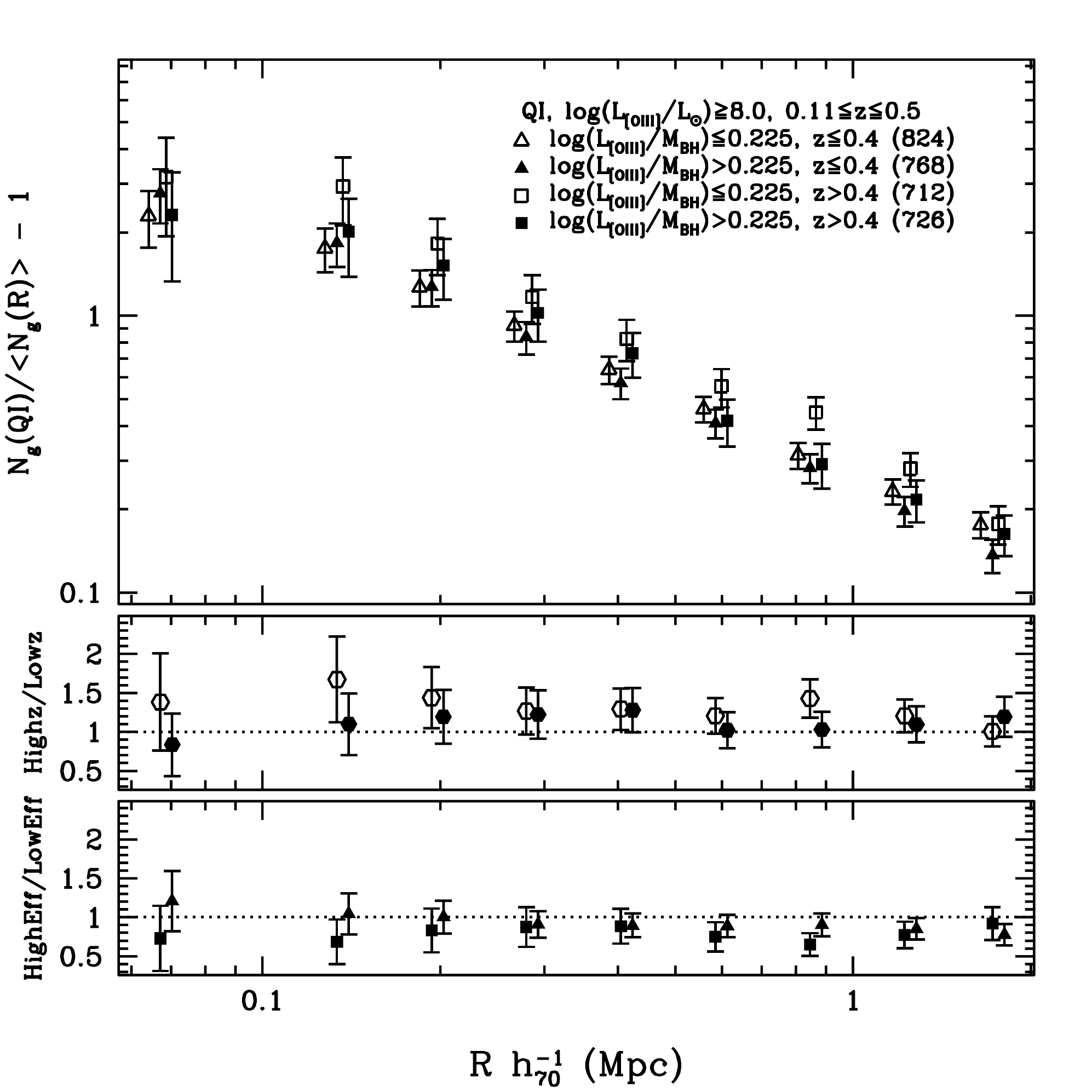}
\caption{\emph{Upper panel:} Mean cumulative overdensity vs. scale and $\LOIII$/$\MBH$ ratio for QIs with log($\LOIII$/$L_{\odot}$)$\geq8.0$ in two different redshift bins.  The dividing values are chosen for direct comparison to other figures.  \emph{Middle panel:} Ratio of environment overdensity of high redshift QIs to that of dim QIs for high efficiency (solid points) and low efficiency (open points) QIs.  \emph{Lower panel:} Ratio of high efficiency overdensity to low efficiency overdensity for high redshift (square) and low redshift (triangle) QIs.
\label{scale_OIIIMBHratio_zsplit_QIptext_extcorr}}
\end{figure}

We first observe that the environment overdensities of high efficiency and low efficiency QIs are similar in Figure~\ref{scale_OIIIMBHratio_QIptext_extcorr}.  The dividing value of log($\LOIII$/$\MBH$)$=0.225$ is at the center of the distribution of efficiency ratios for the sample and gives approximately equal numbers of objects in each bin.  If we use the calibration of $L_{bol}/\LOIII\approx3500$ \citep{Heckman2004}, this ratio corresponds to the QI accreting at approximately 18\% of the Eddington accretion rate (i.e., the rate at which the inward gravitational pull on infalling material balances the outward radiation pressure caused by the accretion process).  On the outermost scales $R\gtrsim500\kpchseventy$, however, it appears that lower efficiency QIs have an environment that is a slightly more overdense compared to the higher efficiency QIs.  In Figure~\ref{scale_OIIIMBHratio_zsplit_QIptext_extcorr}, we divide the accretion efficiency bins into two redshift bins at $z=0.4$.  We see that at all scales there is no significant evidence for redshift evolution (middle panel), and that the ratio of environment overdensities for high efficiency QIs to low efficiency QIs is consistent between the two redshift bins.  Thus, any small amount of redshift evolution has negligible effect on environment overdensity when it is compared in terms of accretion efficiency.  We tabulate the overdensity ratios of high efficiency to low efficiency QIs at scale $R\approx500\kpchseventy$ in Table~\ref{table_overdensityratios_LOIIIMBHratio}.  

\begin{figure}
\plotone{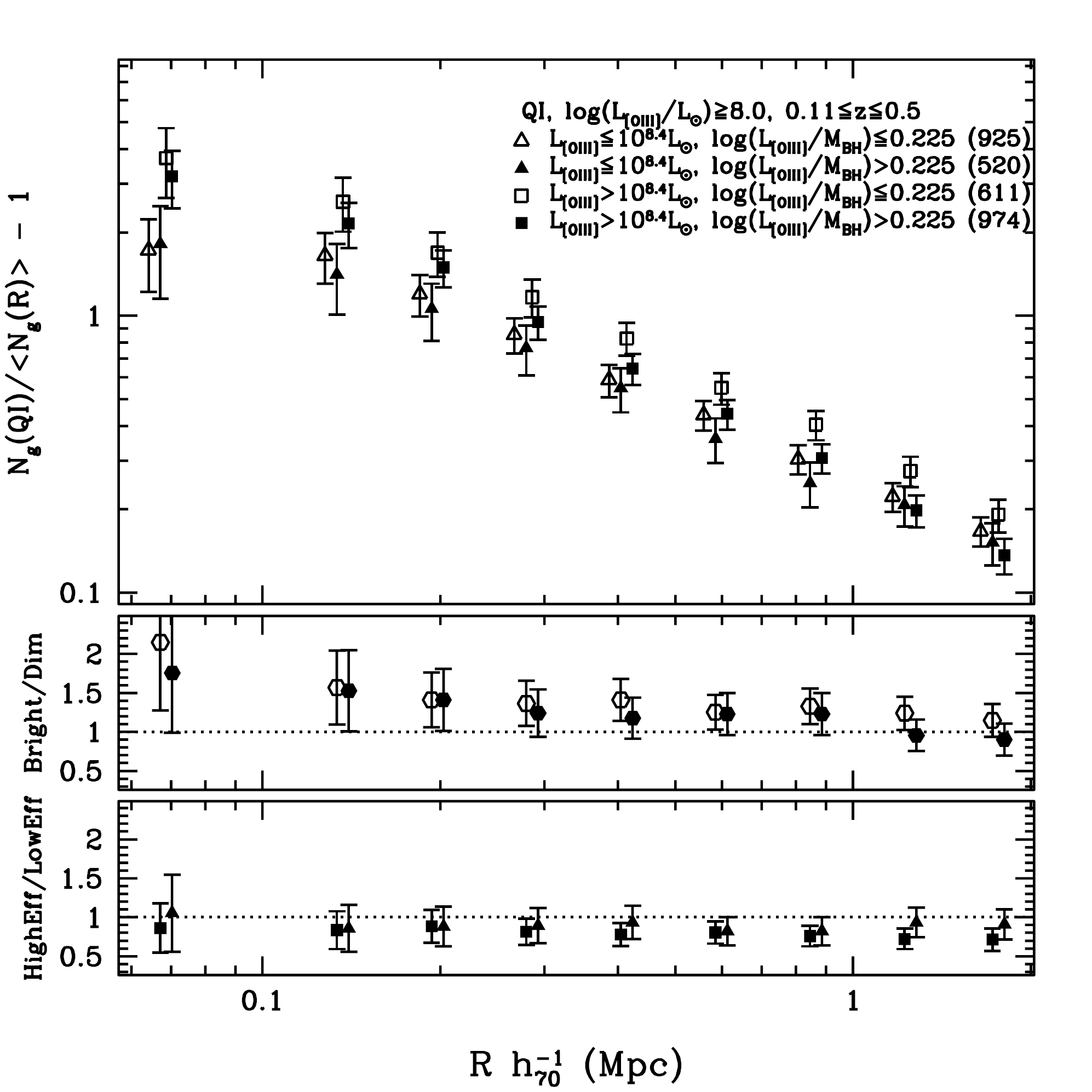}
\caption{\emph{Upper panel:} Mean cumulative overdensity vs. scale and $\LOIII$/$\MBH$ ratio for QIs with log($\LOIII$/$L_{\odot}$)$\geq8.0$ in two different $\LOIII$ bins.  The dividing values are chosen for direct comparison to other figures.  \emph{Middle panel:} Ratio of environment overdensity of bright QIs to that of dim QIs for high efficiency (solid points) and low efficiency (open points) QIs.  \emph{Lower panel:} Ratio of high efficiency overdensity to low efficiency overdensity for bright (square) and dim (triangle) QIs.
\label{scale_OIIIMBHratio_OIIIsplit_QIptext_extcorr}}
\end{figure}

The accretion efficiency as determined by $\LOIII$/$\MBH$ is high with bright $\LOIII$ or with low $\MBH$.  In an attempt to isolate these conditions, we first divide the sample on $\LOIII$ and calculate the overdensities for high- and low-efficiency QIs in each $\LOIII$ bin, as is shown in Figure~\ref{scale_OIIIMBHratio_OIIIsplit_QIptext_extcorr}.  Here we see that low-luminosity, high efficiency QIs have the highest environment overdensity.  Brighter QIs have higher environment overdensity than dimmer QIs, whether they have high or low accretion efficiency (middle panel).  This effect is scale dependent, increasing with decreasing scale, just as we saw in Section~\ref{OIIIlumchapter_typesection}.  Low-efficiency QIs have a slightly higher bright/dim overdensity ratio compared to high-efficiency QIs:  at $R\approx500\kpchseventy$, the bright/dim ratio for low (high) efficiency QIs is $1.41\pm0.27$ ($1.18\pm0.26$) with significance $\approx1.5\sigma$ ($<1\sigma$).  At all scales (and with no scale dependency), both bright and dim low-efficiency QIs have environments with slightly higher overdensity than high-efficiency QIs (lower panel).  

\begin{figure}
\plotone{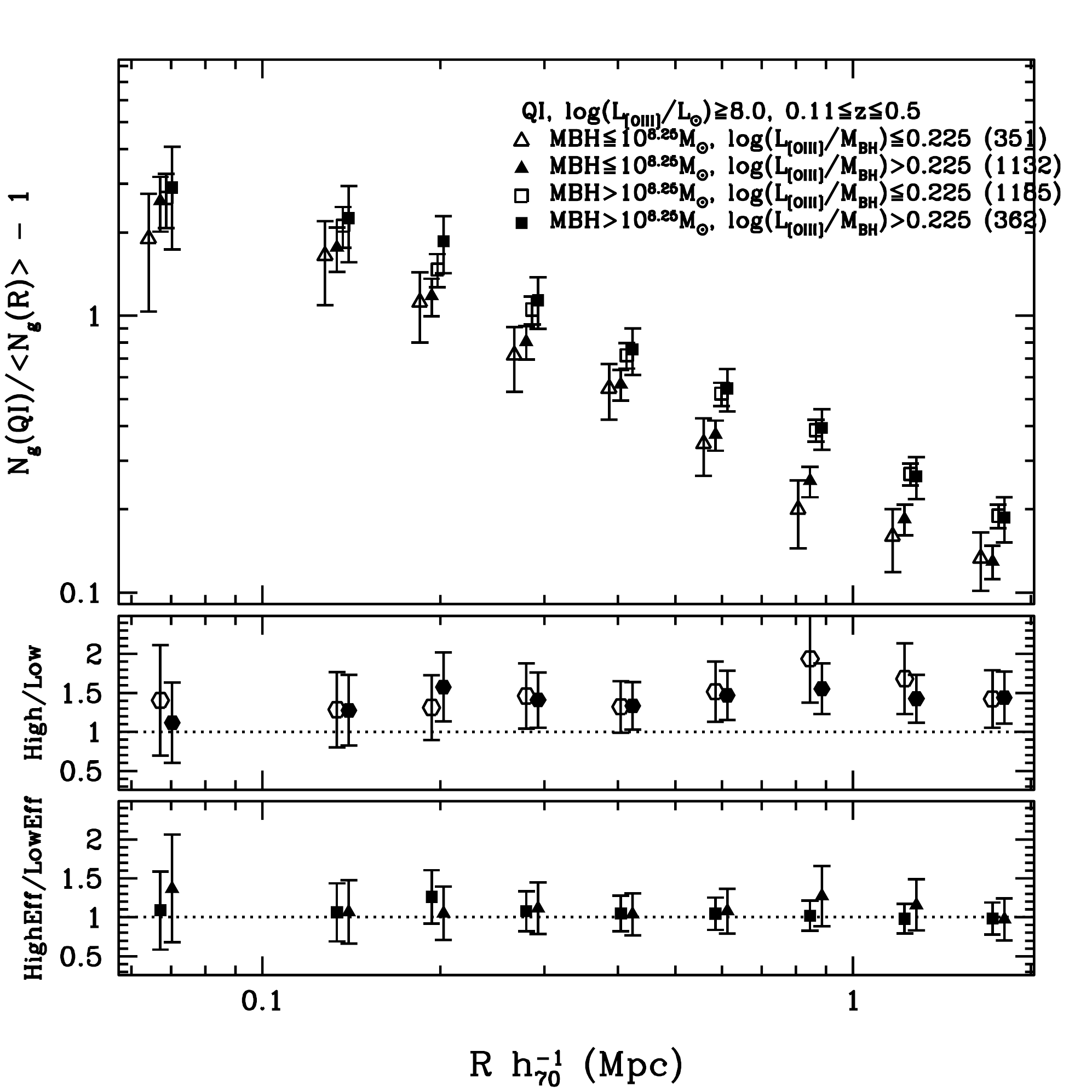}
\caption{\emph{Upper panel:} Mean cumulative overdensity vs. scale and $\LOIII$/$\MBH$ ratio for QIs with log($\LOIII$/$L_{\odot}$)$\geq8.0$ in two $\MBH$ bins.  The dividing values are chosen for direct comparison to other figures.  \emph{Middle panel:} Ratio of environment overdensity of high-mass QIs to that of low-mass QIs for high efficiency (solid points) and low efficiency (open points) QIs.  \emph{Lower panel:} Ratio of high efficiency overdensity to low efficiency overdensity for high mass (square) and low mass (triangle) QIs.  
\label{scale_OIIIMBHratio_BHmasssplit_QIptext_extcorr}}
\end{figure}

Next, we divide the sample on $\MBH$ and compare the overdensities around high- and low-efficiency QIs in Figure~\ref{scale_OIIIMBHratio_BHmasssplit_QIptext_extcorr}.  At all scales, the overdensity ratio of high-efficiency to low-efficiency environments is consistent with unity regardless of mass (lower panel).  However, there is evidence that the environment of high-mass QIs compared to low-mass QIs increases with increasing scale (middle panel); this trend, too, is regardless of the efficiency category of the QI.  

\begin{figure}
\plotone{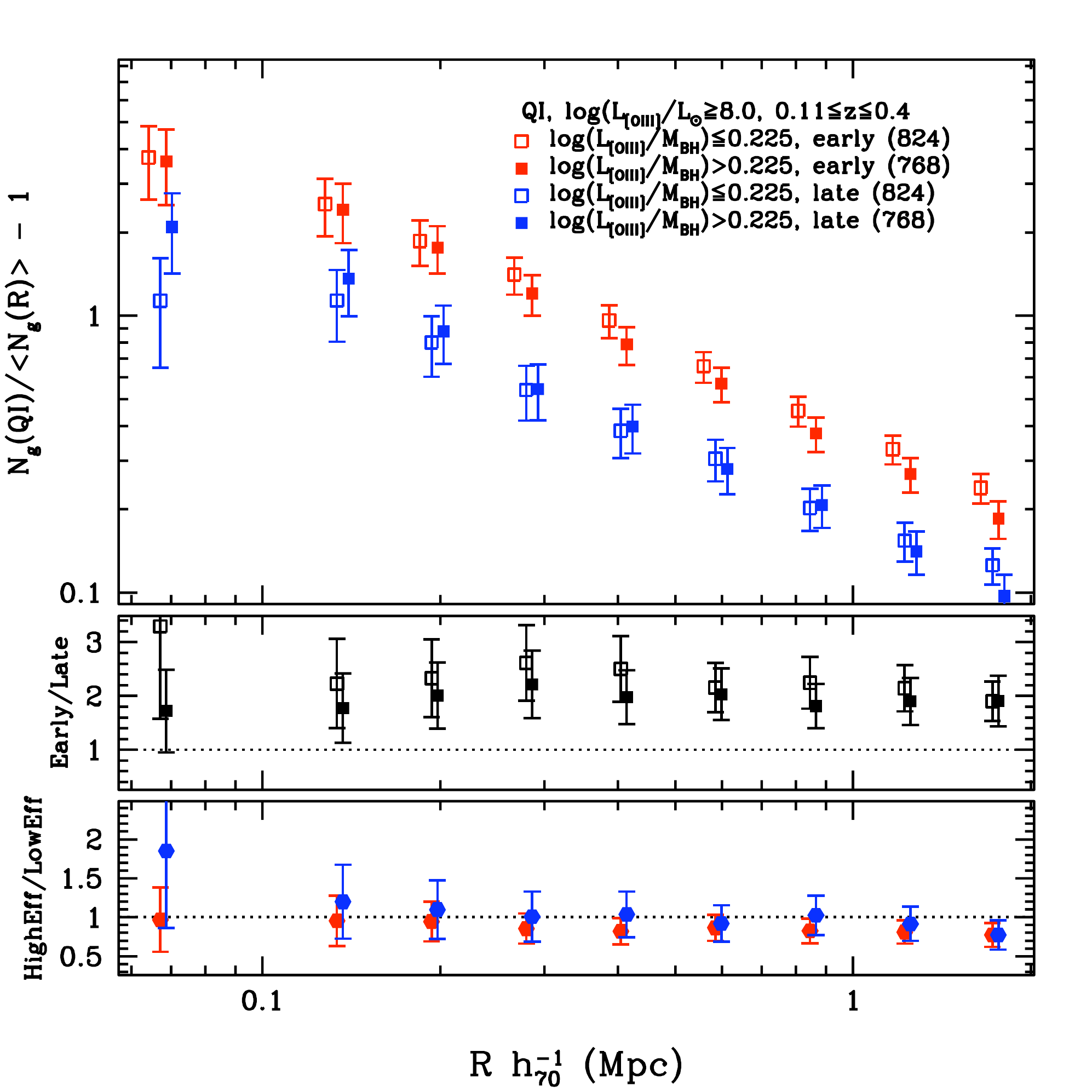}
\caption{\emph{Upper panel:} Mean cumulative overdensity vs. scale, $\LOIII$/$\MBH$ ratio, and environment galaxy type for QIs with log($L_{[OIII]}$/$L_{\odot}$)$\geq8.0$ and $0.11\leq z\leq0.4$.  \emph{Middle panel:} Ratio of early-type environment overdensity to late-type environment overdensity for high efficiency (solid points) and low efficiency (open points) QIs.  \emph{Lower panel:} Ratio of high efficiency overdensity to low efficiency overdensity using early-type environments (red) and late type environments (blue) of QIs.  
\label{scale_OIIIMBHratio_envtgaltype_QIptext_extcorr}}
\end{figure}

When we separate the background galaxies by type in Figure~\ref{scale_OIIIMBHratio_envtgaltype_QIptext_extcorr} (again restricting the sample to have $0.11\leq z\leq0.4$), we see that the early-type overdensity in the environments of both high- and low-efficiency QIs is consistent with unity on all scales except the largest measures, and here the early-type overdensity is slightly higher around low-efficiency QIs with a significance of $\approx1.5\sigma$ (lower panel).  The late-type overdensity ratio is also consistent with unity, though at the innermost scales, high-efficiency QIs may have an increased overdensity of late-type galaxies in their environments.  On all scales, the ratios of early-type to late-type galaxies around high- and low-efficiency quasars are consistent within the error bars (middle panel).  At $R\approx500\kpchseventy$, the ratio around high-efficiency QIs is $1.98\pm0.5$ (significance of $\approx2\sigma$), and the ratio around low-efficiency QIs is $2.51\pm0.61$ (significance of $\approx2.5\sigma$).  

\section{CONCLUSIONS}
We are not surprised to see that there is an increased overdensity of galaxies in the vicinity of more massive black holes.  This result is consistent with the $M_{\rm DMH}-\MBH$ relationship \citep[e.g.,][]{FerrMerr, Gebhardt, Tremaine} in which there is a correlation between the mass of a galaxy's black hole and the mass of the dark matter halo containing the quasar's host galaxy.  Since clustering of halos increases with increasing halo mass, we see higher mass objects located in environments with higher galaxy density.  The fact that we observe QIs with higher black hole mass in environments that are more overdense than those of their lower mass counterparts to larger radii may imply that QIs of different black hole mass are located in galaxy clusters of varying richness, where QIs with higher $\MBH$ may be located in richer clusters with higher halo mass.  Indeed, we see that the lower-mass QIs have an increased overdensity of late-type galaxies in their environments at smaller scales compared to the environments of higher-mass QIs.  This could be the signature of the morphology-density relation seen in local clusters: regions of increased galaxy density have a higher fraction of early-type galaxies and a correspondingly decreased fraction of late-type galaxies \citep[e.g.,][]{Dressler}.  

When we classify the QIs based on accretion efficiency using the ratio $\LOIII$/$\MBH$, we see that low-efficiency QIs have a slightly higher environment overdensity compared to high-efficiency QIs.  This effect is likely due to quenching:  outflows from the central engine of the QI heat the surrounding intergalactic medium (IGM), which prevents the IGM from condensing to form galaxies \citep{ScannapiecoOh}.  QIs with more efficient accretion will have more powerful outflows; therefore, the galaxy overdensity is decreased around these sources compared to their lower-efficiency counterparts.  

\begin{table} \begin{minipage}{140mm}
\caption[Overdensity ratios: High/Low MBH]{Ratio of high mass black hole environment galaxy overdensity to low mass black hole environment galaxy overdensity for QIs at scale $R\approx500\kpchseventy$.  Targets have $0.11\leq z\leq0.5$ and log($\LOIII/L_{\odot}$)$\geq8.0$.  \label{table_overdensityratios_MBH}}
\begin{tabular}{c c c}\hline\hline
Sample&high/low&significance\\ \hline
QI&$1.30\pm0.186$&$1.61\sigma$\\
QI,log($\LOIII/L_{\odot}$)$>8.4$&$1.37\pm0.278$&$1.34\sigma$\\
QI,log($\LOIII/L_{\odot}$)$\leq8.4$&$1.13\pm0.246$&$<1\sigma$\\
QI,$z>0.4$&$1.03\pm0.259$&$<1\sigma$\\
QI,$z\leq0.4$&$1.40\pm0.242$&$1.65\sigma$\\
QI,$z\leq0.4$, early&$1.58\pm0.334$&$1.72\sigma$\\
QI,$z\leq0.4$, late&$1.13\pm0.323$&$<1\sigma$
\end{tabular} \end{minipage}
\end{table}


\begin{table} \begin{minipage}{140mm}
\caption[Fit parameters for overdensity vs. $\MBH$ at four different scales]{Linear least-squares fit parameters for overdensity vs. $\MBH$ at four different scales using QIs with log($\LOIII$/$L_{\odot}$)$\geq8.0$ (see also Figure~\ref{vsMBH_4bins_OIIIgte8_extcorr_remdup} and Figure~\ref{vsMBH_4bins_OIIIsplit_OIIIgte8_extcorr_remdup}).  \label{table_overdensityVSMBH_4scales}}
\begin{tabular}{c | c c c }\hline\hline
\multicolumn{1}{c}{scale}&\multicolumn{3}{c}{}\\
($\Mpchseventy$)&slope&intercept&$\chi^2$\\ \hline
\multicolumn{4}{c}{All QIs}\\ \hline
2.0&$0.052\pm0.033$&$-0.278\pm0.277$&$3.80$\\
1.0&$0.142\pm0.052$&$-0.873\pm0.429$&$2.70$\\
0.5&$0.156\pm0.064$&$-0.656\pm0.535$&$0.862$\\
0.15&$0.564\pm0.074$&$-2.76\pm0.616$&$0.053$\\ \hline
\multicolumn{4}{c}{QIs log($\LOIII$/$L_{\odot}$)$>8.4$}\\ \hline
2.0&$0.082\pm0.042$&$-0.535\pm0.357$&$3.12$\\
1.0&$0.173\pm0.034$&$-1.12\pm0.283$&$0.563$\\
0.5&$0.270\pm0.004$&$-1.57\pm0.037$&$0.002$\\
0.15&$0.657\pm0.288$&$-3.25\pm2.41$&$0.315$\\ \hline
\multicolumn{4}{c}{QIs log($\LOIII$/$L_{\odot}$)$\leq8.4$}\\ \hline
2.0&$0.030\pm0.039$&$-0.088\pm0.318$&$2.17$\\
1.0&$0.096\pm0.091$&$-0.510\pm0.746$&$3.67$\\
0.5&$-0.034\pm0.145$&$0.845\pm1.19$&$2.12$\\
0.15&$0.121\pm0.179$&$0.560\pm1.47$&$0.182$
\end{tabular} \end{minipage}
\end{table}


\begin{table} \begin{minipage}{140mm}
\caption[Overdensity ratios: High/Low Accretion Efficiency]{Ratio of high efficiency environment galaxy overdensity to low efficiency environment galaxy overdensity for QIs at scale $R\approx500\kpchseventy$.  Targets have $0.11\leq z\leq0.5$ and log($\LOIII/L_{\odot}$)$\geq8.0$.  \label{table_overdensityratios_LOIIIMBHratio}}
\begin{tabular}{c c c}\hline\hline
Sample&highEff/lowEff&significance\\ \hline
QI&$0.895\pm0.127$&$<1\sigma$\\
QI, log($\LOIII/L_{\odot}$)$>8.4$&$0.778\pm0.145$&$1.53\sigma$\\
QI, log($\LOIII/L_{\odot}$)$\leq8.4$&$0.934\pm0.212$&$<1\sigma$\\
QI, log($\MBH/M_{\odot}$)$>8.25$&$1.05\pm0.230$&$<1\sigma$\\
QI, log($\MBH/M_{\odot}$)$\leq8.25$&$1.04\pm0.270$&$<1\sigma$\\
QI, $z>0.4$&$0.886\pm0.222$&$<1\sigma$\\
QI, $z\leq0.4$&$0.895\pm0.154$&$<1\sigma$\\
QI, $z\leq0.4$, early&$0.819\pm0.169$&$1.07\sigma$\\
QI, $z\leq0.4$, late&$1.04\pm0.295$&$<1\sigma$
\end{tabular} \end{minipage}
\end{table}





\clearpage

\chapter{The Dependence on Multiwavelength Properties}

\section{OVERVIEW}

By investigating the environment overdensity of QIs with different multiwavelength properties, we can begin to determine whether the local environment of QIs is related to the conditions that produce radio or X-ray emission in quasars.  In this chapter, we take advantage of the radio and X-ray information that has been cross-matched to QIs in the DR5 catalog \citep{Schneider}.  

X-ray emission from quasars is tied to the process of material being accreted onto the central black hole.  Thus, X-ray emission originates close to the black hole, and QIs with high X-ray luminosity must have enough material at hand to support the voracious accretion of their central sources.  Radio emission can originate from the accretion process as well.  But more often, radio emission is associated with synchrotron emission from jets produced by material thrown off from the accretion disk, collimated by magnetic field lines twisted by the rotation of the accretion disk.  Radio emission in these lobes can extend far from the central source, even to Mpc scales \citep[e.g., 3C 236;][]{WillisStrom}, and therefore we should expect to see its influence on the local environment \citep[e.g.,][]{Croton}.  Additionally, due to the geometry of radio jets, the orientation at which the quasar is observed will influence the radio-loudness measured.  

\section{RADIO QUASARS}
The DR5 quasar catalog reports the FIRST peak flux density at 20 cm as an AB magnitude.  
The peak flux density is given by \citep{Ivezic}
\begin{equation}
f_{\nu} = (3631 Jy)\cdot10^{\frac{ABmag}{-2.5}}
\end{equation}
Then the luminosity density is calculated using
\begin{equation}
L_{2cm} = f_{\nu}\cdot \frac{4\pi\cdot D_{L}^{2}}{1+z} = f_{\nu}\cdot4\pi\cdot(1+z)D_{M}^{2}
\end{equation}

\begin{figure}
\plotone{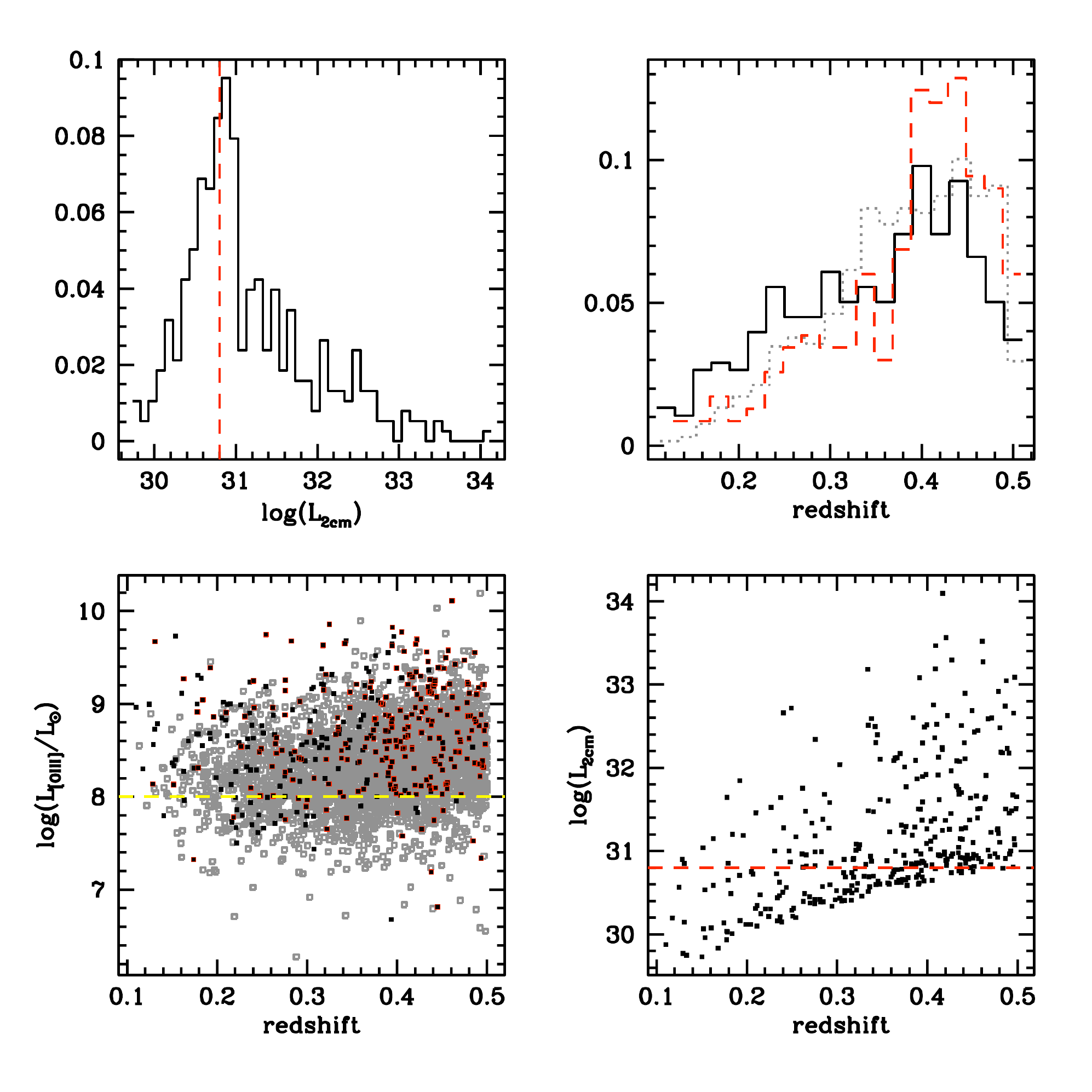}
\caption{\emph{Upper left:}  distribution of radio luminosity for 378 QIs with FIRST detection and $0.11\leq z\leq0.5$. The dashed red line shows the radio luminosity volume-limit cutoff.  \emph{Upper right:}  redshift distribution for 378 QIs with FIRST detection (solid black line) and 3253 QIs with no FIRST detection, but that are still in the FIRST survey area (dotted black line).  The redshift distribution of the ``volume-limited" sample, with log($L_{2cm}$)$\geq30.8$ is shown with the dashed red line; 215 QIs.  \emph{Lower left:}  $\LOIII$ vs. redshift for radio undetected QIs (grey); radio detected QIs (black), where those objects satisfying the radio volume-limit are outlined in red.  The yellow dashed line shows the $\LOIII$ volume-limit that is applied. \emph{Lower right:}  Radio luminosity as a function of redshift for QIs with FIRST detection with log($\LOIII$/$L_{\odot}$)$\geq8.0$.  The dashed red line shows the radio luminosity volume-limit cutoff.  
\label{radiodistr}}
\end{figure}

Figure~\ref{radiodistr} shows the radio luminosity density and redshift distributions of the QI sample with $0.11\leq z\leq0.5$.   In order to ensure that our samples are consistent, we first limit both the radio-detected and radio-undetected samples to $\LOIII$ limit of log($\LOIII$/$L_{\odot}$)$\geq8.0$ (yellow dashed line in the lower left panel).  The obvious flux limit in lower right panel necessitates an volume-limit in radio luminosity.  Our limit of log($L_{2cm}$)$\geq30.8$ (shown by the red dashed lines in the upper left and lower right panels) corresponds to the flux limit of the FIRST survey \citep{Ivezic} at $z=0.5$.  The final radio QI sample contains 215 sources, and the corresponding radio-undetected sample contains 2,569 QIs.  

\begin{figure}
\plotone{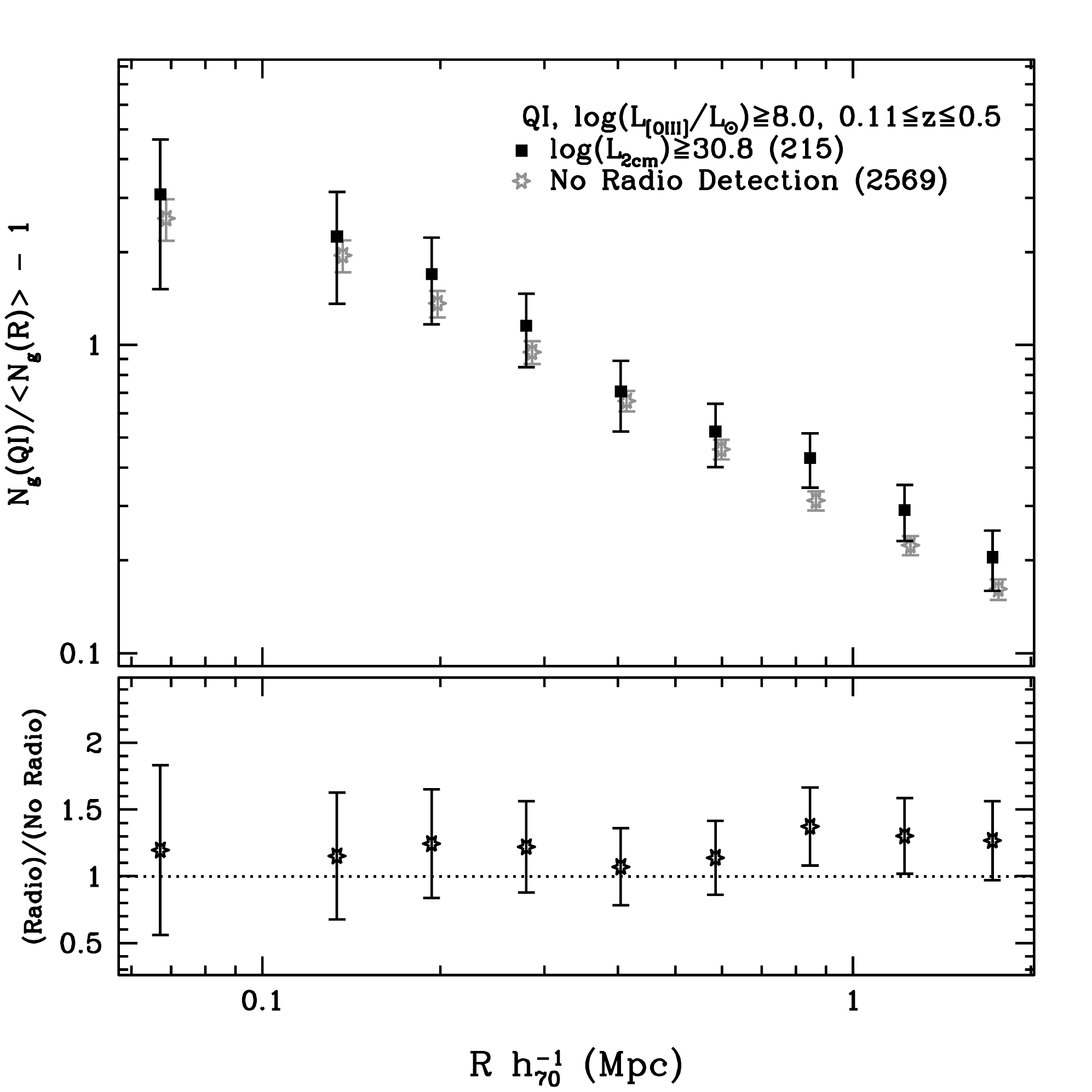}
\caption{\emph{Upper panel:}  Mean cumulative overdensity vs. scale for log($\LOIII$/$L_{\odot}$)$\geq8.0$ and $0.11\leq z\leq0.5$ QIs with and without detections in FIRST (radio detected QIs have log($L_{2cm}$)$\geq30.8$).  \emph{Lower panel:}  Overdensity ratio of FIRST-detected QIs to undetected QIs.  
\label{scale_radio_ptextcombine_q1vollim308}}
\end{figure}

Figure~\ref{scale_radio_ptextcombine_q1vollim308} compares the overdensity of radio detected QIs to undetected QIs that are located in the FIRST survey area.  In addition, overdensity values for these two samples are compared to the overdensity value for the entire QI sample combined at scale $R\approx500\kpchseventy$ in Table~\ref{table_multiwvlgthoverdensitycompare}.  We see very slightly increased overdensity for radio detected QIs compared to undetected QIs with significance $\gtrsim1\sigma$ only at scales $R\gtrsim1.0\Mpchseventy$; at $R\approx1.0\Mpchseventy$, the ratio of overdensity in the environment of radio-detected QIs to that in the environment of radio-undetected QIs is $1.37\pm0.29$ (significance $1.3\sigma$).  

\begin{figure}
\plotone{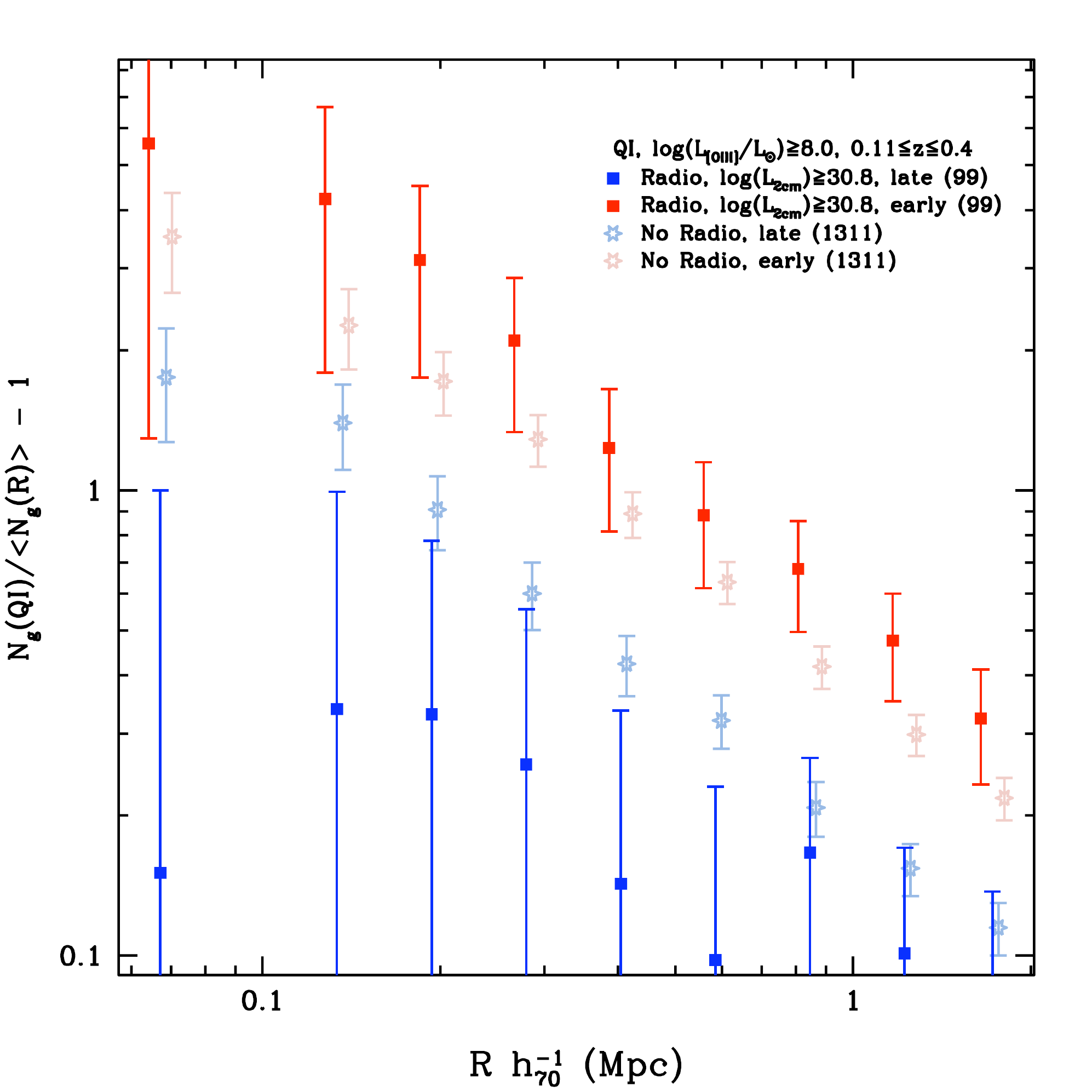}
\caption{Mean cumulative overdensity vs. scale and environment galaxy type for log($\LOIII$/$L_{\odot}$)$\geq8.0$ and $0.11\leq z\leq0.4$ QIs with and without detections in FIRST (radio detected QIs have log($L_{2cm}$)$\geq30.8$).  
\label{scale_radio_ptextcombine_earlylate_q1vollim308}}
\end{figure}

Although the error bars in Figure~\ref{scale_radio_ptextcombine_earlylate_q1vollim308} are very large, we do see clear evidence that the radio detected QIs have a higher overdensity of early-type environment galaxies (at $R\approx500\kpchseventy$, overdensity$=1.23\pm0.418$) and a lower overdensity of late-type environment galaxies (overdensity$=0.143\pm0.194$) compared to the environments of QIs without radio detection (at the same scale, early- and late-type overdensities for undetected QIs are $0.890\pm0.101$ and $0.424\pm0.063$, respectively).  The higher incidence of early-type galaxies in the environments of radio-loud QIs supports the theory that they are located in richer cluster environments \citep[e.g.,][]{Ellingson, Wold2000, McLureDunlop, Best}.  

\section{X-RAY QUASARS}
The DR5 quasar catalog contains the logarithm of the ROSAT All-Sky Survey (RASS) count rate (photons s$^{-1}$) in the broad band (0.1 - 2.4 keV) for each object.  
This count rate must be converted into rest-frame luminosity density for our analysis.  


We use the PIMMS (v3.9b) software to determine the observed flux $f_{0.1-2.4keV}$ from the count rate in the $0.1-2.4$ keV band with $\Gamma= 2.0$, $\alpha= -1.0$ and the corresponding nH (also given in the DR5 quasar catalog) for each object \citep{Vignali}.  


Once the flux is calculated for the $0.1-2.4$ keV band, we calculate the flux density at 2 keV $f_{2keV}$ again using $\alpha= -1.0$:
\begin{equation}\label{flux2fluxdensity}
f_{2keV} = f_{0.1-2.4keV}\cdot\frac{\nu_{2keV}^{\alpha}}{(log(\nu_{2.4keV})-log(\nu_{0.1keV}))}
\end{equation}

We calculate the rest frame luminosity density $L_{2keV}$ from $f_{2keV}$ as follows:

\begin{equation}
L_{2keV} =  \frac{f_{2keV}\cdot4\pi\cdot D_{L}^{2}}{(1+z)}=f_{2keV}\cdot4\pi\cdot(1+z)D_{M}^{2}
\end{equation}
where $D_{L}$ is the luminosity distance and $D_{M}$ is the transverse comoving distance.  

\begin{figure}
\plotone{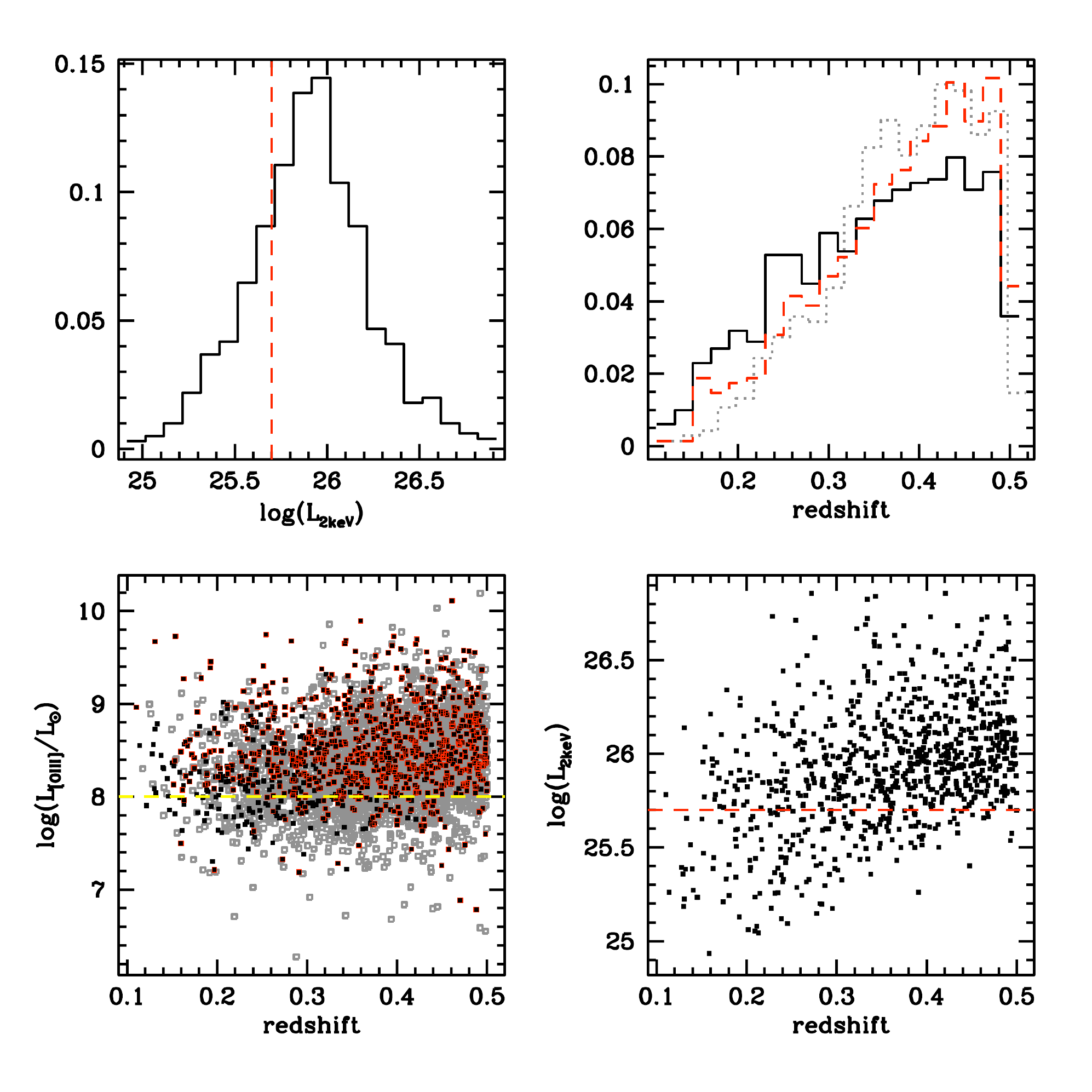}
\caption{\emph{Upper left:}  distribution of X-ray luminosity for 1003 QIs with RASS detection and $0.11\leq z\leq0.5$.  The dashed red line shows the X-ray luminosity volume-limit cutoff.  \emph{Upper right:}  redshift distribution for 1003 QIs with RASS detection (solid black line) and 2790 QIs with no RASS detection (dotted black line).  The redshift distribution of the ``volume-limited" sample, with log($L_{2keV}$)$\geq25.7$ is shown with the dashed red line; 747 QIs.  \emph{Lower left:}  $\LOIII$ vs. redshift for X-ray undetected QIs (grey); X-ray detected QIs (black), where those objects satisfying the X-ray volume-limit are outlined in red.  The yellow dashed line shows the $\LOIII$ volume-limit that is applied.  \emph{Lower right:} X-ray luminosity as a function of redshift for QIs with RASS detection and log($\LOIII$/$L_{\odot}$)$\geq8.0$.  The dashed red line shows the X-ray luminosity volume-limit cutoff.  
\label{xraydistr}}
\end{figure}

Figure~\ref{xraydistr} shows the X-ray luminosity density and redshift distribution of the QI quasar sample with $0.11\leq z\leq0.5$.  Again, we impose a lower $\LOIII$ limit of log($\LOIII$/$L_{\odot}$)$\geq8.0$ on both the X-ray detected and X-ray undetected samples.  Based on the minimum count rate required for inclusion in the RASS \citep[e.g.][]{Voges}, also we impose a lower luminosity density limit of log($L_{2keV}$)$\geq25.7$ for a volume-limited X-ray QI sample.  There are 682 X-ray detected QIs and 2,168 X-ray undetected QIs in our final samples.  

\begin{figure}
\plotone{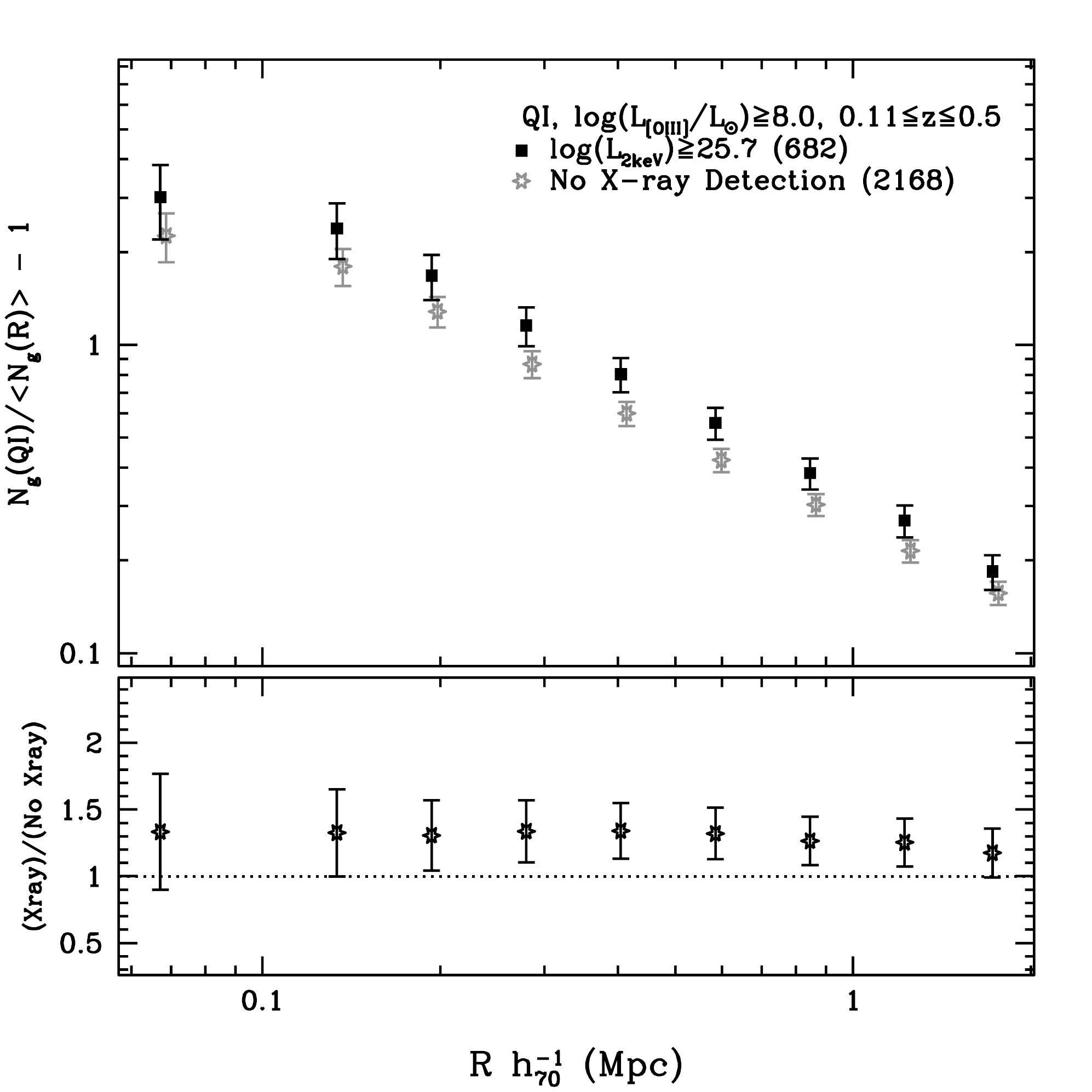}
\caption{\emph{Upper panel:}  Mean cumulative overdensity vs. scale for log($\LOIII$/$L_{\odot}$)$\geq8.0$ and $0.11\leq z\leq0.5$ QIs with and without detections in RASS (X-ray detected QIs have log($L_{2keV}$)$\geq25.7$).  \emph{Lower panel:}  Overdensity ratio of RASS-detected QIs to undetected quasars.  
\label{scale_xray_ptextcombine_q1vollim257}}
\end{figure}

\begin{figure}
\plotone{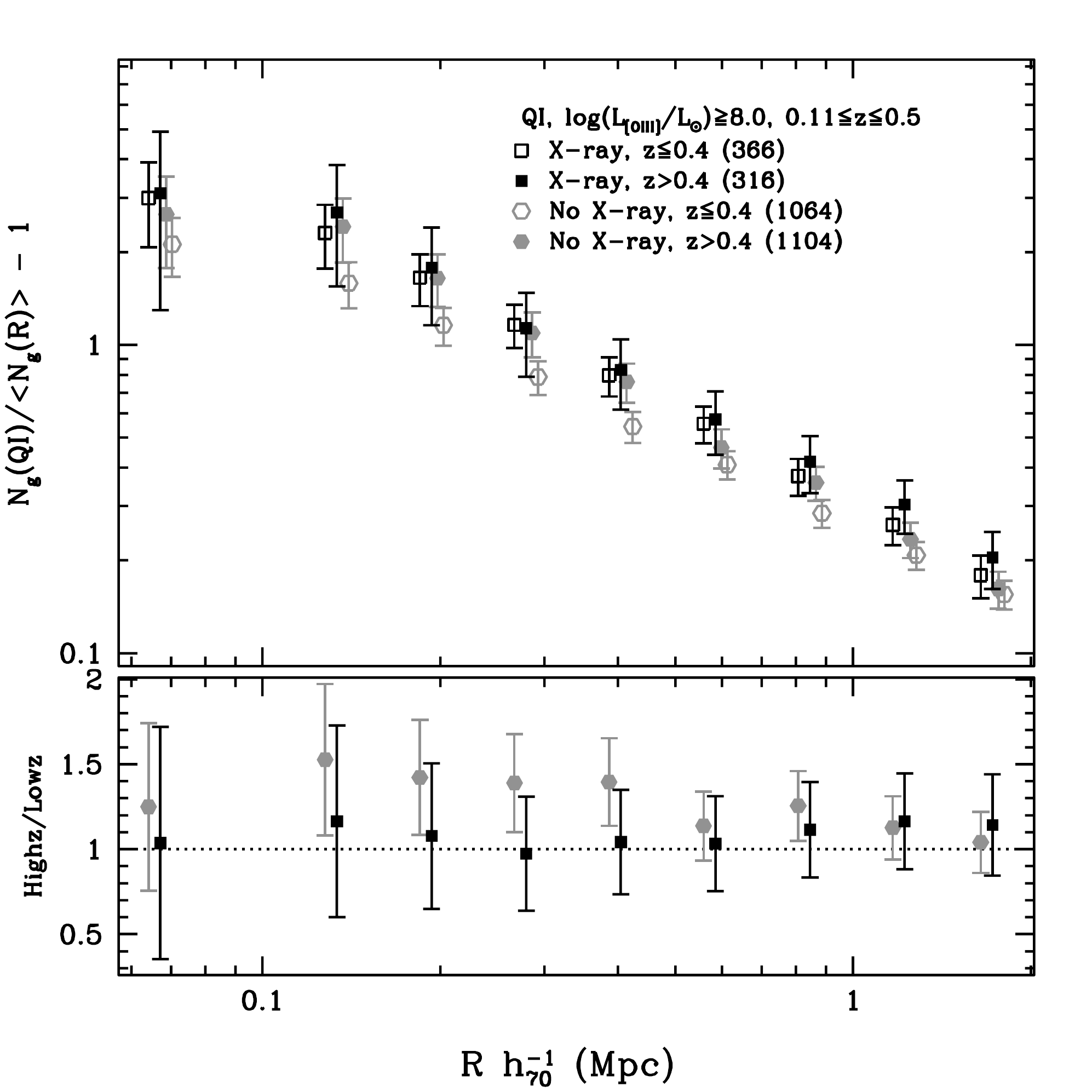}
\caption{Upper panel:  Mean cumulative overdensity vs. scale and redshift for log($\LOIII$/$L_{\odot}$)$\geq8.0$ and $0.11\leq z\leq0.5$ QIs with and without detections in RASS (RASS detected QIs have log($L_{2keV}$)$\geq25.7$).  Lower panel:  Overdensity ratio of $z>0.4$ to $z\leq0.4$ QIs
\label{scale_xray_ptextcombine_zbins_q1vollim257}}
\end{figure}

We compare the environment overdensity of X-ray detected QIs and non-detected QIs in Figure~\ref{scale_xray_ptextcombine_q1vollim257}.  At all scales, X-ray detected QIs are located in environments slightly more overdense than non-detected QIs ($1.34\pm0.21$ times more overdense with significance $1.62\sigma$ at $R\approx500\kpchseventy$).  As shown in Figure~\ref{scale_xray_ptextcombine_zbins_q1vollim257}, there does not appear to be strong redshift evolution of the X-ray detected QI environments, since the ratio of high-redshift X-ray QI overdensity to low-redshift X-ray QI overdensity is consistent with unity on all scales.  We note that there is again evidence for slight redshift evolution in the X-ray undetected QIs, similar to that seen in Chapter~\ref{paper1chapter}.  

\begin{figure}
\plotone{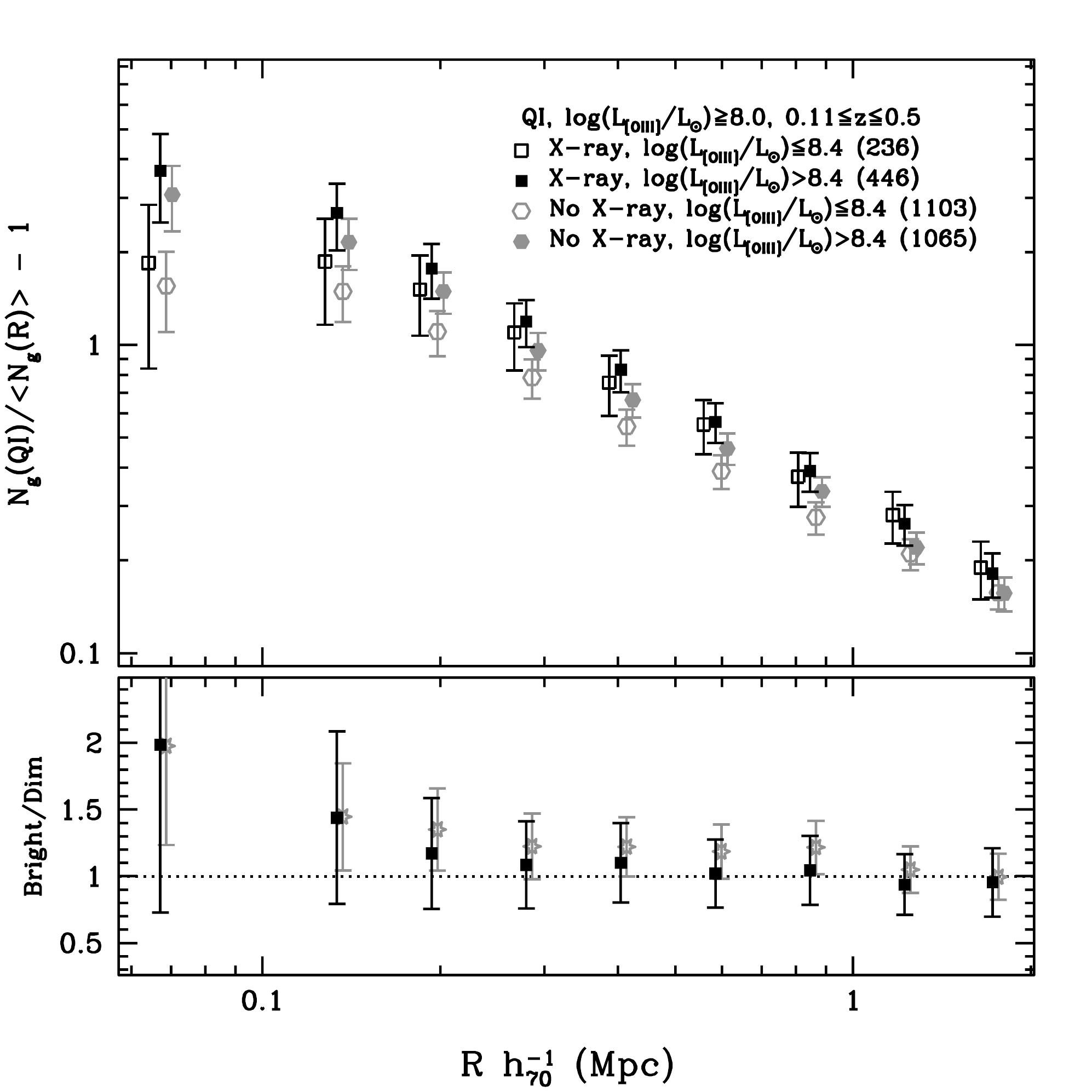}
\caption{\emph{Upper panel:}  Mean cumulative overdensity vs. scale and $\LOIII$ for log($\LOIII$/$L_{\odot}$)$\geq8.0$ and $0.11\leq z\leq0.5$ QIs with and without detections in RASS (X-ray detected QIs have log($L_{2keV}$)$\geq25.7$).  \emph{Lower panel:}  Overdensity ratio of bright $\LOIII$ to dim $\LOIII$ for X-ray detected and X-ray undetected QIs.   
\label{scale_xray_ptextcombine_OIIIsplit_q1vollim257}}
\end{figure}

The X-ray detected and undetected samples are divided into two bins by [OIII] luminosity at log($\LOIII/L_{\odot}$)$=8.4$, the value used in all previous figures.  We plot overdensity vs. scale and $\LOIII$ for these samples in Figure~\ref{scale_xray_ptextcombine_OIIIsplit_q1vollim257}.  We see that although the X-ray detected QIs have slightly higher overdensities at all scales than the undetected QIs (which is consistent with Figure~\ref{scale_xray_ptextcombine_q1vollim257}), the ratio of [OIII] bright QIs to [OIII] dim QIs is about the same within the error bars for both the X-ray detected and undetected samples.  

\begin{figure}
\plotone{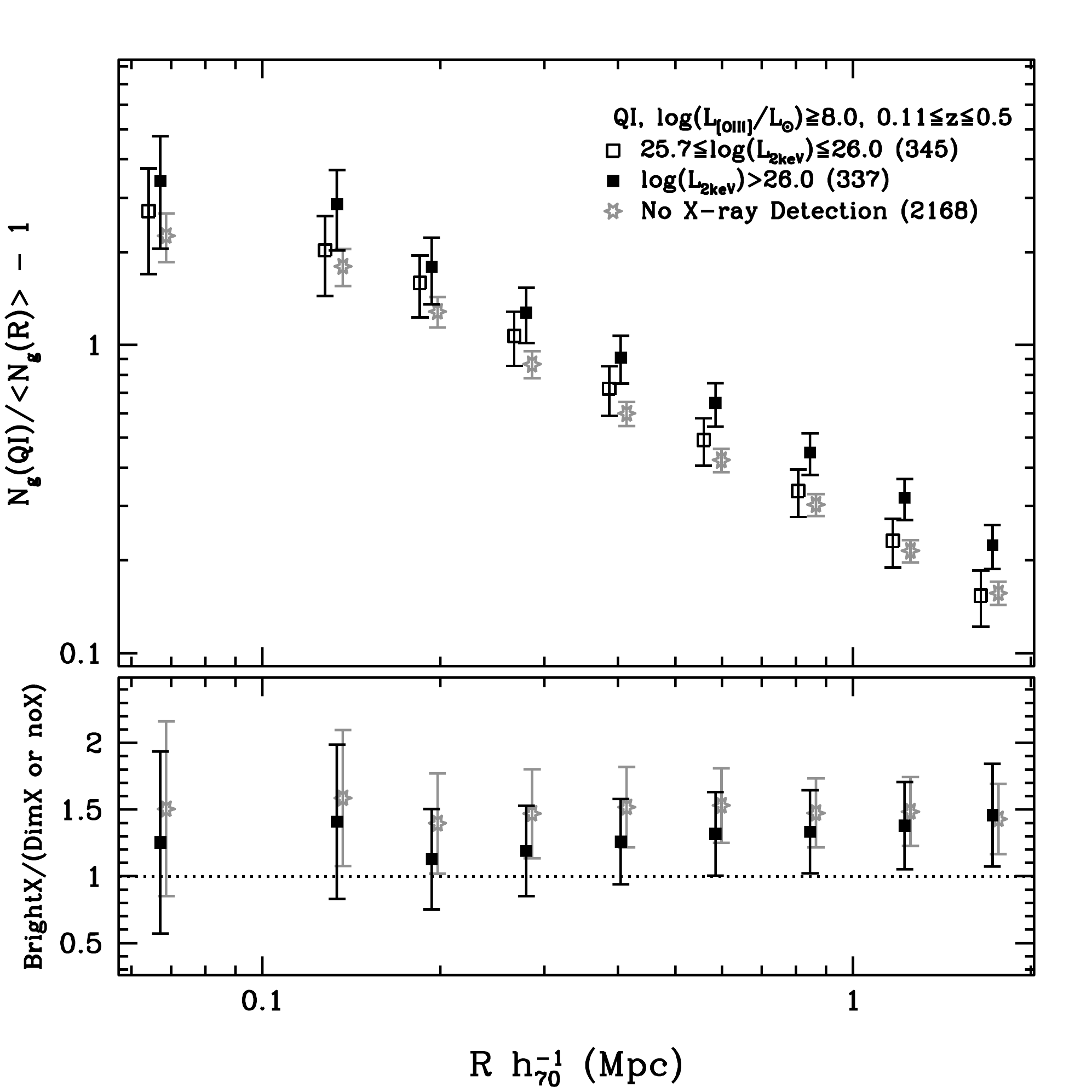}
\caption{\emph{Upper panel:}  Mean cumulative overdensity vs. scale and X-ray luminosity for log($\LOIII$/$L_{\odot}$)$\geq8.0$ and $0.11\leq z\leq0.5$ X-ray detected QIs volume-limited to log($L_{2keV}$)$\geq25.7$).  The gray starred points show the overdensity around QIs not detected by RASS for comparison.  \emph{Lower panel:}  Solid black square points give the overdensity ratio of brighter X-ray QIs to dimmer X-ray QIs.  The grey starred points show the ratio of brighter X-ray QIs to QIs undetected in RASS.  
\label{scale_xray2bins_ptextcombine_q1vollim257}}
\end{figure}

In Figure~\ref{scale_xray2bins_ptextcombine_q1vollim257} we plot mean cumulative overdensity vs. scale for two bins in X-ray luminosity density.  The mean cumulative overdensity around quasars with no RASS detection is plotted in gray for comparison.  The dividing $L_{2keV}$ value is chosen because it gives approximately equal numbers of X-ray QIs in each bin.  On all scales, the brightest X-ray QIs have the highest environment overdensity, though the result is not statistically significant; at $R\approx500\kpchseventy$, the ratio of overdensity for X-ray bright QIs to X-ray dim QIs is $1.26\pm0.320$, with significance $<1\sigma$. However, the overdensities of both X-ray QI samples are higher than the X-ray undetected QI overdensities on most scales.  The brightest X-ray QIs have an overdensity $1.52\pm0.301$ times the overdensity of the undetected QIs with a significance $1.7\sigma$ at the scale $R\approx500\kpchseventy$.  

\begin{figure}
\plotone{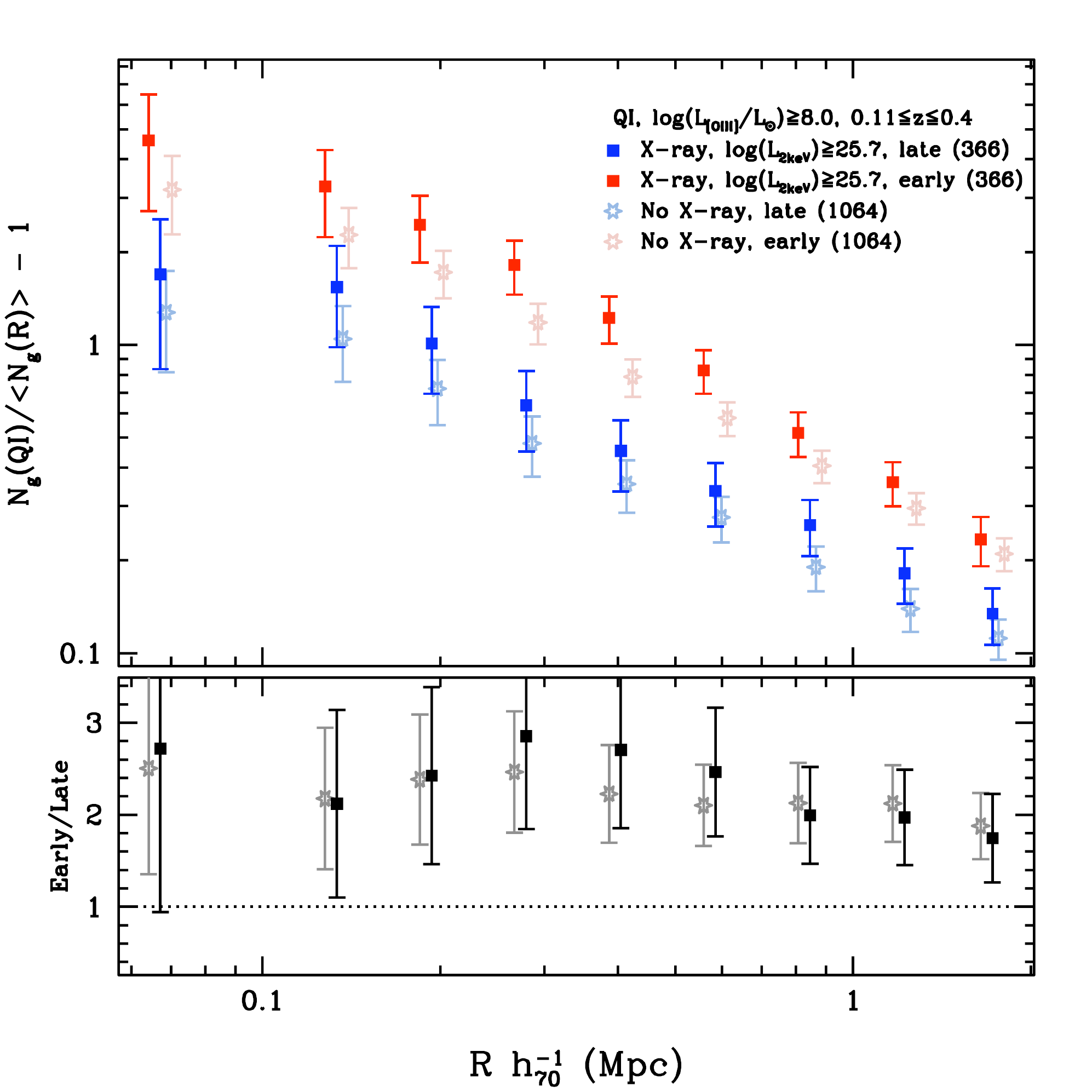}
\caption{\emph{Upper panel:}  Mean cumulative overdensity vs. scale and environment galaxy type for log($\LOIII$/$L_{\odot}$)$\geq8.0$ and $0.11\leq z\leq0.4$ QIs with and without detections in RASS (X-ray detected QIs have log($L_{2keV}$)$\geq25.7$).  \emph{Lower panel:}  Overdensity ratio of RASS-detected QIs to undetected QIs.  
\label{scale_xray_ptextcombine_earlylate_q1vollim257}}
\end{figure}

Finally, Figure~\ref{scale_xray_ptextcombine_earlylate_q1vollim257} shows the early-type and late-type galaxy overdensity around X-ray detected and undetected QIs.  In Table~\ref{table_multiwvlgthoverdensitycompare}, we compare overdensity values at scale $R\approx500\kpchseventy$ for X-ray detected QIs to X-ray undetected QIs.  Although the overdensity values for both early- and late-type environment galaxies are higher on all scales for X-ray detected QIs, it appears that the early- to late-type environment overdensity ratio is about the same and is scale-independent within the error bars.   

\section{CONCLUSIONS}
We have shown some evidence for increased overdensity around radio QIs, but larger samples are required for greater statistical significance.  We cannot make strong claims about the relationship between radio QIs and their environments due to our lack of knowledge of their radio morphology.   This morphology information could allow us to the differentiate radio emission due to accretion from radio emission due to jet activity.  

The fact that QIs with higher X-ray luminosity are shown to have more densely populated environments is not surprising, given our previous results showing that TIs with bright $\LOIII$ are in the most overdense environments, since [OIII]$~\rm\lambda5007$ emission line luminosity and hard ($2-10\rm{KeV}$) X-ray luminosity have been shown to be strongly correlated \citep{Heckman2005}.  We reiterate the need for larger samples of X-ray QIs in order to draw stronger conclusions about their relationship with their environments.  


\begin{table} \begin{minipage}{140mm}
\caption{Comparison of environment overdensity for various samples of QIs with log($\LOIII/L_{\odot}$)$\geq8.0$ at scale $R\approx500\kpchseventy$.  \label{table_multiwvlgthoverdensitycompare}}
\begin{tabular}{c c c c}\hline\hline
Sample&$z\leq0.5$&$z\leq0.4$, early&$z\leq0.4$, late\\ \hline
all QI&$0.644\pm0.045$&$0.877\pm0.090$&$0.391\pm0.056$\\
radio QI&$0.706\pm0.182$&$1.23\pm0.418$&$0.143\pm0.194$\\
no radio QI&$0.659\pm0.051$&$0.890\pm0.101$&$0.424\pm0.063$\\
X-ray QI&$0.803\pm0.102$&$1.22\pm0.214$&$0.453\pm0.118$\\
no X-ray QI&$0.599\pm0.055$&$0.788\pm0.109$&$0.354\pm0.069$
\end{tabular} \end{minipage}
\end{table}


\clearpage

\clearpage

\chapter{Conclusions \& Future Work}

Our results support recently proposed merger models of quasar origin \citep{Hopkins2008}.  On small scales, we measure an increased overdensity around the brightest sources that is scale-dependent, providing evidence that the fueling mechanisms for bright quasars are merger-related \citep{Djorgovski, Hennawi, Myers07b, LaceyCole}.  Lower luminosity AGN do not show the increasing overdensity at smaller scales, leading us to conclude that they are fueled by mechanisms less dependent on the local galaxy environment \citep{Li2006, Li2008}.  On $\sim$Mpc scales, bright and dim quasars have approximately equal overdensities, which implies that they occupy dark matter halos of similar mass.  This result is consistent with the predictions of the merger models that quasars of different luminosities are mainly similar objects radiating at different times in their evolution \citep{Lidz, Hopkins2005a}.  

In addition, we investigate the nature of the quasar and AGN environments themselves.  We find that type I sources are located in environments with a higher overdensity of early-type galaxies than late-type galaxies, and that brighter TIs have an increasing environment overdensity of early-type galaxies on small scales.  A higher overdensity of early-type galaxies on small scales could be evidence that feedback affects nearby galaxies outside the host galaxy, quenching or suppressing star formation \citep{Croton}.  Type II sources have a lower overdensity of early-type galaxies in their environments but similar overdensities of late-type galaxies.  We speculate that TIIs are being observed at a different stage of their evolution compared to TIs and/or are in poorer clusters  than TIs.  

Additional information in our data set allows us to focus on the QIs alone and measure the relationship of QI black hole mass with environment overdensity.  The increased overdensity with higher mass is consistent with the $M_{\rm DMH}-\MBH$ relationship \citep{FerrMerr, Gebhardt, Tremaine}.  The change in the early-to-late type environment galaxy ratio for QIs supports the conclusion that QIs with higher $\MBH$ are located in richer clusters \citep{Zehavi2005}.  Additionally, we find that QIs with low accretion efficiency have a slightly higher environment overdensity compared to high efficiency QIs, which may be evidence of quenching due to accretion feedback \citep{ScannapiecoOh}.  

Finally, our exploration of QI environments and their relationship to QI multiwavelength properties enforced the need for larger samples with deeper radio and X-ray observations, which would allow us to sample a larger range of the radio and X-ray luminosity functions.  Additionally, valuable insight about the relationship of radio sources with their environments could be gained with detailed radio morphology information.  When we measure marginal evolution in the QI environment overdensity with redshift in the range $0.11\leq z\leq0.5$, we determine that it is not the primary factor influencing the increased overdensity of QIs compared to the lower luminosity AGNs.  However, whether there is significant evolution in quasar environments at higher redshifts can be determined by studying the environments of higher redshift quasars using data from deeper surveys such as the Cosmic Evolution Survey (COSMOS).  The two-square-degree field studied by COSMOS has been imaged in the optical as well as across other wavelengths using a combination of space-based and ground-based telescopes.  

Measurements of additional quasar properties will allow us to hone in even deeper on the links between quasars and their environments.  For example, the Large Synoptic Survey Telescope (LSST) will measure quasar variability, which is known to be linked to physical properties of quasars and their fueling mechanisms \citep[e.g.,][]{Wilhite}.  Studying how quasar variability is liked to quasar environments would help us to understand in more detail how the environment influences quasar fueling or vice versa.  

Historically, investigations of quasar and AGN environments have been performed using galaxy-counting techniques similar to those we employ in this dissertation.  However, now that large data sets are available, and we have the promise of even larger ones, it is becoming possible to transition studies of environments to use correlation and marked correlation analyses \citep[e.g.,][]{KauffmannHaehnelt, AdelbergerSteidel2005, WhitePadman}.  The advantage of correlation and cross-correlation measurements is that they can be more directly interpreted via models such as halo occupation distribution models \citep[e.g.,][]{CooraySheth} to constrain theories of structure formation.  

\appendix
\chapter{Documentation for Bincounting Code Package}\label{bincountDoc}
\section{OVERVIEW}
This appendix details the Bincounting Code (BC) package (written in C) developed as part of this dissertation to study the environments of AGNs and quasars in the Sloan Digital Sky Survey.  This package includes code to make a binary file from environment galaxy data or negative random position data (for masking), mask target data using binary files containing negative random position data, and count environment galaxies within defined (comoving) radii of target objects using the binary environment galaxy data files.  

\section{HEADER FILE: bincountsfunctions.h}
The header file contains definitions for all of the functions used in the BC package as well as global variable definitions.  
\begin{itemize}
\item H0 : value of $h_{0}$, such that the Hubble constant $H_{0}=100~h_{0} \kms \rm Mpc^{-1}$.  Default value $= 0.7$.
\item MIND : minimum distance in units of $\Mpch$ from the target coordinate within which objects will be counted.  Default value $=0.025$.
\item MAXD : maximum distance in units of $\Mpch$ from the target coordinate within which obejcts will be counted.  Default value $=10.0$.  
\item NUMBINS :  number of bins in which the objects are (differentially) counted.  This value is used in the binDistance function to select the correct list of maximum bin edges.  Default value $= 14$ for data used in overdensity vs. scale plots.  
\item T1MAX : photometric galaxy type classification cut (see Section~\ref{4types} for details).  Galaxies with spectral-type parameter $t < 0.02$ correspond to the CWW Ell template \citep{Budavari}.  Used for subdividing environment galaxies beyond the early/late division.  Default value $= 0.02$.  
\item T2MAX:  photometric galaxy type classification cut (see Section~\ref{4types} for details).  Galaxies with spectral-type parameter $0.02 < t < 0.3$ approximately correspond to the CWW Sbc template \citep{Budavari}.  Used for subdividing environment galaxies beyond the early/late division.  Default value $= 0.3$.    
\item T3MAX:  photometric galaxy type classification cut (see Section~\ref{4types} for details).  Galaxies with spectral-type parameter $0.3 < t < 0.65$ approximately correspond to the CWW Scd template, and galaxies with $t > 0.65$ approximately correspond to the CWW Irr template \citep{Budavari}.  Used for subdividing environment galaxies beyond the early/late division.  Default value $= 0.65$.  

\item radians(x) : function to convert $x$ to radians
\item degrees(x) : function to convert $x$ to degrees
\end{itemize}

We also define new object types that are used throughout the code:
\begin{itemize}
\item Coord : struct to contain only right ascension and declination of an object
\item zphotCoord :  struct to contain information about photometric galaxies.  Contains right ascension and declination of object, as well as photometric redshift and spectral type parameter as defined by \citet{Budavari}.  
\item zCoord : struct to contain information about targets.  Contains right ascension, declination, and redshift of object.  Also contains comoving distance corresponding to the object's redshift and a string that identifies the object, as well as lists in which (differential) bincounts are stored.  
\item DecDic : ``declination dictionary" created to speed up object searching in the bincounting code.  See Section~\ref{algorithm} for details of its usage.  
\end{itemize}

\section{GENERAL FUNCTIONS: bincountsfunctions.c}
This section contains functions that are used throughout the BC package.  
\begin{itemize}
\item mysort\_ra : given two Coord objects, sort them (Boolean) by their right ascension.  Returns int: 1 if first coordinate is greater than second coordinate, 0 if coordinates are equal, and -1 if second coordinate is greater than first coordinate.  
\item mysort\_dec : given two Coord objects, sort them (Boolean) by their declination.  Returns int (see mysort\_ra).
\item mysort\_dec\_photoz : given two zphotCoord objects, sort (Boolean) by their declination.  Returns int (see mysort\_ra).
\item mysort\_zphot\_photoz : given two two zphotCoord objects, sort (Boolean) by their photometric redshift.  Returns int (see mysort\_ra).
\item minimum : given two doubles, return double of lesser value.  
\item angularDistance : given right ascension and declination (doubles) of two points, use Haversine formula to calcuate angular distance between them.  Coordinates should be entered in degrees; output is in radians (double).  
\item angDiamDistance : given a zCoord object, calculate the angular diameter distance (double) corresponding to its comoving distance.  
\item angle2Mpc : given an angle in radians (double) and a zCoord object, returns the comoving distance in Mpc (double) corresponding to that angle at the redshift of the zCoord object.  
\item Mpc2angle : given a comoving distance in Mpc (double) and a zCoord object, returns the angle in radians (double) corresponding to that comoving distance at the redshift of the zCoord object.  
\item binDistance : given a physical distance (double) between a target and a photometric galaxy, find the bin (bin maxima correspond to value of NUMBINS global variable, see below) in which the galaxy falls and increment the counts in that bin (list of bincounts given as the second argument).  
\item binZ : function to create N(z) for photometric galaxy samples.  
\end{itemize}

The binDistance function uses the value of the NUMBINS global variable to determine the bin maxima to be used in the bincounting.  The binEdges lists are defined at the beginning of the bincountsfunctions.c file.  

\section{CREATE BINARY DATA FILES: \\sortbydec.c and sortbydec\_zphot.c}\label{sort}
\subsection{Negative Random Data for Masking}\label{sort_negran}
In order to mask target data using the BC package, binary files with negative random position data must be created.  These negative random data are points (right ascension, declination) that correspond to ``bad" areas of the sky.  The sortbydec.c program opens an ASCII file (no header line or other columns should appear in the file) with the negative random positions and sorts them by declination.  The code outputs a binary file containing the exact number of objects in the input file and a binary file containing the negative random data sorted by declination.\\

$Command~line~usage:$
\begin{verbatim}
>> gcc -o sortbydec bincountsfunctions.c sortbydec.c -lm
>> sortbydec <catalogname> <binarydatafilename> 
             <binarylinecountname> <approx#ofobjects>
\end{verbatim}

The final argument is an approximate number of objects (long) contained in the catalog (must be greater than actual number) used to allocate enough memory for the code.  

\subsection{Environment Galaxy Data for Bincounting}
The bincounting code also requires binary files containing the environment galaxy data.  The input ASCII file of environment galaxy data must have the columns (and only these columns, with no header) right ascension, declination, redshift, and spectral type parameter.  The code then outputs a binary file containing the exact number of objects in the input file and a binary file containing the environment galaxy data sorted by declination.\\

$Command~line~usage:$
\begin{verbatim}
>> gcc -o sortbydec_zphot bincountsfunctions.c 
                  sortbydec_zphot.c -lm
>> sortbydec <catalogname> <binarydatafilename> 
             <binarylinecountname> <approx#ofobjects>
\end{verbatim}

The final argument is an approximate number of objects (long) contained in the catalog (must be greater than actual number) used to allocate enough memory for the code.  

\section{MASK TARGET DATA: maskforbincount.c}\label{mask}
Target data can be masked using the binary files of negative random data with maskforbincount.c.  This code reads in the entire catalog of negative randoms, then opens the file of targets and reads them in one at a time, determining whether there are negative random points within the maximum radius (as determined by the binDistance function) and writes out only those targets with zero negative random positions within that maximum radius.\\

$Command~line~usage:$
\begin{verbatim}
>> gcc -o maskforbincount bincountsfunctions.c 
               maskforbincount.c -lm
>> maskforbincount <binarylinecountname> <binarydatafilename> 
               <targetcatalogname> <outputfilename> 
               <approx#oftargets>
\end{verbatim}

The final argument is an approximate number of objects (long) contained in the target catalog (must be greater than actual number).  

\section{COUNT ENVIRONMENT GALAXIES: \\bincount\_zphotandztype.c and bincount\_zphot4types.c}
\subsection{Bincounting with All Galaxies and Red-Blue Galaxy Separation}\label{redblue}
The code bincount\_zphotandztype.c is used to count the number of environment objects within a specified distance and redshift interval from the target.  There are three output files:  the first gives the overall bincounts; the second, bincounts of only early-type galaxies; and the third, bincounts of only late-type galaxies.  (Summing the bincounts given in the second two output files should give the counts in the first output file!)\\

$Command~line~usage:$
\begin{verbatim}
>> gcc -o bincount_zphotandztype bincountsfunctions.c 
               bincount_zphotandztype.c -lm
>> bincount_zphotandztype <binarylinecountname> 
               <binarydatafilename> <targetcatalogname> 
               <outputfilename> <outputfilename_earlytype> 
               <outputfilename_latetype> <approx#oftargets> 
               <cumoption> <deltaz> <typesplit>
\end{verbatim}

Command line arguments:
\begin{itemize}
\item linecount filename: binary file produced by sortbydec\_zphot.c corresponding to environment object data file
\item environment object filename: binary file produced by sortbydec\_zphot.c
\item target object filename: ASCII columns id, ra, dec, redshift, comovingdistance, no header line
\item total output filename: will have columns id, bincountMaxR-thru-bincountMinR
\item early-type output filename: will have columns id, bincountMaxR-thru-bincountMinR
\item late-type output filename: will have columns id, bincountMaxR-thru-bincountMinR
\item number of targets: integer (can be approx but must be $>$ actual number)
\item cum option:  cum = 1 if cumulative counts, cum = 0 if differential counts, best if 0
\item deltaz value: the environment objects must have a redshift within this separation from the target redshift to be counted in a radius bin
\item type value split:  value at which early and late type galaxies are to be separated.  According to 
\citet{Budavari}, the best value for this split is $0.3$.  
\end{itemize}

\subsection{Bincounting with \citet{Budavari} Galaxy Subgroups}\label{4types}
The process is almost identical to that of Section~\ref{redblue}, except there are further divisions imposed on the photometric galaxies in the environment.  Additionally, the output files \emph{do not} include a file that gives the total number of galaxies within each counting radius.  

Galaxies with spectral-type parameter $t < 0.02$ (T1) correspond to the CWW Ell template; $0.02 < t < 0.3$ (T2) approximately correspond to the CWW Sbc template; $0.3 < t < 0.65$ (T3) approximately correspond to the CWW Scd template, and galaxies with $t > 0.65$ (T4) approximately correspond to the CWW Irr template \citep{Budavari}.  \\

$Command~line~usage:$
\begin{verbatim}
>> gcc -o bincount_zphot4ztypes bincountsfunctions.c 
                  bincount_zphot4ztypes.c -lm
>> bincount_zphot4ztypes <binarylinecountname> 
                  <binarydatafilename> <targetcatalogname> 
                  <T1outputfilename> <T2outputfilename> 
                  <T3outputfilename> <T4outputfilename>
                  <approx#oftargets> <cumoption> <deltaz> 
\end{verbatim}

Command line arguments:
\begin{itemize}
\item linecount filename: binary file produced by sortbydec\_zphot.c corresponding to environment object data file
\item environment object filename: binary file produced by sortbydec\_zphot.c
\item target object filename: ASCII columns id, ra, dec, redshift, comovingdistance, no header line
\item total output filename: will have columns id, bincountMaxR-thru-bincountMinR
\item T1 output filename: will have columns id, bincountMaxR-thru-bincountMinR
\item T2 output filename: will have columns id, bincountMaxR-thru-bincountMinR
\item T3 output filename: will have columns id, bincountMaxR-thru-bincountMinR
\item T4 output filename: will have columns id, bincountMaxR-thru-bincountMinR
\item number of targets: integer (can be approx but must be $>$ actual number)
\item cum option:  cum = 1 if cumulative counts, cum = 0 if differential counts, best if 0
\item deltaz value: the environment objects must have a redshift within this separation from the target redshift to be counted in a radius bin
\end{itemize}

\section{BINCOUNTING ALGORITHM}\label{algorithm}

Rather than do a brute-force run through the galaxies in order to find those that are in the defined nearby neighborhood of the target position, the bincounting code creates what we call a ``Dec Dictionary" to index the objects by declination.  

The pre-sorted environment galaxy binary data file (and corresponding linecount file) is read into the code.  The galaxy list is divided into fifty equal increments, and the ``Dec Dictionary" holds the index and declination of the object at the end of each of the increments.  Thus, when a target object is read from the file, the code jumps to the increment closest to and less than the target's declination minus the maximum distance set by MAXD.  

The targets are read from the target data file one-by-one and a minimum and maximum RA and dec are calculated based on the maximum distance set by MAXD.  Beginning at the object selected using the Dec Dictionary, each potential environment galaxy is tested to find if it falls within the maximum and minimum coordinate box.  If it does, the full distance calculation is invoked and that distance is sent to the binDistance function for bincounting.  The counts in each bin are printed to a file with the object's ID, redshift, and coordinates.  

\chapter{Photometric Redshift Estimation for SDSS Quasars} 

\section{INTRODUCTION}     

Electromagnetic radiation from a distant object loses energy in its transit through the universe due to the universe's expansion.  This results in a net shift of the radiation toward the red end of the spectrum, an effect known as cosmological redshift.  The most accurate way to find an object's redshift is to acquire its spectrum and measure how far known spectral lines are shifted from their vacuum wavelengths.  However, it can be time-consuming to take spectra for large samples of objects, whereas photometric (or broad-band) measurements can be made much more efficiently.  Therefore, large samples of objects with photometric redshift data can be available for use in cosmological calculations that until now have been restricted by sample size.  For instance, large scale structure in the universe has been studied by using galaxies with photometric redshift information.  Because quasars are more easily observed at higher redshifts than galaxies, accurate photometric redshift information for quasars is extremely valuable for probing large scale structure closer to the earliest times in the universe.  

\subsection{Photometry and Colors}
Broadband filters on a telescope measure the observed flux within a certain wavelength range.  Our data comes from the Sloan Digital Sky Survey (SDSS), which uses a dedicated telescope located at the Apache Point Observatory in New Mexico.  The telescope measures flux through five broadband filters: $u$, $g$, $r$, $i$, and $z$ with average weighted filter wavelengths of 3551, 4686, 6165, 7481, and 8931 Angstroms, respectively \citep{York2000}.  A magnitude is calculated by taking a logarithm of the flux measured in a specific wavelength band \citep{LuptonGunnSzalay}.  Colors are formally defined as the difference between two magnitudes, or equivalently, by the ratio of fluxes through the two different bands; this ratio is independent of redshift.  

\subsection{Previous Photometric Redshift Work: Galaxies}
Photometric redshift estimation techniques were pioneered with galaxies and are particularly effective for galaxies because of distinctive breaks in their spectral energy distributions (SEDs), such as the $4000\rm\AA$~break.  The concept was first introduced by \citet{Baum}, who measured multiple colors of galaxies and used them to estimate the location of the $4000\rm\AA$~break, thereby determining the galaxy's redshift, and \citet{Koo} and \citet{LohSpillar} expanded the technique to larger samples of galaxies.  More recently, work by \citet{Brunner} and \citet{Connolly} matched broad-band galaxy colors to colors estimated using galaxy SEDs generated from theoretical models with excellent results.  

\subsection{Previous Photometric Redshift Work: Quasars} 
The average lower-redshift quasar spectrum can be approximated by a power law, which alone would give an uninformative color-redshift relation.  High redshift quasars have a distinctive break in their optical spectra at the Lyman-$\rm\alpha$ forest, but this feature does not affect the quasar spectrum enough to use for redshift estimation until $z\gtrsim3.0$.  Fortunately, quasars have prominent emission and absorption features that affect the colors measured by broadband photometry \citep[see detailed discussion in][]{RichardsA} enough to give the color-redshift relations shapes that are very often nondegenerate in our four-dimensional color space. \citet{RichardsA} demonstrated that the empirical color-redshift relation (CZR) for quasars using four colors determined from the five SDSS broadband filters has structure that can be exploited to understand quasars themselves, as well as to predict photometric redshifts.    
   
   Photometric redshift estimation techniques for quasars have improved significantly from about 50\% of quasars with photometric redshifts estimated within 0.3 of the actual spectroscopic redshift \citep{Hatzi} to nearly 70\% of quasars having estimated photometric redshifts within 0.2 of the spectroscopic redshift  using a $\chi^2$ minimization technique between observed colors and colors predicted by a CZR \citep{RichardsB}.  Using a probability density function (PDF) derived from a $\chi^2$ distribution, \citet{Weinstein} reports a success rate of over 77\% of quasars within 0.2 of the spectroscopic redshift.  A benefit of this probability density function method is that it also gives the probability that the estimated redshift is correct \citep{Weinstein}.  However, even with these improved results, there is need for still more improvement before these methods produce reliable photometric redshifts suitable for cosmological applications.  
        
   This chapter discusses our implementation of a number of improvements on the photometric redshift determination algorithm developed by \cite{Weinstein} in an attempt to achieve higher accuracy photometric redshift estimates.  
  
\section{DATA SAMPLE}

\begin{figure}
\plotone{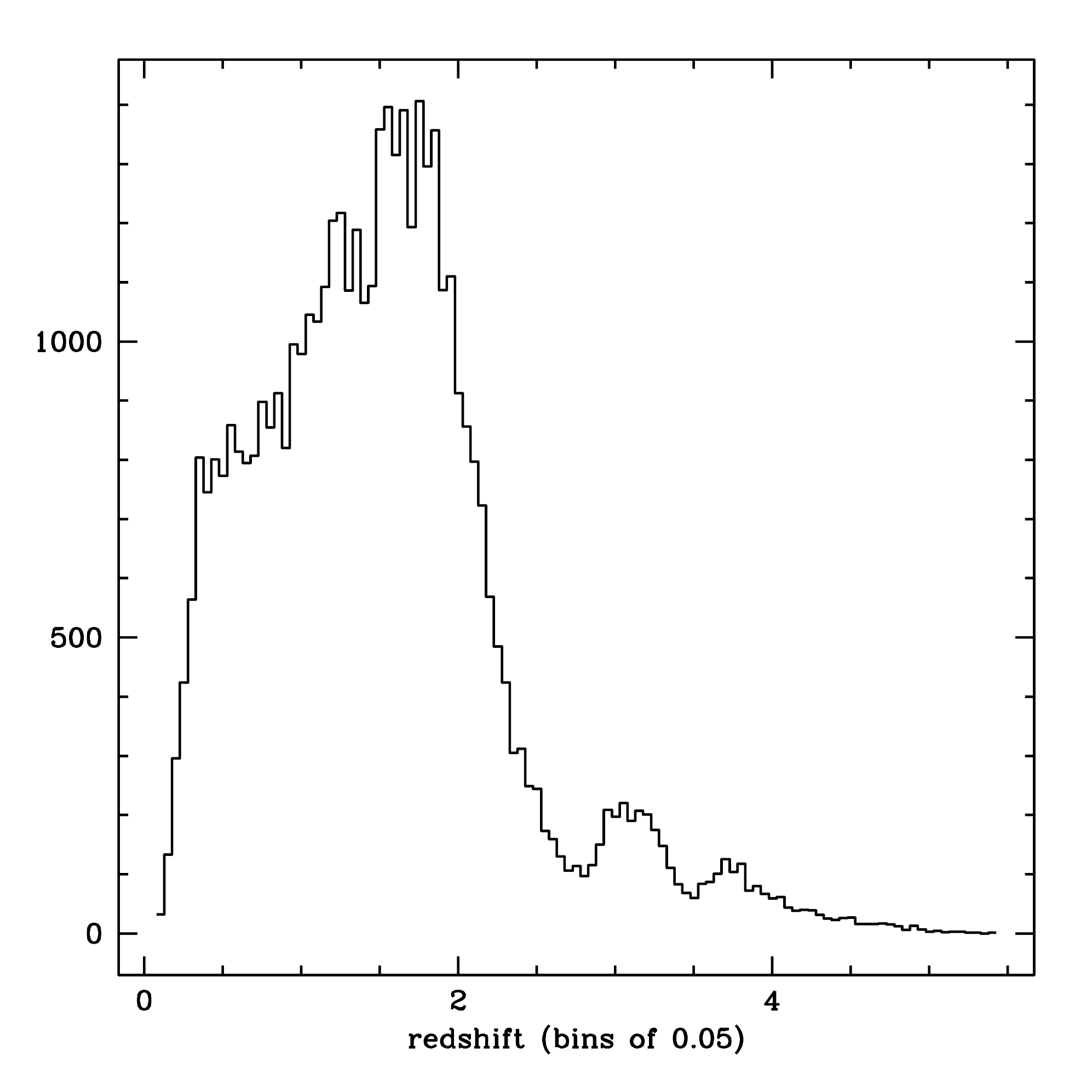}
\caption{Redshift distribution of the SDSS DR3 quasar sample.  
\label{zhist_SDSSDR3qsos}}
\end{figure}

    We use the SDSS Third Data Release \citep[DR3; ][]{Abazajian} Quasar Catalog created by \citet{SchneiderDR3}.  The 46,420 quasars in the catalog were selected from DR3 on the basis of their spectra containing at least one broad emission line or their being unquestionably broad absorption line quasars.  The objects have luminosities $M_{i}>-22.0$ and a bright limit in the $i$-band of 15.0.  Spectroscopic redshifts of quasars in the catalog range from 0.078 to 5.4135.  Figure~\ref{zhist_SDSSDR3qsos} shows the redshift distribution of the sample.  
    
     The catalog contains only point-spread function (PSF) magnitudes, so we matched the quasars to the DR3 SpecPhotoAll Database by plate and fiber numbers to obtain the model magnitudes, fiber magnitudes, and Petrosian magnitudes for each of the objects in the catalog (these magnitude types will be discussed in more detail in Section~\ref{magtypesection}.  One object from the DR3 Quasar Catalog does not have a match in the SpecPhotoAll database and is excluded from our calculations.  While there are PSF magnitudes and errors measured in all bands for all remaining 46,419 objects,  there are a few objects that have ``null" or ``-9999" values for one or more magnitude and/or error measurements in the other magnitude types.  Objects with any invalid field are therefore excluded from our calculations of color-redshift relations when we use those magnitudes.  
    
\section{TECHNIQUE}
\subsection{Recreation of \citet{Weinstein} Algorithm}
Following the matrix method outlined in \citet[][hereafter W04 algorithm]{Weinstein}, we calculate the color-redshift relation (CZR) and its associated ``covariance matrices".  The method is summarized in the following subsections.     

\subsubsection{Removing reddened quasars}\label{RedQsoSection}
     We follow the scheme of \cite{WeinsteinThesis} to determine (and subsequently discard from the sample used to calculate the CZR) those quasars in the input sample which are anomalously red.  First, the quasars are binned by redshift, and the discrete median colors in each bin are calculated.  Next, the ``normalized colors" \citep[][W04]{RichardsA} for $u-g$ and $g-r$ are calculated for every quasar in the input sample.  The normalized color is defined as
      \begin{equation}  
      (u-g)_{norm} = (u-g)_{measured} - (u-g)_{median} 
      \end{equation}
      and
      \begin{equation}
      (g-r)_{norm} = (g-r)_{measured} - (g-r)_{median}
      \end{equation}
where the median color is calculated at the quasar's redshift.  Since the quasar will likely not be exactly at a redshift bin center, a linear interpolation is made between the two bin centers that flank the quasar's redshift to determine the median color used in the formula above.  

When the normalized colors have been calculated for all the quasars in the sample, they are sorted from bluest to reddest (smallest to largest normalized color value).  The ``red limit" in $u-g$ and $g-r$ is defined as the 2.5th percentile of the $u-g$ normalized color ($u-g$)$_{2.5}$ and the 50th percentile of the $g-r$ normalized color ($g-r$)$_{50}$.  A quasar is considered anomalously reddened if both of the following conditions are true \citep{WeinsteinThesis}:
      \begin{equation}  
      (u-g)_{norm} > -(u-g)_{2.5} 
      \end{equation}
      and
      \begin{equation}
      (g-r)_{norm} > (g-r)_{50}
      \end{equation}
The DR3 quasar sample is plotted by color as a function of redshift in Figure~\ref{qsoCZR_red}, and we show the distribution of reddened quasars.  

\subsubsection{Calculating CZR and Covariance Matrix Values}
Once the reddened quasars have been identified and removed from the inital data set, the CZR and corresponding covariance matrices are calculated from the remaining non-reddened quasars.  Following W04, this means calculating the mean color in each redshift bin $z$:
\begin{equation}\label{meanCZReqn}
M_{z}^{j} = \frac{1}{N_{z}}\sum_{q=1}^{N_{z}}x_{j,q}
\end{equation}
where $x_{j}$ are the colors $u-g$, $g-r$, $r-i$, $i-z$ for each quasar $q$ in the bin.  Figure~\ref{czr_psf_O2_mean} shows the mean CZR for the DR3 quasar data.  (Later, we explore the use of the median or mode colors of each redshift bin to form the CZR.)  

The components \emph{j,k} of the CZR covariance matrix in each redshift bin $z$ are (W04):

\begin{equation}\label{CZRcovmatrixeqn}
V_{z}^{jk} = \frac{1}{N_{z} - 1}\sum_{q=1}^{N_{z}}(x_{j,q}-M_{z}^{j})(x_{k,q}-M_{z}^{k})
\end{equation}
where $N_{z}$ is the number of quasars in the redshift bin; $i$, $j$ and $k$ are colors; $M_{z}$ is the vector of CZR colors in the redshift bin $z$ as calculated in Equation~\ref{meanCZReqn}, and $x_{q}$ is the vector of colors for quasar $q$ in the bin.

\subsubsection{Calculating Photometric Redshifts}
A photometric redshift for an observed quasar is found using the CZRs and associated covariance matrices by first calculating the $\chi^2$ distribution between the colors of the observed quasar and the colors of the CZR.  

A color covariance matrix for the observed quasar is calculated from its measured magnitude errors:
\begin{equation}\label{qsocovmatrixeqn}
V_{o} = \left(\begin{array}{cccc}\sigma_{u}^{2}+\sigma_{g}^{2} & -\sigma_{g}^{2} & 0 & 0 \\-\sigma_{g}^{2} & \sigma_{g}^{2}+\sigma_{r}^{2} & -\sigma_{r}^{2} & 0 \\0 & -\sigma_{r}^{2} & \sigma_{r}^{2}+\sigma_{i}^{2} & -\sigma_{i}^{2} \\0 & 0 & -\sigma_{i}^{2} & \sigma_{i}^{2}+\sigma_{z}^{2}\end{array}\right)
\end{equation}

Then, the value of the $\chi^{2}$ distribution in the redshift bin \emph{z} is given by:
\begin{equation}\label{chisquaredcalc}
\chi_{z}^{2} = (X_{o}-M_{z})^{T}(V_{o}+V_{z})^{-1}(X_{o}-M_{z})
\end{equation}
where $X_{o}$ is the vector of color values for the observed quasar and $M_{z}$ is the vector of CZR mean, median, or mode color values in bin $z$.  

The probability distribution function (PDF) is found from this $\chi^2$ distribution and the photometric redshift corresponds to the maximum of the PDF, which is calculated as follows:
    \begin{equation}
    \label{pdfeqn}
      P'_{n} =\frac{W_{z}\rm{exp}[-\frac{\chi ^{2}_{n}}{2}]}{4\pi^{2}|V_{o}+V_{z}|^{1/2}}
     \end{equation}
The normalized PDF is
     \begin{equation}
     P_{n} =\frac{P'_{n}}{\sum_{n = 1}^{N}P'_{n}}
     \end{equation}

Here, \emph{W$_{z}$} is a redshift-dependent weighting parameter.  The weighting function used by W04 is the number distribution $N(z)$ of the CZR training dataset with redshift.  

\section{VARIATIONS ON W04 ALGORITHM}
In this section, we discuss the variations we made to improve the W04 algorithm.  In order to refine the CZR, we varied the binning and tested the use of median and mode statistics instead of the mean; explored the use of different magnitude types and adding additional bands; and imposed magnitude and/or magnitude error limits on the training data.  We also explored the use of different weighting functions for calculating the PDF, as well as combining multiple PDFs to improve the redshift estimations.  Table~\ref{variationSummary} summarizes the variations that are made to the original W04 algorithm.  The following subsections discuss the variations in more detail.  

\subsection{Binning}
    The W04 algorithm, as well as earlier work by \citet{RichardsB}, uses an overlapping binning scheme of the CZR training sample in which the bin centers are located at redshift intervals of 0.05.  Bins with centers $z\leq2.15$ have width $0.075$, bins with centers $2.15<z\leq2.5$ have width 0.2, and bins with centers at $z>2.5$ have width 0.5. They chose to use overlapping bins because their training data set was small; the larger bins increase the number of quasars, thus reducing noise.  Because the sample from the DR3 Quasar Catalog is more than fifteen times larger than the sample used by \citet{RichardsB}, there is no need for overlapping bins.  In addition, we can decrease the size of the bins and have smaller bins to higher redshifts, which increases the accuracy of the estimated photometric redshifts.  With this in mind, we have developed a set of variations of bin widths.  The binning schemes are summarized in Table~\ref{binningtable}; note that the edge of the first bin always starts with $z= 0$.  The first variation is based on the binning scheme of \citet{RichardsB}, where there are two different bin widths in two different redshift intervals (the change occurs at ``division2"), but without overlap.  In the next three variations, we continue to use only two redshift intervals, but change the bin width and the value of division2.  In the last four variations, we increase the number of redshift intervals to three (the first change occurs at ``division1").  This allows us to use the bin width of 0.05 below division1, where there is the smallest number of quasars in the training sample, and use even smaller bins in the interval between division1 and division2.  
        
     \note{I have not yet determined how to properly reference tables in separate sections:  In the case of results-summarizing tables, they are in Appendix A}Table~\ref{PDFsimpleWFremovered_table_mean} gives a comparison of results using different binning schemes.  We isolate the low-redshift and high-redshift quasars (less than or greater than division2, respectively), in Tables~\ref{PDFsimpleWFremovered_lowtable_mean} and~\ref{PDFsimpleWFremovered_hightable_mean}.  There is no drastic improvement in the estimation of quasar redshifts: any increased accuracy in the redshift estimate from decreasing the size of the bins is likely offset by a smaller number of training quasars in those bins, which will cause a decrease in accuracy of the CZR.  
To summarize, we find that a non-overlapping binning scheme does only slightly better than the O2 binning scheme.  
  
\subsection{Statistic}
    The W04 algorithm uses the mean of each color $u-g$, $g-r$, $r-i$, and $i-z$ within each redshift bin to form the CZR.  Here we describe our results when a median \citep{RichardsA, RichardsB}, and mode \citep{HopkinsCZR} of each color is used instead.     
    
\subsubsection{Discrete Statistics}
The ``discrete" median for each color is calculated by ordering the list of quasars by each color (smallest value to largest value) and choosing the central color in each of $u-g$, $g-r$, $r-i$, and $i-z$.  A mode can be calculated from this median and the mean of the bin using the formula \citep{Lupton}:
    \begin{equation}
    \rm{mode} = 3\times \rm{median} - 2\times \rm{mean}
    \end{equation}
    Hereafter, we refer to these statistics as the \emph{discrete median} and \emph{discrete mode}, respectively.
\subsubsection{``Gaussian" Statistics}    
    Since there are inherent errors in the color values (due to the magnitude errors in the measured data), we can alternatively treat the distribution of color in each bin as a continuum.  The color distribution for the quasars in each redshift bin is calculated by approximating each quasarÕs color as a Gaussian centered at the measured color with standard deviation equal to the color error \citep[c.f.][]{HopkinsCZR}.  For a color \emph{i} - \emph{j}, where the measurement error for each magnitude is given by $\delta$m$_{i}$ and $\delta$m$_{j}$, respectively, we define the color error for a single quasar as
    \begin{equation}\label{colorerreqn}
    \delta (m_{i} - m_{j})^{2} \equiv (\delta m_{i})^{2} + (\delta m_{j})^{2} 
    \end{equation}
    The Gaussians for all the quasars in the bin are summed to determine the color distribution.  The median of the distribution is at the 50th percentile; hereafter this statistic is referred to as the \emph{Gaussian median}.  The \emph{Gaussian mode} is the color value at which the color distribution is at its maximum \citep{HopkinsCZR}.  (In Section~\ref{addcorr}, we discuss the effects of including correlation between SDSS magnitude bands.)  Figure~\ref{czr_meanmedianmode} shows a comparison of CZRs using the mean, Gaussian median, and Gaussian mode statistic.   
    
    Tables~\ref{PDFsimpleWFremovered_table_gmedian} (\ref{PDFsimpleWFremovered_lowtable_gmedian}, \ref{PDFsimpleWFremovered_hightable_gmedian})~and~\ref{PDFsimpleWFremovered_table_gmode} (\ref{PDFsimpleWFremovered_lowtable_gmode}, \ref{PDFsimpleWFremovered_hightable_gmode}) can be compared to Table~\ref{PDFsimpleWFremovered_table_mean} (\ref{PDFsimpleWFremovered_lowtable_mean}, \ref{PDFsimpleWFremovered_hightable_mean}) to show the results using different statistics to calculate the CZR.  We find that using a mean statistic does slightly better than gmedian or gmode, with gmode giving the worst results.  
    
\begin{figure}
\plotone{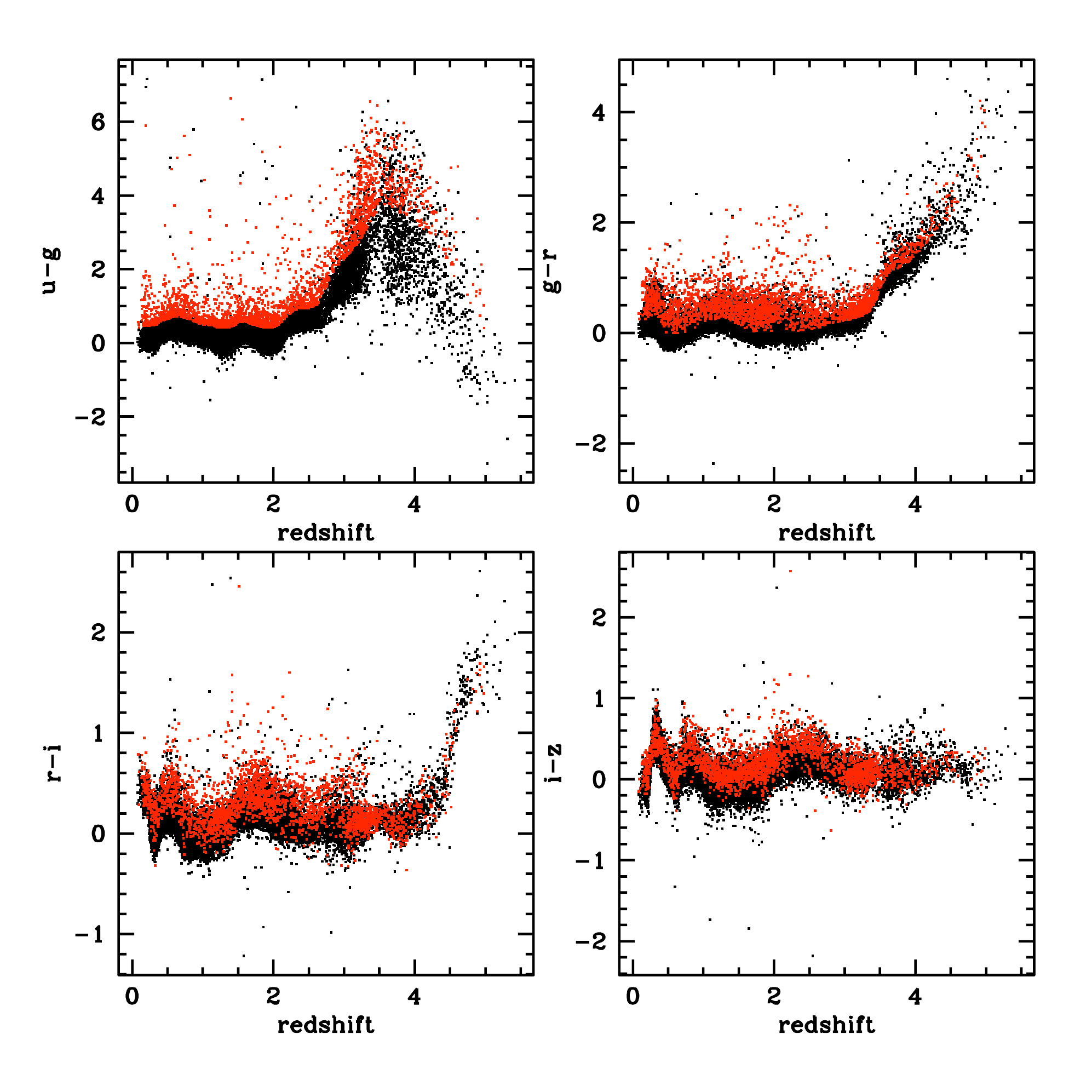}
\caption{Colors of SDSS DR3 quasars plotted as a function of redshift.  The red points are those quasars that are considered reddened according to the scheme presented in Section~\ref{RedQsoSection} (using the O2 binning scheme) and removed from the training set for the CZR (black points).  
\label{qsoCZR_red}}
\end{figure}

\begin{figure}
\plotone{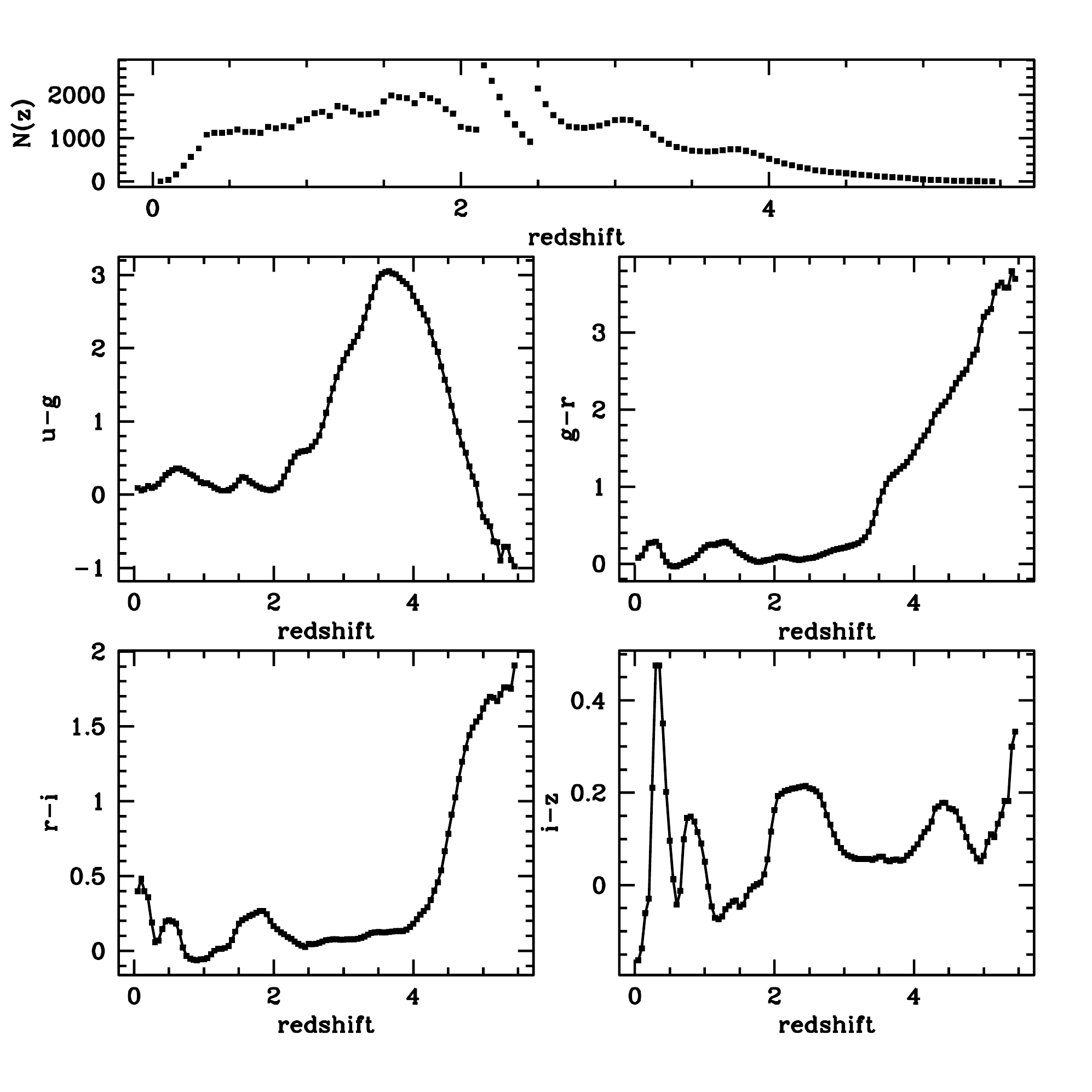}
\caption{Color-redshift relation (CZR) for the SDSS DR3 quasars according to the W04 scheme.  The top panel shows the number distribution of quasars as a function of redshift.  
\label{czr_psf_O2_mean}}
\end{figure}

\begin{figure}
\plotone{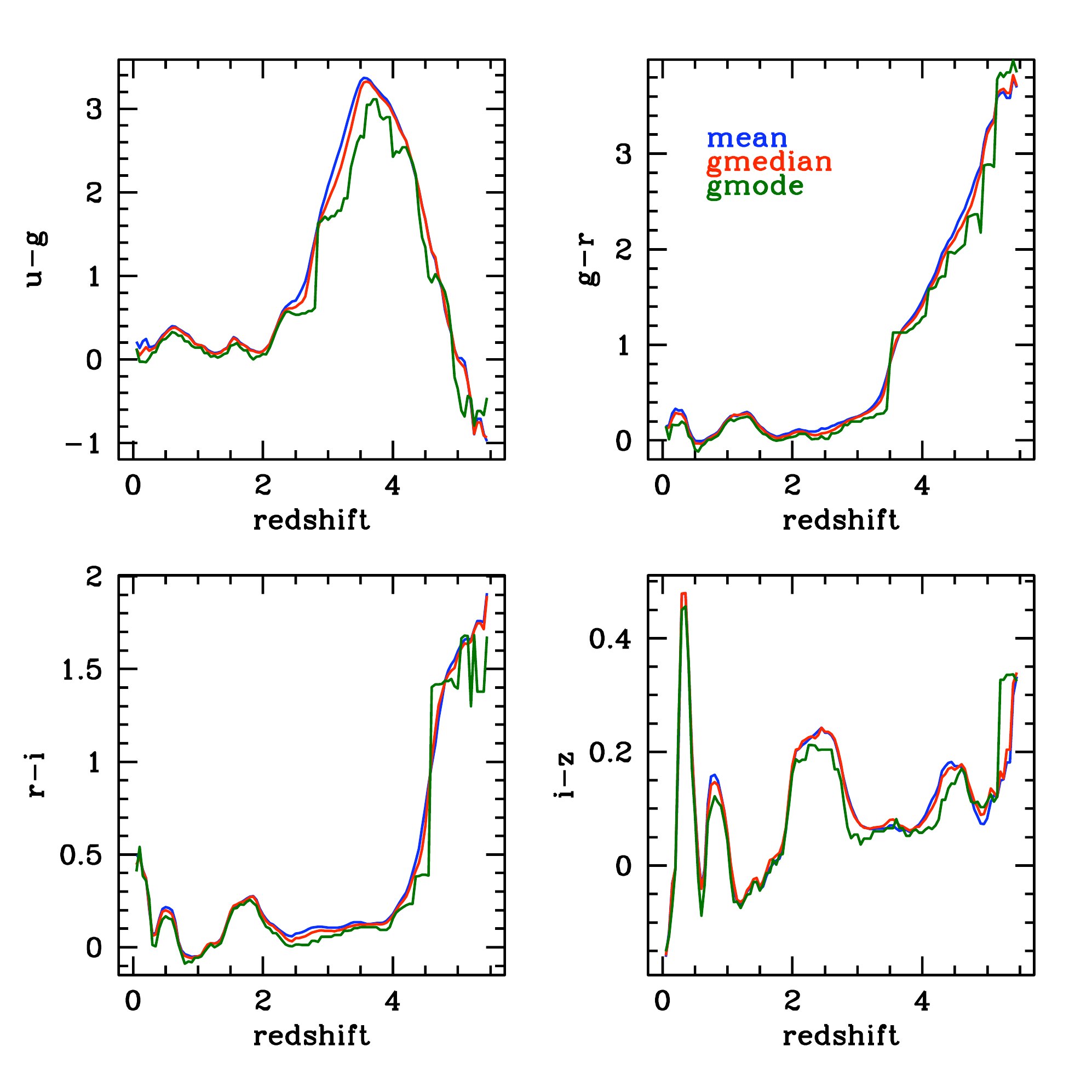}
\caption{Comparison of mean, gmedian, and gmode CZRs for the SDSS DR3 quasars using O2 binning.  
\label{czr_meanmedianmode}}
\end{figure}

    The idea behind using a CZR to predict redshifts is that in a small enough redshift interval, we will have quasars that are very similar to each other.  From those quasars, we choose the colors that best represent that small sample of quasars as the ``ideal".  However, as can be seen in Figure~\ref{qsoCZR_red}, even when the reddened quasars are removed, there is still a noticeable spread in the colors at each redshift.  If the distribution of colors is non-trivial (for instance, the distribution is double-peaked or flat), simply using a median or mode may not be enough to characterize the quasars in that bin.  The mean statistic gives equal weight to each quasar that contributes to the bin, and therefore most consistently gives the best characterization.  
       
\subsection{Magnitude Type}\label{magtypesection}
     SDSS measures four different types of magnitudes: fiber, model, Petrosian, and point-spread function (PSF) for each object.  We investigate how using different magnitude types to create CZRs could improve the photometric redshift estimates.  Below, for completeness, we give a brief discussion of the SDSS magnitude types.  
     
     \begin{itemize}
\item \emph{Point-spread function (psf) magnitudes:}  A point-spread function model is fit to the object to determine the best measure of the total flux.  This magnitude measure is best for isolated point sources.  These sources are unresolved, and the PSF magnitude is unbiased because the image is consistent with a point-spread function.  

\item \emph{Fiber magnitudes:}  Fiber magnitudes are calculated from the flux in each band contained within the three arcsecond aperture of a spectroscopic fiber of the SDSS telescope.  

\item \emph{Model magnitudes:}  Two models, a pure deVaucouleurs profile and a pure exponential profile, are fit to the two-dimensional image of the object in each band.  In the $r$-band, the best-fit model is chosen and applied to the other bands to determine the model magnitudes.  Because the same aperture is used in all bands, the flux measurements are appropriate for determining colors, especially for extended sources such as galaxies.  

\item \emph{Petrosian magnitudes:} Petrosian magnitudes are typically used for galaxy photometry.  From the azimuthally averaged light profile of the galaxy, a radius is determined for a circular aperture in which galaxy fluxes are measured.  The $r$-band profile of the galaxy determines the radius used in all bands to calculate the aperture in which the fluxes are measured. 
\end{itemize}
 
Tables~\ref{PDFsimpleWFremovered_table_mean} through \ref{PDFsimpleWFremovered_hightable_gmode} also compare the photometric redshift estimation results using the different magnitude types.   Using other magnitudes (than PSF) singly usually does worse, but in some cases can do only slightly better.  \note{LOOK AT THE TABLES TO SEE WHAT THIS MEANS}.  It is not surprising that the PSF magnitude CZRs give the most accurate redshift estimates, because this magnitude is optimized for measurement of point sources, which most of the quasars are.  The other magnitude types, especially the fiber magnitude, which measure flux in fixed apertures, may not give as accurate of flux measurements if extraneous flux is allowed through the aperture.   

\subsubsection{Multiple PDFs}
In addition to using different magnitude types separately, we convolve PDFs generated from two, three, and four magnitude types to create a composite PDF from which we estimate a photometric redshift.  Each PDF is weighted equally in the convolution, but giving certain PDFs, such as the PDF generated from PSF magnitudes because it is the most successful alone, greater weight in future work may improve results.  

Tables~\ref{PDFmultimagsO2} through \ref {PDFmultimagsO2_hightable} show that combining 2, 3, and 4 PDFs from different magnitude types does slightly better than the Weinstein standard: by combining the information from several magnitude types, we find that our characterization of quasar colors is improved. 

\subsection{Reddened Quasars}
     In following the W04 algorithm, we remove quasars that we have classified as ``reddened,"  but we also investigate the effect on photometric redshifts if these reddened quasars are not removed from the sample used to calculate the CZR.  \cite{RichardsA} studied the nature of the outliers and found that the scatter is more pronounced on the red end of the distribution and increases for fainter magnitudes.  It was also observed that this reddening effect is independent of redshift, and therefore thought to be due to processes intrinsic to the quasar itself, rather than contamination from the host galaxy or intergalactic extinction.  

By comparing the results shown in Tables~\ref{PDFsimpleWFnoremovered_table_mean} through \ref{PDFsimpleWFnoremovered_hightable_gmode}, it is evident that removing reddened quasars gives slightly better results than leaving reddened quasars in the sample, confirming the method used by W04.  Future work could include a new definition for reddened quasars.  

\subsection{Magnitude and Magnitude Error Limits}\label{maglimsection}
We hypothesize that by restricting the CZR training data to bright objects by imposing magnitude limits, or to objects with good photometry by imposing magnitude error limits, we can improve our estimates.  Tables~\ref{PDFvaryerror} through \ref{PDFvaryerrorhightable} summarize the results using magnitude error limits, while Tables~\ref{PDFvarymags} through \ref{PDFvarymagshightable} summarize the results using magnitude limits.  

We apply magnitude error limits (in all bands) of 1.0, 0.5, 0.2, and 0.1 to the CZR training quasars and found that the limits made essentially no difference in the photometric estimate results, whether or not the same error limits were imposed on the quasars for which photoZs were calculated.  The spread in quasar colors in each redshift interval is much larger than the spread due to including objects with less precise photometry, therefore it is not surprising that we see no effect.  
  
We also imposed magnitude limits (in all bands) of 25.0, 21.0, and 20.0 to build CZRs.  When the same magnitude limits were imposed on the quasars for which photoZs were calculated, a magnitude limit of 20.0 gave a several-percent improvement compared to our recreation of the Weinstein results.  Brighter quasars will have more consistent colors, whereas dimmer quasars are likely to be reddened (though not enough to be removed with the reddening cut described above) and have colors that differ more from the mean colors.  However, if the magnitude limit was not applied to the photoZ data, a CZR-data magnitude limit of 25.0 gave the best results, which were slightly better than the results for the W04 recreation.  A CZR calculated from bright quasars will be a more accurate representation for the bright subset of objects for which photometric redshifts are to be calculated, but the dimmer quasars, with their more anomalous colors, will not fit the bright model as well.  
  
 \subsection{Adding additional bands}\label{addBandssection}
 \subsubsection{Simulation with SDSS colors}
 All of the work thus far has been done using the five SDSS bands ($u$, $g$, $r$, $i$, $z$) and four colors derived by comparing the flux in consecutive bands.  In order to simulate whether there would be an improvement in photometric redshift estimation, we generate CZRs and photometric redshifts using only combinations of three consecutive SDSS bands (two colors) and four consecutive SDSS bands (three colors).     We also investigate whether there is an improvement in photometric redshift estimates if non-consecutive bands are used to calculate 3 colors, i.e., $u-r$, $g-i$, $r-z$.  

Note that for the work in this section, reddened quasars were not removed: since the red limits are only defined in $u-g$ and $g-r$, and these colors are not always used in the CZRs calculated from fewer bands.  Therefore, reported results may look worse, so the second row in the table shows results using the W04 parameters without removing reddened quasars, for a more accurate comparison.  

From the ``simulation" of adding additional bands (results are summarized in Tables~\ref{PDFfewercolorstable} - \ref{PDFfewercolors_hightable}), it is clear that using additional colors improves the photometric redshift estimates.  Using non-adjacent colors yields worse results than the W04 method.  

  \subsubsection{Verification with GALEX data}
 We continue to verify that redshift estimates are improved when additional bands are included in the CZR by using ultraviolet data from GALEX that is matched to SDSS objects.  Using a dataset of 7,642 SDSS DR5 quasars that have matches in GALEX, we calculated the mean CZRs and corresponding covariance matrices with 4 SDSS (psf magnitude) colors + 1 GALEX color.  We continue to use the same redshift binning scheme as W04 without removing reddened quasars.  The PDF for photometric redshift estimation is weighted according to a N(z) weighting scheme based on the input sample.  Tables~\ref{sdssGalex_alltable} - \ref{sdssGalex_hightable} compare the results of using just the 4 colors for this dataset to using 5 colors.  It is obvious that using an additional color improves the redshift estimates by at least 1\%.  We also performed N-fold testing on these results to ensure that we were not overfitting since we have a smaller sample, and find that there is no overfitting.  We caution that some of the improvement may be due to the fact that only bright quasars will have GALEX matches, and we showed in Section~\ref{maglimsection} that redshift estimates are more accurate for brighter objects.  However, because of our ``simulation" above, we conclude that the addition of more colors will be the main factor that improves photometric redshift estimates.  

\subsection{Weighting Functions}
 As stated above, have been using a simple \emph{N(z)} weighting function in which each redshift bin is weighted by the fraction of the total quasars in the training sample within that bin \citep{Weinstein}.  When we use use no weighting function (i.e., $W_{i}$ in equation~\ref{pdfeqn} is always equal to 1/N, where N is the number of redshift bins), we see that the results are typically worse (see Tables~\ref{PDFnoWFremovered_table_mean} - \ref{PDFnoWFnoremovered_hightable_gmode}, which summarize the results when no weighting function is used and binning, magnitude, and statistic are varied with red quasars removed and not removed).  Therefore, we conclude that a weighting function is an effective way to improve the photometric redshift estimates and explore the use of others.  
   
   \subsubsection{Apparent Magnitude Weighting Function}\label{apparMagWFsection}
   We create an apparent magnitude weighting function \emph{N(m, z)} by dividing the entire training sample (the sample used to create the CZR) into bins by apparent magnitude. 
   We then subdivide each of these magnitude bins into bins by spectroscopic redshift, thus defining the boundaries for an apparent magnitude weight function ``matrix."  We count the number of objects in each cell of this matrix and normalize them by the total number of objects \emph{N(m)} within that cell's magnitude shell.  The weighting function $N(m,z)$ used for a given quasar is the matrix row corresponding to the apparent magnitude bin in which that quasar falls.  
    
The redshift binning corresponds to the redshift binning for the CZR (for obvious reasons when calculating photometric redshifts).  The CZR code allows the user to input the magnitude binning.  The user either gives a set apparent magnitude ranges for each bin, or chooses from a set of lists of magnitude-bin minima-- which enables the use of non-uniform width magnitude bins.  
   
   \subsubsection{Absolute Magnitude Weighting Function}\label{NofMfromfilesection}
   From the original training dataset, we bin quasars by absolute magnitude, and create an \emph{N(M)} weighting function in which each absolute magnitude bin is weighted by the fraction of the total quasars within that bin.  The weighting function for an individual quasar is calculated by converting absolute magnitude bin edges to redshift via the distance modulus.  The distance modulus is calculated from the difference between the quasar's apparent magnitude and the absolute magnitude of the bin edge.  Since the distance modulus depends only on redshift and not on magnitude, we take the redshift value corresponding to the distance modulus as the bin redshift.  We use the resulting interpolated \emph{N(z)} as the weighting function.  
   
   \subsubsection{Luminosity-based Weighting Functions}\label{lumWFsection}
We also have developed weighting functions based on the DR3 Quasar Luminosity Function \citep{Richards06}:  
    \begin{equation}\label{phieqn}
  \Phi(M, z) = \Phi^{\ast}10^{A_{1}\mu}
   \end{equation}
  where
    \begin{equation}\label{mueqn}
    \mu = M - (M^{\ast} + B_{1}\xi + B_{2}\xi^{2} + B_{3}\xi^{3})
    \end{equation}
    and 
    \begin{equation}\label{xieqn}
    \xi = \log{\frac{1 + z}{1 + z_{ref}}}
    \end{equation}
The fitting parameters for the luminosity function determined by \cite{Richards06} used in the above equations are listed in Table~\ref{LumFconstants}.  
   
   We have created a set of weighting functions based on this luminosity function by first defining the boundaries for a luminosity weightfunction ``matrix" $L$ by dividing the absolute magnitude range of the sample into bins, then dividing each absolute magnitude bin into redshift bins over the redshift range of the sample.  In each cell of this matrix, we evaluate the value of the luminosity function as 
   
   \begin{equation}\label{lumMatrixeqn}
   L_{M,z} = \int_{M_{min}}^{M_{max}}\int_{z_{min}}^{z_{max}}\Phi(M,z) dV_{c}dM
   \end{equation}
   where $\Phi$ is given by Equation~\ref{phieqn} with apparent magnitude substituted for absolute magnitude using the relation
   \begin{equation}\label{apparmageqn}
    M = m - K_{2} - \frac{c}{H_{o}}5log[(1+z)\int_{0}^{z}\frac{dz'}{\sqrt{\Omega_{M}(1+z')^{3} + \Omega_{\Lambda}}}] + 25
   \end{equation}
   
  \emph{$K_{2}$} is the two-parameter \emph{K}-correction as defined by \citet{Wisotzki}:
   \begin{equation}\label{Kcorreqn}
   K_{2}(z) = K(0) - 2.5(1+\alpha)log(1+z)
   \end{equation}
   where \emph{$K(0)$} is given as -0.42 and $\alpha$ is given as -0.45.  
   
   The comoving volume element is given by \citep{Hogg}
   \begin{equation}\label{dVeqn}
   dV_{c} = 4\pi\frac{c}{H_{o}}\frac{(1+z)^{2}D_{A}^{2}}{\sqrt{\Omega_{M}(1+z)^{3} + \Omega_{\Lambda}}}dz
   \end{equation}
   and \emph{$D_{A}$} is the angular diameter distance from now until redshift \emph{z} given by \cite{Hogg}
   \begin{equation}\label{DAeqn}
   D_{A} = \frac{c}{H_{o}(1+z)}\int_{0}^{z}\frac{dz'}{\sqrt{\Omega_{M}(1+z')^{3} + \Omega_{\Lambda}}}
   \end{equation}

   The matrix cell values are normalized by the probability of finding a quasar from the training set in the bin (M, z) of the matrix: (N$_{M, z}$)/N. The \emph{collapsed matrix N(z)} weight function is calculated by summing the matrix along the magnitude axis.  The \emph{matrix N(z)} weight function is calculated for an individual quasar by using its apparent magnitude to calculating the corresponding absolute magnitude at the center of each redshift bin so that the $i$th value of \emph{N(z)} corresponds to the ($M_{i}$, $z_{i}$) element of the matrix.  
   
        Tables~\ref{PDFapparMagWFtable} - \ref{PDFapparMagWF_hightable} show a comparison of photometric redshift estimation results using each of the above weighting functions.  Using an N(z) weighting function does better than no weighting function by about 1\%.  The apparent magnitude weighting function improves redshift estimates up to about 1\%.  We conclude that a weighting function adds additional information about the training data to the information already contained in the CZR and is necessary for accurate photometric redshift estimates.  
        
\subsection{Covariance between SDSS magnitude bands}\label{addcorr}
    \citet{Scranton} has shown that photometry in the five bands of the SDSS can be highly correlated; therefore covariance between the bands is non-negilgible and should be taken into account when calculating the color error.  The color error for an individual quasar is then expressed as
    \begin{equation}
    \delta (m_{i} - m_{j})^{2} \equiv (\delta m_{i})^{2} + (\delta m_{j})^{2} - 2c_{ij}\delta m_{i}\delta m_{j}
    \end{equation}
where \emph{c$_{ij}$} is the correlation coefficient between the two bands, as determined by \citet{Scranton}.  Table~\ref{corrcoefftable} gives a list of correlation coefficient values used.  
The quasar's color covariance matrix then becomes:
\begin{equation}\label{qsocovmatrixeqn_withcorr}
V_{o} = \left(\begin{array}{cccc}\sigma_{u}^{2}+\sigma_{g}^{2} - c_{ug}\sigma_{u}\sigma_{g}& uggr & ugri & ugiz \\
uggr & \sigma_{g}^{2}+\sigma_{r}^{2} -c_{gr}\sigma_{g}\sigma_{r}& grri & griz \\
ugri & grri & \sigma_{r}^{2}+\sigma_{i}^{2} -c_{ri}\sigma_{r}\sigma_{i} & riiz\\
ugiz & griz & riiz & \sigma_{i}^{2}+\sigma_{z}^{2}-c_{iz}\sigma_{i}\sigma_{z}\end{array}\right)
\end{equation}
where
\begin{equation}\label{matrixelement_uggr}
uggr = -\sigma_{g}^{2} +c_{ug}\sigma_{u}\sigma_{g}+c_{gr}\sigma_{g}\sigma_{r}-c_{ur}\sigma_{u}\sigma_{r}
\end{equation}

\begin{equation}\label{matrixelement_ugri}
ugri = c_{ur}\sigma_{u}\sigma_{r}+c_{gi}\sigma_{g}\sigma_{i}-c_{ui}\sigma_{u}\sigma_{i}-c_{gr}\sigma_{g}\sigma_{r}
\end{equation}

\begin{equation}\label{matrixelement_ugiz}
ugiz = c_{ui}\sigma_{u}\sigma_{z} + c_{gz}\sigma_{g}\sigma_{z} - c_{gi}\sigma_{g}\sigma_{i} - c_{uz}\sigma_{u}\sigma_{z}
\end{equation}

\begin{equation}\label{matrixelement_grri}
grri = -\sigma_{r}^{2} + c_{gr}\sigma_{g}\sigma_{r} + c_{ri}\sigma_{r}\sigma_{i} - c_{gi}\sigma_{g}\sigma_{i}
\end{equation}

\begin{equation}\label{matrixelement_griz}
griz = c_{gi}\sigma_{g}\sigma_{i} + c_{rz}\sigma_{r}\sigma_{z} - c_{ri}\sigma_{r}\sigma_{i} - c_{gz}\sigma_{g}\sigma_{z}
\end{equation}

\begin{equation}\label{matrixelement_riiz}
riiz = -\sigma_{i}^{2} + c_{ri}\sigma_{r}\sigma_{i} + c_{rz}\sigma_{r}\sigma_{z}- c_{iz}\sigma_{i}\sigma_{z}
\end{equation}

\begin{figure}
\plotone{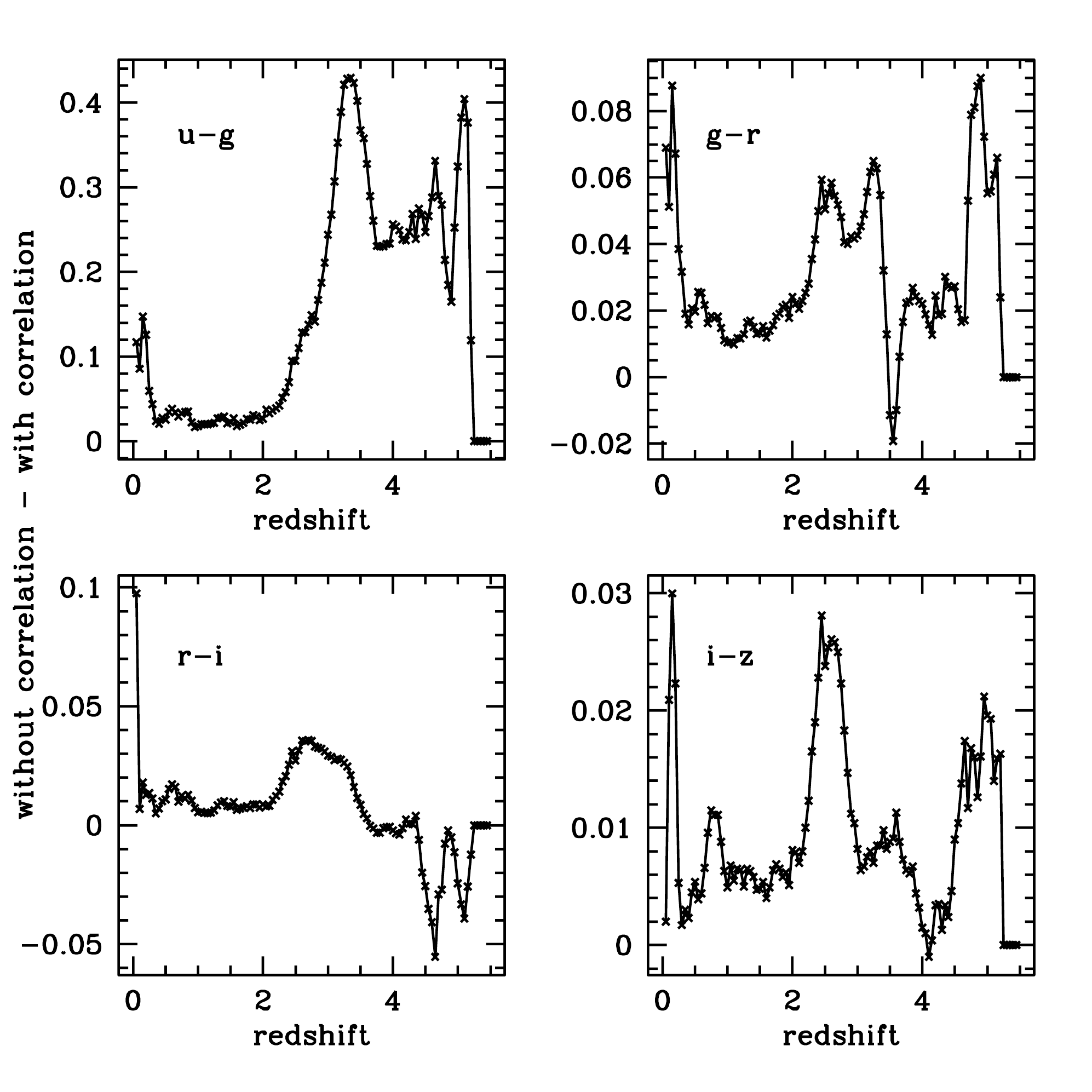}
\caption[Difference between mean CZRs with and without covariance]{The difference between mean CZRs that do and do not include covariance between the SDSS magnitude bands for the four SDSS colors (O2 binning).  
\label{corrCZRcomparefigure_mean}}
\end{figure}

    When the correlation coefficients are included, the changes to the CZRs are small (Figure~\ref{corrCZRcomparefigure_mean}).   The redshift estimate results in Tables~\ref{corr_alltable}-~\ref{corr_hightable} also show that the contribution of correlation coefficients gives only negligible improvement.  
 
 \subsection{CZRs from SDSS DR3 Quasar Subpopulations}
\begin{figure}
\plotone{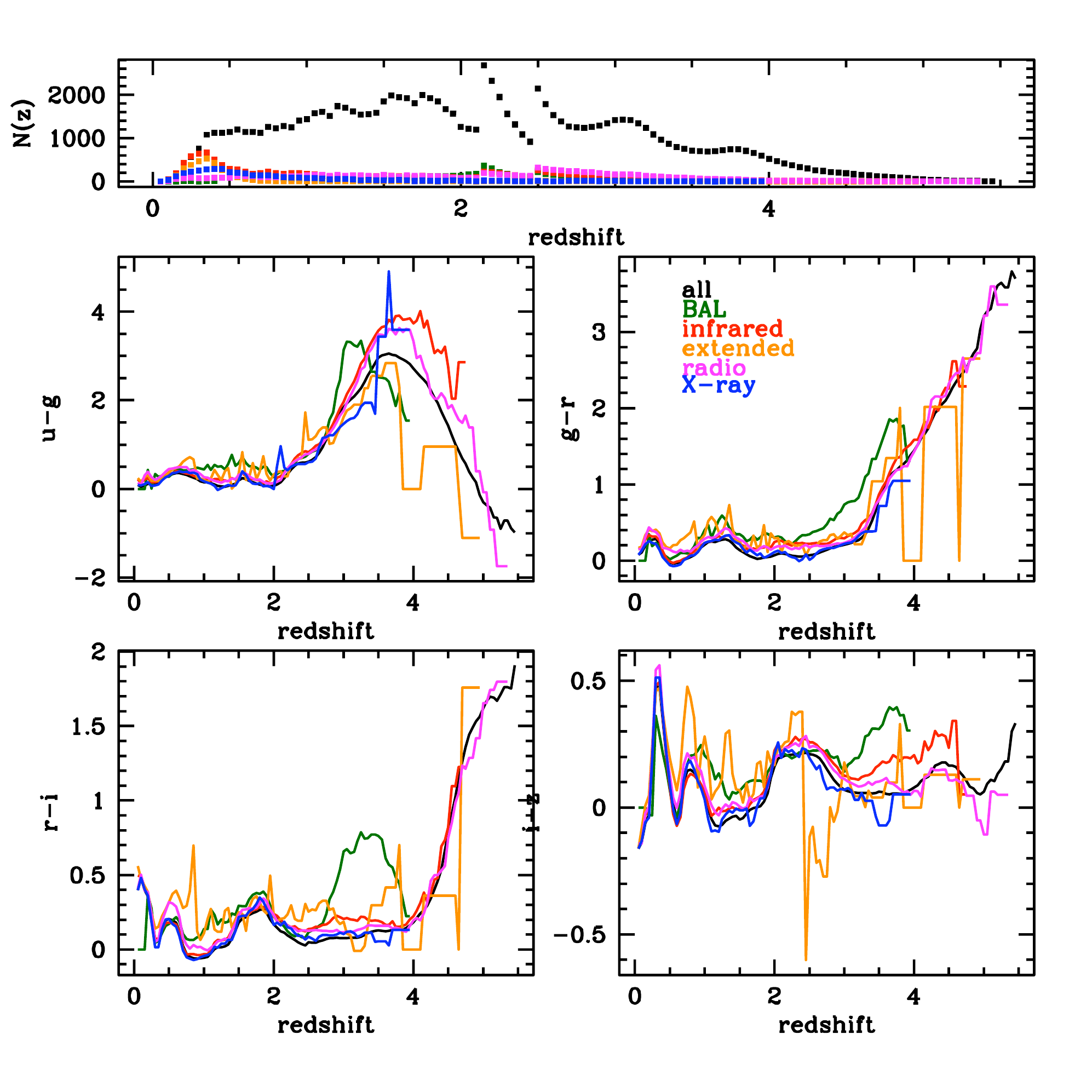}
\caption{Comparison of CZRs calculated according to the W04 scheme.   The top panel shows the number distribution of quasars in each subsample as a function of redshift.  
\label{CZRcompare_subpops}}
\end{figure}

 The CZR technique can be applied to subsets of data from the training sample used above.  We isolated DR3 quasars classified as extended objects, quasars with multiwavelength measurements (e.g. radio, IR, X-ray), or quasars with broad absorption features to generate CZRs for these subpopulations (see Figure~\ref{CZRcompare_subpops}).  Tables~\ref{sdsssubpops_alltable}-~\ref{sdsssubpops_hightable} show results when these CZRs are applied to the subsets to estimate photometric redshifts.  The colors of a subpopulation (especially one based on multiwavelength properties) are more likely to be similar, which will result in a CZR that more accurately characterizes the subpopulation.  When calculating a photometric redshift for an object with a particular attribute, it may improve the estimate to convolve the PDF from the subpopulation CZR with the more general PDF.  
 
 \section{CATASTROPHIC FAILURE ANALYSIS}
     A catastrophic failure is defined to be a photometric redshift estimate for which $|z_{spec}-z_{phot}|\geq0.5$.   This section details the analysis of the 6,821 catastrophic failure objects that occurred in the recreation of the algorithm developed by \citet{Weinstein}.  By analyzing the properties of the objects that failed and searching for similarities and patterns, we hope improve our algorithm to better estimate photometric redshifts.  

\begin{figure}
\plotone{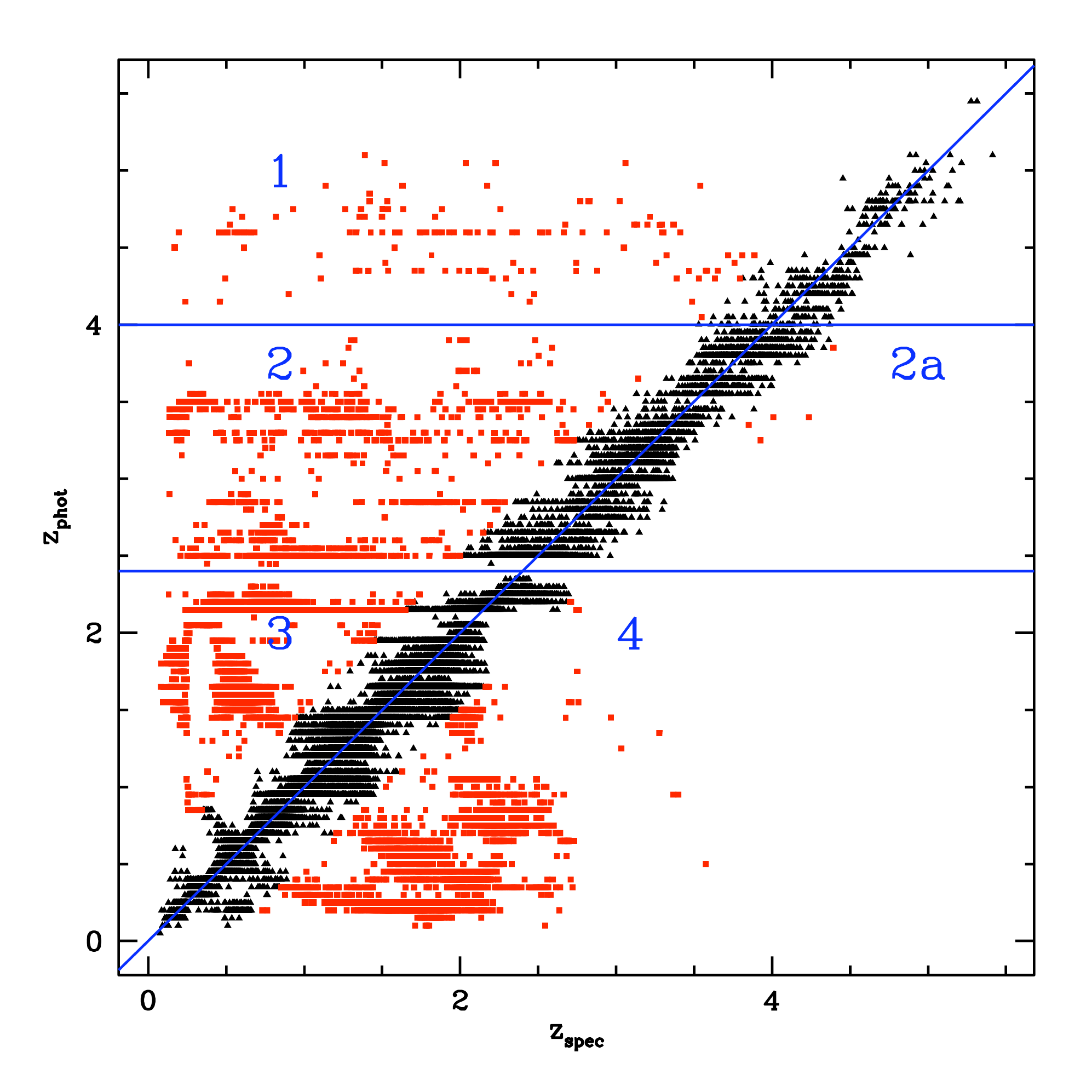}
\caption{Photometric redshift estimates are plotted against SDSS spectroscopic redshift for recreation of W04 algorithm.  Catastrophic failures, where $|$$z_{spec}$ - $z_{phot}$$|$ $\geqslant$ 0.5, are shown in red.  Black points have $|$$z_{spec}$ - $z_{phot}$$|$ $<$ 0.5.  Divisions for the four catastrophic failure sections (in z$_{phot}$) are shown in blue. 
\label{zspeczphot_withdivisions}}
\end{figure}

     It is evident from the $z_{phot}$ vs. $z_{spec}$ plot in Figure~\ref{zspeczphot_withdivisions} that the catastrophic failure objects can be separated into sections, and we have investigated four distinct sections of catastrophic failures.  The sections are described in Table~\ref{CFsectionstable}.  In Section~\ref{CFcolorssection} we investigate whether a color cut may be used to identify catastrophic failure objects, and in Sections~\ref{CFspectrasection} and \ref{CFpdfsection} we give the details from our inspection of the spectra and PDFs for the catastrophic failures, respectively.  
       
     \subsection{Colors of catastrophic failure objects}\label{CFcolorssection}
\begin{figure}
\plotone{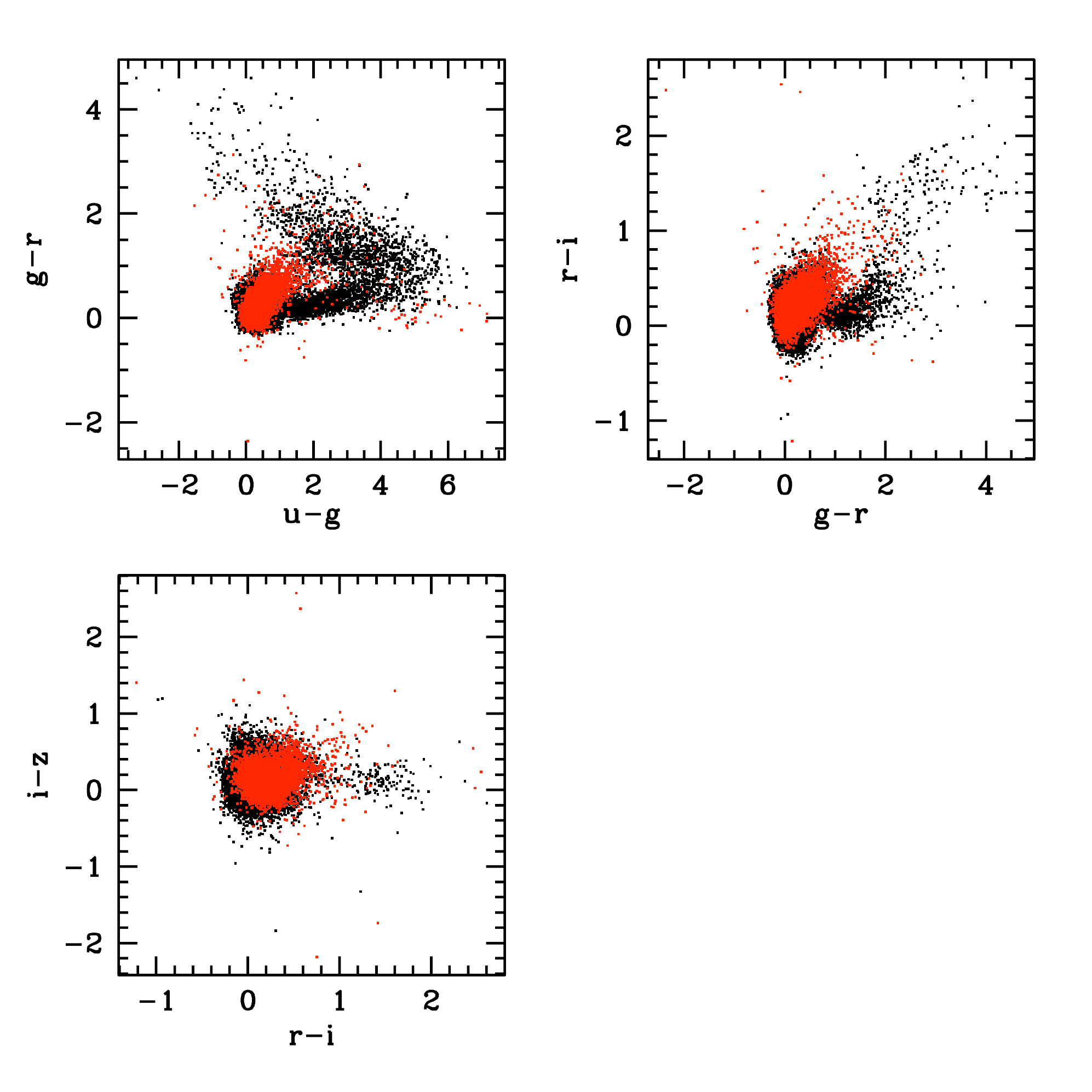}
\caption{Color-color plots for all 6,821 catastrophic failures (red points) superimposed on the entire quasar sample (black points) 
\label{colorcolor_allCF}}
\end{figure}

\begin{figure}
\plotone{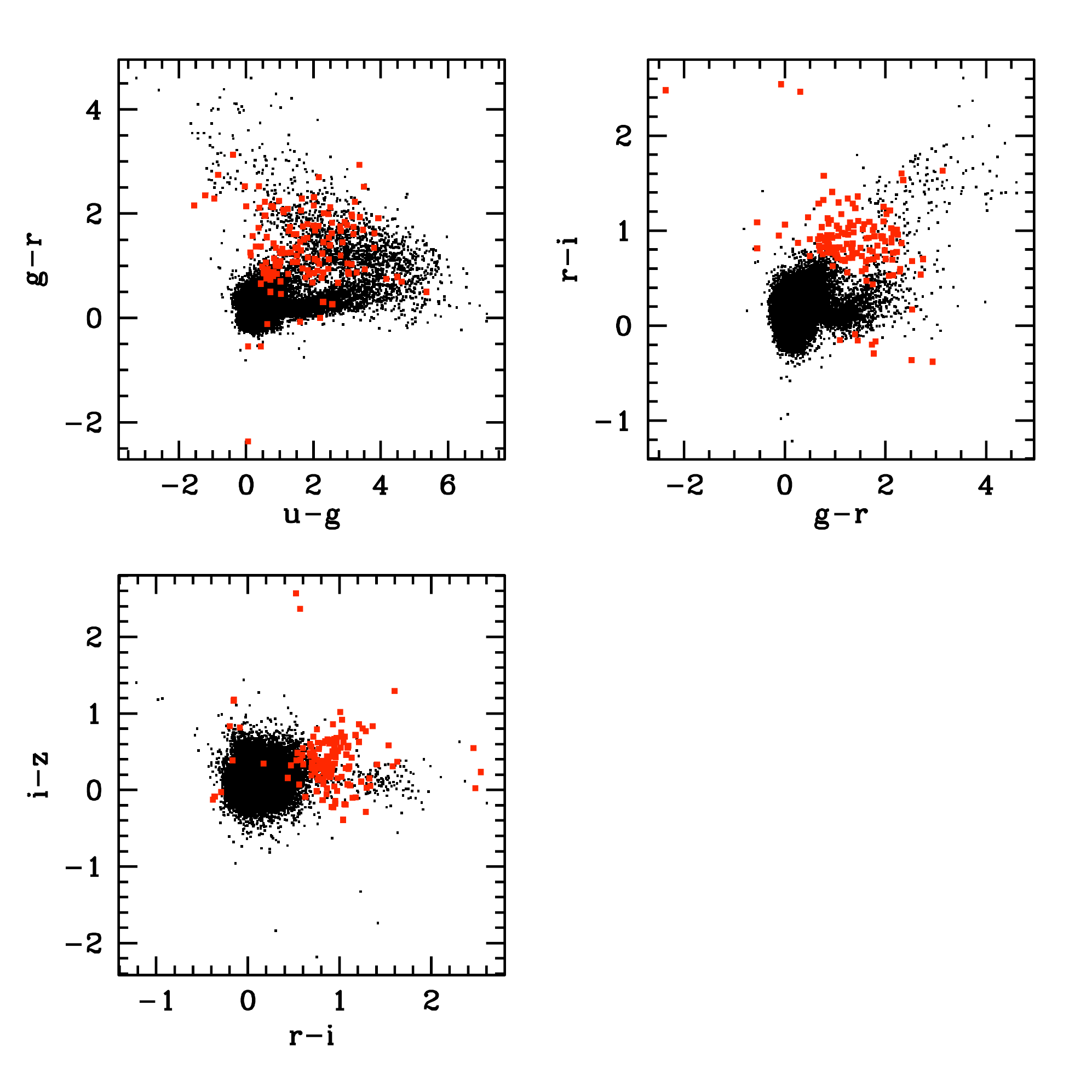}
\caption{Color-color plots for 143 quasars in catastrophic failure section 1 (red points). Since there are so few points in this section, the points have been enlarged for easy viewing.  Black points show all quasars in the sample (including other catastrophic failures. 
\label{colorcolor_sec1CF}}
\end{figure}

\begin{figure}
\plotone{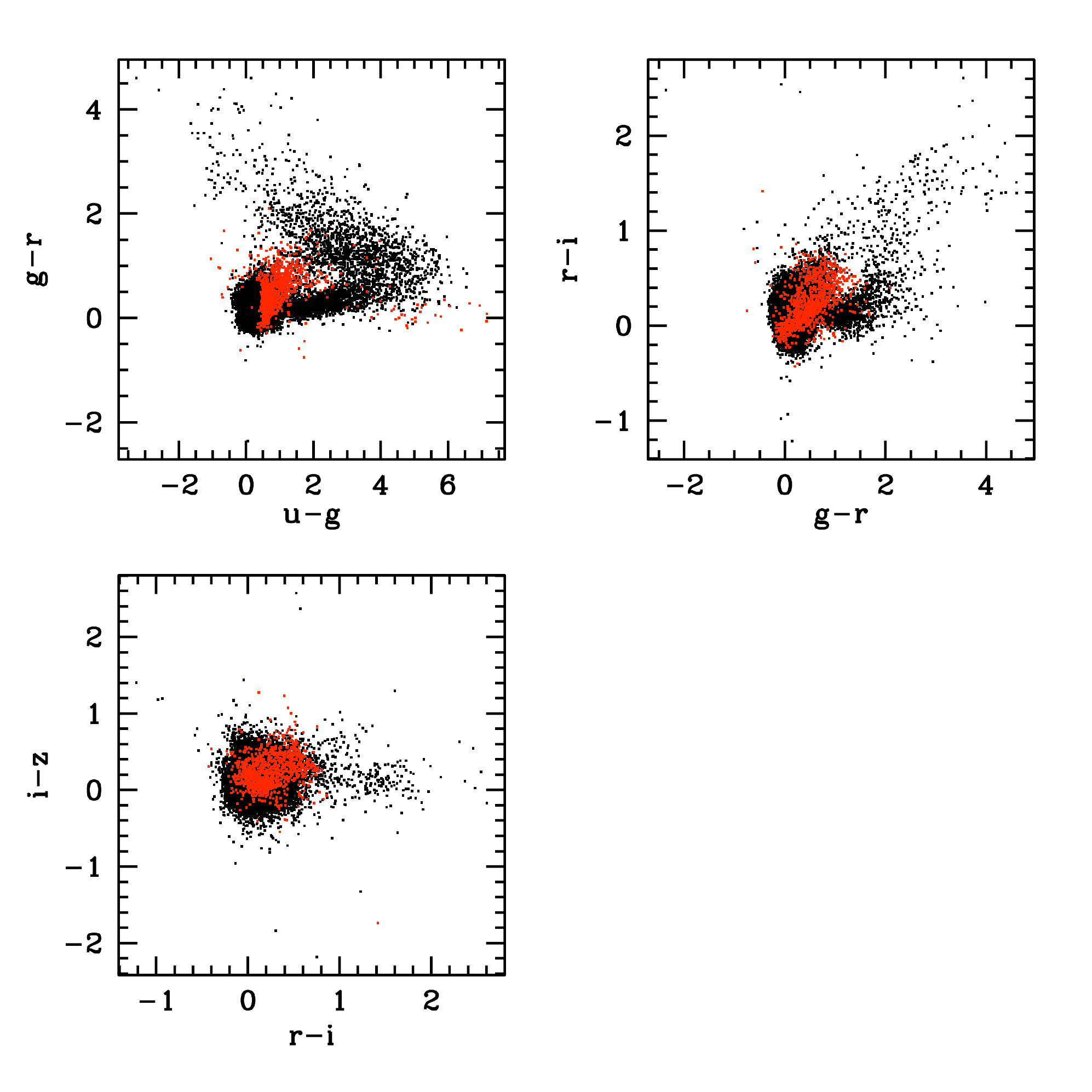}
\caption{Color-color plots for 1,078 quasars in catastrophic failure section 2 (red points). Black points show all quasars in the sample (including other catastrophic failures. 
\label{colorcolor_sec2CF}}
\end{figure}

\begin{figure}
\plotone{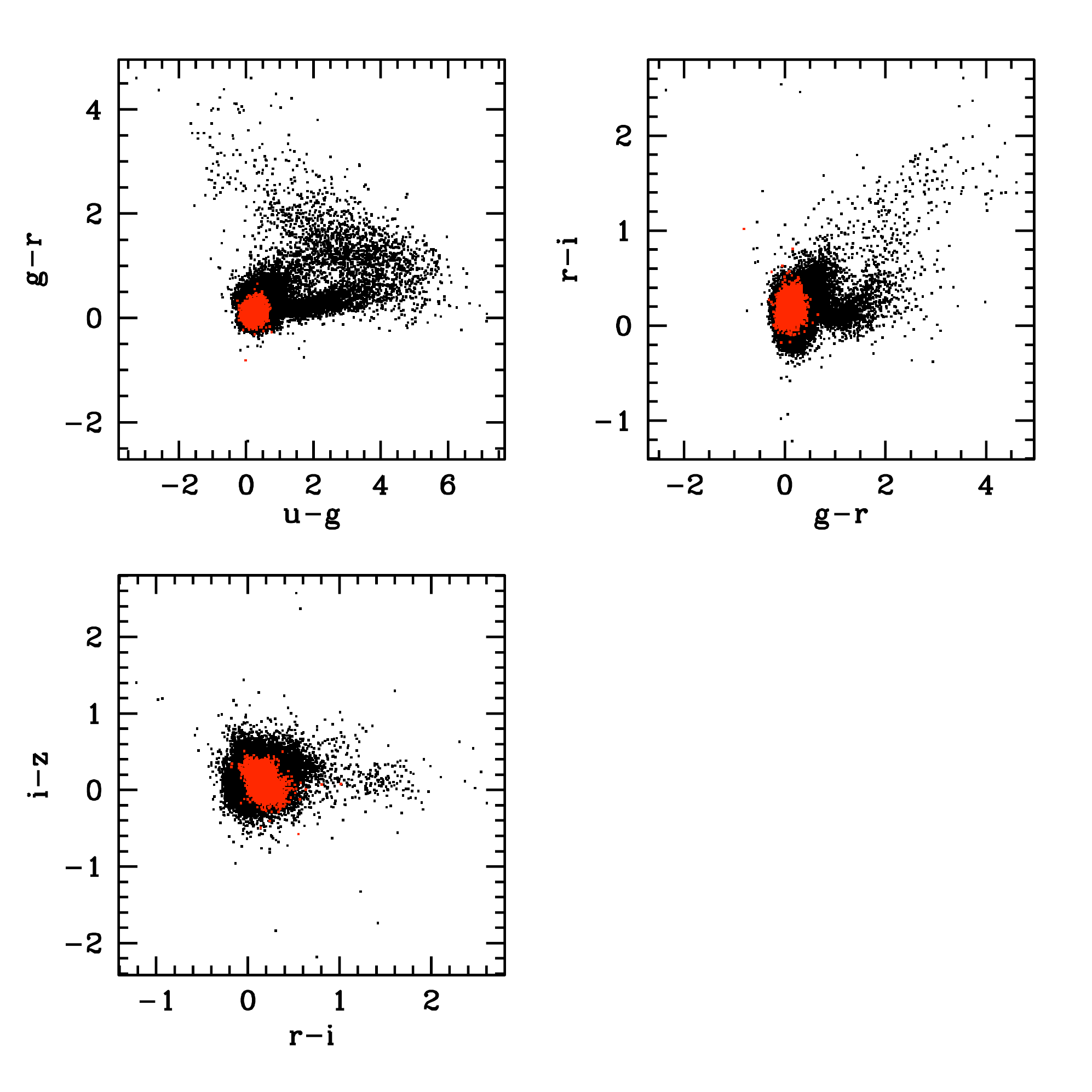}
\caption{Color-color plots for 3,374 quasars in catastrophic failure section 3 (red points). Black points show all quasars in the sample (including other catastrophic failures. 
\label{colorcolor_sec3CF}}
\end{figure}

\begin{figure}
\plotone{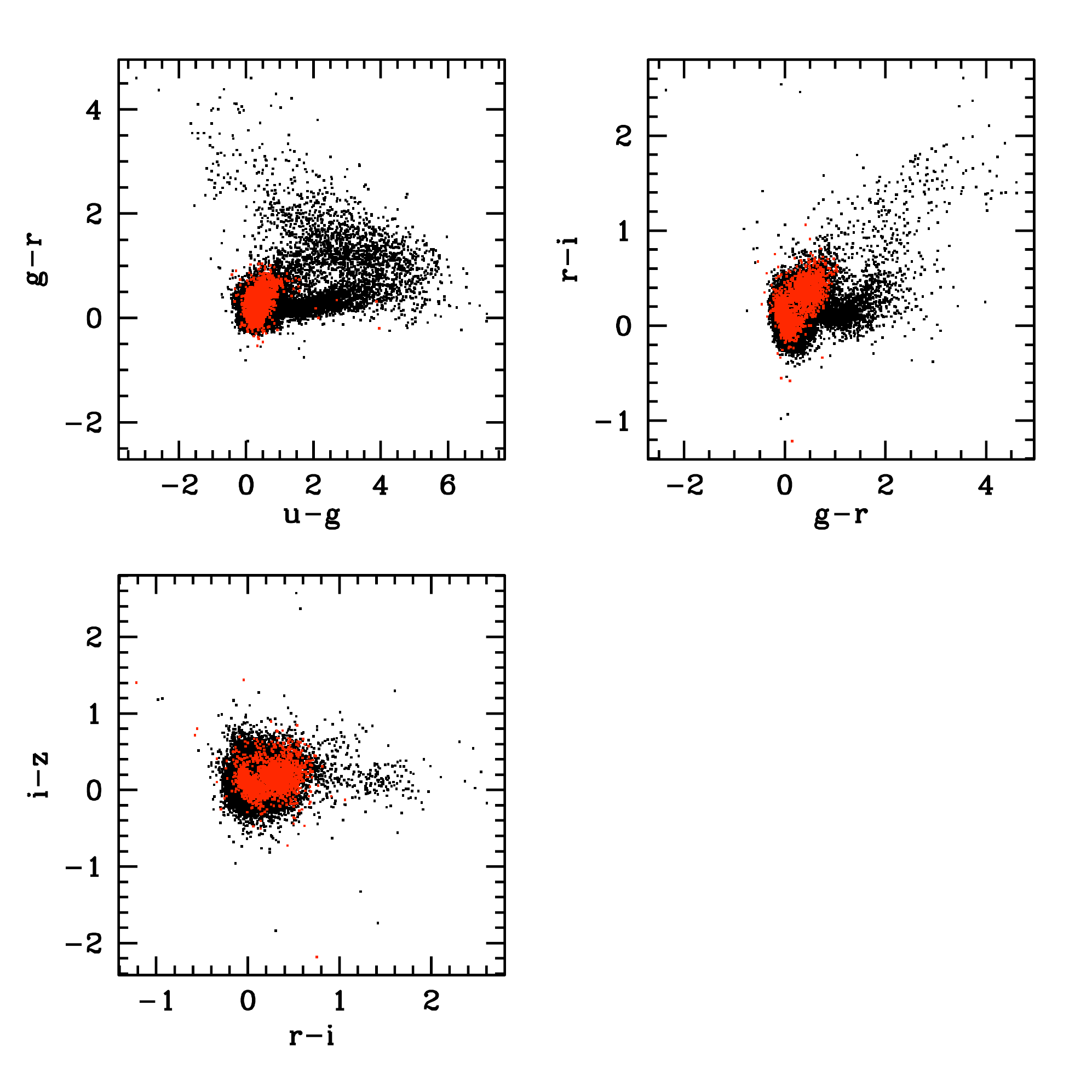}
\caption{Color-color plots for 2,221 quasars in catastrophic failure section 4 (red points). Black points show all quasars in the sample (including other catastrophic failures. 
\label{colorcolor_sec4CF}}
\end{figure}

     In general, it is impossible to separate all catastrophic failure-type objects from non-catastrophic failure objects by a simple color cut, as can be seen in the color-color plots of all catastrophic failures superimposed on the color-color plot for the entire quasar sample in Figure~\ref{colorcolor_allCF}.  However, by looking at the color-color plots for each of the four sections of catastrophic failures (Figures~\ref{colorcolor_sec1CF} to \ref{colorcolor_sec4CF}), we see that objects in Section 1 have noticeably redder colors than the main sample of quasars.  
    
     \subsubsection{Reddened quasars}\label{CFredqsossection}
\begin{figure}
\plotone{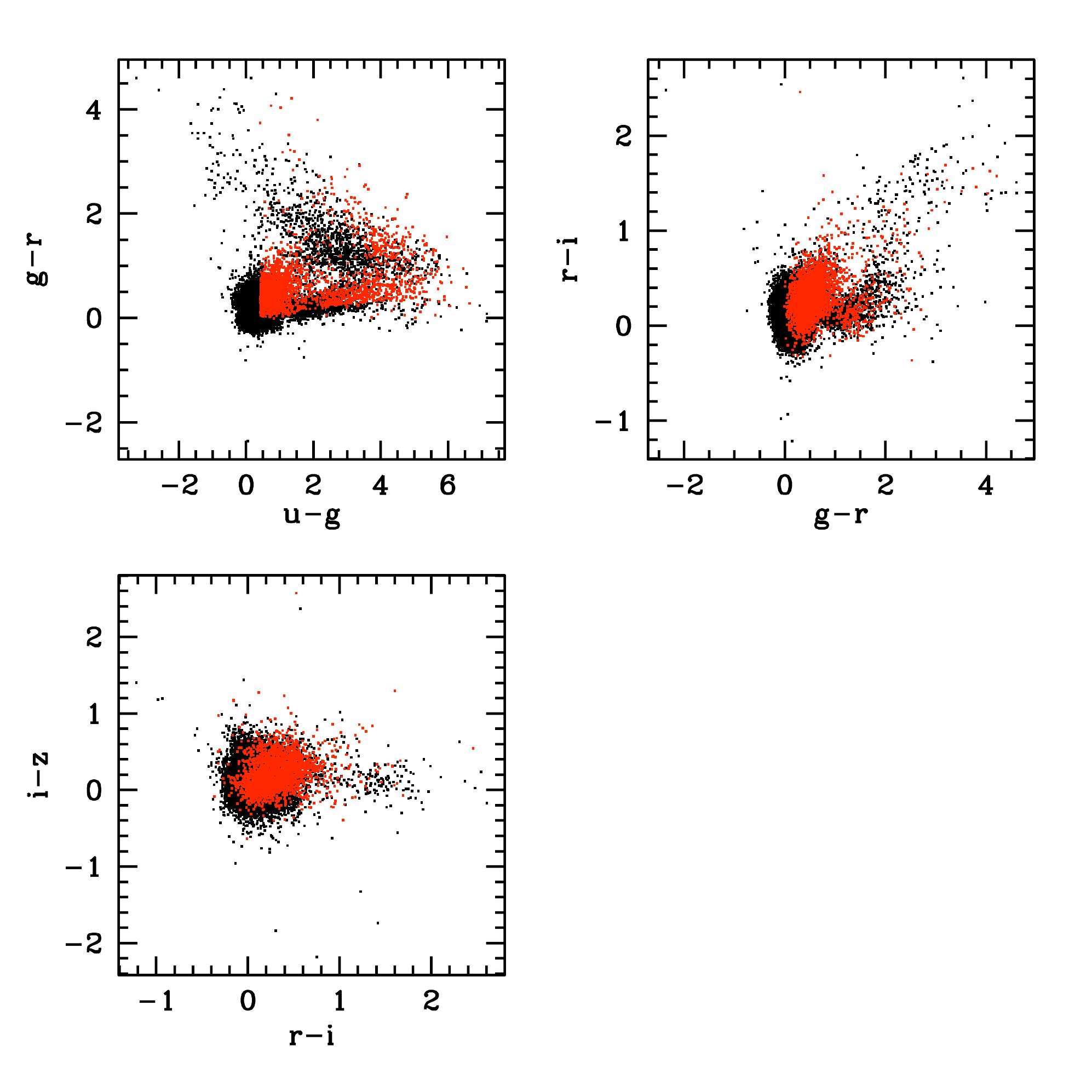}
\caption{Color-color plots for 2,771 reddened quasars classified according to the W04 algorithm.  Black points show all quasars in the sample. 
\label{colorcolor_redqsos}}
\end{figure}

     Another possible source of catastrophic failures is the set of ``reddened" quasars that were excluded from the sample used to generate the CZR.  As discussed in Section~\ref{RedQsoSection}, these quasars are defined as ``red" by a color cut in each (spectroscopic) redshift bin.  A color-color plot superimposing the reddened quasars over the whole sample is shown in Figure~\ref{colorcolor_redqsos}.  When calculating photometric redshifts for the same quasars as were included in the original training data, we find that 1,320 of the 2,771 (~47.6\%) of the reddened quasars are catastrophic failures and 98 of the 143 (~68.5\%) Section 1 objects are considered ``reddened" quasars.  
     
     We generated a CZR from the sample of ``reddened" quasars, and calculated the photometric redshifts for the Section1 quasars.  However, the results are poor.  A rough color cut for ``red" quasars cuts far too many quasars out of the main sample (over 3 times the number of red quasars as eliminated by the algorithm in Section~\ref{RedQsoSection}, and this ``red" CZR is unsuccessful at estimating photometric redshifts.  
     
     \subsection{Spectra of catastrophic failure objects}\label{CFspectrasection}
    We hand-inspected the spectra of the 6,821 catastrophic failures that occurred in the recreation of the results given in \cite{Weinstein} to identify trends that might cause catastrophic failures.  
     \subsubsection{Broad absorption line quasars}\label{CFbalqsosection}
\begin{figure}
\plotone{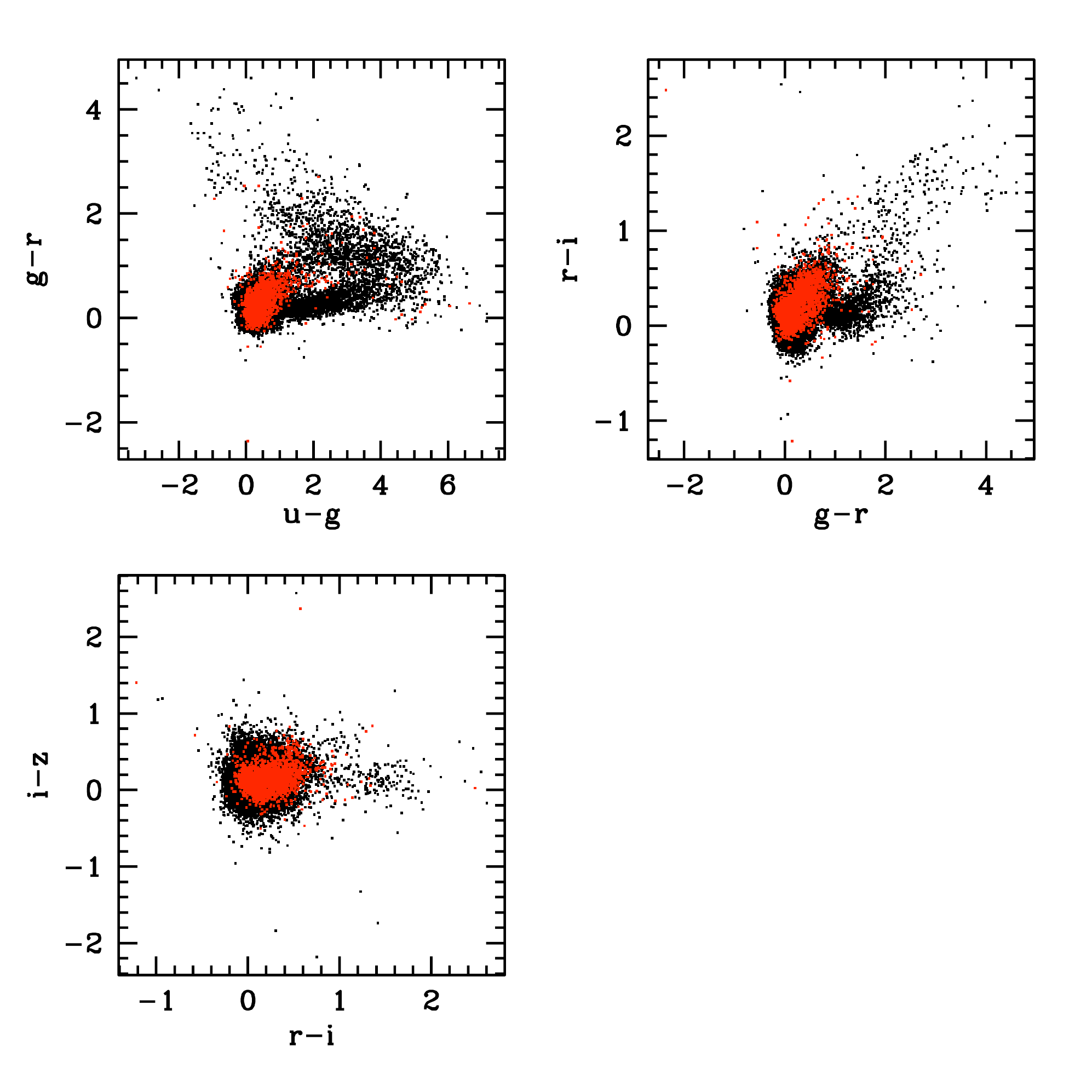}
\caption{Color-color plots for 2,302 BALQSOs that are catastrophic failures (red points). Black points show all other quasars in the sample (including other catastrophic failures. 
\label{colorcolor_BALqsos}}
\end{figure}

     We also have found that a larger-than-average fraction of the catastrophically failing objects are broad absorption line quasars \citep[BALQSOs;][]{Weymann}.  BALQSOs typically constitute about 15\% of all quasars \citep[see, e.g.,][]{Hall}, but we have found that about 34\% of the catastrophic failure objects are BALQSOs.  We suspect that the wide absorption troughs that are seen in BALQSO spectra affect the broadband flux enough that the colors for these objects deviate from the colors seen for normal quasars at the same redshift, giving faulty photometric redshift estimates.  However, we don't see significant color differences of BALQSOs compared to the main sample (see Figure~\ref{colorcolor_BALqsos}).  Therefore, isolating BALQSOs using a simple color cut is not a reasonable way to improve their CZR-based photometric redshift estimates.  
     
     \subsubsection{Quasars with unusual spectra}
     A small subsample of the catastrophic failures (3.2\%) have spectra where the continuum spectrum increases with wavelength.  Therefore, these objects are found to be unusually red compared to normal quasars at the same redshift.  189 of the 219 (~86.3\%) objects with increasing continuum are considered ``reddened" quasars by our definition.  
     
     \subsection{PDFs of catastrophic failure objects}\label{CFpdfsection}
\begin{figure}
\plottwo{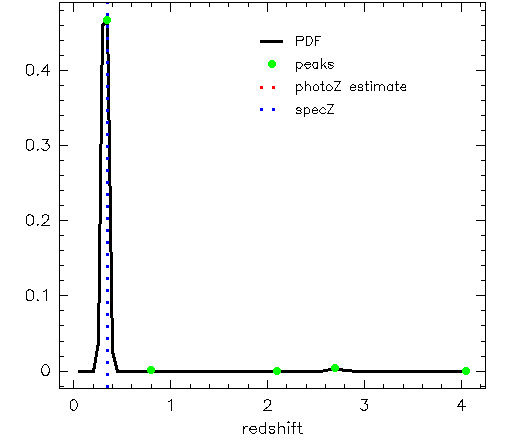}{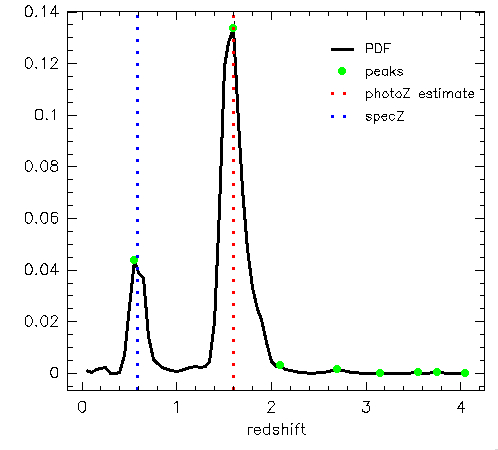}
\caption{\emph{Left:} PDF in which the peak corresponds to the correct redshift.  \emph{Right:} PDF in which the second-highest peak corresponds to the correct redshift.
\label{PDFexamples}}
\end{figure}

  Finally, I have visually inspected the shapes of the PDFs of the objects that have catastrophically failed.  In over 60\% of the PDFs for objects in the extreme failure range, there is no peak at all at the location of the spectroscopic redshift.  However, in CF sections 3 and 4, an overwhelming majority of the PDFs have a peak at the position of the spectroscopic redshift, but a larger peak is present at a different redshift position and therefore being selected by the W04 algorithm (see Figure~\ref{PDFexamples} for example PDFs).  Therefore, changing the weighting used in the calculation of the PDF and/or convolving multiple PDFs may result in selection of the correct peak in the PDF: this could be one of the reasons our variations are improving photometric redshift estimates.  
     
\section{CONCLUSIONS}
Very briefly, when all other things are held constant except the variation being discussed unless otherwise noted, and compared to the Weinstein standard:

\begin{itemize}
  \item A non-overlapping binning scheme does only slightly better than the O2 binning scheme.  
  
  \item Using other magnitudes (than PSF) singly usually gives worse results than the W04 scheme, but in some cases can do only slightly better.  Combining 2, 3, and 4 PDFs from different magnitude types (CZRs generated without removing reddened quasars) does slightly better than the Weinstein standard.  
  
  \item Using a mean statistic does slightly better than gmedian or gmode, with gmode giving the worst results.  
  
  \item Removing reddened quasars gives slightly better results than leaving reddened quasars in the sample.  
  
  \item We applied magnitude error limits (for all bands) on the training data used to build CZRs and found that they made little-to-no difference in the photometric estimate results, whether or not the same error limits were imposed on the test data for which photometric redshifts were calculated.  
  
  \item We imposed magnitude limits (in all bands) of increasing brightness to the traning data used to build CZRs.  When the same magnitude limits were imposed on the test dataset, it is not surprising that a magnitude limit of 20.0 gave a several-percent improvement compared to my recreation of the Weinstein results.  However, if the magnitude limit was not applied to the test data, a CZR-data magnitude limit of 25.0 gave the best results, which were slightly better than the results for my Weinstein recreation. 
  
  \item Using an N(z) weighting function does better than no weighting function by about 1\%.  The apparent magnitude weighting function improves redshift estimates by about 1\%.  
  
  \item From the ``simulation" of adding additional bands, it is clear that using additional colors improves the photometric redshift estimates, though using non-adjacent colors yields worse results than the W04 method. When we use additional non-optical bands to increase the number of CZRs, the photometric redshift estimation success rate increases, but this may be due to the fact that only the brightest quasars with good photometry are present in the non-optical sample, which as we have shown yields a better CZR.  
\end{itemize}

Table~\ref{otherstats} gives the mean statistics for a few of the best photometric redshift estimation techiques.  The best photometric redshift estimates are made when the most parameters are included (i.e., as many bands as possible used to create the CZR, and results from multiple CZRs convolved to calculate the PDF).  We have applied extensive modifications to the W04 CZR algorithm, but found that there is no simple way to meaningfully improve the current success rate with the CZR technique; even the most effective modifications give only marginal improvements on the redshift estimate success rate.  A parallel photometric redshift technique using machine learning algorithms has been developed within our group \citep{Ball2007, Ball2008} with promising results. 

 \begin{table} \centering  \begin{minipage}{140mm}
    \caption{Summary of CZR and Photometric Redshift Algorithm Variations \label{variationSummary}}
    \begin{tabular}{r | l}\hline\hline
    \multicolumn{1}{r |}{Parameter}&\multicolumn{1}{l}{Possible Variations}\\ \hline
    Binning&\emph{overlapping}\\
    &non-overlapping\\
    \hline
    Statistic&\emph{mean}\\
    &discrete median\\
    &Gaussian median\\
    &discrete mode\\
    &Gaussian mode\\
    \hline
    SDSS Magnitude Type&\emph{PSF}\\
    &model\\
    &fiber\\
    &Petrosian\\
    &convolve PDFs for multiple magnitude types\\
    \hline
    Treatment of Reddened Quasars&\emph{removed}\\
    &not removed\\
    &redefined(?)\\
    \hline
    Error and Magnitude Limits&\emph{not explicity imposed}\\
    &impose magnitude limits\\
    &impose error limits\\
    &impose combination of limits\\
    \hline
    PDF Weighting&no weight function\\
    &\emph{simple N(z)}\\
    &apparent magnitude N(m, z)\\
    &luminosity function\\
    \hline
    Additional Bands&\nodata
    \footnotetext{NOTE. -- The parameter variation used by W04 is shown in italics.}
    \end{tabular} \end{minipage}
    \end{table}
    
    \begin{table} \centering \begin{minipage}{140mm}
    \caption{Redshift binning schemes.  \label{binningtable}}
    \begin{tabular}{c | c c | c c c}\hline\hline
    \multicolumn{1}{c}{}&\multicolumn{2}{c}{Divisions}&\multicolumn{3}{c}{$z$-bin Width}\\ 
    Variation&division1&division2&$z\leq$division1&division1$<z\leq$division2&$z>$division2\\ \hline
    A&0.0&2.1&\nodata&0.05&0.2\\
    B&0.0&2.1&\nodata&0.05&0.1\\
    C&0.0&2.25&\nodata&0.05&0.2\\
    D&0.0&2.25&\nodata&0.05&0.1\\
    E&0.1&2.25&0.05&0.025&0.2\\
    F&0.1&2.25&0.05&0.025&0.1\\
    G&0.1&2.25&0.05&0.0125&0.2\\
    H&0.1&2.25&0.05&0.0125&0.1
    \footnotetext{NOTE. -- The (non-overlapping) bin schemes listed in the table above are based on what is referred to in this paper as the O2 (overlapping) binscheme.  This scheme is used by \citet{RichardsB} and \citet{Weinstein}:  bins are centered at intervals in redshift of 0.05; bins with centers \emph{z} $\leq$ 2.15 have width 0.075, bins with centers at 2.15 $<$ \emph{z} $\leq$ 2.5 have width 0.2, and bins with centers at \emph{z} $>$ 2.5 have width 0.5.} 
    \end{tabular} \end{minipage}
    \end{table}

    \begin{table} \centering \begin{minipage}{140mm}
    \caption{Parameters for Luminosity Function. \label{LumFconstants}}
    \begin{tabular}{c c}\hline\hline
    Fitting Parameter&Value\\ \hline
    $A_{1}$&0.78\\
    $B_{1}$&0.10\\
    $B_{2}$&27.35\\
    $B_{3}$&19.27\\
    $M^{*}$&-26\\
    $z_{ref}$&2.45\\
    log($\Phi^{*}$)&-5.75
    \footnotetext{NOTE. -- The values of these parameters are from \citet{Richards06} using the ``Fixed Power Law" form of the luminosity function they derived.  $\Phi^{*}$ has units Mpc$^{-3}$mag$^{-1}$}
    \end{tabular} \end{minipage}
    \end{table}

    \begin{table} \centering \begin{minipage}{140mm}
    \caption{Correlation coefficients between SDSS filter bands.  \label{corrcoefftable}}
    \begin{tabular}{c c}\hline\hline
    Bands&$c_{ij}$\\ \hline
    c$_{ug}$&0.3\\
    c$_{ur}$&0.2\\
    c$_{ui}$&0.1\\
    c$_{uz}$&0.05\\
    c$_{gr}$&0.3\\
    c$_{gi}$&0.2\\
    c$_{gz}$&0.05\\
    c$_{ri}$&0.3\\
    c$_{rz}$&0.05\\
    c$_{iz}$&0.05
    \footnotetext{NOTE. -- The values in this table are estimated from Figure 7 in \citet{Scranton}}  
    \end{tabular} \end{minipage}
    \end{table}

    \begin{table} \centering \begin{minipage}{140mm}
    \caption{Catastrophic Failure Sections for Analysis.  \label{CFsectionstable}}
    \begin{tabular}{c c c c c}\hline\hline
   Section&Range&\# CF objects&\# BALQSOs&\% of section\\ \hline
    1&$z_{phot}>4.0$ and $z_{phot}>z_{spec}$&143&45&31.5\\
    2&$2.4<z_{phot}<4.0$ and $z_{phot}>z_{spec}$&1078&396&36.7\\
    2a&$2.4<z_{phot}<4.0$ and $z_{phot}<z_{spec}$&5&1&20.0\\
    3&$z_{phot}<2.4$ and $z_{phot}>z_{spec}$&3374&586&17.4\\
    4&$z_{phot}<2.4$ and $z_{phot}<z_{spec}$&2221&1273&57.4\\ \hline
    \end{tabular} \end{minipage}
    \end{table}

\clearpage

\clearpage

\clearpage

\clearpage

\clearpage

\clearpage

\clearpage

\clearpage

\clearpage

\chapter{Tables of Results for CZR Based Photometric Redshift Estimator}

\clearpage
    \begin{landscape}
    \begin{table} \begin{minipage}{140mm}
     \caption[PDF Photometric Redshift Estimates:  Mean, SimpleWF, Remove Red]{PDF Photometric Redshift Estimates using mean, Simple Weighting Function and removing reddened quasars.  \label{PDFsimpleWFremovered_table_mean}}
\begin{tabular}{c c c c c || c c c c | c c | c c | c c | c }\hline\hline
     \multicolumn{5}{c}{CZR Parameters}&\multicolumn{4}{c}{$|z_{spec}-z_{phot}|$\footnotemark[1]}&\multicolumn{7}{c}{Catastrophic Failures} \\
     Mag&Bins&Stat&Red?&Weighting&$<0.05$&$<0.1$&$<0.2$&$<0.3$&CF&\% CF&Common\footnotemark[2]&\%&Diff\footnotemark[3]&\%&not CF\footnotemark[4]\\ \hline
psf&O2&mean&yes&SimpleWF&0.3835&0.5889&0.7648&0.8232&6821&0.1469&\nodata&\nodata&\nodata&\nodata&\nodata\\
\hline
psf&A&mean&yes&SimpleWF&0.3878&0.5933&0.7612&0.8192&6822&0.1470&5645&0.8275&693&0.1016&1176\\
psf&C&mean&yes&SimpleWF&0.3975&0.6072&0.7700&0.8258&6524&0.1405&5189&0.7954&875&0.1341&1632\\
psf&E&mean&yes&SimpleWF&0.3967&0.5983&0.7671&0.8265&6503&0.1401&5093&0.7832&942&0.1449&1728\\
psf&G&mean&yes&SimpleWF&0.3930&0.5905&0.7569&0.8175&6872&0.1480&4926&0.7168&1459&0.2123&1895\\
model&O2&mean&yes&SimpleWF&0.3940&0.5988&0.7697&0.8271&6690&0.1443&5533&0.8271&689&0.1030&1288\\
\end{tabular} \end{minipage}
     \end{table}
     \end{landscape}

    \begin{table} \centering \begin{minipage}{140mm}
     \caption[PDF Photometric Redshift Estimates:  Mean, SimpleWF, Remove Red for low-$z$ quasars]{PDF Photometric Redshift Estimates for low-redshift quasars using mean, Simple Weighting Function and removing reddened quasars.  \label{PDFsimpleWFremovered_lowtable_mean}}    
     \begin{tabular}{c c c c c || c c c c } \hline\hline
     \multicolumn{5}{c}{CZR Parameters} & \multicolumn{4}{c}{$z_{spec}\leq$division2} \\ 
     Mag&Bins&Stat&Red?&Weighting&$<0.05$&$<0.1$&$<0.2$&$<0.3$\\ \hline 
psf&O2&mean&yes&SimpleWF&0.3947&0.6036&0.7689&0.8196\\
\hline
psf&A&mean&yes&SimpleWF&0.4065&0.6063&0.7648&0.8155\\
psf&C&mean&yes&SimpleWF&0.4177&0.6222&0.7748&0.8215\\
psf&E&mean&yes&SimpleWF&0.4153&0.6076&0.7660&0.8165\\
psf&G&mean&yes&SimpleWF&0.4102&0.5969&0.7520&0.8039\\
model&O2&mean&yes&SimpleWF&0.4066&0.6152&0.7761&0.8252\\
model&A&mean&yes&SimpleWF&0.4148&0.6121&0.7667&0.8148\\
model&C&mean&yes&SimpleWF&0.4302&0.6314&0.7807&0.8250\\
model&E&mean&yes&SimpleWF&0.4331&0.6219&0.7712&0.8212\\
model&G&mean&yes&SimpleWF&0.4278&0.6090&0.7603&0.8112\\
fiber&O2&mean&yes&SimpleWF&0.3891&0.5935&0.7588&0.8086\\
fiber&A&mean&yes&SimpleWF&0.3911&0.5809&0.7369&0.7868\\
fiber&C&mean&yes&SimpleWF&0.4097&0.6055&0.7610&0.8081\\
fiber&E&mean&yes&SimpleWF&0.4119&0.5982&0.7523&0.8032\\
fiber&G&mean&yes&SimpleWF&0.4018&0.5855&0.7432&0.7942\\
petro&O2&mean&yes&SimpleWF&0.3275&0.5184&0.6956&0.7586\\
petro&A&mean&yes&SimpleWF&0.3351&0.5180&0.6839&0.7439\\
petro&C&mean&yes&SimpleWF&0.3498&0.5377&0.7043&0.7626\\
petro&E&mean&yes&SimpleWF&0.3547&0.5286&0.6885&0.7532\\
petro&G&mean&yes&SimpleWF&0.3418&0.5155&0.6773&0.7400\\ \hline
     \end{tabular} \end{minipage}
     \end{table}

     \begin{table} \centering \begin{minipage}{140mm}
     \caption[PDF Photometric Redshift Estimates:  mean, SimpleWF, Remove Red for high-$z$ quasars]{PDF Photometric Redshift Estimates for high-redshift quasars using mean, Simple Weighting Function and removing reddened quasars.  \label{PDFsimpleWFremovered_hightable_mean}}
     \begin{tabular}{c c c c c || c c c c }\hline\hline
     \multicolumn{5}{c}{CZR Parameters} & \multicolumn{4}{c}{$z_{spec}>$division2} \\
     Mag&Bins&Stat&Red?&Weighting&$<0.05$&$<0.1$&$<0.2$&$<0.3$\\ \hline 
psf&O2&mean&yes&SimpleWF&0.3303&0.5189&0.7458&0.8403\\
\hline
psf&A&mean&yes&SimpleWF&0.2990&0.5318&0.7438&0.8369\\
psf&C&mean&yes&SimpleWF&0.2642&0.5079&0.7382&0.8544\\
psf&E&mean&yes&SimpleWF&0.2740&0.5368&0.7741&0.8927\\
psf&G&mean&yes&SimpleWF&0.2791&0.5482&0.7887&0.9076\\
model&O2&mean&yes&SimpleWF&0.3343&0.5208&0.7389&0.8359\\
model&A&mean&yes&SimpleWF&0.3000&0.5295&0.7388&0.8327\\
model&C&mean&yes&SimpleWF&0.2656&0.5075&0.7373&0.8553\\
model&E&mean&yes&SimpleWF&0.2743&0.5349&0.7723&0.8925\\
model&G&mean&yes&SimpleWF&0.2792&0.5454&0.7856&0.9066\\
fiber&O2&mean&yes&SimpleWF&0.3098&0.4901&0.7272&0.8348\\
fiber&A&mean&yes&SimpleWF&0.2998&0.5243&0.7334&0.8199\\
fiber&C&mean&yes&SimpleWF&0.2656&0.5117&0.7400&0.8516\\
fiber&E&mean&yes&SimpleWF&0.2765&0.5387&0.7768&0.8932\\
fiber&G&mean&yes&SimpleWF&0.2808&0.5488&0.7909&0.9083\\
petro&O2&mean&yes&SimpleWF&0.2605&0.4233&0.6536&0.7760\\
petro&A&mean&yes&SimpleWF&0.2611&0.4619&0.6582&0.7653\\
petro&C&mean&yes&SimpleWF&0.2379&0.4613&0.6759&0.7931\\
petro&E&mean&yes&SimpleWF&0.2510&0.4961&0.7245&0.8476\\
petro&G&mean&yes&SimpleWF&0.2576&0.5149&0.7480&0.8741\\ \hline
     \end{tabular} \end{minipage}
     \end{table}

\clearpage
    \begin{landscape}
    \begin{table} \begin{minipage}{140mm}
     \caption[PDF Photometric Redshift Estimates:  GMedian, SimpleWF, Remove Red]{PDF Photometric Redshift Estimates using gmedian, Simple Weighting Function and removing reddened quasars.  \label{PDFsimpleWFremovered_table_gmedian}}
     \small 
     \begin{tabular}{c c c c c || c c c c | c c | c c | c c | c }\hline\hline
     \multicolumn{5}{c}{CZR Parameters}&\multicolumn{4}{c}{$|z_{spec}-z_{phot}|$}&\multicolumn{7}{c}{Catastrophic Failures} \\
     Mag&Bins&Stat&Red?&Weighting&$<0.05$&$<0.1$&$<0.2$&$<0.3$&CF&\% CF&Common&\%&Diff&\%&not CF\\ \hline
psf&O2&mean&yes&SimpleWF&0.3835&0.5889&0.7648&0.8232&6821&0.1469&\nodata&\nodata&\nodata&\nodata&\nodata\\
\hline
psf&O2&gmedian&yes&SimpleWF&0.3787&0.5840&0.7577&0.8193&6933&0.1494&6062&0.8744&382&0.0551&759\\
psf&A&gmedian&yes&SimpleWF&0.3857&0.5892&0.7576&0.8172&6856&0.1477&5636&0.8221&727&0.1060&1185\\
psf&C&gmedian&yes&SimpleWF&0.3950&0.6023&0.7661&0.8235&6600&0.1422&5247&0.7950&885&0.1341&1574\\
psf&E&gmedian&yes&SimpleWF&0.3958&0.5962&0.7632&0.8228&6634&0.1429&5127&0.7728&1038&0.1565&1694\\
psf&G&gmedian&yes&SimpleWF&0.3906&0.5879&0.7535&0.8149&6961&0.1500&4927&0.7078&1537&0.2208&1894\\
\hline
     \end{tabular} \end{minipage}
     \end{table}
     \end{landscape}
     
    \begin{table} \centering \begin{minipage}{140mm}
     \caption[PDF Photometric Redshift Estimates:  gmedian, SimpleWF, Remove Red for low-$z$ quasars]{PDF Photometric Redshift Estimates for low-redshift quasars using gmedian, Simple Weighting Function and removing reddened quasars.  \label{PDFsimpleWFremovered_lowtable_gmedian}}    
     \begin{tabular}{c c c c c || c c c c } \hline\hline
     \multicolumn{5}{c}{CZR Parameters} & \multicolumn{4}{c}{$z_{spec}\leq$division2} \\ 
     Mag&Bins&Stat&Red?&Weighting&$<0.05$&$<0.1$&$<0.2$&$<0.3$\\ \hline 
psf&O2&mean&yes&SimpleWF&0.3947&0.6036&0.7689&0.8196\\
\hline
psf&O2&gmedian&yes&SimpleWF&0.3899&0.5982&0.7614&0.8157\\
psf&A&gmedian&yes&SimpleWF&0.4043&0.6019&0.7610&0.8136\\
psf&C&gmedian&yes&SimpleWF&0.4147&0.6160&0.7694&0.8184\\
psf&E&gmedian&yes&SimpleWF&0.4141&0.6048&0.7609&0.8120\\
psf&G&gmedian&yes&SimpleWF&0.4075&0.5937&0.7479&0.8009\\
model&O2&gmedian&yes&SimpleWF&0.3989&0.6045&0.7646&0.8167\\
model&A&gmedian&yes&SimpleWF&0.4172&0.6137&0.7660&0.8156\\
model&C&gmedian&yes&SimpleWF&0.4277&0.6263&0.7736&0.8198\\
model&E&gmedian&yes&SimpleWF&0.4333&0.6192&0.7670&0.8166\\
model&G&gmedian&yes&SimpleWF&0.4248&0.6060&0.7555&0.8072\\
fiber&O2&gmedian&yes&SimpleWF&0.3884&0.5884&0.7533&0.8068\\
fiber&A&gmedian&yes&SimpleWF&0.3913&0.5802&0.7366&0.7875\\
fiber&C&gmedian&yes&SimpleWF&0.4070&0.6004&0.7559&0.8041\\
fiber&E&gmedian&yes&SimpleWF&0.4101&0.5968&0.7487&0.8005\\
fiber&G&gmedian&yes&SimpleWF&0.4025&0.5851&0.7404&0.7916\\
petro&O2&gmedian&yes&SimpleWF&0.3249&0.5126&0.6879&0.7543\\
petro&A&gmedian&yes&SimpleWF&0.3378&0.5210&0.6894&0.7498\\
petro&C&gmedian&yes&SimpleWF&0.3418&0.5262&0.6940&0.7526\\
petro&E&gmedian&yes&SimpleWF&0.3504&0.5231&0.6822&0.7468\\
petro&G&gmedian&yes&SimpleWF&0.3410&0.5104&0.6723&0.7364\\ \hline
     \end{tabular} \end{minipage}
     \end{table}

     \begin{table} \centering \begin{minipage}{140mm}
     \caption[PDF Photometric Redshift Estimates:  gmedian, SimpleWF, Remove Red for high-$z$ quasars]{PDF Photometric Redshift Estimates for high-redshift quasars using gmdedian, Simple Weighting Function and removing reddened quasars.  \label{PDFsimpleWFremovered_hightable_gmedian}}
     \begin{tabular}{c c c c c || c c c c }\hline\hline
     \multicolumn{5}{c}{CZR Parameters} & \multicolumn{4}{c}{$z_{spec}>$division2} \\
     Mag&Bins&Stat&Red?&Weighting&$<0.05$&$<0.1$&$<0.2$&$<0.3$\\ \hline 
psf&O2&mean&yes&SimpleWF&0.3303&0.5189&0.7458&0.8403\\
\hline
psf&O2&gmedian&yes&SimpleWF&0.3255&0.5165&0.7399&0.8367\\
psf&A&gmedian&yes&SimpleWF&0.2974&0.5286&0.7412&0.8345\\
psf&C&gmedian&yes&SimpleWF&0.2652&0.5115&0.7441&0.8575\\
psf&E&gmedian&yes&SimpleWF&0.2752&0.5397&0.7785&0.8939\\
psf&G&gmedian&yes&SimpleWF&0.2796&0.5495&0.7910&0.9071\\
model&O2&gmedian&yes&SimpleWF&0.3310&0.5204&0.7337&0.8301\\
model&A&gmedian&yes&SimpleWF&0.3001&0.5300&0.7383&0.8318\\
model&C&gmedian&yes&SimpleWF&0.2661&0.5078&0.7393&0.8561\\
model&E&gmedian&yes&SimpleWF&0.2743&0.5331&0.7718&0.8907\\
model&G&gmedian&yes&SimpleWF&0.2786&0.5445&0.7867&0.9066\\
fiber&O2&gmedian&yes&SimpleWF&0.3113&0.4965&0.7291&0.8348\\
fiber&A&gmedian&yes&SimpleWF&0.2978&0.5212&0.7315&0.8172\\
fiber&C&gmedian&yes&SimpleWF&0.2631&0.5106&0.7390&0.8523\\
fiber&E&gmedian&yes&SimpleWF&0.2724&0.5359&0.7727&0.8900\\
fiber&G&gmedian&yes&SimpleWF&0.2765&0.5454&0.7861&0.9045\\
petro&O2&gmedian&yes&SimpleWF&0.2590&0.4235&0.6510&0.7689\\
petro&A&gmedian&yes&SimpleWF&0.2548&0.4479&0.6446&0.7544\\
petro&C&gmedian&yes&SimpleWF&0.2364&0.4589&0.6790&0.7972\\
petro&E&gmedian&yes&SimpleWF&0.2484&0.4930&0.7260&0.8501\\
petro&G&gmedian&yes&SimpleWF&0.2553&0.5114&0.7496&0.8764\\ \hline
     \end{tabular} \end{minipage}
     \end{table}
     
\clearpage
    \begin{landscape}
    \begin{table} \begin{minipage}{140mm}
     \caption[PDF Photometric Redshift Estimates:  GMode, SimpleWF, Remove Red]{PDF Photometric Redshift Estimates using gmode, Simple Weighting Function and removing reddened quasars.  \label{PDFsimpleWFremovered_table_gmode}}
     \begin{tabular}{c c c c c || c c c c | c c | c c | c c | c }\hline\hline
     \multicolumn{5}{c}{CZR Parameters}&\multicolumn{4}{c}{$|z_{spec}-z_{phot}|$}&\multicolumn{7}{c}{Catastrophic Failures} \\
     Mag&Bins&Stat&Red?&Weighting&$<0.05$&$<0.1$&$<0.2$&$<0.3$&CF&\% CF&Common&\%&Diff&\%&not CF\\ \hline
psf&O2&mean&yes&SimpleWF&0.3835&0.5889&0.7648&0.8232&6821&0.1469&\nodata&\nodata&\nodata&\nodata&\nodata\\
\hline
psf&O2&gmode&yes&SimpleWF&0.3675&0.5696&0.7451&0.8105&7224&0.1556&5795&0.8022&924&0.1279&1026\\
psf&A&gmode&yes&SimpleWF&0.3745&0.5743&0.7434&0.8058&7214&0.1554&5489&0.7609&1209&0.1676&1332\\
psf&C&gmode&yes&SimpleWF&0.3866&0.5924&0.7586&0.8181&6722&0.1448&5283&0.7859&955&0.1421&1538\\
psf&E&gmode&yes&SimpleWF&0.3831&0.5824&0.7526&0.8144&6984&0.1505&5113&0.7321&1374&0.1967&1708\\
psf&G&gmode&yes&SimpleWF&0.3725&0.5706&0.7428&0.8043&7388&0.1592&4933&0.6677&1920&0.2599&1888\\
\hline
     \end{tabular} \end{minipage}
     \end{table}
     \end{landscape}

    \begin{table} \centering \begin{minipage}{140mm}
     \caption[PDF Photometric Redshift Estimates:  gmode, SimpleWF, Remove Red for low-$z$ quasars]{PDF Photometric Redshift Estimates for low-redshift quasars using gmode, Simple Weighting Function and removing reddened quasars.  \label{PDFsimpleWFremovered_lowtable_gmode}}    
     \begin{tabular}{c c c c c || c c c c } \hline\hline
     \multicolumn{5}{c}{CZR Parameters} & \multicolumn{4}{c}{$z_{spec}\leq$division2} \\ 
     Mag&Bins&Stat&Red?&Weighting&$<0.05$&$<0.1$&$<0.2$&$<0.3$\\ \hline 
psf&O2&mean&yes&SimpleWF&0.3947&0.6036&0.7689&0.8196\\
\hline
psf&O2&gmode&yes&SimpleWF&0.3745&0.5808&0.7469&0.8026\\
psf&A&gmode&yes&SimpleWF&0.3910&0.5852&0.7435&0.7975\\
psf&C&gmode&yes&SimpleWF&0.4045&0.6035&0.7589&0.8094\\
psf&E&gmode&yes&SimpleWF&0.3995&0.5893&0.7492&0.8023\\
psf&G&gmode&yes&SimpleWF&0.3869&0.5746&0.7363&0.7890\\
model&O2&gmode&yes&SimpleWF&0.3838&0.5868&0.7460&0.7988\\
model&A&gmode&yes&SimpleWF&0.3932&0.5846&0.7381&0.7890\\
model&C&gmode&yes&SimpleWF&0.4104&0.6078&0.7606&0.8100\\
model&E&gmode&yes&SimpleWF&0.4091&0.5974&0.7556&0.8069\\
model&G&gmode&yes&SimpleWF&0.3921&0.5769&0.7377&0.7901\\
fiber&O2&gmode&yes&SimpleWF&0.3686&0.5680&0.7327&0.7868\\
fiber&A&gmode&yes&SimpleWF&0.3748&0.5677&0.7297&0.7859\\
fiber&C&gmode&yes&SimpleWF&0.3806&0.5737&0.7356&0.7890\\
fiber&E&gmode&yes&SimpleWF&0.3809&0.5686&0.7278&0.7826\\
fiber&G&gmode&yes&SimpleWF&0.3638&0.5448&0.7128&0.7683\\
petro&O2&gmode&yes&SimpleWF&0.3074&0.4884&0.6619&0.7366\\
petro&A&gmode&yes&SimpleWF&0.3137&0.4909&0.6572&0.7266\\
petro&C&gmode&yes&SimpleWF&0.3239&0.5043&0.6722&0.7403\\
petro&E&gmode&yes&SimpleWF&0.3251&0.4951&0.6643&0.7331\\
petro&G&gmode&yes&SimpleWF&0.3038&0.4718&0.6405&0.7076\\ \hline
     \end{tabular} \end{minipage}
     \end{table}

     \begin{table} \centering \begin{minipage}{140mm}
     \caption[PDF Photometric Redshift Estimates:  gmode, SimpleWF, Remove Red for high-$z$ quasars]{PDF Photometric Redshift Estimates for high-redshift quasars using gmode, Simple Weighting Function and removing reddened quasars.  \label{PDFsimpleWFremovered_hightable_gmode}}
     \begin{tabular}{c c c c c || c c c c }\hline\hline
     \multicolumn{5}{c}{CZR Parameters} & \multicolumn{4}{c}{$z_{spec}>$division2} \\
     Mag&Bins&Stat&Red?&Weighting&$<0.05$&$<0.1$&$<0.2$&$<0.3$\\ \hline 
psf&O2&mean&yes&SimpleWF&0.3303&0.5189&0.7458&0.8403\\
\hline
psf&O2&gmode&yes&SimpleWF&0.3342&0.5167&0.7367&0.8482\\
psf&A&gmode&yes&SimpleWF&0.2961&0.5226&0.7427&0.8451\\
psf&C&gmode&yes&SimpleWF&0.2685&0.5188&0.7571&0.8750\\
psf&E&gmode&yes&SimpleWF&0.2750&0.5368&0.7753&0.8943\\
psf&G&gmode&yes&SimpleWF&0.2778&0.5443&0.7854&0.9060\\
model&O2&gmode&yes&SimpleWF&0.3292&0.5081&0.7321&0.8435\\
model&A&gmode&yes&SimpleWF&0.2980&0.5204&0.7353&0.8389\\
model&C&gmode&yes&SimpleWF&0.2676&0.5183&0.7547&0.8774\\
model&E&gmode&yes&SimpleWF&0.2717&0.5368&0.7760&0.8989\\
model&G&gmode&yes&SimpleWF&0.2738&0.5408&0.7816&0.9062\\
fiber&O2&gmode&yes&SimpleWF&0.3127&0.4928&0.7335&0.8461\\
fiber&A&gmode&yes&SimpleWF&0.2872&0.5065&0.7244&0.8250\\
fiber&C&gmode&yes&SimpleWF&0.2621&0.5135&0.7482&0.8661\\
fiber&E&gmode&yes&SimpleWF&0.2669&0.5304&0.7694&0.8897\\
fiber&G&gmode&yes&SimpleWF&0.2701&0.5364&0.7770&0.8986\\
petro&O2&gmode&yes&SimpleWF&0.2691&0.4331&0.6602&0.7850\\
petro&A&gmode&yes&SimpleWF&0.2616&0.4580&0.6582&0.7703\\
petro&C&gmode&yes&SimpleWF&0.2431&0.4791&0.7045&0.8219\\
petro&E&gmode&yes&SimpleWF&0.2533&0.5050&0.7383&0.8595\\
petro&G&gmode&yes&SimpleWF&0.2578&0.5157&0.7513&0.8746\\ \hline
     \end{tabular} \end{minipage}
     \end{table}

\clearpage
     \begin{landscape}
     \begin{table} \begin{minipage}{140mm}
     \caption{PDF Photometric Redshift Estimates using combinations of PDFs from different magnitudes without removing reddened quasars.  \label{PDFmultimagsO2}}
     \footnotesize
     \begin{tabular}{c c c c c || c c c c | c c | c c | c c | c }\hline\hline
     \multicolumn{5}{c}{CZR Parameters} & \multicolumn{4}{c}{$|z_{spec}-z_{phot}|$} & \multicolumn{7}{c}{Catastrophic Failures} \\
     Mag&Bins&Stat&Red?&Weighting&$<0.05$&$<0.1$&$<0.2$&$<0.3$&\# CF&\% CF&\# Common&\%&\# Diff&\%&\# not CF\\ \hline
psf&O2&mean&yes&SimpleWF&0.3835&0.5889&0.7648&0.8232&6821&0.1469&\nodata&\nodata&\nodata&\nodata&\nodata\\
\hline
psfmodel&O2&mean&no&SimpleWF&0.3691&0.5775&0.7693&0.8273&6699&0.1445&4832&0.7213&1411&0.2106&1989\\
psffiber&O2&mean&no&SimpleWF&0.3688&0.5769&0.7697&0.8272&6744&0.1455&4780&0.7088&1495&0.2217&2041\\
psfpetro&O2&mean&no&SimpleWF&0.3444&0.5469&0.7495&0.8148&7094&0.1530&4646&0.6549&1972&0.2780&2175\\
modelfiber&O2&mean&no&SimpleWF&0.3689&0.5755&0.7693&0.8273&6759&0.1458&4694&0.6945&1603&0.2372&2127\\
modelpetro&O2&mean&no&SimpleWF&0.3425&0.5437&0.7457&0.8114&7255&0.1565&4616&0.6363&2176&0.2999&2205\\
fiberpetro&O2&mean&no&SimpleWF&0.3303&0.5278&0.7303&0.7994&7619&0.1644&4511&0.5921&2596&0.3407&2310\\
\hline
psfmodelfiber&O2&mean&no&SimpleWF&0.3707&0.5789&0.7712&0.8286&6689&0.1443&4790&0.7161&1439&0.2151&2031\\
psfmodelpetro&O2&mean&no&SimpleWF&0.3579&0.5615&0.7594&0.8215&6919&0.1493&4691&0.6780&1763&0.2548&2130\\
psffiberpetro&O2&mean&no&SimpleWF&0.3539&0.5564&0.7564&0.8193&6968&0.1503&4634&0.6650&1866&0.2678&2187\\
modelfiberpetro&O2&mean&no&SimpleWF&0.3530&0.5540&0.7537&0.8176&7068&0.1525&4602&0.6511&2001&0.2831&2219\\
\hline
allmags&O2&mean&no&SimpleWF&0.3612&0.5644&0.7621&0.8236&6855&0.1479&4679&0.6826&1714&0.2500&2142\\
allmags&O2&mean&no&NoWF&0.3752&0.5752&0.7507&0.8089&7448&0.1607&4021&0.5399&2944&0.3953&2800\\ \hline
     \end{tabular} \end{minipage}
     \end{table}
     \end{landscape}
     
     \begin{table} \centering \begin{minipage}{140mm}
     \caption{PDF Photometric Redshift Estimates for low-redshift quasars using combinations of PDFs from different magnitudes without removing reddened quasars. \label{PDFmultimagsO2lowtable}}
     \begin{tabular}{c c c c c || c c c c }\hline\hline
     \multicolumn{5}{c}{CZR Parameters} & \multicolumn{4}{c}{$z_{spec}\leq$division2} \\
     Mag&Bins&Stat&Red?&Weighting&$<0.05$&$<0.1$&$<0.2$&$<0.3$\\ \hline
psf&O2&mean&yes&SimpleWF&0.3947&0.6036&0.7689&0.8196\\
\hline
psfmodel&O2&mean&no&SimpleWF&0.3843&0.6006&0.7861&0.8387\\
psffiber&O2&mean&no&SimpleWF&0.3844&0.5988&0.7845&0.8362\\
psfpetro&O2&mean&no&SimpleWF&0.3580&0.5668&0.7646&0.8238\\
modelfiber&O2&mean&no&SimpleWF&0.3840&0.5968&0.7842&0.8365\\
modelpetro&O2&mean&no&SimpleWF&0.3559&0.5631&0.7607&0.8206\\
fiberpetro&O2&mean&no&SimpleWF&0.3427&0.5460&0.7434&0.8061\\
\hline
psfmodelfiber&O2&mean&no&SimpleWF&0.3858&0.6003&0.7860&0.8378\\
psfmodelpetro&O2&mean&no&SimpleWF&0.3717&0.5819&0.7747&0.8308\\
psffiberpetro&O2&mean&no&SimpleWF&0.3675&0.5755&0.7698&0.8269\\
modelfiberpetro&O2&mean&no&SimpleWF&0.3665&0.5731&0.7675&0.8255\\
\hline
allmags&O2&mean&no&SimpleWF&0.3756&0.5848&0.7768&0.8322\\
allmags&O2&mean&no&NoWF&0.4104&0.6189&0.7918&0.8465\\ \hline
     \end{tabular} \end{minipage}
     \end{table}
     
     \begin{table} \centering \begin{minipage}{140mm}
     \caption{PDF Photometric Redshift Estimates for high-redshift quasars using combinations of PDFs from different magnitudes without removing reddened quasars. \label{PDFmultimagsO2_hightable}}
     \begin{tabular}{c c c c c || c c c c }\hline\hline
     \multicolumn{5}{c}{CZR Parameters} & \multicolumn{4}{c}{$z_{spec}>$division2} \\
     Mag&Bins&Stat&Red?&Weighting&$<0.05$&$<0.1$&$<0.2$&$<0.3$\\ \hline
psf&O2&mean&yes&SimpleWF&0.3947&0.6036&0.7689&0.8196\\
\hline
psfmodel&O2&mean&no&SimpleWF&0.2966&0.4681&0.6890&0.7729\\
psffiber&O2&mean&no&SimpleWF&0.2944&0.4727&0.6990&0.7844\\
psfpetro&O2&mean&no&SimpleWF&0.2797&0.4522&0.6777&0.7718\\
modelfiber&O2&mean&no&SimpleWF&0.2972&0.4738&0.6987&0.7831\\
modelpetro&O2&mean&no&SimpleWF&0.2790&0.4512&0.6744&0.7681\\
fiberpetro&O2&mean&no&SimpleWF&0.2715&0.4415&0.6677&0.7675\\
\hline
psfmodelfiber&O2&mean&no&SimpleWF&0.2990&0.4774&0.7010&0.7850\\
psfmodelpetro&O2&mean&no&SimpleWF&0.2922&0.4649&0.6870&0.7774\\
psffiberpetro&O2&mean&no&SimpleWF&0.2890&0.4653&0.6924&0.7828\\
modelfiberpetro&O2&mean&no&SimpleWF&0.2890&0.4634&0.6879&0.7801\\
\hline
allmags&O2&mean&no&SimpleWF&0.2926&0.4672&0.6923&0.7829\\
allmags&O2&mean&no&NoWF&0.2076&0.3672&0.5553&0.6303\\ \hline
     \end{tabular} \end{minipage}
     \end{table}
     
\clearpage          
     \begin{landscape}
     \begin{table} \begin{minipage}{140mm}
     \caption[PDF Photometric Redshift Estimates:  Mean, SimpleWF, Red not removed]{PDF Photometric Redshift Estimates using mean, Simple Weighting Function, without removing reddened quasars.  \label{PDFsimpleWFnoremovered_table_mean}}
     \begin{tabular}{c c c c c || c c c c | c c | c c | c c | c }\hline\hline
     \multicolumn{5}{c}{CZR Parameters}&\multicolumn{4}{c}{$|z_{spec}-z_{phot}|$}&\multicolumn{7}{c}{Catastrophic Failures} \\
     Mag&Bins&Stat&Red?&Weighting&$<0.05$&$<0.1$&$<0.2$&$<0.3$&CF&\% CF&Common&\%&Diff&\%&not CF\\ \hline
psf&O2&mean&yes&SimpleWF&0.3835&0.5889&0.7648&0.8232&6821&0.1469&\nodata&\nodata&\nodata&\nodata&\nodata\\
\hline
psf&O2&mean&no&SimpleWF&0.3648&0.5736&0.7660&0.8242&6800&0.1465&4883&0.7181&1446&0.2126&1938\\
psf&A&mean&no&SimpleWF&0.3596&0.5638&0.7580&0.8198&6815&0.1468&4905&0.7197&1445&0.2120&1916\\
psf&C&mean&no&SimpleWF&0.3669&0.5737&0.7593&0.8168&6959&0.1499&4195&0.6028&2303&0.3309&2626\\
psf&E&mean&no&SimpleWF&0.3736&0.5799&0.7683&0.8284&6372&0.1373&4354&0.6833&1565&0.2456&2467\\
psf&G&mean&no&SimpleWF&0.3828&0.5890&0.7667&0.8274&6410&0.1381&4445&0.6934&1518&0.2368&2376\\
\hline
     \end{tabular} \end{minipage}
     \end{table}
     \end{landscape}

     \begin{table} \centering \begin{minipage}{140mm}
     \caption[PDF Photometric Redshift Estimates:  Mean, SimpleWF, Red not removed for low-$z$ quasars]{PDF Photometric Redshift Estimates for low-redshift quasars using mean, Simple Weighting Function without removing reddened quasars.  \label{PDFsimpleWFnoremovered_lowtable_mean}}
     \begin{tabular}{ c c c c c || c c c c }\hline\hline
     \multicolumn{5}{c}{CZR Parameters} & \multicolumn{4}{c}{$z_{spec}\leq$division2} \\
     Mag&Bins&Stat&Red?&Weighting&$<0.05$&$<0.1$&$<0.2$&$<0.3$\\ \hline 
psf&O2&mean&yes&SimpleWF&0.3947&0.6036&0.7689&0.8196\\
\hline
psf&O2&mean&no&SimpleWF&0.3808&0.5970&0.7830&0.8359\\
psf&A&mean&no&SimpleWF&0.3794&0.5823&0.7729&0.8284\\
psf&C&mean&no&SimpleWF&0.3852&0.5885&0.7696&0.8198\\
psf&E&mean&no&SimpleWF&0.3901&0.5895&0.7710&0.8229\\
psf&G&mean&no&SimpleWF&0.3993&0.5962&0.7646&0.8164\\
model&O2&mean&no&SimpleWF&0.3849&0.5992&0.7853&0.8383\\
model&A&mean&no&SimpleWF&0.3832&0.5824&0.7700&0.8252\\
model&C&mean&no&SimpleWF&0.3918&0.5928&0.7704&0.8201\\
model&E&mean&no&SimpleWF&0.3984&0.5951&0.7730&0.8253\\
model&G&mean&no&SimpleWF&0.4141&0.6072&0.7713&0.8237\\
fiber&O2&mean&no&SimpleWF&0.3731&0.5832&0.7703&0.8254\\
fiber&A&mean&no&SimpleWF&0.3664&0.5603&0.7439&0.7993\\
fiber&C&mean&no&SimpleWF&0.3831&0.5824&0.7605&0.8113\\
fiber&E&mean&no&SimpleWF&0.3886&0.5850&0.7587&0.8112\\
fiber&G&mean&no&SimpleWF&0.3920&0.5858&0.7524&0.8059\\
petro&O2&mean&no&SimpleWF&0.2884&0.4756&0.6790&0.7547\\
petro&A&mean&no&SimpleWF&0.2955&0.4784&0.6745&0.7467\\
petro&C&mean&no&SimpleWF&0.3088&0.4963&0.6856&0.7541\\
petro&E&mean&no&SimpleWF&0.3205&0.5018&0.6836&0.7514\\
petro&G&mean&no&SimpleWF&0.3299&0.5135&0.6860&0.7495\\ \hline
     \end{tabular} \end{minipage}
     \end{table}

     \begin{table} \centering \begin{minipage}{140mm}
     \caption[PDF Photometric Redshift Estimates:  mean, SimpleWF, Red not removed for high-$z$ quasars]{PDF Photometric Redshift Estimates for high-redshift quasars using mean, Simple Weighting Function without removing reddened quasars.  \label{PDFsimpleWFnoremovered_hightable_mean}}
     \begin{tabular}{c c c c c || c c c c }\hline\hline
     \multicolumn{5}{c}{CZR Parameters} & \multicolumn{4}{c}{$z_{spec}> $division2} \\
     Mag&Bins&Stat&Red?&Weighting&$<0.05$&$<0.1$&$<0.2$&$<0.3$\\ \hline 
psf&O2&mean&yes&SimpleWF&0.3303&0.5189&0.7458&0.8403\\
\hline
psf&O2&mean&no&SimpleWF&0.2887&0.4624&0.6848&0.7687\\
psf&A&mean&no&SimpleWF&0.2655&0.4758&0.6869&0.7790\\
psf&C&mean&no&SimpleWF&0.2462&0.4760&0.6907&0.7967\\
psf&E&mean&no&SimpleWF&0.2644&0.5170&0.7499&0.8650\\
psf&G&mean&no&SimpleWF&0.2744&0.5414&0.7807&0.8999\\
model&O2&mean&no&SimpleWF&0.3008&0.4711&0.6894&0.7730\\
model&A&mean&no&SimpleWF&0.2636&0.4763&0.6822&0.7749\\
model&C&mean&no&SimpleWF&0.2482&0.4773&0.6947&0.7997\\
model&E&mean&no&SimpleWF&0.2643&0.5140&0.7454&0.8600\\
model&G&mean&no&SimpleWF&0.2735&0.5385&0.7780&0.8965\\
fiber&O2&mean&no&SimpleWF&0.2892&0.4628&0.6934&0.7825\\
fiber&A&mean&no&SimpleWF&0.2648&0.4741&0.6821&0.7771\\
fiber&C&mean&no&SimpleWF&0.2478&0.4737&0.6942&0.7978\\
fiber&E&mean&no&SimpleWF&0.2678&0.5204&0.7571&0.8705\\
fiber&G&mean&no&SimpleWF&0.2738&0.5395&0.7830&0.8992\\
petro&O2&mean&no&SimpleWF&0.2268&0.3823&0.5957&0.7016\\
petro&A&mean&no&SimpleWF&0.2131&0.3883&0.5754&0.6886\\
petro&C&mean&no&SimpleWF&0.2088&0.4068&0.6038&0.7097\\
petro&E&mean&no&SimpleWF&0.2321&0.4562&0.6705&0.7884\\
petro&G&mean&no&SimpleWF&0.2443&0.4901&0.7181&0.8414\\ \hline
     \end{tabular} \end{minipage}
     \end{table}

\clearpage          
     \begin{landscape}
     \begin{table}  \begin{minipage}{140mm}
     \caption[PDF Photometric Redshift Estimates:  gmedian, SimpleWF, Red not removed]{PDF Photometric Redshift Estimates using gmedian, Simple Weighting Function, without removing reddened quasars.  \label{PDFsimpleWFnoremovered_table_gmedian}}
     \small
     \begin{tabular}{c c c c c || c c c c | c c | c c | c c | c }\hline\hline
     \multicolumn{5}{c}{CZR Parameters}&\multicolumn{4}{c}{$|z_{spec}-z_{phot}|$}&\multicolumn{7}{c}{Catastrophic Failures} \\
     Mag&Bins&Stat&Red?&Weighting&$<0.05$&$<0.1$&$<0.2$&$<0.3$&CF&\% CF&Common&\%&Diff&\%&not CF\\ \hline
psf&O2&mean&yes&SimpleWF&0.3835&0.5889&0.7648&0.8232&6821&0.1469&\nodata&\nodata&\nodata&\nodata&\nodata\\
\hline
psf&O2&gmedian&no&SimpleWF&0.3633&0.5702&0.7601&0.8212&6917&0.1490&4854&0.7017&1580&0.2284&1967\\
psf&A&gmedian&no&SimpleWF&0.3580&0.5593&0.7518&0.8164&6897&0.1486&4841&0.7019&1585&0.2298&1980\\
psf&C&gmedian&no&SimpleWF&0.3646&0.5684&0.7545&0.8149&7003&0.1509&4214&0.6017&2315&0.3306&2607\\
psf&E&gmedian&no&SimpleWF&0.3730&0.5762&0.7635&0.8253&6516&0.1404&4382&0.6725&1679&0.2577&2439\\
psf&G&gmedian&no&SimpleWF&0.3816&0.5837&0.7613&0.8231&6558&0.1413&4439&0.6769&1670&0.2547&2382\\
\hline
     \end{tabular} \end{minipage}
     \end{table}
     \end{landscape}

     \begin{table} \centering \begin{minipage}{140mm}
     \caption[PDF Photometric Redshift Estimates:  gmedian, SimpleWF, Red not removed for low-$z$ quasars]{PDF Photometric Redshift Estimates for low-redshift quasars using gmedian, Simple Weighting Function without removing reddened quasars.  \label{PDFsimpleWFnoremovered_lowtable_gmedian}}
     \begin{tabular}{ c c c c c || c c c c }\hline\hline
     \multicolumn{5}{c}{CZR Parameters} & \multicolumn{4}{c}{$z_{spec}\leq$division2} \\
     Mag&Bins&Stat&Red?&Weighting&$<0.05$&$<0.1$&$<0.2$&$<0.3$\\ \hline 
psf&O2&mean&yes&SimpleWF&0.3947&0.6036&0.7689&0.8196\\
\hline
psf&O2&gmedian&no&SimpleWF&0.3789&0.5912&0.7745&0.8309\\
psf&A&gmedian&no&SimpleWF&0.3772&0.5763&0.7648&0.8236\\
psf&C&gmedian&no&SimpleWF&0.3815&0.5810&0.7618&0.8158\\
psf&E&gmedian&no&SimpleWF&0.3887&0.5841&0.7636&0.8180\\
psf&G&gmedian&no&SimpleWF&0.3973&0.5892&0.7570&0.8108\\
model&O2&gmedian&no&SimpleWF&0.3836&0.5943&0.7761&0.8330\\
model&A&gmedian&no&SimpleWF&0.3813&0.5807&0.7668&0.8235\\
model&C&gmedian&no&SimpleWF&0.3861&0.5860&0.7626&0.8143\\
model&E&gmedian&no&SimpleWF&0.3955&0.5898&0.7660&0.8206\\
model&G&gmedian&no&SimpleWF&0.4108&0.5997&0.7627&0.8173\\
fiber&O2&gmedian&no&SimpleWF&0.3709&0.5756&0.7617&0.8208\\
fiber&A&gmedian&no&SimpleWF&0.3640&0.5580&0.7395&0.7961\\
fiber&C&gmedian&no&SimpleWF&0.3777&0.5759&0.7522&0.8049\\
fiber&E&gmedian&no&SimpleWF&0.3858&0.5778&0.7506&0.8061\\
fiber&G&gmedian&no&SimpleWF&0.3873&0.5759&0.7427&0.7990\\
petro&O2&gmedian&no&SimpleWF&0.2849&0.4700&0.6694&0.7480\\
petro&A&gmedian&no&SimpleWF&0.2985&0.4833&0.6738&0.7481\\
petro&C&gmedian&no&SimpleWF&0.3027&0.4878&0.6709&0.7416\\
petro&E&gmedian&no&SimpleWF&0.3139&0.4885&0.6670&0.7387\\
petro&G&gmedian&no&SimpleWF&0.3252&0.5041&0.6745&0.7411\\ \hline
     \end{tabular} \end{minipage}
     \end{table}

     \begin{table} \centering \begin{minipage}{140mm}
     \caption[PDF Photometric Redshift Estimates:  gmedian, SimpleWF, Red not removed for high-$z$ quasars]{PDF Photometric Redshift Estimates for high-redshift quasars using gmedian, Simple Weighting Function without removing reddened quasars.  \label{PDFsimpleWFnoremovered_hightable_gmedian}}
     \begin{tabular}{c c c c c || c c c c }\hline\hline
     \multicolumn{5}{c}{CZR Parameters} & \multicolumn{4}{c}{$z_{spec}> $division2} \\
     Mag&Bins&Stat&Red?&Weighting&$<0.05$&$<0.1$&$<0.2$&$<0.3$\\ \hline 
psf&O2&mean&yes&SimpleWF&0.3303&0.5189&0.7458&0.8403\\
\hline
psf&O2&gmedian&no&SimpleWF&0.2892&0.4704&0.6917&0.7750\\
psf&A&gmedian&no&SimpleWF&0.2670&0.4784&0.6900&0.7818\\
psf&C&gmedian&no&SimpleWF&0.2526&0.4855&0.7060&0.8088\\
psf&E&gmedian&no&SimpleWF&0.2690&0.5242&0.7625&0.8735\\
psf&G&gmedian&no&SimpleWF&0.2781&0.5469&0.7897&0.9047\\
model&O2&gmedian&no&SimpleWF&0.2962&0.4739&0.6927&0.7764\\
model&A&gmedian&no&SimpleWF&0.2668&0.4763&0.6849&0.7765\\
model&C&gmedian&no&SimpleWF&0.2532&0.4863&0.7086&0.8118\\
model&E&gmedian&no&SimpleWF&0.2697&0.5234&0.7600&0.8728\\
model&G&gmedian&no&SimpleWF&0.2778&0.5449&0.7879&0.9034\\
fiber&O2&gmedian&no&SimpleWF&0.2907&0.4755&0.7026&0.7898\\
fiber&A&gmedian&no&SimpleWF&0.2651&0.4769&0.6836&0.7784\\
fiber&C&gmedian&no&SimpleWF&0.2526&0.4838&0.7052&0.8085\\
fiber&E&gmedian&no&SimpleWF&0.2713&0.5293&0.7669&0.8785\\
fiber&G&gmedian&no&SimpleWF&0.2772&0.5449&0.7886&0.9033\\
petro&O2&gmedian&no&SimpleWF&0.2252&0.3819&0.5910&0.6922\\
petro&A&gmedian&no&SimpleWF&0.2155&0.3870&0.5762&0.6864\\
petro&C&gmedian&no&SimpleWF&0.2129&0.4108&0.6102&0.7142\\
petro&E&gmedian&no&SimpleWF&0.2385&0.4694&0.6904&0.8068\\
petro&G&gmedian&no&SimpleWF&0.2486&0.4988&0.7309&0.8519\\ \hline
     \end{tabular} \end{minipage}
     \end{table}

\clearpage          
     \begin{landscape}
     \begin{table} \begin{minipage}{140mm}
     \caption[PDF Photometric Redshift Estimates:  gmode, SimpleWF, Red not removed]{PDF Photometric Redshift Estimates using gmode, Simple Weighting Function, without removing reddened quasars.  \label{PDFsimpleWFnoremovered_table_gmode}}
     \begin{tabular}{c c c c c || c c c c | c c | c c | c c | c }\hline\hline
     \multicolumn{5}{c}{CZR Parameters}&\multicolumn{4}{c}{$|z_{spec}-z_{phot}|$}&\multicolumn{7}{c}{Catastrophic Failures} \\
     Mag&Bins&Stat&Red?&Weighting&$<0.05$&$<0.1$&$<0.2$&$<0.3$&CF&\% CF&Common&\%&Diff&\%&not CF\\ \hline
psf&O2&mean&yes&SimpleWF&0.3835&0.5889&0.7648&0.8232&6821&0.1469&\nodata&\nodata&\nodata&\nodata&\nodata\\
\hline
psf&O2&gmode&no&SimpleWF&0.3546&0.5597&0.7510&0.8184&6890&0.1484&4821&0.6997&1576&0.2287&2000\\
psf&A&gmode&no&SimpleWF&0.3502&0.5502&0.7391&0.8070&7138&0.1538&4775&0.6690&1871&0.2621&2046\\
psf&C&gmode&no&SimpleWF&0.3607&0.5655&0.7507&0.8138&6901&0.1487&4239&0.6143&2192&0.3176&2582\\
psf&E&gmode&no&SimpleWF&0.3646&0.5692&0.7538&0.8174&6765&0.1457&4343&0.6420&1947&0.2878&2478\\
psf&G&gmode&no&SimpleWF&0.3650&0.5673&0.7474&0.8117&6967&0.1501&4398&0.6313&2085&0.2993&2423\\
\hline
     \end{tabular} \end{minipage}
     \end{table}
     \end{landscape}

     \begin{table} \centering \begin{minipage}{140mm}
     \caption[PDF Photometric Redshift Estimates:  gmode, SimpleWF, Red not removed for low-$z$ quasars]{PDF Photometric Redshift Estimates for low-redshift quasars using gmode, Simple Weighting Function without removing reddened quasars.  \label{PDFsimpleWFnoremovered_lowtable_gmode}}
     \begin{tabular}{ c c c c c || c c c c }\hline\hline
     \multicolumn{5}{c}{CZR Parameters} & \multicolumn{4}{c}{$z_{spec}\leq$division2} \\
     Mag&Bins&Stat&Red?&Weighting&$<0.05$&$<0.1$&$<0.2$&$<0.3$\\ \hline 
psf&O2&mean&yes&SimpleWF&0.3947&0.6036&0.7689&0.8196\\
\hline
psf&O2&gmode&no&SimpleWF&0.3681&0.5788&0.7618&0.8207\\
psf&A&gmode&no&SimpleWF&0.3657&0.5627&0.7448&0.8049\\
psf&C&gmode&no&SimpleWF&0.3757&0.5754&0.7547&0.8107\\
psf&E&gmode&no&SimpleWF&0.3782&0.5752&0.7520&0.8081\\
psf&G&gmode&no&SimpleWF&0.3779&0.5703&0.7416&0.7977\\
model&O2&gmode&no&SimpleWF&0.3607&0.5686&0.7484&0.8103\\
model&A&gmode&no&SimpleWF&0.3597&0.5538&0.7383&0.7951\\
model&C&gmode&no&SimpleWF&0.3739&0.5722&0.7533&0.8062\\
model&E&gmode&no&SimpleWF&0.3756&0.5727&0.7554&0.8103\\
model&G&gmode&no&SimpleWF&0.3813&0.5715&0.7460&0.8047\\
fiber&O2&gmode&no&SimpleWF&0.3535&0.5583&0.7415&0.8012\\
fiber&A&gmode&no&SimpleWF&0.3513&0.5498&0.7328&0.7947\\
fiber&C&gmode&no&SimpleWF&0.3561&0.5556&0.7352&0.7930\\
fiber&E&gmode&no&SimpleWF&0.3593&0.5507&0.7320&0.7905\\
fiber&G&gmode&no&SimpleWF&0.3531&0.5405&0.7188&0.7765\\
petro&O2&gmode&no&SimpleWF&0.2732&0.4561&0.6557&0.7387\\
petro&A&gmode&no&SimpleWF&0.2821&0.4598&0.6503&0.7328\\
petro&C&gmode&no&SimpleWF&0.2910&0.4727&0.6586&0.7376\\
petro&E&gmode&no&SimpleWF&0.2975&0.4693&0.6521&0.7301\\
petro&G&gmode&no&SimpleWF&0.3032&0.4760&0.6500&0.7214\\ \hline
     \end{tabular} \end{minipage}
     \end{table}

     \begin{table} \centering \begin{minipage}{140mm}
     \caption[PDF Photometric Redshift Estimates:  gmode, SimpleWF, Red not removed for high-$z$ quasars]{PDF Photometric Redshift Estimates for high-redshift quasars using gmode, Simple Weighting Function without removing reddened quasars.  \label{PDFsimpleWFnoremovered_hightable_gmode}}
     \begin{tabular}{c c c c c || c c c c }\hline\hline
     \multicolumn{5}{c}{CZR Parameters} & \multicolumn{4}{c}{$z_{spec}> $division2} \\
     Mag&Bins&Stat&Red?&Weighting&$<0.05$&$<0.1$&$<0.2$&$<0.3$\\ \hline 
psf&O2&mean&yes&SimpleWF&0.3303&0.5189&0.7458&0.8403\\
\hline
psf&O2&gmode&no&SimpleWF&0.2901&0.4691&0.6995&0.8072\\
psf&A&gmode&no&SimpleWF&0.2768&0.4911&0.7119&0.8171\\
psf&C&gmode&no&SimpleWF&0.2613&0.5002&0.7238&0.8339\\
psf&E&gmode&no&SimpleWF&0.2749&0.5299&0.7654&0.8786\\
psf&G&gmode&no&SimpleWF&0.2799&0.5471&0.7859&0.9038\\
model&O2&gmode&no&SimpleWF&0.2942&0.4706&0.7050&0.8101\\
model&A&gmode&no&SimpleWF&0.2681&0.4792&0.6941&0.8042\\
model&C&gmode&no&SimpleWF&0.2627&0.5055&0.7304&0.8386\\
model&E&gmode&no&SimpleWF&0.2756&0.5365&0.7734&0.8901\\
model&G&gmode&no&SimpleWF&0.2809&0.5478&0.7880&0.9057\\
fiber&O2&gmode&no&SimpleWF&0.2975&0.4880&0.7319&0.8364\\
fiber&A&gmode&no&SimpleWF&0.2651&0.4696&0.6881&0.7954\\
fiber&C&gmode&no&SimpleWF&0.2583&0.4978&0.7241&0.8310\\
fiber&E&gmode&no&SimpleWF&0.2726&0.5350&0.7745&0.8872\\
fiber&G&gmode&no&SimpleWF&0.2767&0.5436&0.7865&0.9014\\
petro&O2&gmode&no&SimpleWF&0.2230&0.3919&0.6127&0.7260\\
petro&A&gmode&no&SimpleWF&0.2266&0.4060&0.6017&0.7199\\
petro&C&gmode&no&SimpleWF&0.2233&0.4334&0.6469&0.7567\\
petro&E&gmode&no&SimpleWF&0.2458&0.4878&0.7211&0.8424\\
petro&G&gmode&no&SimpleWF&0.2515&0.5035&0.7411&0.8647\\ \hline
     \end{tabular} \end{minipage}
     \end{table}

\clearpage
     \begin{landscape}
     \begin{table} \begin{minipage}{140mm}
     \caption[PDF Photometric Redshift Estimates varying magnitude error limits and removing reddened quasars]{PDF Photometric Redshift Estimates varying magnitude error limits and removing reddened quasars.  \label{PDFvaryerror}}
     \footnotesize
     \begin{tabular}{c c c c c c c || c c c c | c c | c c | c c | c }\hline\hline
     \multicolumn{7}{c}{CZR Parameters} & \multicolumn{4}{c}{$|z_{spec}-z_{phot}|$} & \multicolumn{7}{c}{Catastrophic Failures} \\
     &&&&&CZR&$z_{phot}$&&&&&\#&\%&\#&&\#&&\#\\
     Mag&Bins&Stat&Red?&Weighting&errlim&errlim&$<0.05$&$<0.1$&$<0.2$&$<0.3$&CF&CF&Common&\%&Diff&\%&not CF\\ \hline
psf&O2&mean&yes&SimpleWF&None&None&0.3835&0.5889&0.7648&0.8232&6821&0.1469&\nodata&\nodata&\nodata&\nodata&\nodata\\
\hline
psf&O2&mean&yes&SimpleWF&1.0&1.0&0.3982&0.5973&0.7646&0.8227&6774&0.1478&5847&0.8632&433&0.0639&974\\
psf&O2&mean&yes&SimpleWF&0.5&0.5&0.4081&0.6049&0.7639&0.8164&6935&0.1564&5407&0.7797&1030&0.1485&1414\\
psf&O2&mean&yes&SimpleWF&0.2&0.2&0.4125&0.6071&0.7629&0.8166&6743&0.1560&5123&0.7598&1133&0.1680&1698\\
psf&O2&mean&yes&SimpleWF&0.1&0.1&0.4222&0.6167&0.7642&0.8157&6169&0.1588&4460&0.7230&1281&0.2077&2361\\
\hline
     \end{tabular} \end{minipage}
     \end{table}  
     \end{landscape}   
     
     \begin{table} \centering \begin{minipage}{140mm}
     \caption[PDF Photometric Redshift Estimates for low-$z$ quasars varying magnitude error limits and removing reddened quasars]{PDF Photometric Redshift Estimates for low-redshift quasars varying magnitude error limits and removing reddened quasars. \label{PDFvaryerrorlowtable}}
     \begin{tabular}{c c c c c c c || c c c c }\hline \hline
     \multicolumn{7}{c}{CZR Parameters} & \multicolumn{4}{c}{$z_{spec}\leq $division2} \\
     &&&&&CZR&$z_{phot}$&&&&\\ 
     Mag&Bins&Stat&Red?&Weighting&errlim&errlim&$<0.05$&$<0.1$&$<0.2$&$<0.3$\\ \hline
psf&O2&mean&yes&SimpleWF&10.0&10.0&0.3947&0.6036&0.7689&0.8196\\
\hline
psf&O2&mean&yes&SimpleWF&1.0&1.0&0.4119&0.6131&0.7675&0.8178\\
psf&O2&mean&yes&SimpleWF&0.5&0.5&0.4169&0.6159&0.7660&0.8122\\
psf&O2&mean&yes&SimpleWF&0.2&0.2&0.4217&0.6197&0.7679&0.8143\\
psf&O2&mean&yes&SimpleWF&0.1&0.1&0.4300&0.6284&0.7698&0.8138\\
psf&O2&mean&yes&SimpleWF&0.05&0.05&0.4529&0.6454&0.7773&0.8180\\
psf&O2&mean&yes&SimpleWF&0.02&0.02&0.5013&0.6354&0.7453&0.7909\\
\hline
psf&C&mean&yes&SimpleWF&1.0&1.0&0.4262&0.6229&0.7734&0.8199\\
psf&C&mean&yes&SimpleWF&0.5&0.5&0.4261&0.6181&0.7664&0.8123\\
psf&C&mean&yes&SimpleWF&0.2&0.2&0.4287&0.6209&0.7684&0.8138\\
psf&C&mean&yes&SimpleWF&0.1&0.1&0.4384&0.6309&0.7719&0.8169\\
\hline
psf&O2&mean&yes&SimpleWF&1.0&None&0.4116&0.6126&0.7669&0.8172\\
psf&O2&mean&yes&SimpleWF&0.5&None&0.4158&0.6143&0.7640&0.8101\\
psf&O2&mean&yes&SimpleWF&0.2&None&0.4180&0.6144&0.7614&0.8076\\
psf&O2&mean&yes&SimpleWF&0.1&None&0.4161&0.6131&0.7578&0.8027\\
psf&O2&mean&yes&SimpleWF&0.05&None&0.4124&0.6073&0.7505&0.7960\\
psf&O2&mean&yes&SimpleWF&0.02&None&0.3854&0.5741&0.7269&0.7740\\
psf&O2&mean&yes&SimpleWF&0.01&None&0.2664&0.4386&0.6160&0.6931\\
\hline
psf&C&mean&yes&SimpleWF&1.0&None&0.4259&0.6223&0.7727&0.8191\\
psf&C&mean&yes&SimpleWF&0.5&None&0.4250&0.6164&0.7642&0.8101\\
psf&C&mean&yes&SimpleWF&0.2&None&0.4246&0.6151&0.7616&0.8069\\
psf&C&mean&yes&SimpleWF&0.1&None&0.4232&0.6139&0.7581&0.8038\\ \hline
     \end{tabular} \end{minipage}
     \end{table}     

     \begin{table} \centering \begin{minipage}{140mm}
     \caption[PDF Photometric Redshift Estimates for high-$z$ quasars varying magnitude error limits and removing reddened quasars]{PDF Photometric Redshift Estimates for high-redshift quasars varying magnitude error limits and removing reddened quasars. \label{PDFvaryerrorhightable}}
     \begin{tabular}{c c c c c c c || c c c c }\hline\hline
     \multicolumn{7}{c}{CZR Parameters} & \multicolumn{4}{c}{$z_{spec}> $division2} \\
     &&&&&CZR&$z_{phot}$&&&&\\ 
     Mag&Bins&Stat&Red?&Weighting&errlim&errlim&$<0.05$&$<0.1$&$<0.2$&$<0.3$\\ \hline
psf&O2&mean&yes&SimpleWF&10.0&10.0&0.3947&0.6036&0.7689&0.8196\\
\hline
psf&O2&mean&yes&SimpleWF&1.0&1.0&0.3284&0.5164&0.7496&0.8478\\
psf&O2&mean&yes&SimpleWF&0.5&0.5&0.3531&0.5361&0.7505&0.8426\\
psf&O2&mean&yes&SimpleWF&0.2&0.2&0.3468&0.5171&0.7276&0.8335\\
psf&O2&mean&yes&SimpleWF&0.1&0.1&0.3579&0.5190&0.7172&0.8317\\
psf&O2&mean&yes&SimpleWF&0.05&0.05&0.3881&0.5474&0.7468&0.8725\\
psf&O2&mean&yes&SimpleWF&0.02&0.02&0.4583&0.5417&0.7917&0.8333\\
\hline
psf&C&mean&yes&SimpleWF&1.0&1.0&0.2677&0.5159&0.7513&0.8655\\
psf&C&mean&yes&SimpleWF&0.5&0.5&0.2633&0.5247&0.7652&0.8799\\
psf&C&mean&yes&SimpleWF&0.2&0.2&0.2583&0.5251&0.7550&0.8781\\
psf&C&mean&yes&SimpleWF&0.1&0.1&0.2604&0.5405&0.7528&0.8830\\
\hline
psf&O2&mean&yes&SimpleWF&1.0&None&0.3242&0.5127&0.7516&0.8507\\
psf&O2&mean&yes&SimpleWF&0.5&None&0.3283&0.5205&0.7543&0.8531\\
psf&O2&mean&yes&SimpleWF&0.2&None&0.2979&0.4695&0.6935&0.8143\\
psf&O2&mean&yes&SimpleWF&0.1&None&0.2749&0.4270&0.6505&0.7835\\
psf&O2&mean&yes&SimpleWF&0.05&None&0.2582&0.4000&0.6056&0.7387\\
psf&O2&mean&yes&SimpleWF&0.02&None&0.2017&0.3265&0.5232&0.6808\\
psf&O2&mean&yes&SimpleWF&0.01&None&0.1368&0.2193&0.3529&0.4693\\
\hline
psf&C&mean&yes&SimpleWF&1.0&None&0.2691&0.5161&0.7535&0.8676\\
psf&C&mean&yes&SimpleWF&0.5&None&0.2627&0.5143&0.7636&0.8773\\
psf&C&mean&yes&SimpleWF&0.2&None&0.2416&0.4763&0.7202&0.8452\\
psf&C&mean&yes&SimpleWF&0.1&None&0.2334&0.4567&0.7063&0.8344\\ \hline
     \end{tabular} \end{minipage}
     \end{table}     
     
     \begin{landscape}
     \begin{table} \begin{minipage}{140mm}
     \caption[PDF Photometric Redshift Estimates varying magnitude limits and removing reddened quasars]{PDF Photometric Redshift Estimates varying magnitude limits and removing reddened quasars.  \label{PDFvarymags}}
     \footnotesize
     \begin{tabular}{c c c c c c c || c c c c | c c | c c | c c | c }\hline\hline
     \multicolumn{7}{c}{CZR Parameters} & \multicolumn{4}{c}{$|z_{spec}-z_{phot}|$} & \multicolumn{7}{c}{Catastrophic Failures} \\
     &&&&&CZR&$z_{phot}$&&&&&\#&\%&\#&&\#&&\#\\
     Mag&Bins&Stat&Red?&Weighting&maglim&maglim&$<0.05$&$<0.1$&$<0.2$&$<0.3$&CF&CF&Common&\%&Diff&\%&not CF\\ \hline
psf&O2&mean&yes&SimpleWF&None&None&0.3835&0.5889&0.7648&0.8232&6821&0.1469&\nodata&\nodata&\nodata&\nodata&\nodata\\
\hline
psf&O2&mean&yes&SimpleWF&25.0&25.0&0.3898&0.5963&0.7687&0.8237&6830&0.1485&6055&0.8865&296&0.0433&766\\
psf&O2&mean&yes&SimpleWF&21.0&21.0&0.4172&0.6075&0.7567&0.8127&6839&0.1619&4809&0.7032&1572&0.2299&2012\\
psf&O2&mean&yes&SimpleWF&20.0&20.0&0.4545&0.6505&0.7945&0.8486&4471&0.1296&3148&0.7041&982&0.2196&3673\\
\hline
psf&C&mean&yes&SimpleWF&25.0&25.0&0.3975&0.6074&0.7712&0.8260&6539&0.1422&5121&0.7831&952&0.1456&1700\\
psf&C&mean&yes&SimpleWF&21.0&21.0&0.4263&0.6218&0.7735&0.8252&6228&0.1474&4239&0.6806&1556&0.2498&2582\\
psf&C&mean&yes&SimpleWF&20.0&20.0&0.4615&0.6634&0.8050&0.8531&4287&0.1243&2852&0.6653&1108&0.2585&3969\\
\hline
psf&O2&mean&yes&SimpleWF&25.0&None&0.3882&0.5948&0.7683&0.8243&6857&0.1477&6080&0.8867&296&0.0432&741\\
psf&O2&mean&yes&SimpleWF&21.0&None&0.3890&0.5700&0.7222&0.7861&8227&0.1772&5371&0.6529&2252&0.2737&1450\\
psf&O2&mean&yes&SimpleWF&20.0&None&0.3904&0.5719&0.7156&0.7754&8907&0.1919&5099&0.5725&3179&0.3569&1722\\
\hline
psf&C&mean&yes&SimpleWF&25.0&None&0.3962&0.6064&0.7711&0.8265&6566&0.1415&5144&0.7834&954&0.1453&1677\\
psf&C&mean&yes&SimpleWF&21.0&None&0.3971&0.5834&0.7375&0.7940&8082&0.1741&4770&0.5902&2717&0.3362&2051\\
psf&C&mean&yes&SimpleWF&20.0&None&0.4007&0.5888&0.7336&0.7855&8596&0.1852&4705&0.5473&3288&0.3825&2116\\ \hline
     \end{tabular} \end{minipage}
     \end{table}
     \end{landscape}     
     
     \begin{table} \centering \begin{minipage}{140mm}
     \caption[PDF Photometric Redshift Estimates for low-$z$ quasars varying magnitude limits and removing reddened quasars]{PDF Photometric Redshift Estimates for low-redshift quasars varying magnitude limits and removing reddened quasars. \label{PDFvarymagslowtable}}
     \small
     \begin{tabular}{c c c c c c c || c c c c }\hline\hline
     \multicolumn{7}{c}{CZR Parameters} & \multicolumn{4}{c}{$z_{spec}\leq$division2} \\
     &&&&&CZR&$z_{phot}$&&&&\\
     Mag&Bins&Stat&Red?&Weighting&maglim&maglim&$<0.05$&$<0.1$&$<0.2$&$<0.3$\\ \hline
psf&O2&mean&yes&SimpleWF&None&None&0.3947&0.6036&0.7689&0.8196\\
\hline
psf&O2&mean&yes&SimpleWF&25.0&25.0&0.3977&0.6068&0.7699&0.8181\\
psf&O2&mean&yes&SimpleWF&21.0&21.0&0.4287&0.6238&0.7670&0.8141\\
psf&O2&mean&yes&SimpleWF&20.0&20.0&0.4634&0.6636&0.8021&0.8482\\
\hline
psf&C&mean&yes&SimpleWF&25.0&25.0&0.4158&0.6191&0.7725&0.8189\\
psf&C&mean&yes&SimpleWF&21.0&21.0&0.4373&0.6267&0.7741&0.8218\\
psf&C&mean&yes&SimpleWF&20.0&20.0&0.4706&0.6666&0.8049&0.8501\\
\hline
psf&O2&mean&yes&SimpleWF&25.0&None&0.3976&0.6066&0.7697&0.8178\\
psf&O2&mean&yes&SimpleWF&21.0&None&0.4222&0.6144&0.7558&0.8026\\
psf&O2&mean&yes&SimpleWF&20.0&None&0.4286&0.6253&0.7656&0.8132\\
\hline
psf&C&mean&yes&SimpleWF&25.0&None&0.4156&0.6189&0.7722&0.8186\\
psf&C&mean&yes&SimpleWF&21.0&None&0.4306&0.6175&0.7633&0.8109\\
psf&C&mean&yes&SimpleWF&20.0&None&0.4348&0.6252&0.7654&0.8125\\ \hline
     \end{tabular} \end{minipage}
     \end{table}     

     \begin{table} \centering \begin{minipage}{140mm}
     \caption[PDF Photometric Redshift Estimates for high-$z$ quasars varying magnitude limits and removing reddened quasars]{PDF Photometric Redshift Estimates for high-redshift quasars varying magnitude limits and removing reddened quasars. \label{PDFvarymagshightable}}
     \small
     \begin{tabular}{c c c c c c c || c c c c }\hline\hline
     \multicolumn{7}{c}{CZR Parameters} & \multicolumn{4}{c}{$z_{spec}> $division2} \\
     &&&&&CZR&$z_{phot}$&&&&\\
     Mag&Bins&Stat&Red?&Weighting&maglim&maglim&$<0.05$&$<0.1$&$<0.2$&$<0.3$\\ \hline
psf&O2&mean&yes&SimpleWF&None&None&0.3947&0.6036&0.7689&0.8196\\
\hline
psf&O2&mean&yes&SimpleWF&25.0&25.0&0.3502&0.5437&0.7625&0.8517\\
psf&O2&mean&yes&SimpleWF&21.0&21.0&0.3208&0.4715&0.6710&0.8011\\
psf&O2&mean&yes&SimpleWF&20.0&20.0&0.3669&0.5206&0.7189&0.8520\\
\hline
psf&C&mean&yes&SimpleWF&25.0&25.0&0.2679&0.5238&0.7627&0.8769\\
psf&C&mean&yes&SimpleWF&21.0&21.0&0.2598&0.5473&0.7648&0.8762\\
psf&C&mean&yes&SimpleWF&20.0&20.0&0.2873&0.6016&0.8073&0.9113\\
\hline
psf&O2&mean&yes&SimpleWF&25.0&None&0.3437&0.5385&0.7621&0.8549\\
psf&O2&mean&yes&SimpleWF&21.0&None&0.2310&0.3589&0.5626&0.7072\\
psf&O2&mean&yes&SimpleWF&20.0&None&0.2085&0.3182&0.4776&0.5956\\
\hline
psf&C&mean&yes&SimpleWF&25.0&None&0.2681&0.5238&0.7641&0.8785\\
psf&C&mean&yes&SimpleWF&21.0&None&0.1758&0.3577&0.5667&0.6822\\
psf&C&mean&yes&SimpleWF&20.0&None&0.1754&0.3487&0.5233&0.6066\\ \hline
     \end{tabular} \end{minipage}
     \end{table}     

\clearpage
     \begin{landscape}
     \begin{table} \begin{minipage}{140mm}
     \caption[PDF Photometric Redshift Estimates using fewer colors]{PDF Photometric Redshift Estimates using fewer colors.  \label{PDFfewercolorstable}}
     \small
     \begin{tabular}{c c c c c || c c c c | c c | c c | c c | c }\hline\hline
     \multicolumn{5}{c}{CZR Parameters}&\multicolumn{4}{c}{$|z_{spec}-z_{phot}|$}&\multicolumn{7}{c}{Catastrophic Failures} \\
     Mag&Bins&Stat&Red?&Weighting&$<0.05$&$<0.1$&$<0.2$&$<0.3$&CF&\% CF&Common&\%&Diff&\%&not CF\\ \hline          
psf UGRIZ&O2&mean&yes&SimpleWF&0.3835&0.5889&0.7648&0.8232&6821&0.1469&\nodata&\nodata&\nodata&\nodata&\nodata\\
psf UGRIZ&O2&mean&no&SimpleWF&0.3648&0.5736&0.7660&0.8242&6800&0.1465&4883&0.7181&1446&0.2126&1938\\
\hline
psf UGR&O2&mean&no&SimpleWF&0.1724&0.3130&0.5034&0.5918&14775&0.3183&4948&0.3349&8729&0.5908&1873\\
psf GRI&O2&mean&no&SimpleWF&0.2050&0.3548&0.5282&0.6050&15241&0.3283&4909&0.3221&9261&0.6076&1912\\
psf RIZ&O2&mean&no&SimpleWF&0.2209&0.3768&0.5702&0.6586&13532&0.2915&4711&0.3481&7879&0.5822&2110\\
\hline
psf UGRI&O2&mean&no&SimpleWF&0.2729&0.4577&0.6500&0.7195&10572&0.2278&4971&0.4702&4829&0.4568&1850\\
psf GRIZ&O2&mean&no&SimpleWF&0.3037&0.4844&0.6651&0.7271&10986&0.2367&4979&0.4532&5260&0.4788&1842\\
\hline\hline
psf non-adj&O2&mean&no&SimpleWF&0.2573&0.4440&0.6438&0.7312&8585&0.1849&4680&0.5451&3904&0.4547&2142\\ \hline
     \end{tabular} \end{minipage}
     \end{table}  
     \end{landscape}

     \begin{table} \centering \begin{minipage}{140mm}
     \caption[PDF Photometric Redshift Estimates for low-$z$ quasars using fewer colors]{PDF Photometric Redshift Estimates for low-redshift quasars using fewer colors.  \label{PDFfewercolors_lowtable}}
     \begin{tabular}{c c c c c || c c c c }\hline\hline
     \multicolumn{5}{c}{CZR Parameters} & \multicolumn{4}{c}{$z_{spec}\leq$division2} \\ 
     Mag&Bins&Stat&Red?&Weighting&$<0.05$&$<0.1$&$<0.2$&$<0.3$\\ \hline 
psf UGRIZ&O2&mean&yes&SimpleWF&0.3947&0.6036&0.7689&0.8196\\
psf UGRIZ&O2&mean&no&SimpleWF&0.3808&0.5970&0.7830&0.8359\\
\hline
psf UGR&O2&mean&no&SimpleWF&0.1712&0.3114&0.4943&0.5745\\
psf GRI&O2&mean&no&SimpleWF&0.2171&0.3809&0.5650&0.6453\\
psf RIZ&O2&mean&no&SimpleWF&0.2343&0.4095&0.6236&0.7185\\
\hline
psf UGRI&O2&mean&no&SimpleWF&0.2816&0.4719&0.6595&0.7248\\
psf GRIZ&O2&mean&no&SimpleWF&0.3283&0.5256&0.7128&0.7724\\
\hline\hline
psf non-adj&O2&mean&no&SimpleWF&0.2528&0.4429&0.6395&0.7251\\ \hline
     \end{tabular} \end{minipage}
     \end{table}  

     \begin{table} \centering \begin{minipage}{140mm}
     \caption[PDF Photometric Redshift Estimates for high-$z$ quasars using fewer colors]{PDF Photometric Redshift Estimates for high-redshift quasars using fewer colors.  \label{PDFfewercolors_hightable}}
     \begin{tabular}{c c c c c || c c c c }\hline\hline
     \multicolumn{5}{c}{CZR Parameters} & \multicolumn{4}{c}{$z_{spec}\leq$division2} \\ 
     Mag&Bins&Stat&Red?&Weighting&$<0.05$&$<0.1$&$<0.2$&$<0.3$\\ \hline 
psf&O2&mean&yes&SimpleWF&0.3303&0.5189&0.7458&0.8403\\
psf&O2&mean&no&SimpleWF&0.2887&0.4624&0.6848&0.7687\\
\hline
psf UGR&O2&mean&no&SimpleWF&0.1780&0.3209&0.5466&0.6739\\
psf GRI&O2&mean&no&SimpleWF&0.1479&0.2311&0.3534&0.4137\\
psf RIZ&O2&mean&no&SimpleWF&0.1575&0.2214&0.3164&0.3737\\
\hline
psf UGRI&O2&mean&no&SimpleWF&0.2314&0.3902&0.6045&0.6947\\
psf GRIZ&O2&mean&no&SimpleWF&0.1870&0.2888&0.4386&0.5118\\
\hline\hline
psf non-adj&O2&mean&no&SimpleWF&0.2788&0.4489&0.6645&0.7605\\ \hline
     \end{tabular} \end{minipage}
     \end{table}  

     \begin{table} \centering \begin{minipage}{140mm}
     \caption{Photometric Redshift Estimates: Comparison between using 4 colors and 5 colors for quasars with GALEX matches.  \label{sdssGalex_alltable}}
     \small
     \begin{tabular}{c c c c c || c c c c | c c }\hline\hline
     \multicolumn{5}{c}{CZR Parameters} & \multicolumn{5}{c}{$|z_{spec}-z_{phot}|$} \\
     Mag&Bins&Stat&Red?&Weighting&$<0.05$&$<0.1$&$<0.2$&$<0.3$&CF&\% CF\\ \hline
psf&O2&mean&no&SimpleWF&0.5533&0.7429&0.8609&0.9004&1250&0.0818\\
psfgalex&O2&mean&no&SimpleWF&0.5752&0.7567&0.8731&0.9135&530&0.0694\\ \hline
     \end{tabular} \end{minipage}
     \end{table}  
     
     \begin{table} \centering \begin{minipage}{140mm}
     \caption{Photometric Redshift Estimates for low-redshift quasars: Comparison between using 4 colors and 5 colors for quasars with GALEX matches.  \label{sdssGalex_lowtable}}
     \begin{tabular}{c c c c c || c c c c }\hline\hline
     \multicolumn{5}{c}{CZR Parameters} & \multicolumn{4}{c}{$z_{spec}\leq $2.1} \\
     Mag&Bins&Stat&Red?&Weighting&$<0.05$&$<0.1$&$<0.2$&$<0.3$\\ \hline
psf&O2&mean&no&SimpleWF&0.5647&0.7569&0.8767&0.9162\\
psfgalex&O2&mean&no&SimpleWF&0.5871&0.7700&0.8884&0.9284\\ \hline
     \end{tabular} \end{minipage}
     \end{table}  
     
     \begin{table} \centering \begin{minipage}{140mm}
     \caption{Photometric Redshift Estimates for high-redshift quasars: Comparison between using 4 colors and 5 colors for quasars with GALEX matches.  \label{sdssGalex_hightable}}
     \begin{tabular}{c c c c c || c c c c }\hline\hline
     \multicolumn{5}{c}{CZR Parameters} & \multicolumn{4}{c}{$z_{spec}\leq $2.1} \\
     Mag&Bins&Stat&Red?&Weighting&$<0.05$&$<0.1$&$<0.2$&$<0.3$\\ \hline
psf&O2&mean&no&SimpleWF&0.1834&0.2882&0.3493&0.3886\\
psfgalex&O2&mean&no&SimpleWF&0.1921&0.3275&0.3755&0.4323\\ \hline
     \end{tabular} \end{minipage}
     \end{table}  
     
\clearpage
     \begin{landscape}
     \begin{table} \begin{minipage}{140mm}
     \caption[PDF Photometric Redshift Estimates:  Mean, NoWF, Remove Red]{PDF Photometric Redshift Estimates using mean, No Weighting Function and removing reddened quasars.  \label{PDFnoWFremovered_table_mean}}
     \begin{tabular}{c c c c c c || c c c c | c c | c c | c c }\hline\hline
\multicolumn{5}{c}{CZR Parameters}&\multicolumn{4}{c}{$|z_{spec}-z_{phot}|$}&\multicolumn{7}{c}{Catastrophic Failures} \\
     Mag&Bins&Stat&Red?&Weighting&$<0.05$&$<0.1$&$<0.2$&$<0.3$&CF&\% CF&Common&\%&Diff&\%&not CF\\ \hline
     psf&O2&mean&yes&SimpleWF&0.3835&0.5889&0.7648&0.8232&6821&0.1469&\nodata&\nodata&\nodata&\nodata&\nodata\\
\hline
psf&O2&mean&yes&NoWF&0.3871&0.5890&0.7502&0.8074&7481&0.1612&4831&0.6458&2650&0.3542&1990\\
psf&A&mean&yes&NoWF&0.3885&0.5888&0.7462&0.7988&7814&0.1683&4584&0.5866&3230&0.4134&2237\\
psf&C&mean&yes&NoWF&0.3908&0.5934&0.7481&0.8024&7607&0.1639&4961&0.6522&2646&0.3478&1860\\
psf&E&mean&yes&NoWF&0.3968&0.5905&0.7457&0.8015&7637&0.1645&4801&0.6286&2836&0.3714&2020\\
psf&G&mean&yes&NoWF&0.4030&0.5922&0.7435&0.8011&7690&0.1657&4706&0.6120&2984&0.3880&2115\\
\hline
     \end{tabular} \end{minipage}
     \end{table}
     \end{landscape}

    \begin{table} \centering \begin{minipage}{140mm}
     \caption[PDF Photometric Redshift Estimates:  mean, NoWF, Remove Red for low-$z$ quasars]{PDF Photometric Redshift Estimates for low-redshift quasars using mean, No Weighting Function and removing reddened quasars.  \label{PDFnoWFremovered_lowtable_mean}}    
     \begin{tabular}{c c c c c || c c c c } \hline\hline
     \multicolumn{5}{c}{CZR Parameters} & \multicolumn{4}{c}{$z_{spec}\leq$division2} \\ 
     Mag&Bins&Stat&Red?&Weighting&$<0.05$&$<0.1$&$<0.2$&$<0.3$\\ \hline 
psf&O2&mean&yes&SimpleWF&0.3947&0.6036&0.7689&0.8196\\
\hline
psf&O2&mean&yes&NoWF&0.4105&0.6142&0.7657&0.8173\\
psf&A&mean&yes&NoWF&0.4180&0.6196&0.7670&0.8151\\
psf&C&mean&yes&NoWF&0.4112&0.6096&0.7521&0.7973\\
psf&E&mean&yes&NoWF&0.4181&0.6072&0.7508&0.7974\\
psf&G&mean&yes&NoWF&0.4253&0.6102&0.7495&0.7983\\
model&O2&mean&yes&NoWF&0.4226&0.6245&0.7706&0.8204\\
model&A&mean&yes&NoWF&0.4271&0.6275&0.7712&0.8185\\
model&C&mean&yes&NoWF&0.4259&0.6249&0.7650&0.8099\\
model&E&mean&yes&NoWF&0.4403&0.6259&0.7617&0.8104\\
model&G&mean&yes&NoWF&0.4418&0.6263&0.7599&0.8091\\
fiber&O2&mean&yes&NoWF&0.4018&0.6028&0.7564&0.8070\\
fiber&A&mean&yes&NoWF&0.4012&0.5949&0.7420&0.7907\\
fiber&C&mean&yes&NoWF&0.4019&0.5949&0.7410&0.7881\\
fiber&E&mean&yes&NoWF&0.4145&0.5984&0.7357&0.7850\\
fiber&G&mean&yes&NoWF&0.4176&0.5985&0.7383&0.7869\\
petro&O2&mean&yes&NoWF&0.3440&0.5319&0.6983&0.7599\\
petro&A&mean&yes&NoWF&0.3533&0.5375&0.6960&0.7528\\
petro&C&mean&yes&NoWF&0.3471&0.5264&0.6800&0.7347\\
petro&E&mean&yes&NoWF&0.3552&0.5266&0.6734&0.7338\\
petro&G&mean&yes&NoWF&0.3545&0.5287&0.6733&0.7322\\ \hline
     \end{tabular} \end{minipage}
     \end{table}

    \begin{table} \centering \begin{minipage}{140mm}
     \caption[PDF Photometric Redshift Estimates:  Mean, NoWF, Remove Red for high-$z$ quasars]{PDF Photometric Redshift Estimates for high-redshift quasars using mean, No Weighting Function and removing reddened quasars.  \label{PDFnoWFremovered_hightable_mean}}    
     \begin{tabular}{c c c c c || c c c c } \hline\hline
     \multicolumn{5}{c}{CZR Parameters} & \multicolumn{4}{c}{$z_{spec}>$division2} \\ 
     Mag&Bins&Stat&Red?&Weighting&$<0.05$&$<0.1$&$<0.2$&$<0.3$\\ \hline 
psf&O2&mean&yes&SimpleWF&0.3303&0.5189&0.7458&0.8403\\
\hline
psf&O2&mean&yes&NoWF&0.2762&0.4692&0.6767&0.7601\\
psf&A&mean&yes&NoWF&0.2484&0.4427&0.6476&0.7211\\
psf&C&mean&yes&NoWF&0.2559&0.4862&0.7217&0.8364\\
psf&E&mean&yes&NoWF&0.2565&0.4808&0.7124&0.8292\\
psf&G&mean&yes&NoWF&0.2557&0.4731&0.7037&0.8198\\
model&O2&mean&yes&NoWF&0.2841&0.4729&0.6796&0.7609\\
model&A&mean&yes&NoWF&0.2544&0.4510&0.6529&0.7250\\
model&C&mean&yes&NoWF&0.2573&0.4902&0.7222&0.8369\\
model&E&mean&yes&NoWF&0.2576&0.4833&0.7145&0.8307\\
model&G&mean&yes&NoWF&0.2591&0.4765&0.7068&0.8240\\
fiber&O2&mean&yes&NoWF&0.2484&0.4348&0.6604&0.7440\\
fiber&A&mean&yes&NoWF&0.2272&0.4129&0.6156&0.6903\\
fiber&C&mean&yes&NoWF&0.2600&0.4930&0.7208&0.8308\\
fiber&E&mean&yes&NoWF&0.2603&0.4958&0.7261&0.8415\\
fiber&G&mean&yes&NoWF&0.2577&0.4828&0.7190&0.8380\\
petro&O2&mean&yes&NoWF&0.1965&0.3508&0.5662&0.6591\\
petro&A&mean&yes&NoWF&0.1836&0.3395&0.5261&0.6156\\
petro&C&mean&yes&NoWF&0.2226&0.4290&0.6585&0.7797\\
petro&E&mean&yes&NoWF&0.2200&0.4270&0.6580&0.7831\\
petro&G&mean&yes&NoWF&0.2208&0.4218&0.6552&0.7825\\ \hline
     \end{tabular} \end{minipage}
     \end{table}       

\clearpage
     \begin{landscape}
     \begin{table} \begin{minipage}{140mm}
     \caption[PDF Photometric Redshift Estimates:  gmedian, NoWF, Remove Red]{PDF Photometric Redshift Estimates using gmedian, No Weighting Function and removing reddened quasars.  \label{PDFnoWFremovered_table_gmedian}}
     \small
     \begin{tabular}{c c c c c c || c c c c | c c | c c | c c }\hline\hline
\multicolumn{5}{c}{CZR Parameters}&\multicolumn{4}{c}{$|z_{spec}-z_{phot}|$}&\multicolumn{7}{c}{Catastrophic Failures} \\
     Mag&Bins&Stat&Red?&Weighting&$<0.05$&$<0.1$&$<0.2$&$<0.3$&CF&\% CF&Common&\%&Diff&\%&not CF\\ \hline
     psf&O2&mean&yes&SimpleWF&0.3835&0.5889&0.7648&0.8232&6821&0.1469&\nodata&\nodata&\nodata&\nodata&\nodata\\
\hline
psf&O2&gmedian&yes&NoWF&0.3843&0.5848&0.7458&0.8045&7543&0.1625&4910&0.6509&2633&0.3491&1911\\
psf&A&gmedian&yes&NoWF&0.3864&0.5857&0.7439&0.7971&7841&0.1689&4650&0.5930&3191&0.4070&2171\\
psf&C&gmedian&yes&NoWF&0.3875&0.5889&0.7437&0.7994&7677&0.1654&4986&0.6495&2691&0.3505&1835\\
psf&E&gmedian&yes&NoWF&0.3957&0.5875&0.7413&0.7982&7740&0.1667&4834&0.6245&2906&0.3755&1987\\
psf&G&gmedian&yes&NoWF&0.4003&0.5904&0.7397&0.7977&7793&0.1679&4752&0.6098&3041&0.3902&2069\\
\hline
     \end{tabular} \end{minipage}
     \end{table}
     \end{landscape}

    \begin{table} \centering \begin{minipage}{140mm}
     \caption[PDF Photometric Redshift Estimates:  gmedian, NoWF, Remove Red for low-$z$ quasars]{PDF Photometric Redshift Estimates for low-redshift quasars using gmedian, No Weighting Function and removing reddened quasars.  \label{PDFnoWFremovered_lowtable_gmedian}}    
     \begin{tabular}{c c c c c || c c c c } \hline\hline
     \multicolumn{5}{c}{CZR Parameters} & \multicolumn{4}{c}{$z_{spec}\leq$division2} \\ 
     Mag&Bins&Stat&Red?&Weighting&$<0.05$&$<0.1$&$<0.2$&$<0.3$\\ \hline 
psf&O2&mean&yes&SimpleWF&0.3947&0.6036&0.7689&0.8196\\
\hline
psf&O2&gmedian&yes&NoWF&0.4081&0.6106&0.7621&0.8152\\
psf&A&gmedian&yes&NoWF&0.4144&0.6142&0.7619&0.8115\\
psf&C&gmedian&yes&NoWF&0.4078&0.6046&0.7471&0.7939\\
psf&E&gmedian&yes&NoWF&0.4170&0.6042&0.7459&0.7937\\
psf&G&gmedian&yes&NoWF&0.4228&0.6082&0.7447&0.7933\\
model&O2&gmedian&yes&NoWF&0.4148&0.6147&0.7617&0.8134\\
model&A&gmedian&yes&NoWF&0.4278&0.6240&0.7704&0.8186\\
model&C&gmedian&yes&NoWF&0.4234&0.6169&0.7585&0.8043\\
model&E&gmedian&yes&NoWF&0.4377&0.6201&0.7562&0.8055\\
model&G&gmedian&yes&NoWF&0.4395&0.6194&0.7531&0.8026\\
fiber&O2&gmedian&yes&NoWF&0.4020&0.6019&0.7567&0.8083\\
fiber&A&gmedian&yes&NoWF&0.3996&0.5914&0.7400&0.7899\\
fiber&C&gmedian&yes&NoWF&0.4003&0.5901&0.7373&0.7853\\
fiber&E&gmedian&yes&NoWF&0.4117&0.5950&0.7309&0.7803\\
fiber&G&gmedian&yes&NoWF&0.4150&0.5953&0.7319&0.7809\\
petro&O2&gmedian&yes&NoWF&0.3386&0.5237&0.6918&0.7548\\
petro&A&gmedian&yes&NoWF&0.3551&0.5385&0.7001&0.7575\\
petro&C&gmedian&yes&NoWF&0.3432&0.5202&0.6756&0.7305\\
petro&E&gmedian&yes&NoWF&0.3519&0.5220&0.6665&0.7270\\
petro&G&gmedian&yes&NoWF&0.3499&0.5203&0.6623&0.7208\\ \hline
     \end{tabular} \end{minipage}
     \end{table}

    \begin{table} \centering \begin{minipage}{140mm}
     \caption[PDF Photometric Redshift Estimates:  gmedian, NoWF, Remove Red for high-$z$ quasars]{PDF Photometric Redshift Estimates for high-redshift quasars using gmedian, No Weighting Function and removing reddened quasars.  \label{PDFnoWFremovered_hightable_gmedian}}    
     \begin{tabular}{c c c c c || c c c c } \hline\hline
     \multicolumn{5}{c}{CZR Parameters} & \multicolumn{4}{c}{$z_{spec}>$division2} \\ 
     Mag&Bins&Stat&Red?&Weighting&$<0.05$&$<0.1$&$<0.2$&$<0.3$\\ \hline 
psf&O2&mean&yes&SimpleWF&0.3303&0.5189&0.7458&0.8403\\
\hline
psf&O2&gmedian&yes&NoWF&0.2710&0.4623&0.6687&0.7536\\
psf&A&gmedian&yes&NoWF&0.2535&0.4506&0.6585&0.7289\\
psf&C&gmedian&yes&NoWF&0.2532&0.4848&0.7210&0.8357\\
psf&E&gmedian&yes&NoWF&0.2552&0.4773&0.7107&0.8280\\
psf&G&gmedian&yes&NoWF&0.2513&0.4724&0.7070&0.8262\\
model&O2&gmedian&yes&NoWF&0.2831&0.4690&0.6699&0.7542\\
model&A&gmedian&yes&NoWF&0.2594&0.4582&0.6623&0.7341\\
model&C&gmedian&yes&NoWF&0.2541&0.4845&0.7209&0.8361\\
model&E&gmedian&yes&NoWF&0.2563&0.4789&0.7152&0.8320\\
model&G&gmedian&yes&NoWF&0.2561&0.4737&0.7103&0.8285\\
fiber&O2&gmedian&yes&NoWF&0.2525&0.4376&0.6606&0.7432\\
fiber&A&gmedian&yes&NoWF&0.2310&0.4170&0.6210&0.6940\\
fiber&C&gmedian&yes&NoWF&0.2601&0.4937&0.7203&0.8298\\
fiber&E&gmedian&yes&NoWF&0.2575&0.4938&0.7225&0.8387\\
fiber&G&gmedian&yes&NoWF&0.2569&0.4832&0.7192&0.8413\\
petro&O2&gmedian&yes&NoWF&0.1936&0.3493&0.5667&0.6552\\
petro&A&gmedian&yes&NoWF&0.1874&0.3448&0.5287&0.6158\\
petro&C&gmedian&yes&NoWF&0.2236&0.4278&0.6590&0.7807\\
petro&E&gmedian&yes&NoWF&0.2213&0.4283&0.6603&0.7856\\
petro&G&gmedian&yes&NoWF&0.2216&0.4234&0.6582&0.7856\\\hline
     \end{tabular} \end{minipage}
     \end{table}       

\clearpage
     \begin{landscape}
     \begin{table} \begin{minipage}{140mm}
     \caption[PDF Photometric Redshift Estimates:  gmode, NoWF, Remove Red]{PDF Photometric Redshift Estimates using gmode, No Weighting Function and removing reddened quasars.  \label{PDFnoWFremovered_table_gmode}}
     \begin{tabular}{c c c c c c || c c c c | c c | c c | c c}\hline\hline
\multicolumn{5}{c}{CZR Parameters}&\multicolumn{4}{c}{$|z_{spec}-z_{phot}|$}&\multicolumn{7}{c}{Catastrophic Failures} \\
     Mag&Bins&Stat&Red?&Weighting&$<0.05$&$<0.1$&$<0.2$&$<0.3$&CF&\% CF&Common&\%&Diff&\%&not CF\\ \hline
     psf&O2&mean&yes&SimpleWF&0.3835&0.5889&0.7648&0.8232&6821&0.1469&\nodata&\nodata&\nodata&\nodata&\nodata\\
\hline
psf&O2&gmode&yes&NoWF&0.3757&0.5715&0.7356&0.7989&7717&0.1662&4925&0.6382&2792&0.3618&1896\\
psf&A&gmode&yes&NoWF&0.3719&0.5702&0.7301&0.7848&8297&0.1787&4796&0.5780&3501&0.4220&2025\\
psf&C&gmode&yes&NoWF&0.3723&0.5712&0.7272&0.7849&8248&0.1777&5022&0.6089&3226&0.3911&1799\\
psf&E&gmode&yes&NoWF&0.3774&0.5687&0.7235&0.7816&8492&0.1829&4936&0.5813&3556&0.4187&1885\\
psf&G&gmode&yes&NoWF&0.3785&0.5664&0.7216&0.7830&8354&0.1800&4799&0.5745&3555&0.4255&2022\\
\hline
     \end{tabular} \end{minipage}
     \end{table}
     \end{landscape}
     
    \begin{table} \centering \begin{minipage}{140mm}
     \caption[PDF Photometric Redshift Estimates:  gmode, NoWF, Remove Red for low-$z$ quasars]{PDF Photometric Redshift Estimates for low-redshift quasars using gmode, No Weighting Function and removing reddened quasars.  \label{PDFnoWFremovered_lowtable_gmode}}    
     \begin{tabular}{c c c c c || c c c c } \hline\hline
     \multicolumn{5}{c}{CZR Parameters} & \multicolumn{4}{c}{$z_{spec}\leq$division2} \\ 
     Mag&Bins&Stat&Red?&Weighting&$<0.05$&$<0.1$&$<0.2$&$<0.3$\\ \hline 
psf&O2&mean&yes&SimpleWF&0.3947&0.6036&0.7689&0.8196\\
\hline
psf&O2&gmode&yes&NoWF&0.3964&0.5947&0.7482&0.8037\\
psf&A&gmode&yes&NoWF&0.3955&0.5928&0.7397&0.7890\\
psf&C&gmode&yes&NoWF&0.3918&0.5861&0.7281&0.7746\\
psf&E&gmode&yes&NoWF&0.3971&0.5838&0.7252&0.7727\\
psf&G&gmode&yes&NoWF&0.3998&0.5830&0.7238&0.7745\\
model&O2&gmode&yes&NoWF&0.3954&0.5970&0.7443&0.7968\\
model&A&gmode&yes&NoWF&0.4062&0.5985&0.7452&0.7930\\
model&C&gmode&yes&NoWF&0.4059&0.5979&0.7415&0.7880\\
model&E&gmode&yes&NoWF&0.4152&0.6029&0.7420&0.7927\\
model&G&gmode&yes&NoWF&0.4064&0.5883&0.7316&0.7834\\
fiber&O2&gmode&yes&NoWF&0.3895&0.5884&0.7435&0.7963\\
fiber&A&gmode&yes&NoWF&0.3856&0.5820&0.7337&0.7871\\
fiber&C&gmode&yes&NoWF&0.3736&0.5629&0.7110&0.7619\\
fiber&E&gmode&yes&NoWF&0.3820&0.5662&0.7104&0.7621\\
fiber&G&gmode&yes&NoWF&0.3827&0.5685&0.7102&0.7639\\
petro&O2&gmode&yes&NoWF&0.3169&0.4941&0.6518&0.7191\\
petro&A&gmode&yes&NoWF&0.3376&0.5209&0.6811&0.7461\\
petro&C&gmode&yes&NoWF&0.3299&0.5073&0.6613&0.7234\\
petro&E&gmode&yes&NoWF&0.3315&0.5010&0.6505&0.7131\\
petro&G&gmode&yes&NoWF&0.3255&0.5008&0.6537&0.7156\\ \hline
     \end{tabular} \end{minipage}
     \end{table}

    \begin{table} \centering \begin{minipage}{140mm}
     \caption[PDF Photometric Redshift Estimates:  gmode, NoWF, Remove Red for high-$z$ quasars]{PDF Photometric Redshift Estimates for high-redshift quasars using gmode, No Weighting Function and removing reddened quasars.  \label{PDFnoWFremovered_hightable_gmode}}    
     %
 \end{minipage}
     \end{table}       

     \begin{landscape}
     \begin{table} \begin{minipage}{140mm}
     \caption[PDF Photometric Redshift Estimates:  Mean, NoWF, Red not Removed]{PDF Photometric Redshift Estimates using mean, No Weighting Function, without removing reddened quasars.  \label{PDFnoWFnoremovered_table_mean}}
     %
 \end{minipage}
     \end{table}     
     \end{landscape}

    \begin{table} \centering \begin{minipage}{140mm}
     \caption[PDF Photometric Redshift Estimates:  Mean, NoWF, Red not removed for low-$z$ quasars]{PDF Photometric Redshift Estimates for low-redshift quasars using mean, No Weighting Function and without removing reddened quasars.  \label{PDFnoWFnoremovered_lowtable_mean}}    
     %
 \end{minipage}
     \end{table}

    \begin{table} \centering \begin{minipage}{140mm}
     \caption[PDF Photometric Redshift Estimates:  mean, NoWF, Red not removed for high-$z$ quasars]{PDF Photometric Redshift Estimates for high-redshift quasars using mean, No Weighting Function and without removing reddened quasars.  \label{PDFnoWFnoremovered_hightable_mean}}    
     %
 \end{minipage}
     \end{table}

     \begin{landscape}
     \begin{table} \begin{minipage}{140mm}
     \caption[PDF Photometric Redshift Estimates:  gmedian, NoWF, Red not Removed]{PDF Photometric Redshift Estimates using gmedian, No Weighting Function, without removing reddened quasars.  \label{PDFnoWFnoremovered_table_gmedian}}
     \small
     %
 \end{minipage}
     \end{table}     
     \end{landscape}

    \begin{table} \centering \begin{minipage}{140mm}
     \caption[PDF Photometric Redshift Estimates:  gmedian, NoWF, Red not removed for low-$z$ quasars]{PDF Photometric Redshift Estimates for low-redshift quasars using gmedian, No Weighting Function and without removing reddened quasars.  \label{PDFnoWFnoremovered_lowtable_gmedian}}    
     %
 \end{minipage}
     \end{table}

    \begin{table} \centering \begin{minipage}{140mm}
     \caption[PDF Photometric Redshift Estimates:  gmedian, NoWF, Red not removed for high-$z$ quasars]{PDF Photometric Redshift Estimates for high-redshift quasars using gmedian, No Weighting Function and without removing reddened quasars.  \label{PDFnoWFnoremovered_hightable_gmedian}}    
     %
 \end{minipage}
     \end{table}

     \begin{landscape}
     \begin{table} \begin{minipage}{140mm}
     \caption[PDF Photometric Redshift Estimates:  gmode, NoWF, Red not Removed]{PDF Photometric Redshift Estimates using gmode, No Weighting Function, without removing reddened quasars.  \label{PDFnoWFnoremovered_table_gmode}}
     %
 \end{minipage}
     \end{table}     
     \end{landscape}

    \begin{table} \centering \begin{minipage}{140mm}
     \caption[PDF Photometric Redshift Estimates:  gmode, NoWF, Red not removed for low-$z$ quasars]{PDF Photometric Redshift Estimates for low-redshift quasars using gmode, No Weighting Function and without removing reddened quasars.  \label{PDFnoWFnoremovered_lowtable_gmode}}    
     %
 \end{minipage}
     \end{table}

    \begin{table} \centering \begin{minipage}{140mm}
     \caption[PDF Photometric Redshift Estimates:  gmode, NoWF, Red not removed for high-$z$ quasars]{PDF Photometric Redshift Estimates for high-redshift quasars using gmode, No Weighting Function and without removing reddened quasars.  \label{PDFnoWFnoremovered_hightable_gmode}}    
     %
 \end{minipage}
     \end{table}
     
     \begin{landscape}
     \begin{table} \begin{minipage}{140mm}
     \caption[PDF Photometric Redshift Estimates using other weighting functions]{PDF Photometric Redshift Estimates using other weighting functions.  \label{PDFapparMagWFtable}}
     \scriptsize
     %
 \end{minipage}
\end{table}
\end{landscape}

    \begin{table} \centering \begin{minipage}{140mm}
     \caption[PDF Photometric Redshift Estimates for low-$z$ quasars using other weighting functions]{PDF Photometric Redshift Estimates for low-redshift quasars using other weighting functions.  \label{PDFapparMagWF_lowtable}}
    \footnotesize
     %
 \end{minipage}
\end{table}

     \begin{table} \centering \begin{minipage}{140mm}
     \caption[PDF Photometric Redshift Estimates for high-$z$ quasars using other weighting functions]{PDF Photometric Redshift Estimates for high-redshift quasars using other weighting functions.  \label{PDFapparMagWF_hightable}}
     \footnotesize
     %
 \end{minipage}
\end{table}

\clearpage
\begin{landscape}
    \begin{table} \begin{minipage}{140mm}
     \caption[PDF Photometric Redshift Estimates for quasars in the SDSS DR3 Quasar Catalog including correlation]{PDF Photometric Redshift Estimates for quasars in the SDSS DR3 Quasar Catalog including correlation.   \label{corr_alltable}}
     \small
     %
 \end{minipage}
    \end{table}
    \end{landscape}
    
    \begin{table} \centering \begin{minipage}{140mm}
     \caption[Photometric Redshift Estimates for low-$z$ quasars in the SDSS DR3 Quasar Catalog including correlation]{Photometric Redshift Estimates for low-redshift quasars in the SDSS DR3 Quasar Catalog including correlation.  \label{corr_lowtable}}
     %
 \end{minipage}
     \end{table}
     
     \begin{table} \centering \begin{minipage}{140mm}
     \caption[Photometric Redshift Estimates for high-$z$ quasars in the SDSS DR3 Quasar Catalog including correlation]{Photometric Redshift Estimates for high-redshift quasars in the SDSS DR3 Quasar Catalog including correlation.  \label{corr_hightable}}
     %
 \end{minipage}
     \end{table}

     \clearpage
     
     \begin{landscape}
     \begin{table} \centering \begin{minipage}{140mm}
     \caption{Photometric Redshift Estimates for subpopulations in the SDSS DR3 Quasar Catalog. \label{sdsssubpops_alltable}}
     %
 \end{minipage}
    \end{table}
    \end{landscape}

    \begin{table} \centering \begin{minipage}{140mm}
     \caption[Photometric Redshift Estimates for low-$z$ quasars in subpopulations in the SDSS DR3 Quasar Catalog]{Photometric Redshift Estimates for low-redshift quasars in subpopulations in the SDSS DR3 Quasar Catalog.  \label{sdsssubpops_lowtable}}
     %
 \end{minipage}
     \end{table}

     \begin{table} \centering \begin{minipage}{140mm}
     \caption[Photometric Redshift Estimates for high-$z$ quasars in subpopulations in the SDSS DR3 Quasar Catalog]{Photometric Redshift Estimates for high-redshift quasars in subpopulations in the SDSS DR3 Quasar Catalog.  \label{sdsssubpops_hightable}}
     %
 \end{minipage}
     \end{table}

    \clearpage
     \begin{landscape}
     \begin{table} \begin{minipage}{140mm}
     \caption[Mean Statistics for selected variations for calculating PDF PhotoZs]{Mean Statistics for selected variations for calculating PDF PhotoZs.  \label{otherstats}}
     \scriptsize
     %
 \end{minipage}
     \end{table} 
     \end{landscape}

\backmatter

\chapter{Curriculum Vitae of\\ Natalie Erin Strand}

\noindent\textbf{EDUCATION}
\begin{itemize} 
\item Ph. D. in Physics:  University of Illinois at Urbana-Champaign, October 2009\\
            Thesis Advisor: Professor Robert J. Brunner\\
            Thesis Title:  AGN Environments in the Sloan Digital Sky Survey
 
\item M. S. in Physics:  University of Illinois at Urbana-Champaign, October 2005
 
\item B. S. in Physics \& Minor in Mathematics, The Pennsylvania State University, May 2003\\
            Honors in Physics; Thesis Advisor: Professor Lee Samuel Finn\\
            Thesis Title: Wavelets for LIGO Data Analysis\\
\end{itemize}

\noindent\textbf{TEACHING EXPERIENCE}
\begin{itemize}
\item Graduate Teacher Certificate at the University of Illinois at Urbana Champaign:  Spring 2006
 
\item Discussion Teaching Assistant for PHYS 199M: Enrichment Mechanics, UIUC\\
            Spring 2008**
\item Discussion Teaching Assistant for PHYS 211: Introductory Mechanics, UIUC\\
            Fall 2003\\
            Spring 2004**\\
            Spring 2005*\\
            Fall 2006**\\
            Fall 2007 (Mentor TA, no sections)\\
            Spring 2008**
\item Discussion Teaching Assistant for ASTR 122: Stars and Galaxies, UIUC\\
            Fall 2004*
\end{itemize}

\indent* Rated Excellent on ÒList of Teachers Rated Excellent by their StudentsÓ\\
\indent** Rated Outstanding on ÒList of Teachers Rated Excellent by Their StudentsÓ\\
 
\noindent\textbf{RESEARCH EXPERIENCE}
\begin{itemize}
\item Summer 2004 to present:  Cosmology with AGN\\
      Advisor:  Professor Robert J. Brunner \\
      Investigation of active galactic nuclei environments using Sloan Digital Sky Survey (SDSS) data \\
      Analysis of photometric redshift prediction techniques using optical colors of quasars in SDSS \\
      Wrote software to search for quasar groups in SDSS data \\
      \emph{Tools:}  C, Python, SuperMongo, use of NCSA TeraGrid resources
 
\item Fall 2000 to Spring 2003:  Penn State LIGO Data Analysis Research \\
      Advisor:  Professor Lee Samuel Finn\\
      Wrote wavelet software package to be part of the LIGO Data Analysis System to search for burst gravitational wave signals in LIGO data\\
      \emph{Tools:}  C++\\
\end{itemize}

\noindent\textbf{PUBLICATIONS}
 
    Strand, N. E., Brunner, R. B. and Myers, A. D.  ÒAGN Environments in the Sloan Digital Sky Survey I: Dependence on Type, Redshift, and Luminosity.Ó  The Astrophysical Journal 688 (2008): 180

    Ball, N. M. et al.  ÒRobust Machine Learning Applied to Astronomical Data Sets III: Probabilistic Photometric Redshifts for Galaxies and Quasars in the SDSS and GALEX.Ó  The Astrophysical Journal 683 (2008): 12

    Myers, A. D. et al.  ÒQuasar Clustering at 25 h-1 kpc from a Complete Sample of Binaries.Ó The Astrophysical Journal 678 (2007): 635

    Ball, N. M. et al.  ÒRobust Machine Learning Applied to Astronomical Data Sets II: Quantifying Photometric Redshifts for Quasars Using Instance-based Learning.Ó The Astrophysical Journal 663 (2007): 774

    Abbott, B. et al.  ÒSearch for Gravitational Waves Associated with 39 Gamma-ray Bursts using Data from the Second, Third, and Fourth LIGO Runs.Ó  Physical Review D 77 (2008): 062004

    Abbott, B. et al.  ÒSearches for Periodic Gravitational Waves from Unknown Isolated Sources and Scorpius X-1: Results from the Second LIGO Science Run.Ó  Physical Review D 76 (2007): 082001

    Abbott, B. et al.  ÒJoint LIGO and TAMA300 search for gravitational waves from inspiralling neutron star binaries.Ó  Physical Review D 73 (2006): 102002

    Abbot, B. et al.  ÒSearch for gravitational waves from binary black hole inspirals in LIGO data.Ó  Physical Review D 73 (2006): 062001

    Abbot, B. et al.  ÒUpper limits from the LIGO and TAMA detectors on the rate of gravitational-wave bursts.Ó  Physical Review D 72 (2005): 122004

    Abbot, B. et al.  ÒFirst all-sky upper limits from LIGO on the strength of periodic gravitational waves using the Hough transform.Ó  Physical Review D 72 (2005): 102004

    Abbot, B. et al.  ÒSearch for gravitational waves from primordial black hole coalescences in the galactic halo.Ó  Physical Review D 72 (2005): 0282002

    Abbot, B. et al.  ÒSearch for gravitational waves from galactic and extra-galactic binary neutron stars.Ó Physical Review D 72 (2005): 082001

    Abbott, B. et al.  ÒUpper limits on gravitational wave bursts in LIGO's second science run.Ó  Physical Review D 72 (2005): 062001

    Abbott, B. et al.  ÒLimits on Gravitational-Wave Emission from Selected Pulsars Using LIGO Data.Ó  Physical Review Letters 94 (2005): 181103

    Abbott, B. et al.  ÒSetting upper limits on the strength of periodic gravitational waves from PSR J1939+2134 using the first science data from the GEO 600 and LIGO detectors.Ó  Physical Review D 69 (2004): 082004

    Abbott, B. et al.  ÒSearch for gravitational waves associated with the gamma ray burst GRB030329 using the LIGO detectors.Ó  Physical Review D 72 (2005): 042002

    Abbott, B. et al.  ÒFirst upper limits from LIGO on gravitational wave bursts.Ó  Physical Review D 69 (2004): 102001

    Abbott, B. et al.  ÒAnalysis of LIGO data for gravitational waves from binary neutron stars.Ó  Physical Review D 69 (2004): 122001

   Abbott, B. et al.  ÒAnalysis of first LIGO science data for stochastic gravitational waves.Ó  Physical Review D 69 (2004): 122004\\

\noindent\textbf{AWARDS}
\begin{itemize}
\item Scott Anderson Award, University of Illinois Department of Physics (2009)

\item Selected as GAANN Fellow 2007-2008 by the University of Illinois Department of Physics

\item Phi Beta Kappa Thesis Award (May 2003)

\item Langhorne H. Brickwedde Award for Excellence in Undergraduate Physics, The Pennsylvania State University Department of Physics (May 2003)

\item DeanÕs List, Pennsylvania State University (Spring 2000-Spring 2003)

\item Bookstore Academic Excellence Scholarship, Schreyer Honors College, The Pennsylvania State University (Fall 1999-Spring 2003)

\item John and Elizabeth Holmes Teas Scholarship, The Pennsylvania State University Department of Physics (Fall 2000-Spring 2003)\\
\end{itemize}

\noindent\textbf{HONOR SOCIETY MEMBERSHIPS}
\begin{itemize}
\item Phi Beta Kappa                       

\item Sigma Pi Sigma\\
\end{itemize}

\end{document}